\documentclass[manuscript]{acmart}
\AtBeginDocument{%
  }

\setcopyright{acmlicensed}
\copyrightyear{2026}
\acmYear{2026}
\acmDOI{XXXXXXX.XXXXXXX}

\usepackage{graphicx}
\usepackage{subcaption}
\usepackage{float}
\usepackage{xargs}
\usepackage{booktabs}
\usepackage{makecell}
\usepackage{multirow}
\usepackage{tabularx}
\usepackage{tcolorbox}
\usepackage{todonotes}
\usepackage{footnote}
\usepackage[inline]{enumitem}
\newcolumntype{Y}{>{\centering\arraybackslash}X}

\newcommand{\collection}{{\mathcal{C}}}

\newcommand{\CLS}{\texttt{[CLS]}}
\newcommand{\SEP}{\texttt{[SEP]}}
\newcommand{\SEPone}{\texttt{[SEP1]}}
\newcommand{\SEPtwo}{\texttt{[SEP2]}}
\newcommand{\PAD}{\texttt{[PAD]}}

\newcommand{\llmdim}{\ensuremath{d_\mathcal{T}}}
\newcommand{\headdim}{\ensuremath{{d_\mathcal{H}}}}

\newcommand{\maxatt}[1]{M^{h,\ell}_{#1}}
\newcommand{\patt}[0]{p_{h,\ell}}

\newcommand{\ignore}[1]{}

\newcommand{\result}[1]{%
  \noindent
  \textcolor{red}{\rule{1.2ex}{1.2ex}}~\textbf{#1} \noindent %
}

\usepackage{mdframed}
\newcounter{revision}

\renewcommand{\therevision}{R\arabic{revision}}

\begin{document}

\newenvironment{formulas}{}{}

\title{How BERT-based Cross-Encoders Model Relevance in Information Retrieval?}


\author{Mathias Vast}
\email{mathias.vast@isir.upmc.fr}
\orcid{0009-0007-4612-717X}
\affiliation{%
  \institution{Sinequa by ChapsVision}
    \city{Paris}
  \country{France}
}
\affiliation{%
  \institution{Sorbonne Université, CNRS, ISIR}
  \city{F-75005 Paris}
  \country{France}
}

\author{Basile Van Cooten}
\orcid{0009-0002-0234-917X}
\affiliation{%
  \institution{Sinequa by ChapsVision}
  \city{Paris}
  \country{France}
}
\email{bvancooten@chapsvision.com}

\author{Laure Soulier}
\orcid{0000-0001-9827-7400}
\affiliation{%
  \institution{Sorbonne Université, CNRS, ISIR}
  \city{F-75005 Paris}
  \country{France}
}
\email{laure.soulier@isir.upmc.fr}

\author{Benjamin Piwowarski}
\orcid{0000-0001-6792-3262}
\affiliation{%
  \institution{Sorbonne Université, CNRS, ISIR}
  \city{F-75005 Paris}
  \country{France}
}
\email{benjamin.piwowarski@cnrs.fr}

\renewcommand{\shortauthors}{Vast et al.}
\acmArticleType{Research}

\begin{abstract}
Cross-encoders are the most effective but also the least efficient models in Information Retrieval. Despite their success,  the mechanisms underlying their relevance predictions remain poorly understood, particularly how they process to match queries and documents.
This lack of interpretability limits their optimization, can lead to suboptimal retrieval, especially out-of-domain, and raises efficiency concerns for over-parameterized models.
To address these challenges, this paper investigates two under-explored aspects of cross-encoder behavior. 
First, to understand the relevance prediction process, we trace the flow of information at the different stages of the process and between different input segments. We study the nature of the signals that compose this flow and that allow cross-encoders to predict the relevance of a passage to a query. Our work particularly focuses on the matching performed by the cross-encoders between queries and documents. 
Second, we explore how cross-encoders behave with out-of-domain documents and/or queries. Through a series of interpretability experiments, we evaluate three BERT-based cross-encoder models, varying in size and pre-training strategies, across both in-domain and out-of-domain datasets.
Overall, this study both consolidates what is known about how cross-encoders model relevance and extends it by exploring the role of matching signals.


\end{abstract}

\maketitle

\section{Introduction}\label{section:intro}

With the rise of transformer-based models \cite{transformers}, currently culminating in Large Language Models (LLMs), information retrieval (IR) has changed drastically since the BM25 era \cite{bm25}. 
Current transformer-based IR models can be categorized according to how they model the interaction between queries and documents: implicitly (cross-encoders \cite{monobert, monot5}), explicitly (dense \cite{karpukhin-etal-2020-dense} or sparse encoders \cite{Formal_Lassance_Piwowarski_Clinchant_2021}) or a mix of both (late interactions \cite{khattab2020colbert}). With their ability to model semantics and
produce contextualized representations, these models have overcome the query-document exact match limitation of traditional lexical methods.  This advancement, however, has come with an important trade-off: interpretability. Unlike lexical approaches such as BM25 -- where relevance can be directly traced to term-statistics -- neural models are mostly black-boxes. For end users, this opacity may be acceptable, provided that the retrieved results or the LLM-generated outputs are relevant \cite{lewis2020retrieval}. For researchers and practitioners, however, this opacity hinders both systematic improvement and the diagnosis of failure cases. In particular, cross-encoders are widely used in the re-ranking stage of retrieval pipelines -- whether in search engines or RAG frameworks \cite{rag} -- despite limited understanding of their internal mechanisms and failure modes.  While we know that these models are over-parameterized \cite{xin-etal-2020-early} and could be made more efficient by removing unnecessary layers or attention heads \cite{voita-etal-2019-analyzing,he2024matterstransformersattentionneeded}, our limited comprehension prevents us from designing more efficient and/or effective alternatives.  

This inability to diagnose failure cases (as well as the scientific challenge that understanding these models represents) has motivated a surge in \textit{interpretability} and \textit{explainability} studies, not limited to IR. From attribution methods \cite{LIME_Ribeiro_2016,sundararajan2017axiomatic,shrikumar2017just,GRAD_CAM_Selvaraju_2019} to probing \cite{probing_overview,probing_for_ranking,van_Aken_2019}, many techniques have been developed to interpret and explain the behavior of neural networks. 
In IR, efforts have largely concentrated on elucidating the nature of the signals involved when computing document relevance, though with mixed outcomes \cite{IR_survey}. Some works study the agreement of neural IR models with established IR axioms \cite{chowdhury2024probingrankingllmsmechanistic,lu-etal-2025-pathway} or use probing \cite{probing_overview} to search for specific encoded features \cite{macavaney-etal-2022-abnirml,probing_for_ranking}, while other studies have highlighted the roles of specific attention heads in detecting matching signals \cite{lu-etal-2025-pathway}.

However,  we still lack a \textit{mechanistic} understanding of how these signals are detected or processed, let alone incorporated to form the \textit{relevance prediction}. In addition, few works have attempted to provide a holistic view of the entire relevance prediction process, as most of them generally focus on a specific aspect. This also raises the question of whether any single interpretability technique (probing, ablations, attention) can bridge this gap. Notable exceptions in IR are studies of the processing stages within MonoBERT \cite{zhan2020bertanalysis,jiang-etal-2021-bert,qiao2019understandingbehaviorsbertranking}, but even these lack supporting evidence for the hypotheses they formulate. 
More recently, \citeauthor{lu-etal-2025-pathway}~\cite{lu-etal-2025-pathway} identified a mechanism similar to a semantic variant of BM25 within a MiniLM-based cross-encoder, while \citeauthor{chen2024axiomatic}~\cite{chen2024axiomatic} used mechanistic interpretability to interpret the roles of some attention heads for duplicate tokens inside TAS-B \cite{hofstatter2021efficiently}.  These works provide an end-to-end picture of a specific mechanism inside neural IR models, yet they also highlight another overlooked aspect of interpretability studies: their generalization to other models and datasets.

Cross-encoders stand as the most effective models for relevance estimation, while their internal mechanisms remain poorly understood. This paradox makes them a high-potential target for interpretability research. Better understanding how they work could help overcome their two main limitations in search pipelines: computational inefficiency and dependence on a first-stage retriever.
To address these limitations, our experimental plan combines 1) multiple interpretability techniques, since signals in transformer-based models are entangled \cite{sentiment_neuron,elhage2022superposition}; 2) multiple BERT-based cross-encoders of different sizes and pre-training strategies; and 3) multiple datasets, both from the training domain and from domains unknown to the models. This last point is motivated by the release of the BEIR benchmark \cite{thakur_beir_2021} which highlighted disparities between the in-domain and out-of-domain performance of neural IR models. Although smaller for cross-encoders than for bi-encoders, this gap, together with their reliance on semantic knowledge inherited from pre-training, raises the question of whether cross-encoders behave consistently In-Domain (ID) and Out-of-Domain (OOD).

Thanks to the points 1) and 3) of this experimental plan, i.e., multiple techniques applied across multiple datasets, we further gain the ability to draw stronger conclusions, as we can cross observations.
Our objective is to consolidate and extend previous findings on the role of attention heads and the nature of signals processed by cross-encoders. Beyond unifying these findings, we seek to shed light on the full process through which these models arrive at a relevance prediction. 
Specifically, we address the following research questions:
\begin{itemize}
    \item \textbf{RQ1}: What is the contribution of the different models' components in predicting relevance?
    \item \textbf{RQ2}: How does information flow within the model, particularly through its most important components?
        \item \textbf{RQ3}: Which signals flow through the model and contribute to predicting relevance or non-relevance?
    \item \textbf{RQ4}: How is the relevance signal constructed within the \CLS{} representation of cross-encoders?
\end{itemize}

These main research questions structure our work. We study each one for different models and across different domains to answer a series of orthogonal research questions that strengthen our findings:
\begin{itemize}
    \item \textbf{RQ5}: To what extent are these mechanisms consistent across different sizes of cross-encoders?
    \item \textbf{RQ6}: How robust are these findings to domain shifts?
\end{itemize}

In Section \ref{section:Related_works}, we present an overview of the current state of understanding of how transformer-based models \cite{transformers} work in general (Subsection \ref{subsection:explain_bert}) and more precisely in IR (Subsection \ref{subsection:explain_ir}). Then, we detail our experimental protocol in Section \ref{section:experiment_setup}, including the different models and datasets we used to answer \textbf{RQ5} and \textbf{RQ6}.
Once the setup is explained, we detail in sections \ref{section:NIG} to \ref{section:subspace_analysis} the results of our experiments and how they help us shape answers to our research questions. We summarize below the main highlights of these different sections.

To identify the most crucial components involved in the relevance prediction (\textbf{RQ1}), we follow our previous work in 
~\cite{Vast_2024}, leveraging the Neuron Integrated Gradients (NIG) method \cite{shrikumar2018computationally} (see \autoref{section:NIG}). 
\begin{quote}
\emph{Key highlights.} 
Some components are important \emph{across} domains, while others are important \emph{specifically} in OOD, particularly for large models. In particular, our method highlights the key role of a few attention heads and FFN neurons in the relevance prediction process.
\end{quote}

As a ``naive'' usage of NIG cannot provide enough details on the nature of the contributions from the critical components, we investigate the flow of information inside the model (\textbf{RQ2}). In Section \ref{section:information_flow}, we combine insights drawn from improving the granularity of the original NIG technique, extending the ablation analysis from \citeauthor{zhan2020bertanalysis}~\cite{zhan2020bertanalysis}, and using a variant of the Information Bottleneck Attribution \cite{Schulz2020Restricting}. While NIG and ablations allow for the extraction of the most salient informational directions throughout the model, the Information Bottleneck yields a more holistic view of the information flow. Altogether, they map the pathways through which information flows between input segments.
\begin{quote}
\emph{Key highlights.} We empirically validate the existence of the multi-stage process described in \citeauthor{zhan2020bertanalysis}~\cite{zhan2020bertanalysis} across both ID and OOD, but provide more details: Our results confirm that the relevance prediction process begins with the contextualization of the query and document tokens and, in parallel, query-to-document information transfers (i.e., the document tokens receive information from the query tokens through the attention mechanism). Contrary to \cite{zhan2020bertanalysis}, we observe that document-to-query transfers are equally important at this stage. These transfers last for most of the models' layers, until the later layers. At this point, relevance information shared across document and query tokens is not directly transmitted to the \CLS{} but is first accumulated in the query tokens before being sent to the \CLS{} in the very last few layers. In addition, we confirm the role of the ``no-op'' operation \cite{clark-etal-2019-bert} for \SEP{} and further show that \CLS{} might play a similar role through the early to the middle layers.

Finally, our study reveals two new insights: 1) not all cross-encoders share the same process of accumulating relevance information inside the \CLS{} and 2) MiniLM, contrary to MonoBERT cross-encoders, does not seem to rely on ``relevance'' signals to predict relevance, but instead on ``non-relevance'' signals.
\end{quote}
By answering \textbf{RQ1} and \textbf{RQ2}, we notice the key role of some high-attribution attention heads in query-to-document and document-to-query transfers, a behavior we link to the detection of matching signals across them. 
To derive the nature of these matching signals (\textbf{RQ3}), we analyze in Section \ref{section:understand_attention} the attention patterns of every attention head by contrasting their attributions when exposed to relevant pairs versus non-relevant ones. Our objective here is twofold: firstly, to uncover the nature of the query-document interactions detected by high-attribution heads, and secondly, to discover heads specialized in detecting \emph{relevant} matching signals. 

Furthermore, we extend this experiment with a mechanistic interpretability study (see \autoref{subsection:matching_mechanism}) to better understand the mechanism at stake that allows these heads to match together (contextualized) query tokens and (contextualized) document tokens. 
\begin{quote}
\emph{Key highlights.} Our results confirm that heads specialized in detecting matching signals between queries and documents exist in the model. By analyzing their attention patterns, we further characterize the nature of the matching signals they capture and report two distinct stages. From early to middle layers, these interactions match occurrences of the same token in the query and the document (\emph{syntactic matching}), while in the last layers, matching occurs between semantically-contextualized tokens (\emph{semantic matching}) that are close to each other. These heads are also among the most important components of cross-encoders, which confirms the significance of matching signals, and our mechanistic study explains the behavior of most of them: their Query-Key circuit favors cross-attention (query to document or document to query) over self-attention, and associates highly similar source and target tokens.
\end{quote}


After discovering the nature of the signals that are leveraged by cross-encoders to predict relevance, we look into the process behind the construction of the ``relevance'' signal itself (\textbf{RQ4}). We target key stages of the process, i.e., when the information stored inside the query tokens is transferred into the \CLS{} representation, and investigate the geometry of the hidden state's subspaces to understand if a ``relevance'' subspace exists.
\begin{quote}
\emph{Key highlights.} Our analysis reveals differences between MiniLM, MonoBERT-base and MonoBERT-large, in both the construction and the nature of their respective ``relevance'' subspace. In addition, we report that relevance signals emerge in the \CLS{} token only in the final layers, which supports the view that it performs a ``no-op'' function in earlier layers, akin to the \SEP{} tokens \cite{clark-etal-2019-bert}.
\end{quote}

Taken together, our findings both reinforce and refine prior insights into cross-encoder behavior. In particular, we are able, not only to confirm, but also to specify, the flow of information between input parts inside BERT-based cross-encoders. Crucially, we go beyond the existing work by testing the consistency of this finding across multiple architectures and domains (\textbf{RQ5}, \textbf{RQ6}). In doing so, we help move the field toward a more unified and generalizable understanding of how cross-encoders compute relevance.

\section{Related Work}
\label{section:Related_works}

\subsection{Comparisons of interpretability approaches}
\label{subsection:interpretability_comparison}

To interpret neural models, researchers rely on a large variety of interpretability techniques that can be categorized into distinct families.
In IR, some architectures such as SPLADE \cite{Formal_Lassance_Piwowarski_Clinchant_2021} or ColBERT \cite{khattab2020colbert} are ``more explainable'' than cross-encoders which must rely on the interpretability approaches we describe next.

One of the most widespread ways to access the internal knowledge of neural networks is through probing \cite{probing_overview}. Probing studies rely on training external models from hidden states on auxiliary but related tasks (e.g., for IR, it can  be Named-Entity Recognition, Semantic Similarity or co-reference Resolution \cite{probing_for_ranking, van_Aken_2019}), revealing the specific abilities that the model learned implicitly during its training to solve the task. Probing is popular because it is easy to use: given a dataset for the auxiliary task, one can test whether a model's activations encode the corresponding information. In IR, probing studies often rely on ``diagnostic datasets'' related to specific IR axioms \cite{camaraandhauf,chen2024axiomatic,chowdhury2024probingrankingllmsmechanistic}. However, because probing relies on external classifiers \cite{probing_overview}, its results are only correlational and can be hard to interpret, especially with overly expressive nonlinear probes. In addition, probing methods cannot be used to target specific neurons as they use each hidden representation as a whole and do not provide any explanation of the interplay between these abilities, as well as their relative importance in the model output. 

Other simple introspection techniques include the study of the attention process (e.g., the distribution of the attention weights across the tokens) in transformer-based architectures. Similarly to probing, these techniques have been criticized for providing explanations uncorrelated to the predictions of the model \cite{jain-wallace-2019-attention}. This claim has been challenged since \cite{wiegreffe-pinter-2019-attention}, and researchers now address this issue by combining attention patterns analysis with other interpretability methods to strengthen their conclusions~\cite{zhan2020bertanalysis}.

A way to test the importance of specific components is through ablation studies. By comparing the performance of the model with and without a specific part, one can directly assess its importance. Ablations belong to the extended family of \textit{perturbation-based} attribution approaches \cite{LIME_Ribeiro_2016, Fong_2017, petsiuk2018rise}, as they compare the outputs of the original model with those of the ablated model. Similarly, mechanistic interpretability \cite{cammarata2020thread:} approaches\footnote{Mechanistic interpretability is a line of work that tries to ``\textit{reverse engineer transformers into human-understandable computer programs}'' (quote taken from the Transformer Circuits blog: \url{https://transformer-circuits.pub/})}, such as activation patching \cite{meng_locating_2023,causal_mediation_analysis} and path patching \cite{goldowsky2023localizing}, introduce perturbations inside the computational graph of the model to deduce the importance of specific components.  While they led to significant breakthroughs \cite{elhage2021mathematical,olsson2022context}, including in IR \cite{lu-etal-2025-pathway,chen2024axiomatic}, mechanistic interpretability approaches and perturbation-based ones are hard to scale in an exhaustive study like ours. Even for more straightforward approaches, such as ablations, perturbations need to be conducted in isolation and put into perspective with the perturbations of the surrounding components. On larger models than BERT, this can make them intractable. 

More tractable attribution methods are the \textit{backpropagation-based} ones. They usually have more desirable theoretical properties than perturbation-based methods, such as sensitivity or additivity \cite{sundararajan2017axiomatic, shrikumar2017just, GRAD_CAM_Selvaraju_2019}. Their main drawback is the difficulty of giving a human-understandable meaning to their attribution. Among such techniques, the Integrated Gradients \cite{sundararajan2017axiomatic} and Neuron Integrated Gradients (NIG) \cite{dhamdhere2018important, lundstrom2022rigorous} respectively allow tracing back the contributions of input parts and individual neurons to the predictions of the model. However, these approaches fail to provide insights into the nature of these contributions. Notably, attribution patching \cite{syed2023attribution} alleviates this limitation for mechanistic interpretability approaches by making use of gradient-based approximation (borrowing from NIG-related techniques), but it can lead to the appearance of false negatives \cite{kramr2024atp}, potentially harming its conclusions. 

While attribution methods help to understand the relative importance of input parts or components of the model (at various granularity levels), they fall short of providing a holistic view of information flow without being applied repeatedly across the architecture. In contrast, the Information Bottleneck \cite{Schulz2020Restricting} allows for automatically determining the most important parts in the input. This approach estimates the minimal information flow between an input $X$ and an output $Y$, necessary to preserve $Y$, limiting the components to their most important features. As it acts on all the components at once, it further accounts for their relations, alleviating one of the main limitations of attribution methods -- but at the cost of some potential instability during training. However, its results need to be further refined to understand what makes a part with a high attribution score important.

\begin{table}[t]
\caption{Summary of interpretability approaches in IR and whether we used them to answer our Research Questions (RQ).}
\begin{tabular}{|c|c|c|c||c|}
\hline
\textbf{Method} & Strengths & Weaknesses & Used in & Sections / RQ \\ \hline \hline
Probing & Versatility & \makecell{Results are correlational,\\poor granularity \ignore{(cannot be\\used to study individual neurons)} } &  \cite{chowdhury2024probingrankingllmsmechanistic,choiFindingInverseDocument2022} & None  \\ \hline 
Ablations & Ease-of-use, granularity & \makecell{Needs to be scaled\\for better analysis} & \cite{zhan2020bertanalysis} & \ref{section:information_flow} / RQ2\\ \hline
Attention patterns & \makecell{Ease-of-use,\\can be applied to a few\\samples or a dataset} & \makecell{Results need further\\interpretation} & \cite{zhan2020bertanalysis,choiFindingInverseDocument2022,lu-etal-2025-pathway} & \ref{section:understand_attention} / RQ3\\ \hline
NIG & \makecell{Gives the importance of\\individual neurons} & No info. on their role(s) & \cite{Vast_2024} & \ref{section:NIG},\ref{section:information_flow} / RQ1,2\\ \hline
Information Bottlenecks & \makecell{Get important features\\all at once} & Needs (limited) training & \cite{jiang-etal-2021-bert} & \ref{section:information_flow} / RQ2 \\ \hline
Causal / Path Patching & \makecell{Give the role of\\components} & Hard to scale & \cite{chen2024axiomatic,lu-etal-2025-pathway} & None \\ \hline
\end{tabular}
\label{tab:ir-sota-overview}
\end{table}

These different approaches, along with their associated findings, strengths, and weaknesses, are summarized in Table \ref{tab:ir-sota-overview}. Based on this comparison, several techniques emerge as especially promising for our study. The analysis of attention patterns will be useful for examining the information exchanges between the queries and the documents (or matches) and for determining the nature of these exchanges (\textbf{RQ3}). Both perturbation-based (ablations) and backpropagation-based (NIG) attribution methods are efficient and easy to interpret. They give the contribution of each component (\textbf{RQ1}) and help put other results into perspective, e.g., are the important attention heads those that detect matches? (\textbf{RQ2, RQ3}). Finally, the Information Bottleneck approach, despite involving some training, provides a suitable way to validate the findings of other methods and analyze the information flow (\textbf{RQ2}) by considering all the interactions inside the models at once. Table \ref{tab:ir-sota-overview} also summarizes the Sections where each technique is deployed, as well as the Research Questions it is used to answer. 
In this work, the choice of methods is mainly motivated by the ease of scaling them and interpreting their results. Indeed, as we want to study three models across various datasets (\textbf{RQ5, RQ6}), we cannot afford to use methods that require too many computing resources (like mechanistic interpretability). For the same reasons, we refrain from using probing, as it is hard to link its results to the actual mechanisms of the model. 

In addition to the methods listed above, we further developed our own framework to interpret neural IR models. 
Following our previous work involving NIG to study the most important components inside MonoBERT \cite{Vast_2024} in both in-domain and out-of-domain settings, we repeat each experiment in both setups and cross the results to differentiate the in- and out-of-domain behavior. 
Extending this study, we further develop a method based on NIG, which differentiates the different parts inside the input (the \CLS{}, the query, the passage, and the two \SEP{} tokens) to study the most important directions of the attention inside the model as well as the most important components in the FFN.
To study the roles of the attention heads, we extend the traditional attention patterns analysis by contrasting the distribution of attention scores between query-relevant passage pairs and query-non-relevant passage pairs to obtain the attention heads whose roles are directly related to detecting signals important for the prediction of relevance.
Lastly, to study at which points the relevance signal starts to be encoded inside the \CLS{} token representation (\textbf{RQ4}), we analyze the representational subspace of the different tokens with Linear Discriminant Analysis.

So far, we reviewed methods designed to \textit{interpret} models' behaviors, i.e., decipher their inner mechanisms. By design, interpretability is tied to the studied model. On the other hand, \textit{explainability} refers to a different line of work, whose methods produce post-hoc explanations, are generally model-agnostic, and specifically target the role of the input features \cite{LIME_Ribeiro_2016,shapley}. In IR, notable works include applications of SHAP \cite{heuss2024rankingshap,chowdhury2025rankshap} or LIME \cite{lirme,rank_lime} to explain pointwise predictions and even listwise rankings (for a more comprehensive view, we refer the interested readers to this tutorial \cite{explain_ir}).
We consider this line of work complementary to ours, as its focus is mostly on providing post-hoc explanations of a model's behavior, rather than on its \textit{internal mechanisms}. 

\subsection{Interpreting BERT and transformer-based models}
\label{subsection:explain_bert}

Numerous works investigate the behaviors and tendencies of BERT \cite{bert}, a field of research called BERTology \cite{rogers-etal-2020-primer}. In particular, studies have shown that earlier layers tend to encode syntactic knowledge, such as parts of speech, syntactic structures (chunks, trees, or dependencies) \cite{lin-etal-2019-open,liu-etal-2019-linguistic,hewitt-manning-2019-structural,clark-etal-2019-bert}. BERT models have also been found to encode IDF information inside their parameters, in particular inside their embeddings \cite{choiFindingInverseDocument2022}, and some of their behaviors appear to depend on token frequency, with attention heads in early layers attending primarily to rare tokens \cite{voita-etal-2019-analyzing}. 
In addition to syntax and lexical properties, BERT's embeddings are also known to be highly contextualized \cite{ethayarajh-2019-contextual}, i.e., each token representation encodes information about its surroundings. 
In later layers, contextualization allows the model to capture the words' meaning \cite{wiedemann2019doesbertmakesense}.


Combining these insights, it is possible to draw a high-level structure underlying the different steps of a forward pass in a transformer model. In the first layers, neurons detect surface signals and shallow patterns in the input. At the same time, the input is contextualized, and moving on to the middle layers, neurons specialize in higher-level concepts. Finally, in the last layers, neurons use the signals extracted earlier to take actions and promote features in the residual stream. In the context of BERT-based models specifically, this is encapsulated by the fact that pre-trained BERT models can encode knowledge related to different NLP tasks and organize them dynamically into a pipeline \cite{tenney-etal-2019-bert}.

These insights obviously overlap with IR interpretability studies, and ours in particular, as most observed behaviors and conclusions are transferable across all NLP tasks. Yet, some specificities of IR, like the modeling of relevance, are not properly captured by generic NLP studies. IR thus developed its interpretability field, summarized in the next section.

\subsection{Interpreting neural IR models}
\label{subsection:explain_ir}

Soon after BERT was successfully applied to IR \cite{monobert}, studies aimed at understanding how neural IR models work and how they compare to traditional IR approaches emerged. In particular, \citeauthor{zhan2020bertanalysis}~\cite{zhan2020bertanalysis} and \citeauthor{jiang-etal-2021-bert}~\cite{jiang-etal-2021-bert} both study the attention process \cite{attention} of MonoBERT. \citeauthor{zhan2020bertanalysis} state that MonoBERT predicts relevance by first contextualizing queries and documents, then capturing interaction signals, and finally composing them. With the Information Bottleneck attribution method \cite{Schulz2020Restricting}, \citeauthor{jiang-etal-2021-bert} find that MonoBERT not only relies on lexical matching but also captures semantic relationships between the contextualized representations of the query and the document tokens. They also provide further details on the role of \CLS{}, which starts to gather relevance information in the very last layers of the model.  
On this matter, \citeauthor{zhan2020bertanalysis}~\cite{zhan2020bertanalysis} go even further and show that across most of the model's layers, \CLS{}, \SEP{}, and punctuations attract a large proportion of attention but carry little relevance information, echoing a similar behavior to the ``no-op'' operation \cite{clark-etal-2019-bert}. They further emphasize the importance of document-to-query information transfers while questioning the role of query-to-document transfers. Overall, their study provides various insights into the inner mechanisms of MonoBERT. However, they do not specify the types of interactions detected (our focus in Section \ref{section:understand_attention}) or how these contribute to the relevance signal (studied in Section \ref{section:subspace_analysis}), and their conclusions on attention directionality, between query and document, lack stronger empirical evidence (as illustrated in our Section \ref{section:information_flow}). 

More recently, \citeauthor{lu-etal-2025-pathway}~\cite{lu-etal-2025-pathway} apply path patching \cite{goldowsky2023localizing}, another mechanistic interpretability approach, to a BERT-based cross-encoder and observe that it ``rediscovers'' a semantic variant of BM25, as mentioned in \citeauthor{jiang-etal-2021-bert}~\cite{jiang-etal-2021-bert}. They identify specific attention heads responsible for computing ``soft'' term frequency while accounting for term saturation and document length effects (Matching Heads). Combined with the finding that the IDF is encoded in a low-rank component of the model's embedding matrix \cite{choiFindingInverseDocument2022}, 
\citeauthor{lu-etal-2025-pathway} further show that the model output can be reconstructed (correlation of 0.80) using these signals. 

Aside from BERT-based ranking models, some works have also started to explore the mechanisms of LLMs for ranking. Notably, \citeauthor{chowdhury2024probingrankingllmsmechanistic}~\cite{chowdhury2024probingrankingllmsmechanistic} conduct a probing study to examine activations in fine-tuned LLMs for passage re-ranking, finding that certain human-engineered features, such as the number or ratio of covered query terms, 
are encoded in activations, while others, like BM25 scores, do not seem to be present (no linear correlation). If we further extend the scope to different architectures, other works have also explored the behavior of learned sparse retrieval models \cite{mackenzie2023exploring}, late-interaction models \cite{white_box_colbert_formal}, and bi-encoders \cite{moller_attribution_2023,chen2024axiomatic}, in particular the matching behaviors of these models. For example, \citeauthor{chen2024axiomatic}~\cite{chen2024axiomatic} discover the key roles of some attention heads in detecting duplicate instances of the same tokens while studying TAS-B \cite{hofstatter2021efficiently}, a dual embedding model, using activation patching \cite{meng_locating_2023,causal_mediation_analysis}. However, these studies are architecture-specific and cannot be generalized to the study of cross-encoders.

As we have seen in this section, our understanding of how cross-encoder models work is rapidly growing, fueled by a wide variety of interpretability approaches. Besides the aforementioned limitations, most of these studies, including \cite{lu-etal-2025-pathway}, are conducted in a relatively limited setup, studying a single model and a single dataset. This questions the generalization of their findings across models (architectures, as well as sizes) and across domains. For example, MonoBERT has been shown to exhibit different behaviors with in-domain versus out-of-domain \cite{Vast_2024}, and larger models (such as MonoBERT) have been hypothesized to behave differently than shallower neural rankers \cite{qiao2019understandingbehaviorsbertranking}. Thus, this motivates the need to conduct a more holistic interpretability study. In this work, we conduct our experiments using three cross-encoders of different sizes to compare their mechanisms and evaluate the impact of the number of parameters on their behavior (\textbf{RQ5}). In addition, we use several datasets covering various domains to verify the extent to which the mechanisms they learn can be transferred to out-of-domain data and to strengthen our findings (\textbf{RQ6}).


\section{Experimental setup}
\label{section:experiment_setup}

\subsection{Cross-Encoders Architecture}
\label{subsection:architecture}
\newcommand{\sample}{\mathbf{x}}

\newcommandx{\rep}[2][1=j, 2=\ell]{\ensuremath{\sample_{#1}^{(#2)}}}
\newcommand{\repToken}[2][\ell]{\ensuremath{\sample_{\mathtt{[#2]}}^{(#1)}}}
\newcommandx{\repi}[3][1=j, 2=\ell]{\ensuremath{\sample_{#1}^{(#2,#3)}}}

\newcommandx{\repAr}[2][1=j, 2=\ell]{\ensuremath{\sample_{#1}^{(#2,\tilde a)}}}
\newcommandx{\repA}[2][1=j, 2=\ell]{\ensuremath{\sample_{#1}^{(#2, a)}}}
\newcommandx{\repFFNr}[2][1=j, 2=\ell]{\ensuremath{\sample_{#1}^{(#2,\tilde f)}}}
\newcommandx{\repFFN}[2][1=j, 2=\ell]{\ensuremath{\sample_{#1}^{(#2)}}}

Cross-encoders classify a query-document pair $(q,d)$. 
Note that in this work, we focus exclusively on BERT-based cross-encoders and consider that both the queries and the documents are textual. This excludes other re-rankers such as Qwen3-VL \cite{qwen3vlembedding} which are based on a decoder-only backbone and/or deal with other modalities than text. 
We denote $\sample$ the cross-encoder input, which is composed of the sequence of tokens \texttt{[CLS] $q_1 \ldots q_n$ [SEP1] $d_1 \ldots d_m$ [SEP2]} ($n$ query tokens and $m$ document tokens). 
The transformer architecture updates the representation of the input tokens after each layer $\ell \in [0 \ldots L-1]$ (layers are indexed from 0 throughout the paper, consistently with the figures and tables). We denote $\rep[j][\ell]\in\mathbb{R}^{\llmdim}$ the representation of the token $j$ at the output of layer\footnote{For consistency, we assume $\rep[j][-1]$ is the model input embedding.} $\ell$, where $j$ can be the token \texttt{[CLS]}, a query $q_k$ or document token $d_k$, \texttt{[SEP1]} or \texttt{[SEP2]}. Each layer $\ell$ is composed of two main processing steps:
\begin{enumerate}
    \item \textbf{The attention step}, summarized as: 
    \begin{eqnarray*}    
    \repAr & = & \sum_{h\in \mathcal{H}_\ell} W_O^{h} \ \mbox{SelfAttention}^{h}(\rep[j][\ell-1], \rep[][\ell-1]) \\
    \repA  & = & LN\left(\repi{\tilde a} + \rep[j][\ell-1] \right)
    \end{eqnarray*}
     where $h\in \mathcal{H}_\ell$ is an attention head among the $H$ heads of layer $\ell$, and $\rep[][\ell-1]$ denotes the representations of all the input tokens at layer $\ell-1$. An attention head $h$ works in a subspace of dimension $\headdim = \llmdim / H$, and is computed using the following formula: 
     $$\mbox{SelfAttention}^{h}(\rep[j][\ell-1], \rep[][\ell-1]) = \sum_{i} \mbox{softmax}_i \left( \frac{\left(W_Q^{h} \rep[j][\ell-1]\right)^{\top} \left(W_K^{h} \rep[i][\ell-1]\right)}{\sqrt{\headdim}} \right) \, W_V^{h} \rep[i][\ell-1]$$
     where $W_Q^h,W_K^h,W_V^h \in \mathbb{R}^{\headdim \times \llmdim}$ are the query, key and value matrices, and $W_O^h \in \mathbb{R}^{\llmdim \times \headdim}$ is the output matrix.
     Last, $LN$ is the layer norm~\cite{ba2016layernormalization} that normalizes each vector to zero mean and unit variance across its components, before applying a learned affine transformation.
    
    \item \textbf{The feed-forward network (FFN) step}:
    $$\repFFNr = \sum_{i \in \mathrm{FFN}_\ell} \phi(W^{\ell,\text{in}}_i\cdot\repA + b^{\ell,\text{in}}_i)W^{\ell,\text{out}}_i + b^{\ell,\text{out}}$$
    where $\phi$ is the activation function (GELU), $W^{\ell,\text{in}}_i \in \mathbb{R}^{1\times\llmdim}$ and $W^{\ell,\text{out}}_i \in \mathbb{R}^{\llmdim}$ are respectively the $i$-th row of the input projection and the $i$-th column of the output projection, and the number of FFN neurons $|\mathrm{FFN}_\ell|$ is a multiple (e.g., 4) of $\llmdim$. 
    From this formulation, ``FFN neuron $i$'', used throughout the work, corresponds to the scalar $\phi(\cdot)$. 
\end{enumerate}

For any token $j$, the final representation $\rep[j][\ell]$ on layer $\ell$ is computed with $$\repFFN[j][\ell] = LN(\repFFNr[j][\ell] + \repi[j][\ell]{a})$$

Finally, the classification head of a cross-encoder is a two-layer neural network which is applied to the \CLS{} token representation $\repToken[L-1]{CLS}$ at the last layer $L-1$, formally $$P(\repToken[L-1]{CLS}) = \sigma(w_2 \cdot \tanh(W_1 \repToken[L-1]{CLS} + b_1) + b_2)$$ where $W_1$, $w_2$, $b_1$ and $b_2$ are parameters, $\sigma$ the sigmoid function that ensures that the value is between 0 and 1\footnote{In the case of MonoBERT, the network has two outputs, one for not relevant and another for relevant, and the probability distribution is obtained through a softmax, which is equivalent to a sigmoid. We assume in the remainder of this paper that there is only one output corresponding to the prediction of relevance.}. 
As a shorthand, for an input $\sample$, we denote $P(\sample)$ the output of a cross-encoder when applied to the last layer representation of the token \texttt{[CLS]}, i.e., $\repToken[L-1]{CLS}$. We denote $z(\sample)$ the corresponding logit, i.e., the input of the sigmoid $\sigma$ in $P$.

In this paper, we consider three BERT-based cross-encoders with different numbers of parameters but also pre-training, and study their respective behaviors. More precisely, we study cross-encoders based on MiniLM-L12-v2 \cite{minilmv2}, BERT-base and BERT-large \cite{monobert}, i.e., MonoBERT-base and MonoBERT-large. In the remainder of the paper, we simply refer to the former as MiniLM. While there exist fine-tuned checkpoints for MonoBERT-large and MiniLM, we had to fine-tune a version of MonoBERT-base ourselves, using the training split of MS MARCO\footnote{All models are available on the Hugging Face Hub: \href{https://huggingface.co/castorini/monobert-large-msmarco}{castorini/monobert-large-msmarco} / \href{https://huggingface.co/xpmir/cross-encoder-bert-base-infoNCE/tree/main}{xpmir/cross-encoder-bert-base-infoNCE} / \href{https://huggingface.co/cross-encoder/ms-marco-MiniLM-L12-v2}{cross-encoder/ms-marco-MiniLM-L12-v2}}. We report the training recipe and the performance of our MonoBERT-base in Appendix \ref{appendix:monobert_base_details}, along with the architectural details of the models in Table \ref{tab:models-overview}.

These three models let us vary three factors: size (from 33M to 340M parameters), depth (12 or 24 layers) and pre-training strategy (the classical Masked Language Modeling and Next Sentence Prediction of BERT \cite{bert}, or the deep self-attention distillation of MiniLM \cite{minilmv2}). MonoBERT-base serves as a ``control'' model: it has the depth of MiniLM (12) and the pre-training of MonoBERT-large, so that the influence of each factor can be studied separately (see \textbf{RQ5}).

\subsection{Datasets}
\label{subsection:datasets}

For our experiments, we rely on the Deep Learning tracks from TREC (TREC-DL) between 2019 and 2022 \cite{craswell2019overview,craswell2020overview,craswell2021overview,craswell2022overview} to observe the behavior of the models on in-domain data (ID), as well as on a subset of the datasets composing the BEIR benchmark \cite{thakur_beir_2021} for out-of-domain data (OOD).

TREC-DL datasets complement the MS MARCO dataset \cite{MSMARCO} with manually annotated relevance judgments, with multiple annotated passages per query. It is thus more reliable (fewer false negatives). We restrict ourselves to the queries that have been manually annotated by the NIST assessors (\textit{judged} version of the data on \textit{ir-datasets} \cite{macavaney:sigir2021-irds}). This amounts to 226 topics and more than 400k query relevance assessments along 4 degrees of relevance $r\in [0..3]$. 

For OOD (see \textbf{RQ6}), we carefully select the datasets corresponding to tasks that resemble the most a traditional IR setup in the BEIR benchmark \cite{thakur_beir_2021} according to their classification\footnote{The BEIR benchmark covers a total of 9 tasks, among which 3 can be considered the closest to the IR task: News Retrieval, Question-Answering (minus HotpotQA \cite{hotpotqa} which is a multi-hop QA dataset) and Bio-Medical IR.}: FiQA \cite{fiqa}, TREC-Covid \cite{trec-covid}, TREC-News \cite{trec_news}, NFCorpus \cite{nfcorpus}, BioASQ \cite{bioasq}, Natural Questions\footnote{Note that for Natural Questions, we use the document collection as it was pre-processed in DPR \cite{karpukhin-etal-2020-dense} to benefit from their additional filtering.} \cite{nq} and TREC-Robust04 \cite{robust04}. From this reduced pool of datasets, we further eliminated TREC-News because of the format of the task, as ``queries'' are in fact documents from the corpus. Since some datasets are not directly available in BEIR and \textit{ir-datasets} \cite{macavaney:sigir2021-irds} offers several versions with different pre-processing, we report in Appendix \ref{annex:baselines} the performance of the three models on the versions we use.

As documents labeled as ``not relevant'' could be semantically related (they were retrieved by an IR model, e.g., BM25), we also study the behaviors of the models in the case where the document is irrelevant \emph{ and } not semantically related to the query. To obtain these additional levels of annotations, we randomly sample documents for each query (excluding assessed ones). Following the terminology used in Neural IR training methodology, we call those documents \textit{easy negatives}, as opposed to the \textit{hard negatives}, labeled as non-relevant~\cite {tonellottoLectureNotesNeural2022}. We attribute the relevance level $r=-1$ to these passages and use $r=0$ for hard negatives. For each dataset listed in Table \ref{tab:datasets-passages}, we consider the highest available level of relevance to be ``highly relevant'', or $r=1$, following the annotation in the 4 tracks of TREC-DL. Throughout our work, we focus on these three levels of relevance. In addition to this pre-processing, note that we also applied three different sampling strategies to each dataset's list of query relevance judgments, to meet the requirements of our experiments. These three strategies, called respectively \emph{Top sampling}, \emph{Diverse Sampling} and \emph{BM25 sampling}, are detailed in the Appendix \ref{appendix:sampling_strategies}, Figure \ref{fig:dataset_construction}.

We access the datasets using \textit{ir-datasets} \cite{macavaney:sigir2021-irds}\footnote{For TREC-DL 2022, this implies that the version we use has been filtered down to remove all duplicate documents.} and managed the experimental plan with the \textit{experimaestro} framework \cite{experimaestro,xpmir}. All the code necessary to reproduce the results is public\footnote{Code available at: \url{https://github.com/MathVast/CE-relevance-integration}}.


\section{RQ1: Identifying the contribution of the different components to the prediction of relevance}
\label{section:NIG}

To unveil the different components involved in modeling and predicting relevance in cross-encoders, we first look at characterizing their respective contributions when performing the IR task. Following \citeauthor{Vast_2024}~\cite{Vast_2024}, we use the Neuron Integrated Gradients approach (or NIG) \cite{dhamdhere2018important,lundstrom2022rigorous} to identify a reduced set of core neurons with consistently high attributions. In what follows, we begin by describing NIG, how it can be adapted to fit the IR task requirements, and the specificity of this work compared to the original setup.

\subsection{Background}
\label{subsection:background}
\paragraph{(Neuron) Integrated Gradients}

Originally designed for computing the attributions of the input features (to the output), Integrated Gradients (IG) \cite{sundararajan2017axiomatic} is based on path integrals. IG requires carefully selecting a baseline input $\sample'$ for which the model's prediction is neutral, e.g., in IR $p(relevant|\sample') = p(non-relevant|\sample')$. IG measures how the decision evolves between $\sample$ and $\sample'$, and more precisely along the path defined by $\gamma(t) = \sample' + t(\sample - \sample'), t \in [0, 1]$. 
IG has been generalized into \emph{conductance} \cite{dhamdhere2018important} to obtain the importance of any neuron $y$. In this framework, the importance of an individual neuron is given by the Neuron Integrated Gradients formula \cite{shrikumar2018computationally}:
\begin{equation}
    \label{equ:NIG}
    NIG^y(\sample) = \int_{t=0}^{t=1}\frac{\partial P(\sample'+t(\sample-\sample'))}{\partial \gamma_y(t)}
    \frac{\partial \gamma_y(t)}{\partial t} dt
\end{equation}
where $P$ is the classification model (our cross-encoder in Section~\ref{subsection:architecture}), $\gamma_y(t)$ denotes the activation value of a neuron $y$ at the point $t$ of the path $\gamma$, i.e., the output $y$ of the network, $P_y(\gamma(t))$. Note that this can be efficiently estimated~\cite{shrikumar2018computationally}.

 \begin{figure}[t]
    \centering
    \includegraphics[width=0.5\linewidth]{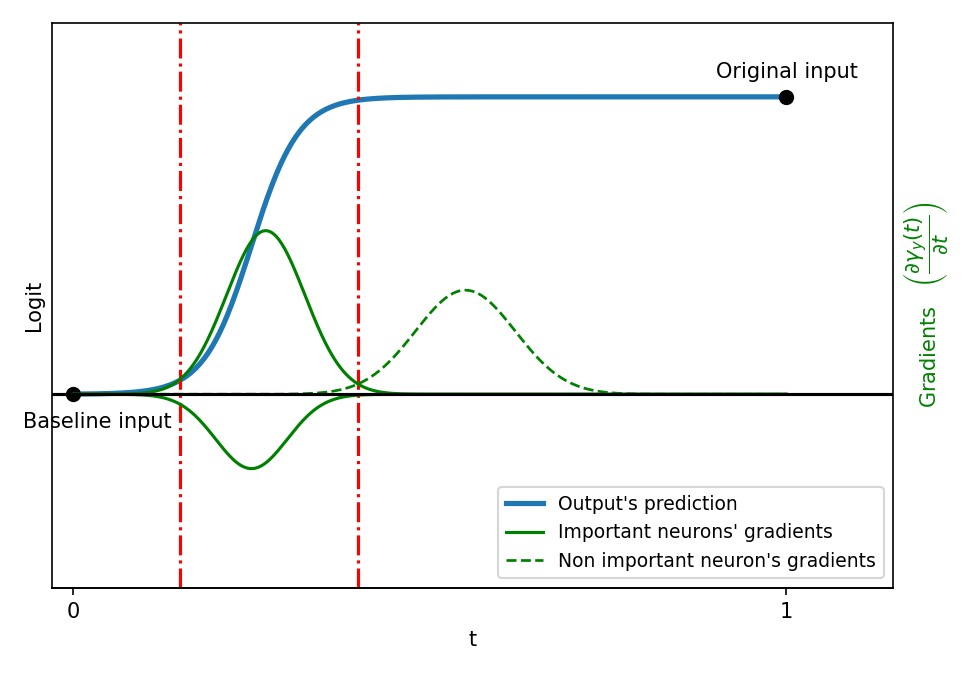}
    \caption{Illustration of the Neuron Integrated Gradients method. The blue curve (left axis) describes the recovery of the model's logit $z(\cdot)$ (i.e., the input of the sigmoid in $P$) as we progress along the path $\gamma(t)$. Green curves (right axis) represent the gradients $\frac{\partial \gamma_y(t)}{\partial t}$ of individual neurons along this path: plain curves correspond to important neurons, whose gradient is non-zero where the logit varies (between the two red lines), and the dashed curve to a non-important neuron.}
    \label{fig:nig}
\end{figure}

To illustrate the intuition behind Neuron Integrated Gradients, let us use Figure \ref{fig:nig} that depicts the evolution along the path $\gamma(t)$, between the ``Baseline'' $\sample'$ and the ``Original'' input $\sample$, of the model's logit (in blue, left axis) and the gradients $\frac{\partial \gamma_y(t)}{\partial t}$ of two types of neurons (in green, right axis). 
In the interval delimited by the two red vertical lines in the figure, a change in the input results in a significant recovery of the original output signal, symbolized by a strong increase in the blue curve. According to NIG, the important neurons are neurons $y$ whose gradient $\frac{\partial \gamma_y(t)}{\partial t}$ evolves with the output $P$, i.e., whose gradient is non-zero inside this interval (plain green curves) -- note that the contribution can be either positive or negative. Conversely, a neuron whose gradient is non-zero only outside of the zone of variation of $P$ (dashed green curve) is not important for the prediction.


Over the years, a series of works have improved the quality of the attributions from Neuron Integrated Gradients. Notably, Integrated Decision Gradients \cite{walker2024integrated} samples mostly in the region where the model decision changes (between the two red vertical lines in Figure \ref{fig:nig}). We use their algorithm to sample the steps and obtain better attributions.

\paragraph{Neurons.} Usually, ``neurons'' correspond to the activation of any layer/block in the model. Previous works used an even more restricted scope and used ``neurons'' to refer to outputs of linear transformations (used to compute keys, queries, values, and inside the feed-forward block) and demonstrated their importance in various mechanisms (e.g., factual recall, performing auxiliary tasks) \cite{probing_for_ranking, van_Aken_2019, geva_transformer_2021, dai2022kn, chen2023journey, wang2022finding}.
While \citeauthor{Vast_2024}~\cite{Vast_2024} followed the same definition and applied NIG to the outputs of the linear transformations in MonoBERT, we argue that not all of them are equally interesting and can provide meaningful interpretations. As pointed out in \cite{brown2023privilegedconvergentbasesneural}, activations in transformer-based models do not necessarily possess a \emph{privileged basis}, which makes them hard to interpret. \citeauthor{elhage2021mathematical}~\cite{elhage2021mathematical} list ``tokens, attention patterns and FFN activations'' as ``the only vectors with privileged bases'' based on their mathematical definitions, but later works discovered that some features in the residual stream are also \emph{basis-aligned} \cite{elhage2023privilegedbasis}. Motivated by these works, we further restrict the scope of vectors and study the attention patterns (i.e., the attention probabilities obtained after the softmax over the query--key products) and the FFN neurons (see Section~\ref{subsection:architecture}).

\subsection{Adapting NIG to identify ``task-related'' neurons}
\label{adapting_nig}

As IR differs from other domains (CV or NLP) in which NIG has been applied, it requires some adaptations.

\paragraph{Comparisons across datasets.} To answer \textbf{RQ6} and draw fair comparisons between attributions obtained for different datasets, we aggregate the contribution of each activation over tokens. Since the same neuron is applied to every token of the query-document pair, we sum its conductance over tokens: a ``neuron'' in our experiments thus includes its corresponding outputs over all tokens. For the FFN neurons, this aggregated conductance can be thought of as the importance of the neuron within the Transformer. For the attention heads, we obtain a score characterizing their importance within the model.

\paragraph{Dependency on the input.} As pointed out by \citeauthor{wang2022finding}~\cite{wang2022finding}, gradient-based methods produce results that are input-dependent. As we want to identify ``IR task-related'' or ``skill'' neurons, i.e., neurons that are important for any input in the IR task, we average NIG results over multiple samples across multiple datasets and only retain the neurons with the highest mean attribution values.

\paragraph{Baseline.} For images, an obvious baseline $\sample^\prime$ is a black image. In IR and NLP in general, there is no obvious baseline, and current works that apply NIG to NLP tasks do so at the level of the embeddings. In the IR task, \citeauthor{zhan2020bertanalysis}~\cite{zhan2020bertanalysis} use Integrated Gradients to study the role of different input parts at different stages of MonoBERT. As a baseline, they replace the part they want to study (for example, the query tokens) with \PAD{} tokens, compute the corresponding embedding of the baseline, and then proceed with the method. We follow the same procedure to build our baselines. 

\paragraph{Better targeting ``relevance'' neurons.} Unlike our previous work~\cite{Vast_2024}, where we applied NIG to query-passage pairs and considered positive passages and negative ones separately, and to~\cite{zhan2020bertanalysis}, who used only relevant query-passage pairs, we work with triplets (query, positive passage, negative passage). In practice, this is closely related to the InfoNCE loss~\cite{infonce}. The rationale is that we want to study the discriminative power of cross-encoders and not only their pointwise decision process. Moreover, this simplifies our analysis compared to \cite{Vast_2024} as we have only a single set of important neurons per dataset instead of two (i.e., one for the positive passages and one for the negatives).
We hence use this version of Eq.~\eqref{equ:NIG}:
\begin{equation}
    \label{equ:NIG_infonce}
    \text{NIG}^y_{\text{InfoNCE}}(\sample_+,\sample_-) = \int_{t=0}^{t=1}\frac{\partial \sigma\left(z(\sample_+'+t(\sample_+-\sample_+'))-z(\sample_-'+t(\sample_--\sample_-'))\right)}{\partial \gamma_y(t)}
    \frac{\partial \gamma_y(t)}{\partial t} dt
\end{equation}
where $\sample_+$ and $\sample_-$ respectively represent a query-relevant passage pair and a query-non-relevant passage pair, $\sample_{\{+,-\}}'$ their respective baselines, and $z(\cdot)$ the model's logit (see Section~\ref{subsection:architecture}). 

\subsection{Experiment description}\label{sec:nig_experimental}

Using the datasets described in Section \ref{subsection:datasets} and the \emph{Top sampling} strategy as described in Appendix \ref{appendix:sampling_strategies}, Figure \ref{fig:dataset_construction}, we obtain one set of NIG scores per dataset and per model, corresponding to the importance of each neuron. While it is interesting to observe the most important neurons for a given sample or even a specific dataset, we are more interested in identifying the most important neurons \emph{in general} for the IR task. Following \citeauthor{Vast_2024}~\cite{Vast_2024}, we achieve this by aggregating the different sets of NIG scores across multiple datasets with: 
\begin{itemize}
    \item \textbf{Intersection}: $\text{I}(t, \collection) = \bigcap_{c \in \collection}\text{Top}\%(t,c)$, where $\collection$ is a pool of datasets and $\text{Top}\%(t, c)$ is the set of the top $t\%$ neurons for a dataset $c$;
    \item \textbf{Fusion}: $\text{F}(t, \collection) = \text{Top}\%\left(t, \left\langle \text{NIG}_c \right\rangle_{c \in \collection}\right)$, where $\left\langle \cdot \right\rangle_{c \in \collection}$ corresponds to the average, neuron-wise, of the NIG scores. 
\end{itemize}

The rationale behind these two different ways of aggregating the results across different datasets is that each allows us to observe a different phenomenon. With the Fusion, we can observe \emph{generic} important neurons, i.e., neurons that are important on average over the datasets. In contrast, the Intersection allows us to observe the important neurons \emph{shared} across the pool, i.e., neurons that are important for each dataset in the pool. Using these operations and for $t \in \{0.1\%, 0.5\%, 1\%\}$, we select the most important neurons from the following sets of datasets:
\begin{itemize}
    \item \textbf{M} corresponds to the 4 TREC-DL datasets \cite{craswell2019overview,craswell2020overview,craswell2021overview,craswell2022overview} all derived from MS MARCO \cite{MSMARCO};
    \item \textbf{O} corresponds to the out-of-domain datasets, i.e., all datasets from Table \ref{tab:datasets-passages} but the 4 TREC-DL datasets; 
    \item \textbf{A} corresponds to all the datasets.
\end{itemize}

By combining a set of datasets, an aggregation operation, and a percentage $t$, we obtain a list of the most important neurons inside a model. We follow \citeauthor{Vast_2024}~\cite{Vast_2024} and refer to this list as an \textit{ablation scheme}, even if we do not perform a complementary ablation study. We denote an ablation scheme by its aggregation operation and its pool of datasets, e.g., $F_M = \text{F}(t, M)$, $I_O = \text{I}(t, O)$ or $I_A = \text{I}(t, A)$; unless stated otherwise, $t=1\%$. We refer to $t$ as the \emph{pruning percentage}.

\begin{figure}[t]
    \centering
    \begin{subfigure}[h]{0.45\textwidth}
        \centering
        \includegraphics[width=\textwidth]{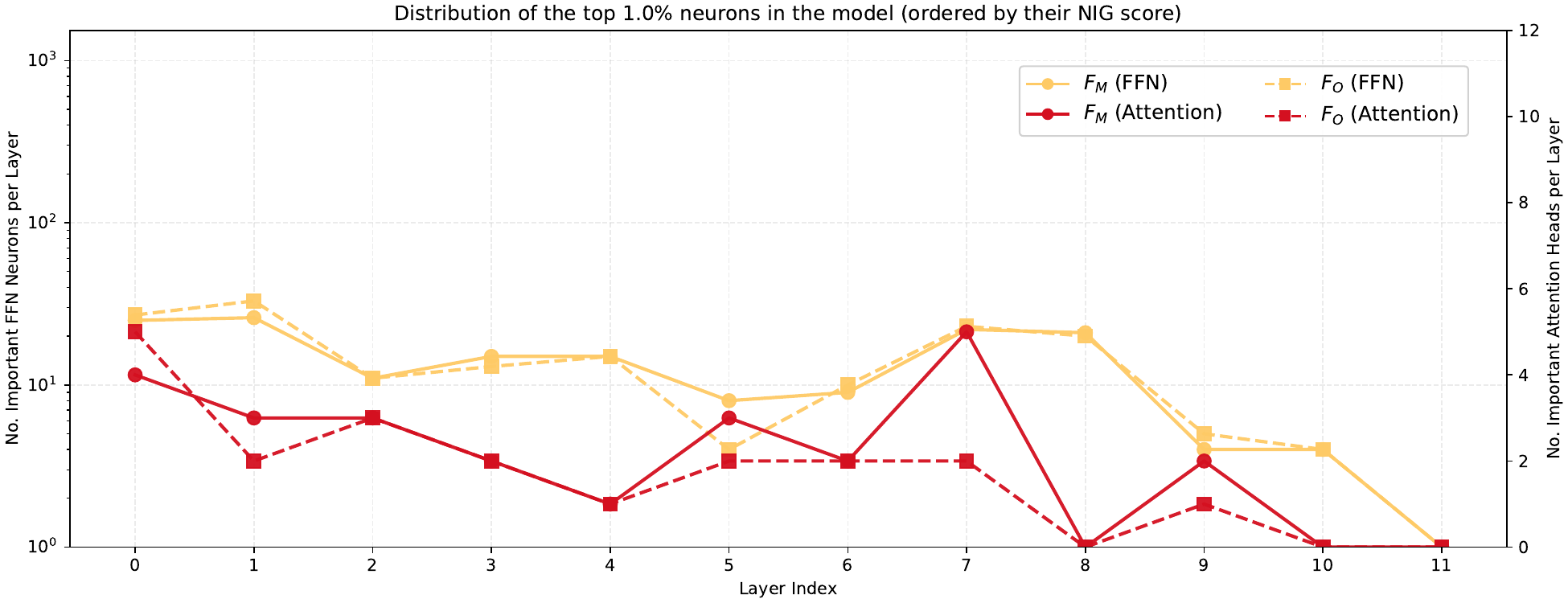}
        \caption{Ablation schemes $F_M$ and $F_O$ for MiniLM.}
    \end{subfigure}
    \hspace{0.02\textwidth} 
    \begin{subfigure}[h]{0.45\textwidth}
        \centering
        \includegraphics[width=\textwidth]{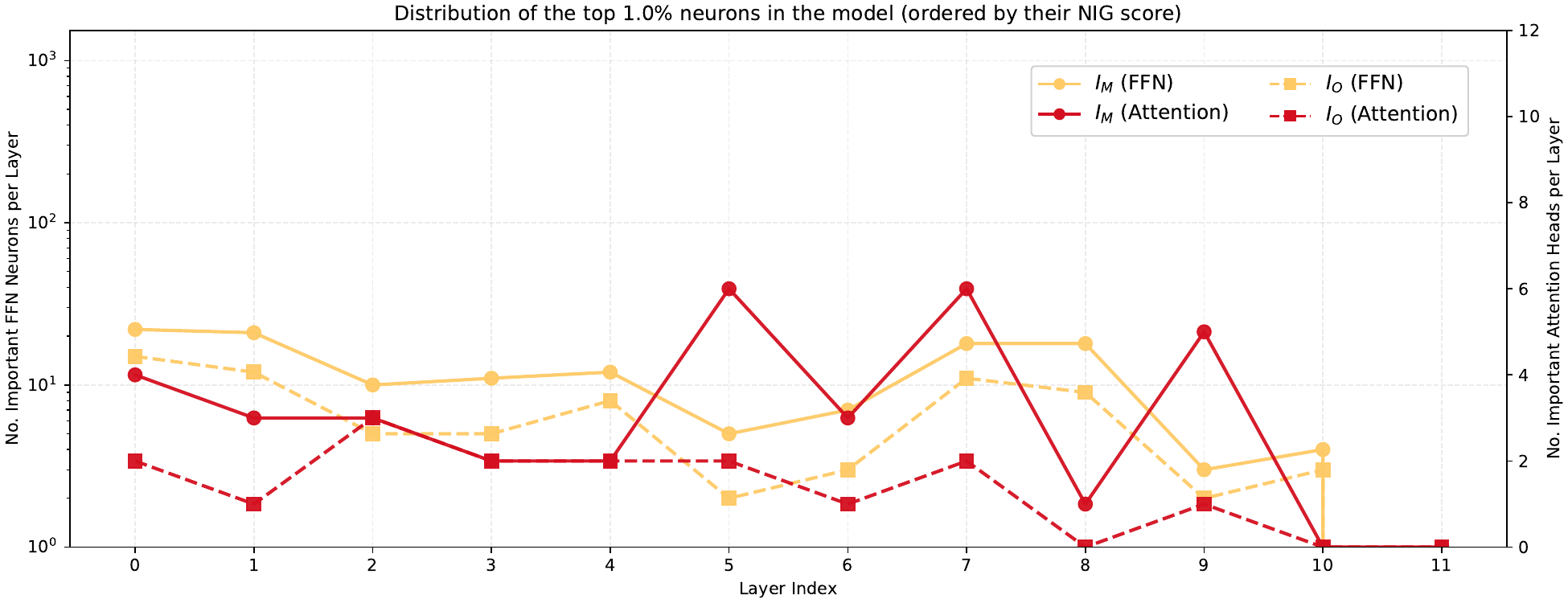}
        \caption{Ablation schemes $I_M$ and $I_O$ for MiniLM.}
    \end{subfigure}
    \begin{subfigure}[h]{0.45\textwidth}
        \centering
        \includegraphics[width=\textwidth]{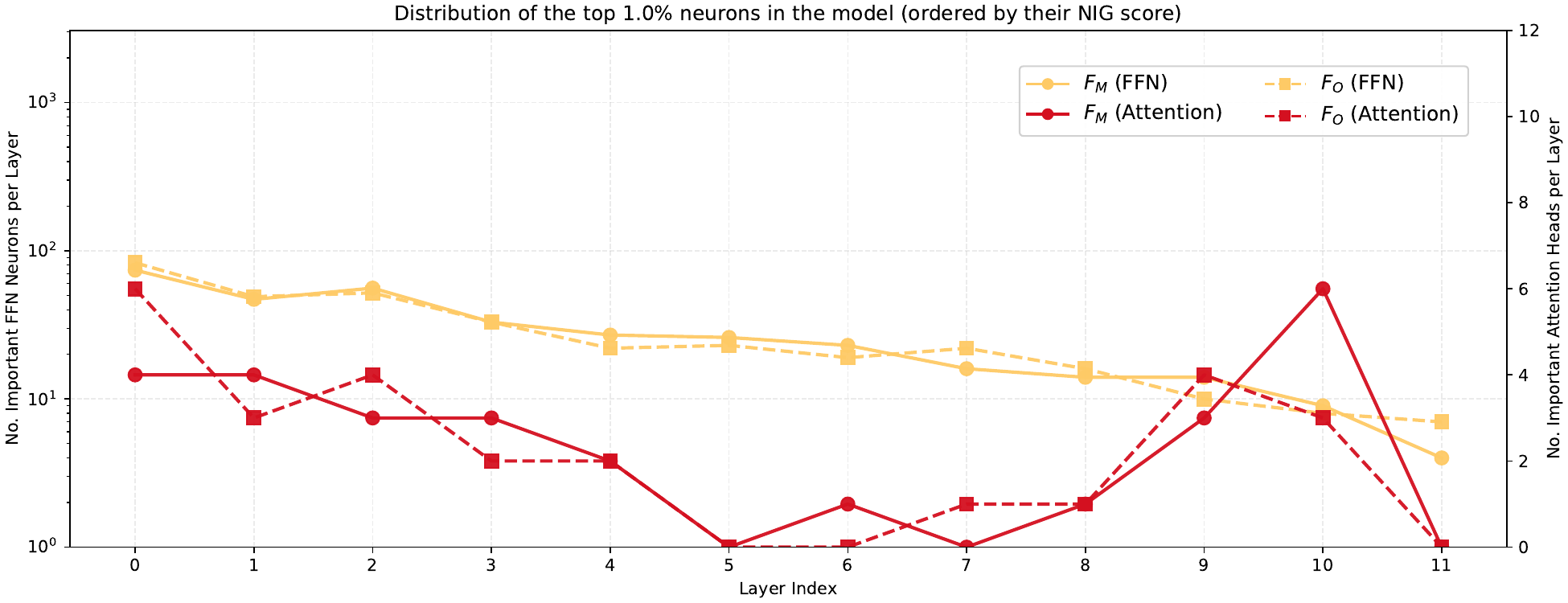}
        \caption{Ablation schemes $F_M$ and $F_O$ for MonoBERT-base.}
        \label{fig:nig_details_fusion_monobert_base}
    \end{subfigure}
    \hspace{0.02\textwidth} 
    \begin{subfigure}[h]{0.45\textwidth}
        \centering
        \includegraphics[width=\textwidth]{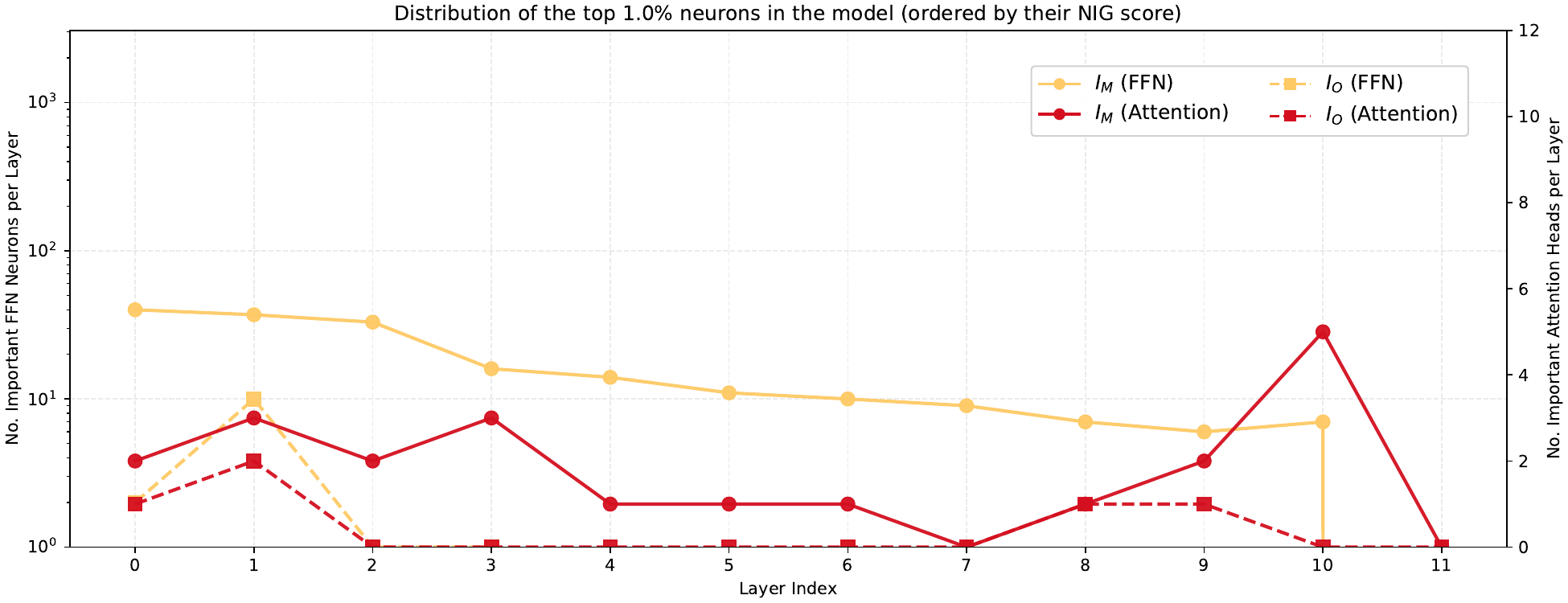}
        \caption{Ablation schemes $I_M$ and $I_O$ for MonoBERT-base.}
        \label{fig:nig_details_intersection_monobert_base}
    \end{subfigure}
    \centering
    \begin{subfigure}[h]{0.45\textwidth}
        \centering
        \includegraphics[width=\textwidth]{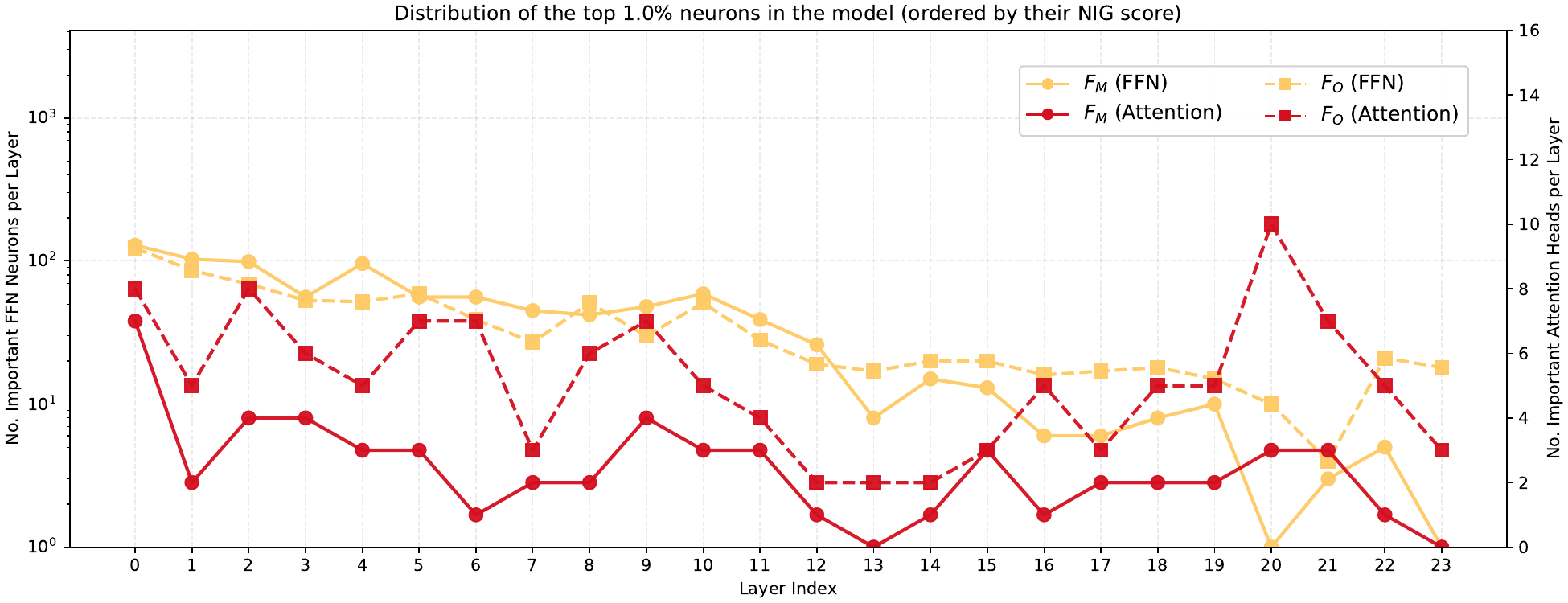}
        \caption{Ablation schemes $F_M$ and $F_O$ for MonoBERT-large.}
        \label{fig:nig_details_fusion_monobert_large}
    \end{subfigure}
    \hspace{0.02\textwidth} 
    \begin{subfigure}[h]{0.45\textwidth}
        \centering
        \includegraphics[width=\textwidth]{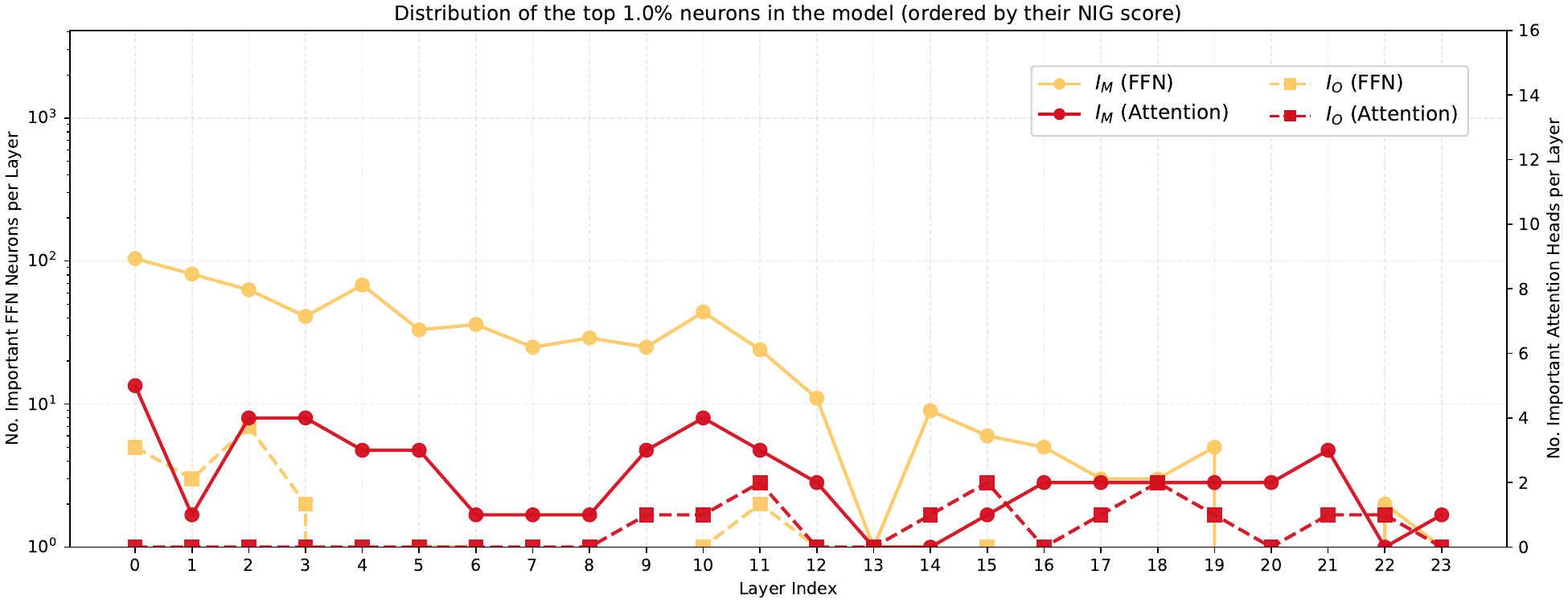}
        \caption{Ablation schemes $I_M$ and $I_O$ for MonoBERT-large.}
        \label{fig:nig_details_intersection_monobert_large}
    \end{subfigure}
    \caption[Number of most important components per layer]{Number of most important components per layer in the models: attention heads (red, right y-axis) and FFN neurons (yellow, left y-axis), following the ablation schemes obtained from the \textbf{Fusion} and the \textbf{Intersection} of the NIG scores for the different pools of datasets, at pruning percentage $t=1\%$. Solid lines correspond to the TREC-DL pool ($M$), dashed lines to the OOD pool ($O$).}
    \label{fig:nig_details}
\end{figure}

\begin{figure}[t]
    \centering
    \begin{subfigure}[h]{\textwidth}
        \centering
        \includegraphics[width=\textwidth]{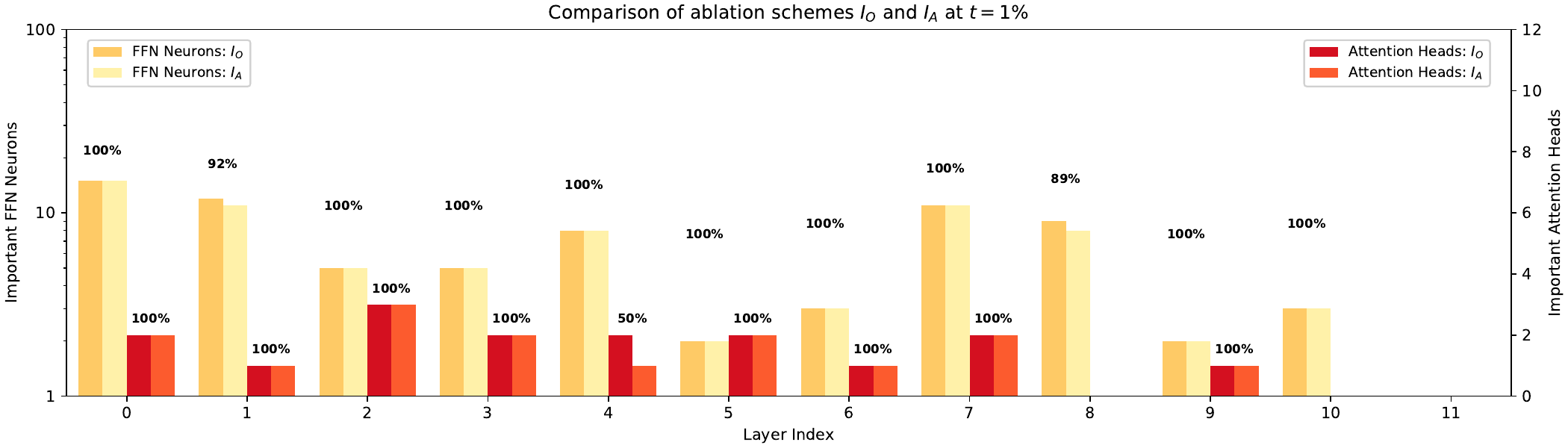}
        \caption{Comparison of the most important neurons and attention heads between ablation schemes $I_O$ and $I_A$ for MiniLM.}
    \end{subfigure}
    \begin{subfigure}[h]{\textwidth}
        \centering
        \includegraphics[width=\textwidth]{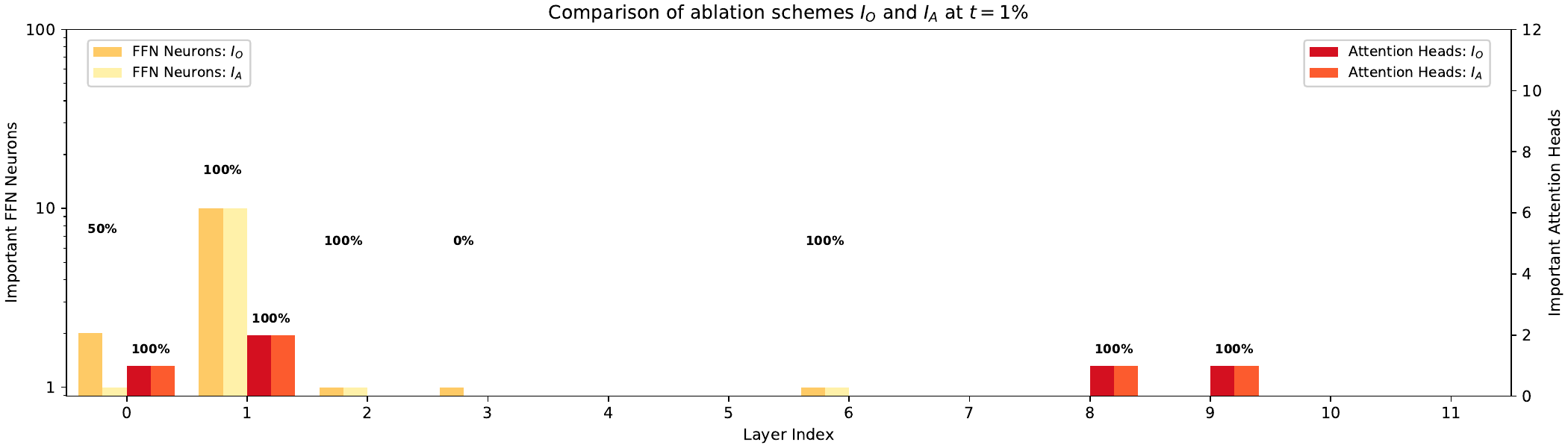}
        \caption{Comparison of the most important neurons and attention heads between ablation schemes $I_O$ and $I_A$ for MonoBERT-base.}
    \end{subfigure}
    \begin{subfigure}[h]{\textwidth}
        \centering
        \includegraphics[width=\textwidth]{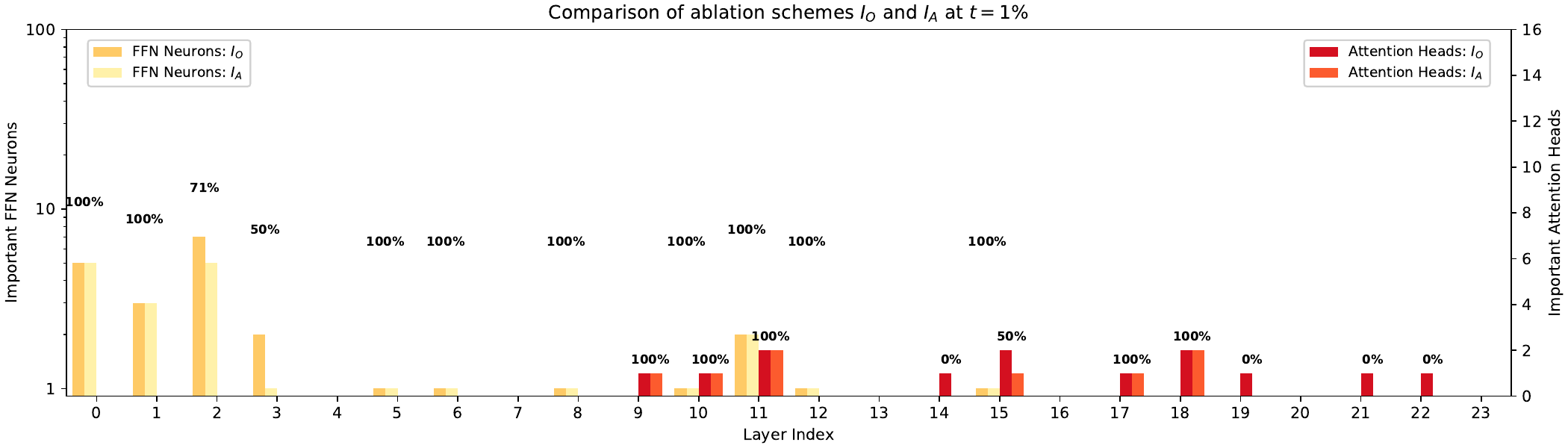}
        \caption{Comparison of the most important neurons and attention heads between ablation schemes $I_O$ and $I_A$ for MonoBERT-large.}
    \end{subfigure}
    \caption[Comparison of the ablation schemes $I_O$ and $I_A$]{Number of most important FFN neurons (yellow, left y-axis) and attention heads (red, right y-axis), compared between ablation schemes $I_O$ and $I_A$ at pruning percentage $t=1\%$. Percentage values above the bars correspond to the Jaccard similarity between the two sets of neurons for each model's component at each layer $\ell$ from the ablation schemes.}
    \label{fig:nig_comparisons}
\end{figure}

\subsection{Results analysis}

\result{Important neurons according to NIG are not located in specific layers or transformer components.} 
Figure~\ref{fig:nig_details} reports the neurons selected with NIG for MiniLM and both versions of MonoBERT. We can see that, in proportion and across ablation schemes, the most important parts of the models tend to be attention heads, but no layer particularly stands out, even though there seems to be a trend where the deeper the layer, the less important its components. 
Given the known roles of the different layers in transformer-based models (see Section \ref{section:Related_works}), this suggests that relevance prediction involves several consecutive stages, in which attention plays a key role by moving information across input parts.

\result{When exposed to domain shifts, MonoBERT tends to rely on different neurons, whereas MiniLM has learned a more consistent mechanism.} 
By construction, $F(t,\cdot)$ always contains $t\%$ of the neurons while $I(t,\cdot)$ contains at most $t\%$, so the gap between the $I$ curve and the $F$ curve measures how much the top-$t\%$ set changes from one dataset to the next. In Figure~\ref{fig:nig_details}, for both MonoBERT models, $I_O$ is nearly empty while $F_O$ is full (see the dashed curves of Figures \ref{fig:nig_details_intersection_monobert_base} and \ref{fig:nig_details_intersection_monobert_large}, versus the dashed curves of Figures \ref{fig:nig_details_fusion_monobert_base} and \ref{fig:nig_details_fusion_monobert_large}), so the top neurons differ across OOD datasets; for MiniLM,  $I_O$ and $F_O$ have comparable cardinalities at each layer, i.e., the same neurons are among the most important for every dataset. In other words, there are comparatively fewer neurons that are \textit{always} important across different situations in MonoBERT-base and large than in MiniLM, which leads us to the conclusion that MiniLM relies on a more consistent mechanism than MonoBERT. 

\result{OOD and ID predictions rely on a disjoint set of neurons.} 
Across models, in Figure \ref{fig:nig_comparisons} we compare the similarities between the sets of the most important neurons shared between the OOD datasets and all the datasets per component (respectively $I_O$ and $I_A$). Although more visible with MonoBERT-base and MonoBERT-large than for MiniLM (see previous remark), we observe that OOD predictions involve additional important neurons, different from those for ID predictions (i.e., the intersection of the most important neurons across all datasets is a subset of the intersection of the most important neurons across OOD datasets). For MiniLM, this is observable at Layers 1 and 8 for FFN neurons and 4 for attention heads and, for MonoBERT-base, at Layers 0 and 3 and for MonoBERT-large, at Layers 2 and 3 for FFN neurons and 14, 15, 19, 21 and 22 for attention heads. \citeauthor{Vast_2024}~\cite{Vast_2024} report the same observation solely for MonoBERT-large, but we specify here that this is mostly true for attention heads. 
Since the OOD-specific components of MonoBERT-large are mostly attention heads in later layers, the differences across domains may lie more in what information is moved (attention) than in how it is transformed (FFN).

This observation, as well as the identification of other important attention heads, motivates the need to better understand what information is moved between the tokens' representations and how it is moved. In the next section, we focus on characterizing the information flow across the attention heads of each cross-encoder. 

\section{RQ2: Identifying the flow of information inside the model}
\label{section:information_flow}

\subsection{Context and notations}

\begin{figure}[t]
    \centering
    \includegraphics[width=0.7\textwidth]{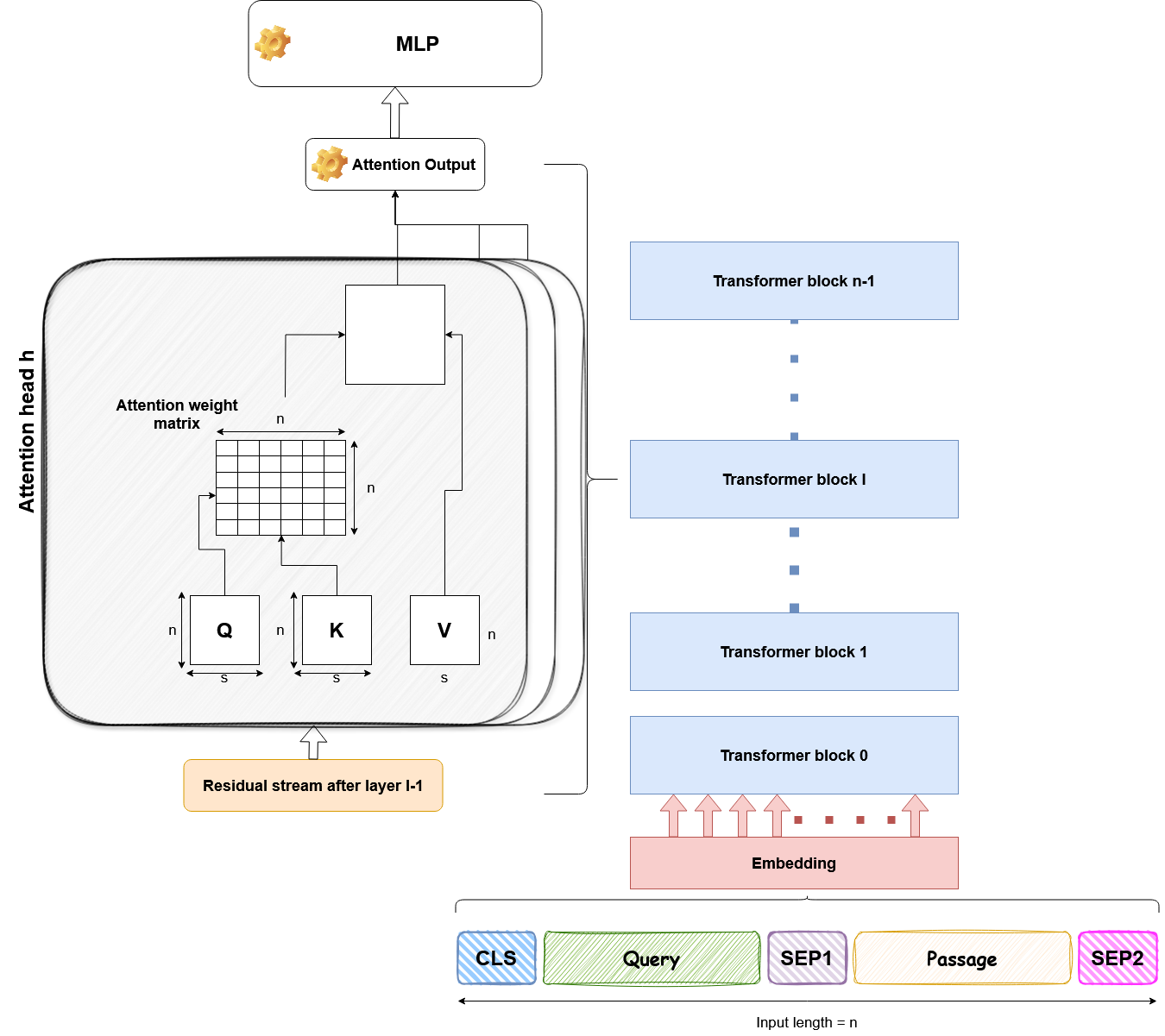}
    \caption[Description of an attention weight matrix]{Description of an attention weight matrix, at the layer $\ell$ for an attention head $h$, in the context of a transformer-based model. Each row inside these matrices is a distribution of probabilities over the entire input, relatively to a given token $t$ (the target token), quantifying the amount of information each token (the sources) will send to $t$.}
    \label{fig:attention_weight_in_context}
\end{figure}

\begin{figure}[h]
    \centering
    \includegraphics[width=0.72\textwidth]{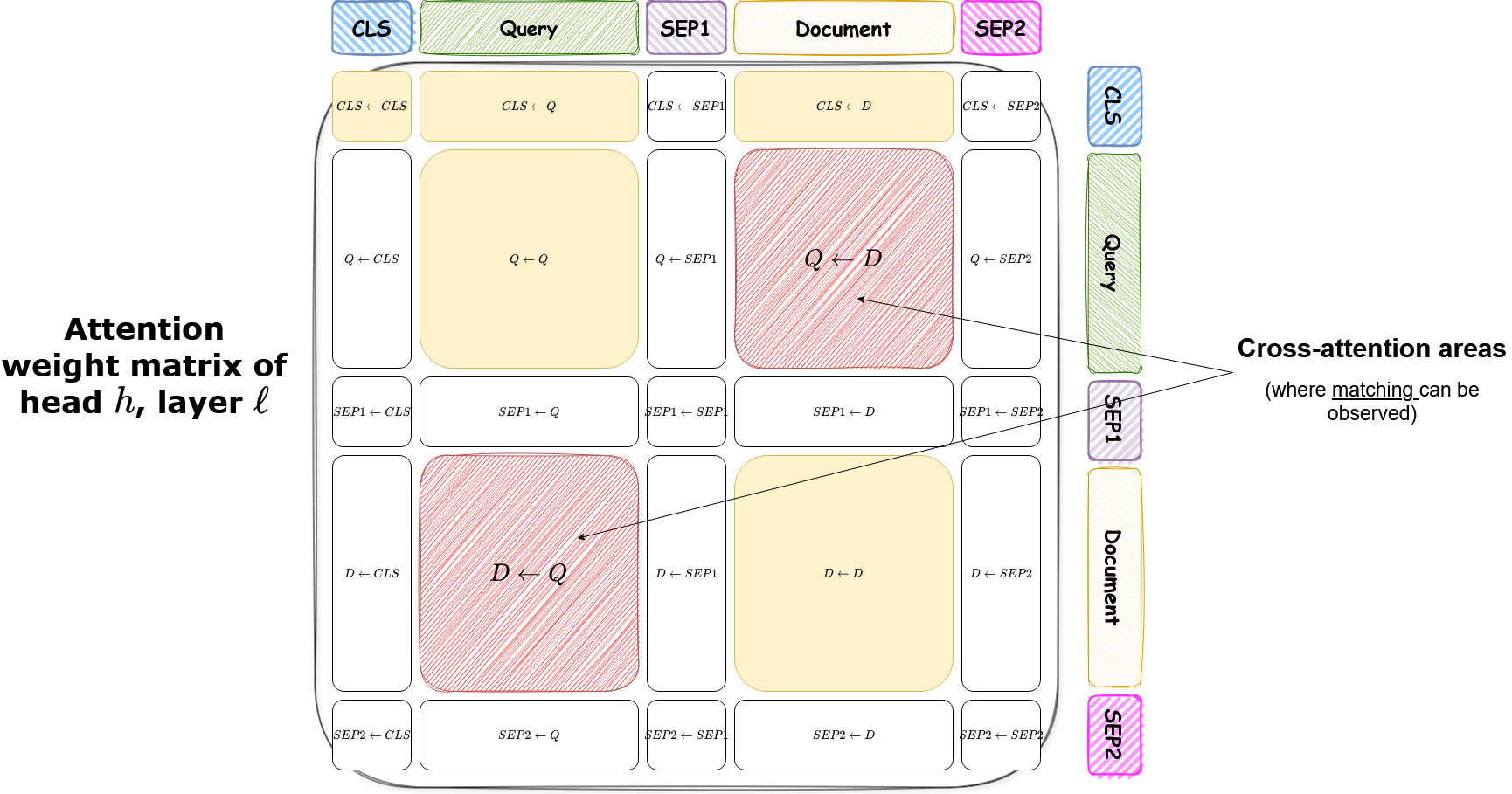}
    \caption[Segmentation of an attention weight matrix between the different input parts]{Segmentation of the attention weight matrices into the 25 parts dictating the information flow between all the input parts $X \leftarrow Y$, $X, Y \in \{\CLS, Q, \SEPone, D, \SEPtwo\}$. Highlighted in yellow are the areas we suspect to be the most important for cross-encoders, and in red, the areas where Query and Document exchange information.}
    \label{fig:attention_weight_segmented}
\end{figure}

In the previous section, we identified the most important components of the cross-encoders for the IR task. Yet, directly using NIG does not unveil their respective roles in modeling relevance. To tackle this issue, we first look to characterize the information flow between these components. To do so, we consider 5 distinct input parts for each query-passage pair: the \CLS{} token, the query $Q$, the first separator token $\SEPone$, the document (or passage) $D$, and the last separator token $\SEPtwo$ (see Figure \ref{fig:attention_weight_in_context}). In transformer-based models, information is moved between token representations thanks to the attention process \cite{attention}. As described in Figure \ref{fig:attention_weight_in_context}, the attention weight matrix dictates the effective information transfers.  
In this section, we focus mostly on these matrices and study the flow of information between input parts. 
Note that the FFN can also be used as a proxy in this study (see Subsection \ref{subsection:nig_information_flow}), given that important transformations on a part of the input necessarily imply that important signals are available there.
We decompose each attention score matrix into different parts, characterizing the information being sent from one input part to another. Figure \ref{fig:attention_weight_segmented} describes these different parts for a given attention head $h$ at a layer $\ell$ of a cross-encoder model. Each one of these input parts can attend to itself or any of the others, amounting to 25 possible \emph{directions} for the information flow. We name \textit{direction} the attention of the input's part $X$ to the input's part $Y$ within the attention weight matrix or equivalently, the information transmitted by the input's part $Y$ to the input's part $X$, also denoted as $X \leftarrow Y$ with $X, Y \in \{\CLS, Q, \SEPone, D, \SEPtwo\}$. We further introduce the notation $X \nleftarrow Y$ to indicate that the transfer of information from the input part $Y$ to the input part $X$ is blocked.

Throughout this section, we use the following notations to refer to these directions:
\begin{formulas}
For a given attention head $h$ at layer $\ell$, we denote $\patt(X_i\leftarrow Y_j)$ the probability that $X_i$, the $i^{th}$ token of $X$, \emph{attends to} $Y_j$, the $j^{th}$ token  of $Y$. 
\end{formulas}

\subsection{Directly characterizing the flow of information in cross-encoders}
\label{subsection:nig_information_flow}

In this Section, we analyze the results obtained by applying two different methods to the attention weight matrix framework described above. We begin this section by introducing these two approaches.

\subsubsection{Extending the NIG approach to study the information flow}
In Section \ref{section:NIG}, we identified the most important components using NIG. As we monitored the raw importance of each component, we obtained a single score per component (attention heads or FFN neurons) by aggregating along the entire input length (see Subsection \ref{adapting_nig}). While this is useful for comparing overall attributions, it discards fine-grained signal about \emph{where} in the input each component's importance is concentrated. We therefore aggregate the conductances (see Subsection \ref{subsection:background}) \emph{per input part} (query, document, \CLS{}, \SEP{}) rather than over the whole input, following the segmentation of Figure~\ref{fig:attention_weight_segmented}. By applying the same experimental protocol described in Section \ref{sec:nig_experimental} at this finer granularity, we obtain different ablation schemes, defined by their aggregation operation (Intersection or Fusion), the set of datasets used ($M$, $O$ or $A$) and a pruning percentage $t$.

\subsubsection{Studying the information flow with the Information Bottleneck approach}\label{sec:ib_details}

In the X-AI literature, the Information Bottleneck \cite{Schulz2020Restricting} approach is helpful to analyze different components of a model simultaneously. This approach has already been applied to IR and re-ranking \cite{jiang-etal-2021-bert}, but never to directly analyze the directions followed by the information inside a cross-encoder. From a high-level perspective, its objective is to restrict as much as possible the amount of transfers between an input and an output (the bottlenecks), while maintaining the quality of the output.

We provide more background on the theory behind the information bottleneck approach in the Appendix \ref{appendix:information_bottleneck}, and further detail there our experimental setup. 
We learn the bottlenecks across the whole model at once under the constraint that predictions are preserved. To verify that this constraint holds, we measured the degradation caused when applying the learned masks on the cross-encoders. Table \ref{tab:mask_eval} reports the score of the models with their masks on the 4 TREC-DL datasets as well as the OOD datasets. From these results, no statistically significant degradation can be observed across all three models. This validates that the masks we learned do not harm the flow of the information of relevance inside the models. 

\begin{table}[t]
\centering
\caption{Performance (in  nDCG@10) of the models evaluated while applying the masks learned with the Information Bottleneck to their attention matrices (see Sec. \ref{sec:ib_details}) on the datasets from TREC-DL and BEIR. No difference with the original model performance is statistically significant at the 5\% level under the two-tailed Student’s t-test.}
\begin{tabular}{l|cccc|cccccc}
\toprule
\textbf{Model} & \textbf{T.-DL 19} & \textbf{20} & \textbf{ 21} & \textbf{22} & \textbf{NFCorpus} & \textbf{FiQA} & \textbf{T.-Covid} & \textbf{Robust-04} & \textbf{NQ} & \textbf{BioASQ}  \\
\midrule
\texttt{MiniLM} & 0.74 & 0.68 & 0.65 & 0.49 & 0.35 & 0.31 & 0.74 & 0.46 & 0.36 & 0.50 \\
\texttt{+ mask} & 0.71 & 0.65 & 0.63 & 0.47 & 0.35 & 0.30 & 0.75 & 0.44 & 0.33 & 0.50 \\
\midrule
\texttt{MonoBERT-base} & 0.73 & 0.68 & 0.64 & 0.49 & 0.35 & 0.30 & 0.73 & 0.47 & 0.36 & 0.50 \\
\texttt{+ mask} & 0.72 & 0.67 & 0.63 & 0.47 & 0.35 & 0.29 & 0.73 & 0.45 & 0.34 & 0.49 \\
\midrule
\texttt{MonoBERT-large} & 0.71 & 0.68 & 0.63 & 0.49 & 0.35 & 0.32 & 0.70 & 0.45 & 0.35 & 0.47 \\
\texttt{+ mask} & 0.70 & 0.66 & 0.61 & 0.47 & 0.35 & 0.32 & 0.70 & 0.44 & 0.34 & 0.45 \\
\bottomrule
\end{tabular}
\label{tab:mask_eval}
\end{table}

\begin{figure}[t]
    \centering
    \begin{subfigure}[t]{0.45\textwidth}
        \centering
        \includegraphics[width=\textwidth]{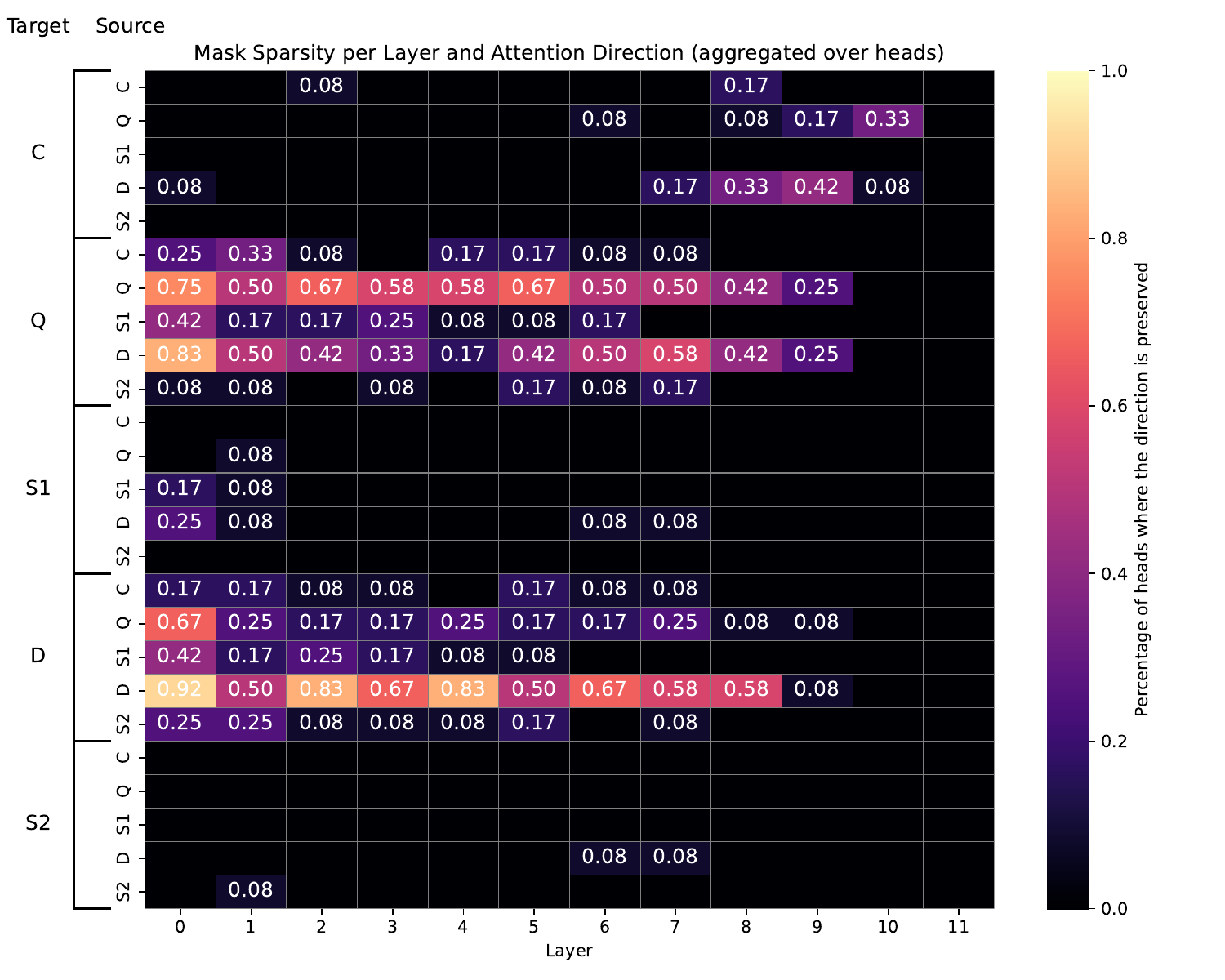}
        \caption{Results of the Information Bottleneck on the attention heads of MiniLM.}
        \label{fig:minilm_mask_overview}
    \end{subfigure}%
    \hspace{0.01\textwidth}
    \begin{subfigure}[t]{0.45\textwidth}
        \centering
        \includegraphics[width=\textwidth]{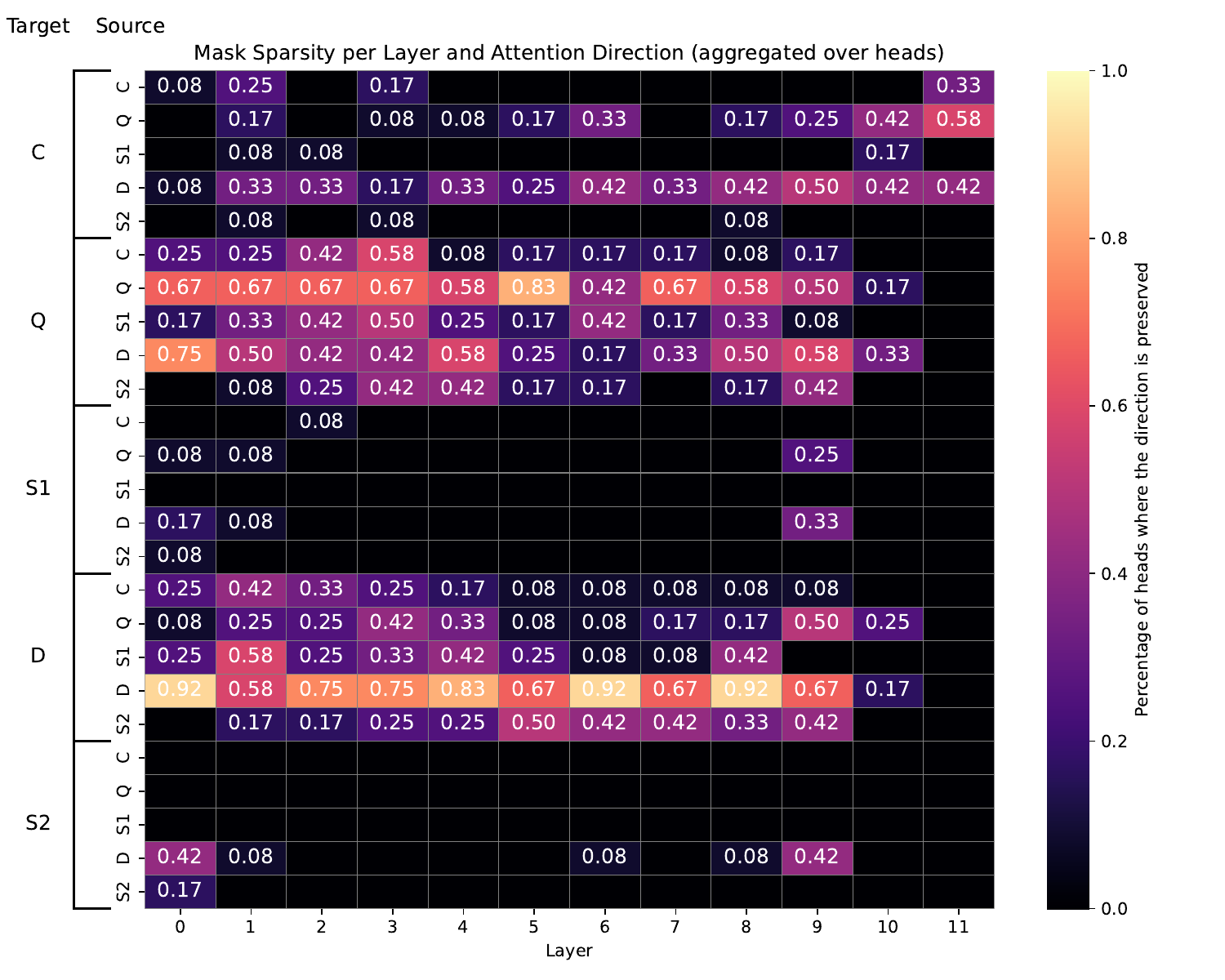}
        \caption{Results of the Information Bottleneck on the attention heads of MonoBERT-base.}
        \label{fig:monobert_base_mask_overview}
    \end{subfigure}%
    \hfill
    \begin{subfigure}[t]{\textwidth}
        \centering
        \includegraphics[width=0.85\textwidth]{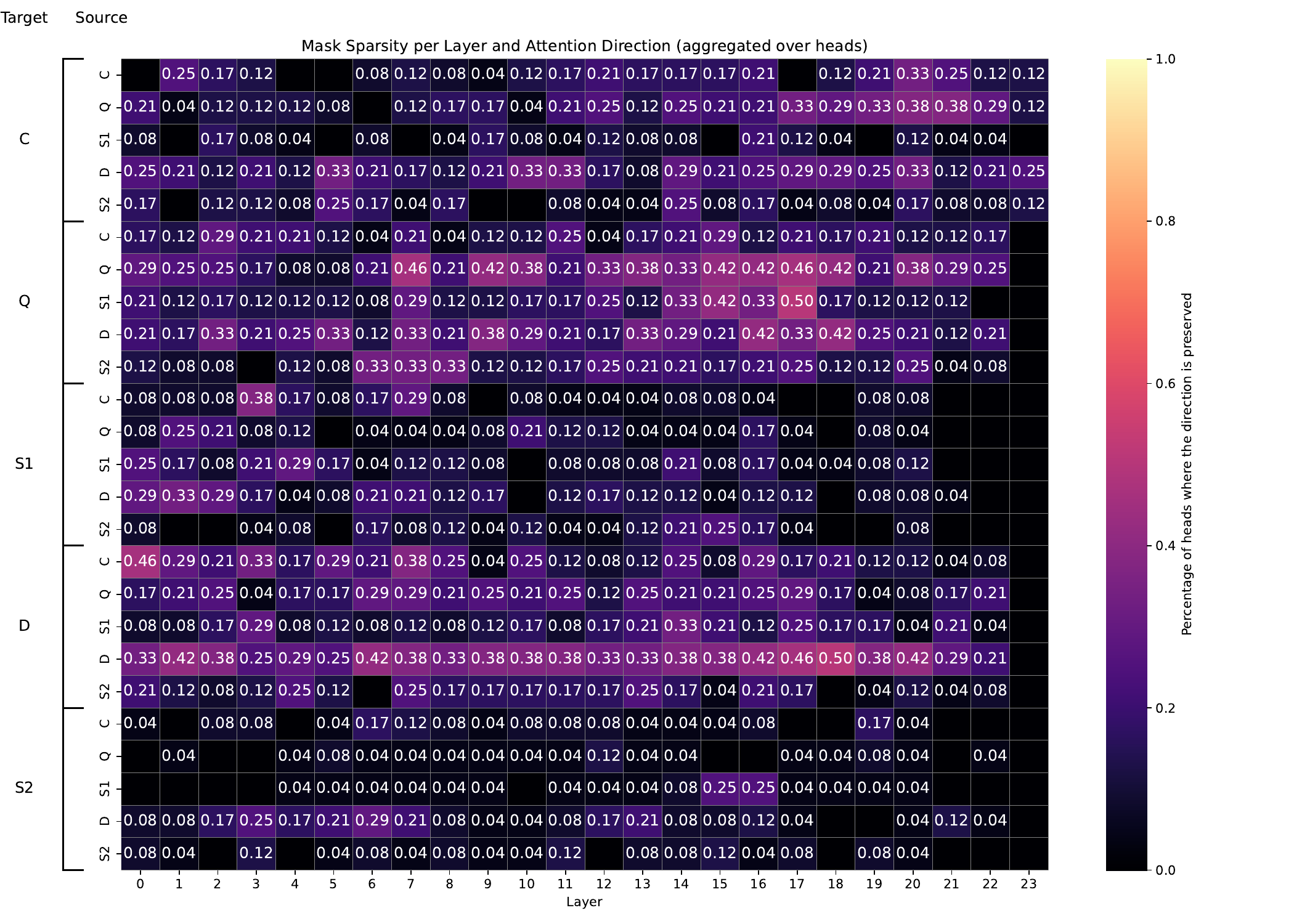}
        \caption{Results of the Information Bottleneck on the attention heads of MonoBERT-large.}
        \label{fig:monobert_mask_overview}
    \end{subfigure}%
    \caption[Information Bottleneck results on the information flow]{Overview of the masking of each direction at each layer in the models, averaged over the attention heads. }
\label{fig:information_bottlenecks}
\end{figure}

\begin{figure}[t]
    \centering
    \begin{subfigure}[h]{0.44\textwidth}
        \centering
        \includegraphics[width=\textwidth]{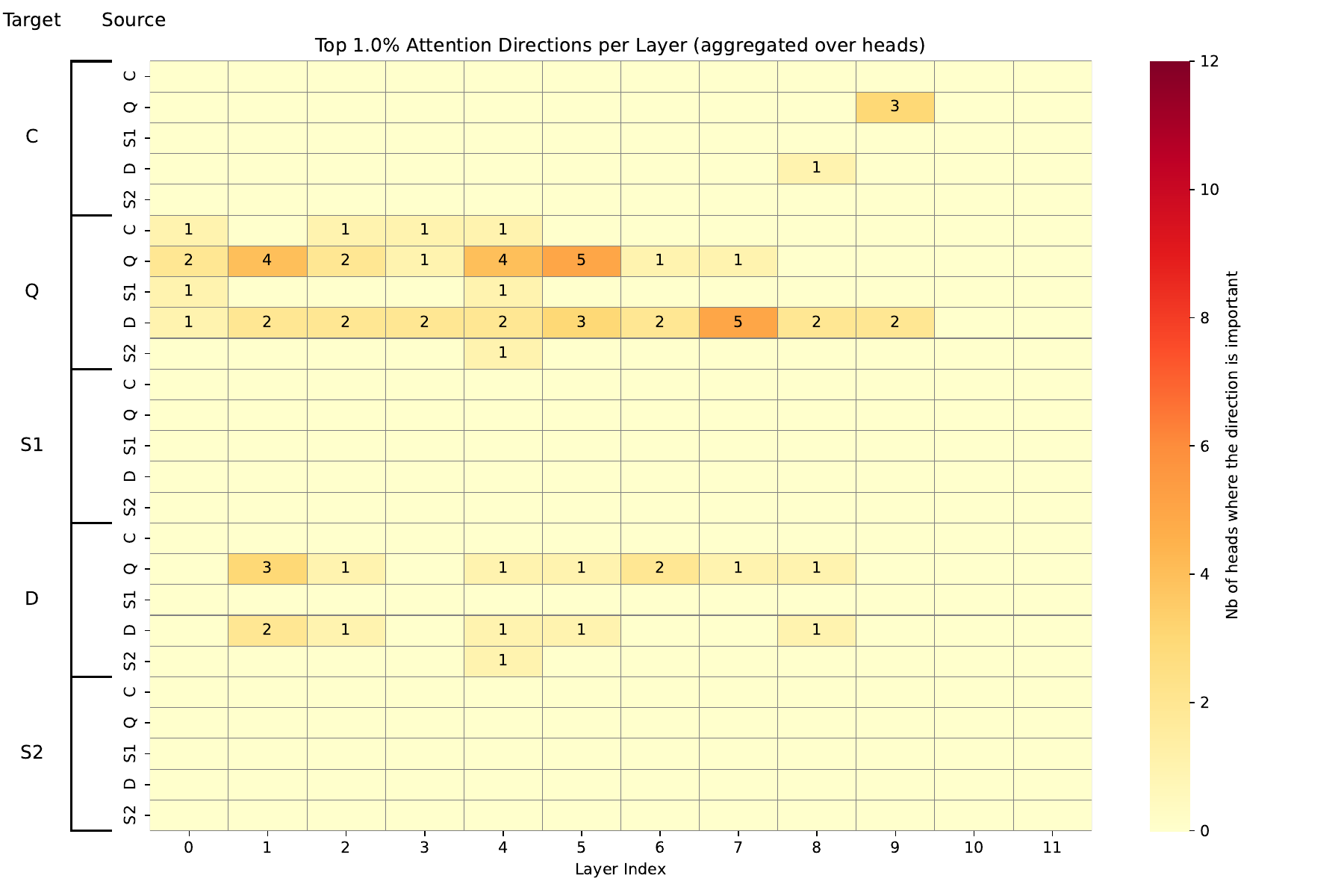}
        \caption{Ablation scheme $I_O$ for MiniLM.}
    \end{subfigure}
    \hspace{0.02\textwidth} 
    \begin{subfigure}[h]{0.44\textwidth}
        \centering
        \includegraphics[width=\textwidth]{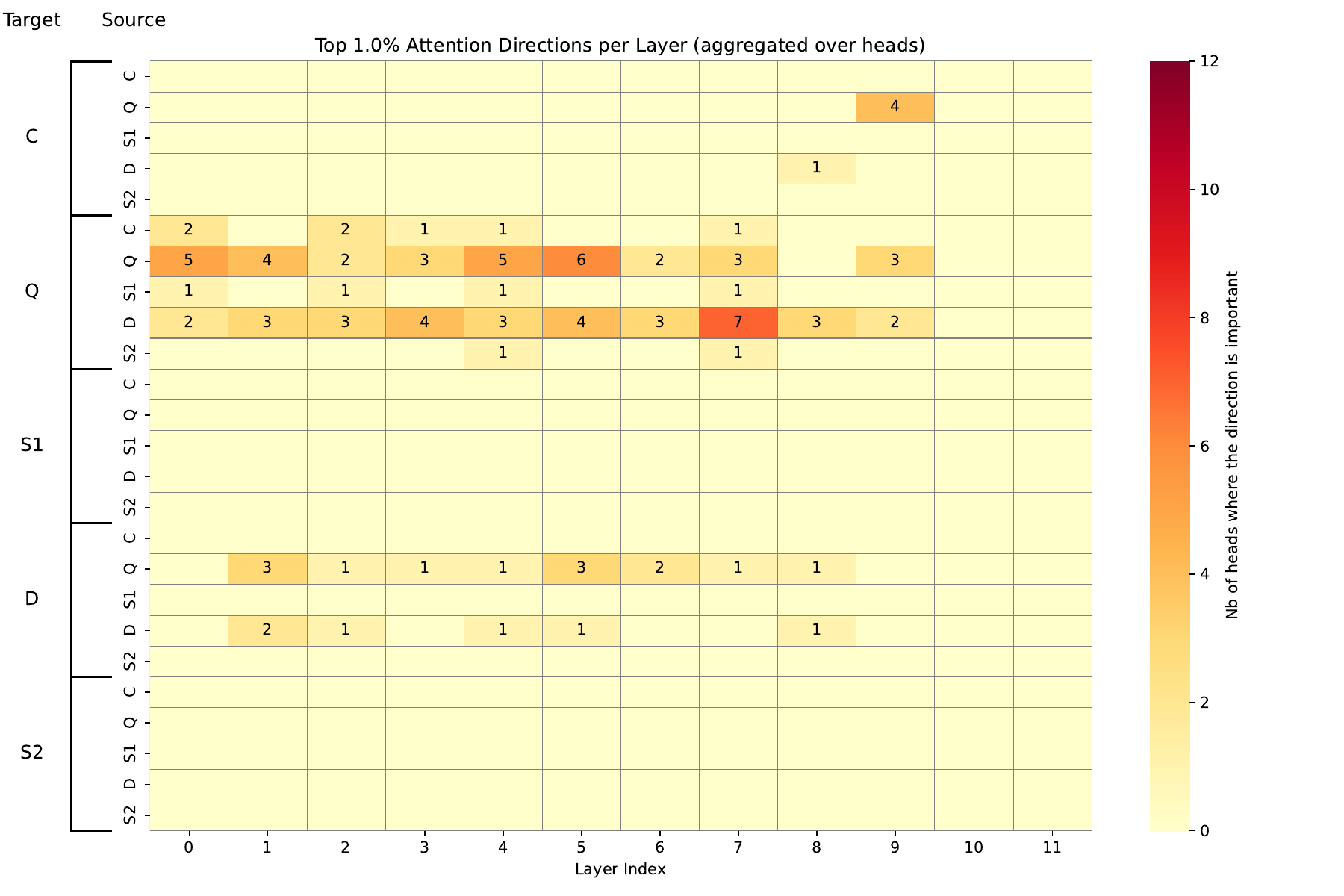}
        \caption{Ablation scheme $F_O$ for MiniLM.}
    \end{subfigure}

    \begin{subfigure}[h]{0.44\textwidth}
        \centering
        \includegraphics[width=\textwidth]{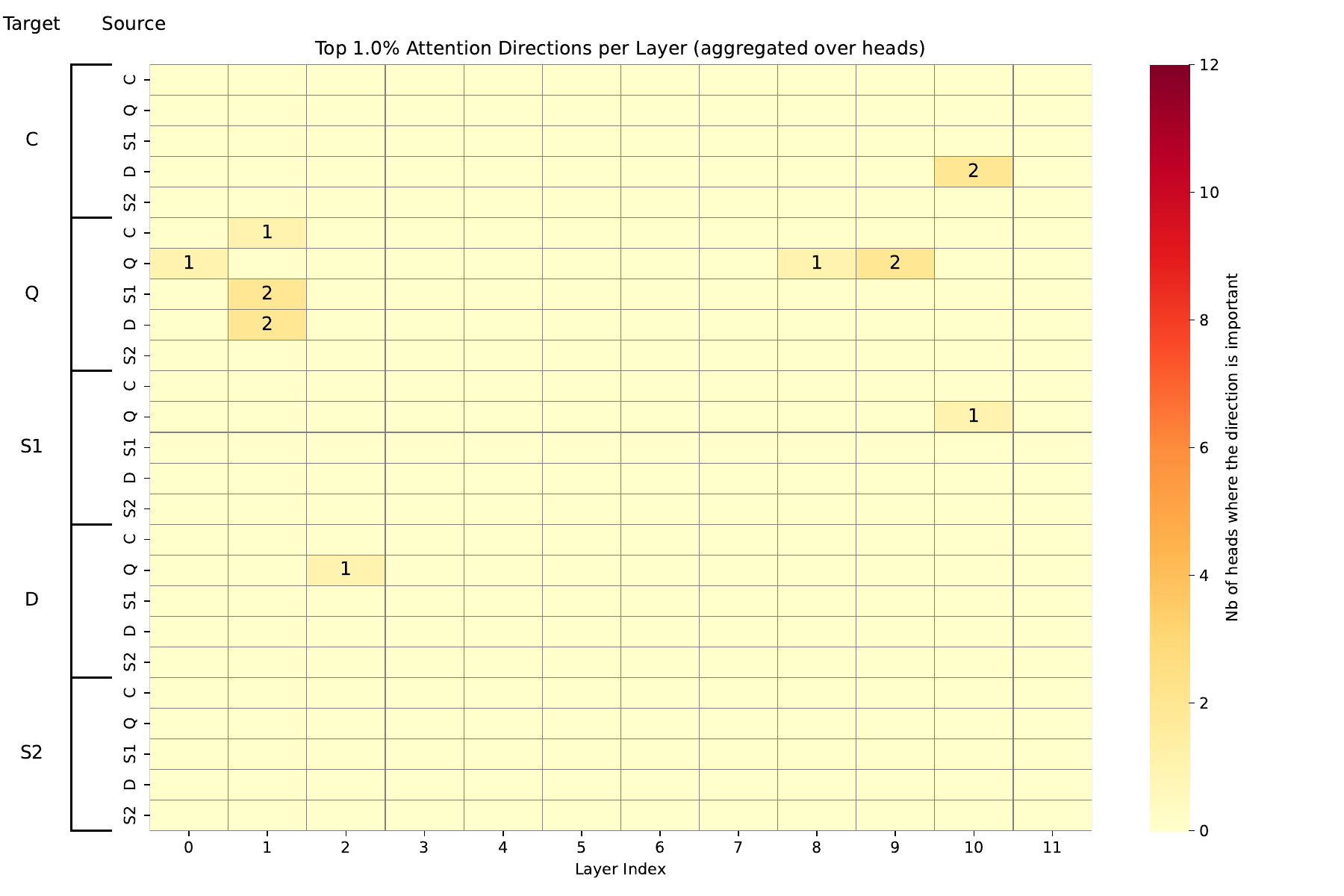}
        \caption{Ablation scheme $I_O$ for MonoBERT-base.}
    \end{subfigure}
    \hspace{0.02\textwidth} 
    \begin{subfigure}[h]{0.44\textwidth}
        \centering
        \includegraphics[width=\textwidth]{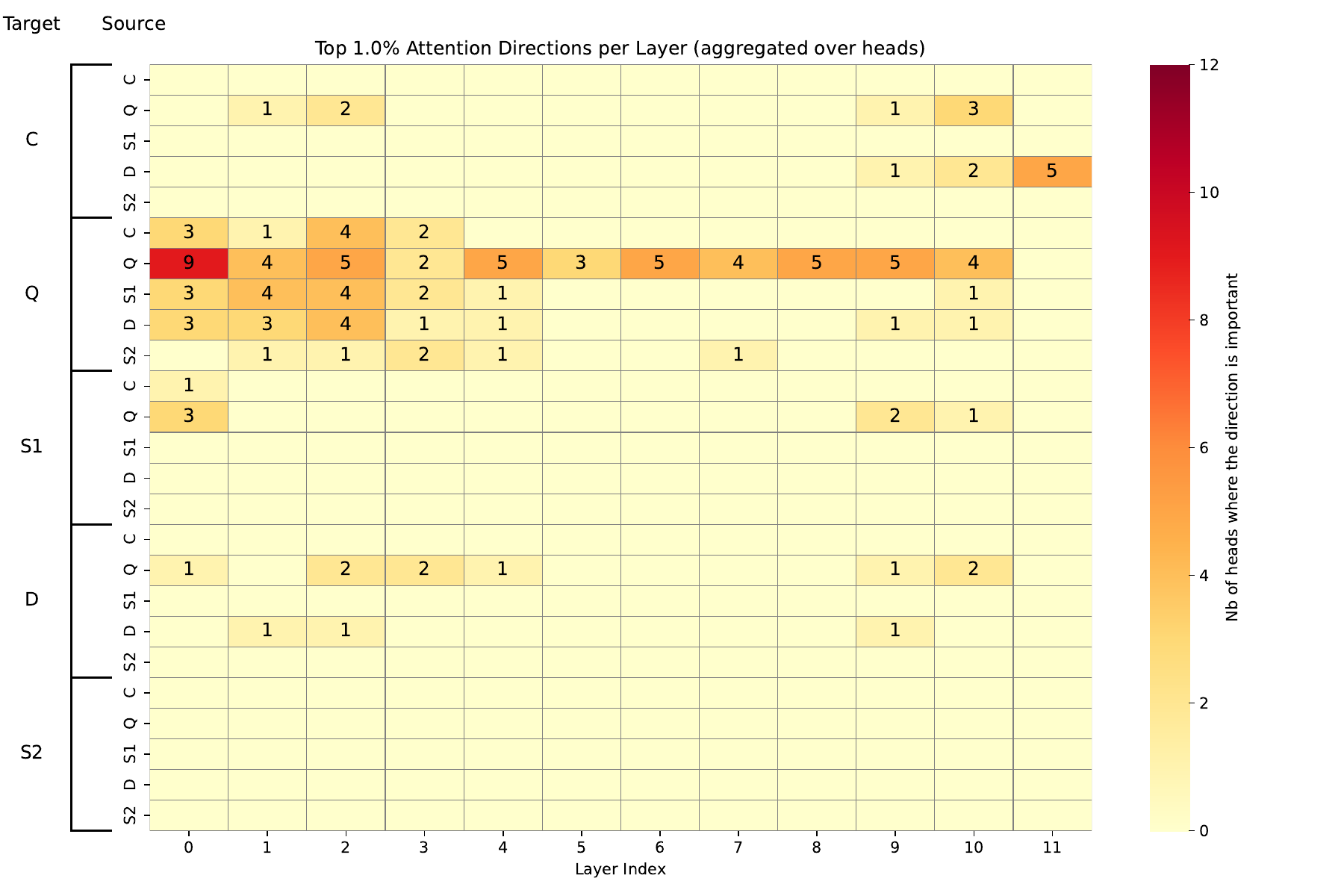}
        \caption{Ablation scheme $F_O$ for MonoBERT-base.}
    \end{subfigure}

    \begin{subfigure}[h]{0.44\textwidth}
        \centering
        \includegraphics[width=\textwidth]{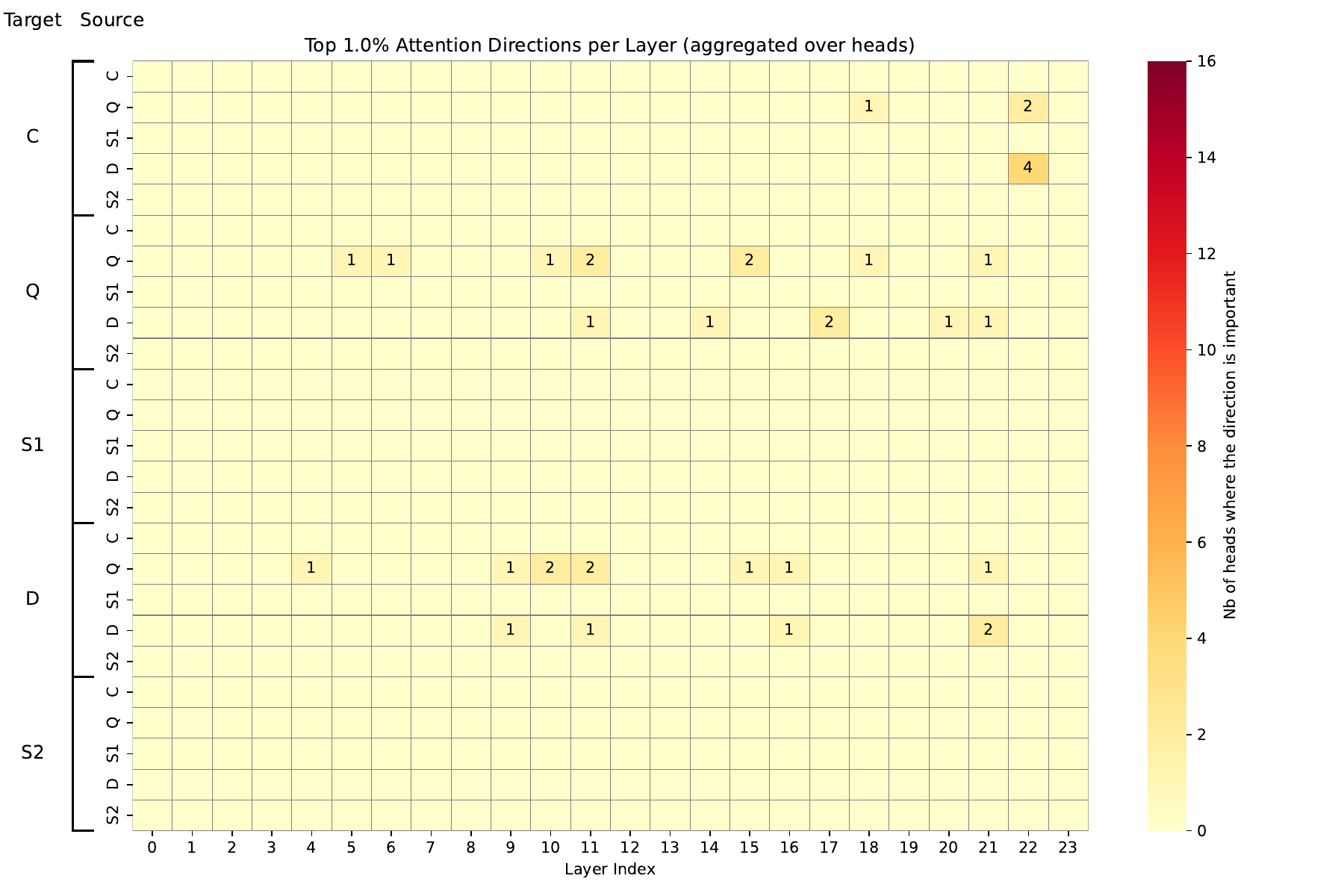}
        \caption{Ablation scheme $I_O$ for MonoBERT-large.}
    \end{subfigure}
    \hspace{0.02\textwidth} 
    \begin{subfigure}[h]{0.44\textwidth}
        \centering
        \includegraphics[width=\textwidth]{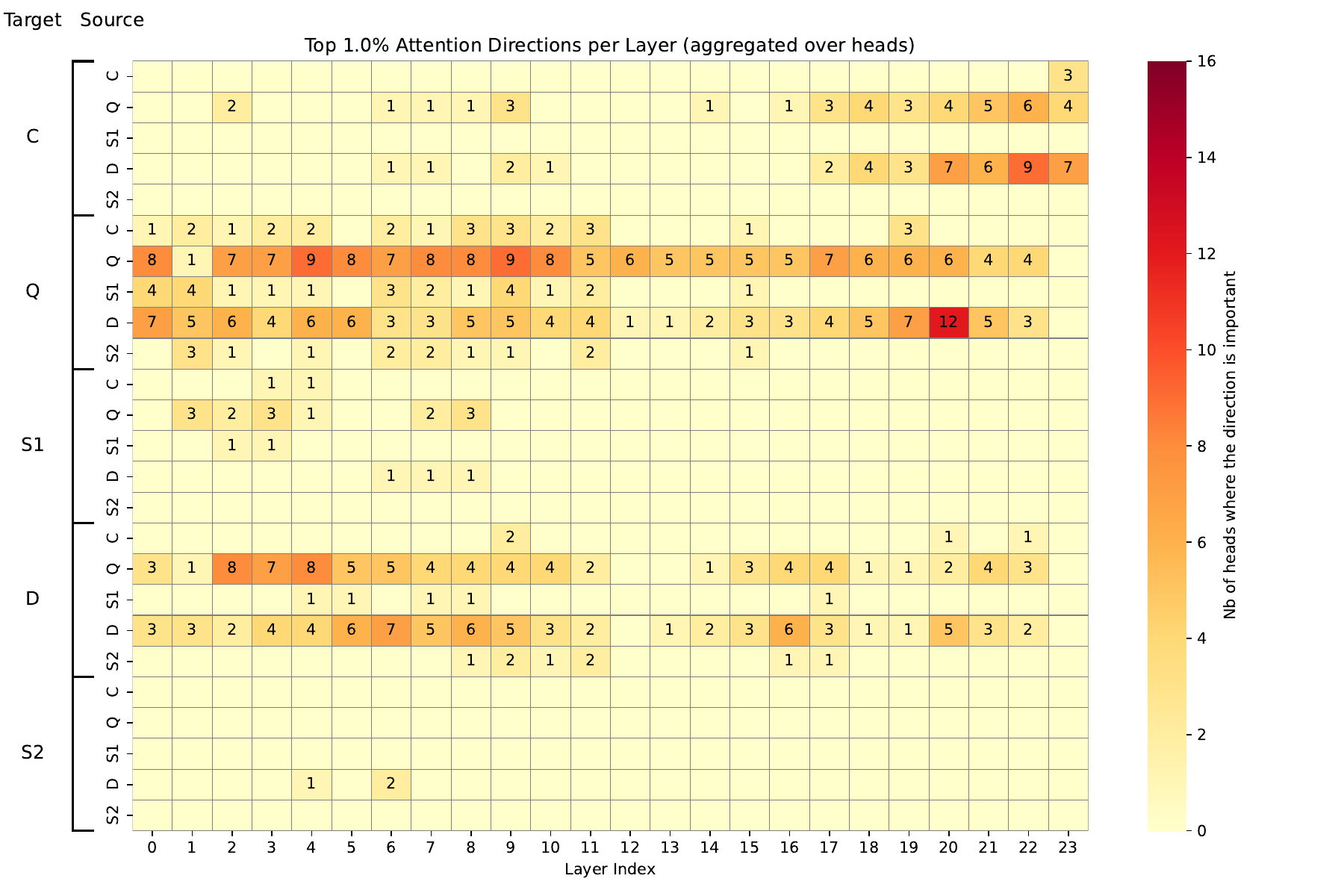}
        \caption{Ablation scheme $F_O$ for MonoBERT-large.}
    \end{subfigure} 
    
    \caption[NIG results on the information flow (attention heads)]{Distribution of the most important attention heads, relative to each input part, per layer, following the ablation schemes $F_O$ and $I_O$ at pruning percentage $t=1\%$.}
    \label{fig:nig_aggregation_attn}
\end{figure}

By combining the results from NIG and IB, we can formulate the following observations:

\result{The information flow inside MiniLM is more consistent across domains than the information flow inside MonoBERT models.}
According to NIG, the most important transfers of information are consistent in all models for ID (see Figure \ref{fig:appendix_nig_attn_msmarco} of the Appendix). Indeed, we observe a high similarity between the patterns reported for the most important neurons in general ($F_M$) and for the most important neurons shared across the four TREC-DL datasets ($I_M$). Meanwhile in OOD, Figure \ref{fig:nig_aggregation_attn} tells a different story: for MiniLM, important transfers (i.e., with a NIG score in the top $t=1\%$) in both $F_O$ and $I_O$ appear to be relatively consistent, especially between the query and the document and toward the \CLS{} token. However, it is not the case for the two MonoBERT models. This is illustrated by the transfers occurring towards the first \SEP{} token in $F_O$ as, surprisingly, $\SEPone \leftarrow Q$ appears to be important for MonoBERT-base and large, but absent from $I_O$. 
Note that the same conclusion can be drawn by comparing the ablation schemes $I_A$ and $F_A$, detailed in Figure \ref{fig:appendix_nig_attn_all} of the Appendix.
The information flow inside MiniLM appears to be more robust to domain shift than that of MonoBERT-base and large, both in terms of the attention heads that convey it (see Section \ref{section:NIG}) and in terms of directions.

\result{MiniLM integrates relevance information inside the \CLS{} for a small set of layers, and marginally refines it once it has been moved. In contrast, MonoBERT-large needs a longer span and refines the information once it has been moved inside the \CLS{} representation. Finally, MonoBERT-base exhibits a ``hybrid'' behavior, with similarities to both MonoBERT-large and MiniLM.}
In addition to the (surprising) presence of the \SEP{} tokens in important directions inside MonoBERT, the other major difference in the information flow compared to MiniLM is in the later layers and how relevance is built inside the \CLS{} token. Figures \ref{fig:nig_aggregation_attn}, \ref{fig:appendix_nig_attn_msmarco} and \ref{fig:appendix_nig_attn_all} (NIG) and \ref{fig:information_bottlenecks} (IB) show that MiniLM moves important information into the \CLS{} in very few layers (1 or 2 according to NIG depending on the scheme, 3 according to IB), and that $\CLS \leftarrow \CLS$ plays little role. In contrast, on the same figures, MonoBERT-large needs multiple layers to achieve the same objective, and, in the very last layer, NIG highlights the direction $\CLS \leftarrow \CLS$ in some domains (see $F_M$, $F_O$ and $F_A$ for MonoBERT-large). For MonoBERT-base, NIG and IB suggest a longer integration inside the \CLS{}, as in MonoBERT-large. Yet the IB masks (Figure \ref{fig:information_bottlenecks}) also preserve query and document self-attention ($Q \leftarrow Q$ and $D \leftarrow D$) more than other directions, as in MiniLM. In addition, the direction $\CLS \leftarrow \CLS$, does not seem to have an important role, just as in MiniLM. A tentative explanation, developed in later sections, is that the \CLS{} token integrates \emph{already refined} relevance information from query/document in MiniLM and marginally in MonoBERT-base, whereas relevance information is still \emph{raw signals} in MonoBERT-large.
Combined with the previous observation on the greater robustness to domain shifts of MiniLM, this suggests that the three models do not capture relevance through the same mechanism.
\begin{figure}[t]
    \centering
    \begin{subfigure}[h]{0.49\textwidth}
        \centering
        \includegraphics[width=\textwidth]{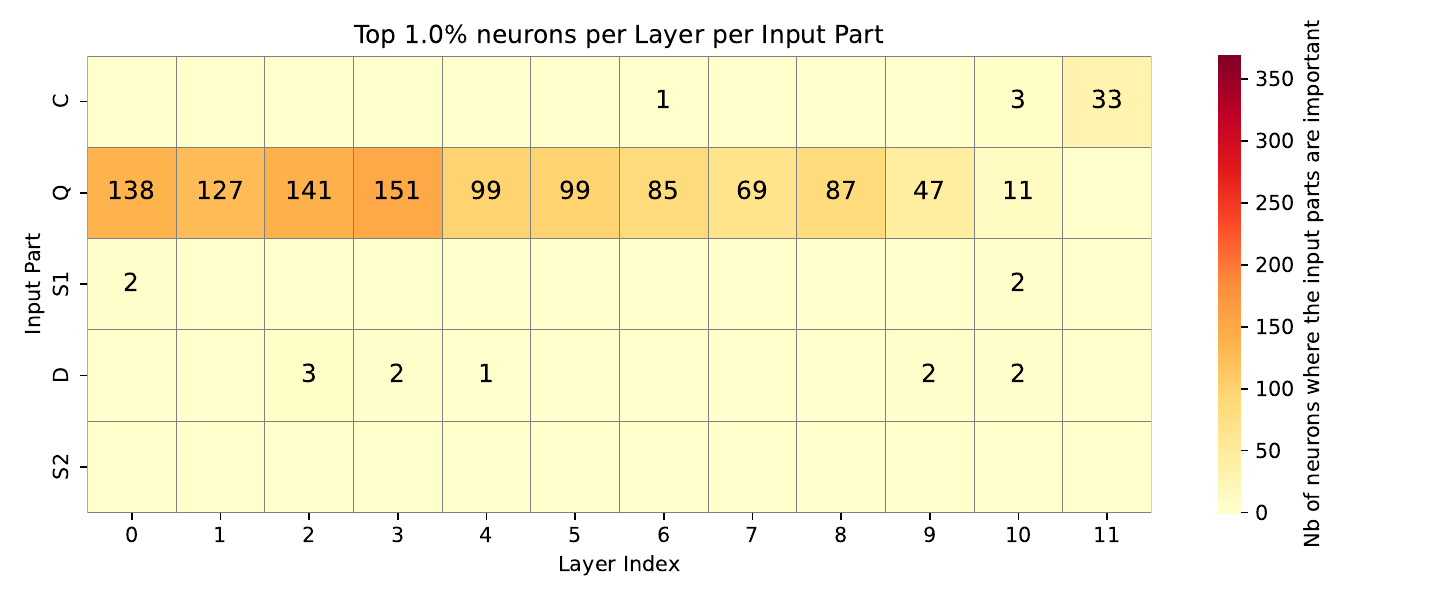}
        \caption{Ablation scheme $I_M$ for MonoBERT-base.}
    \end{subfigure}
    \begin{subfigure}[h]{0.49\textwidth}
        \centering
        \includegraphics[width=\textwidth]{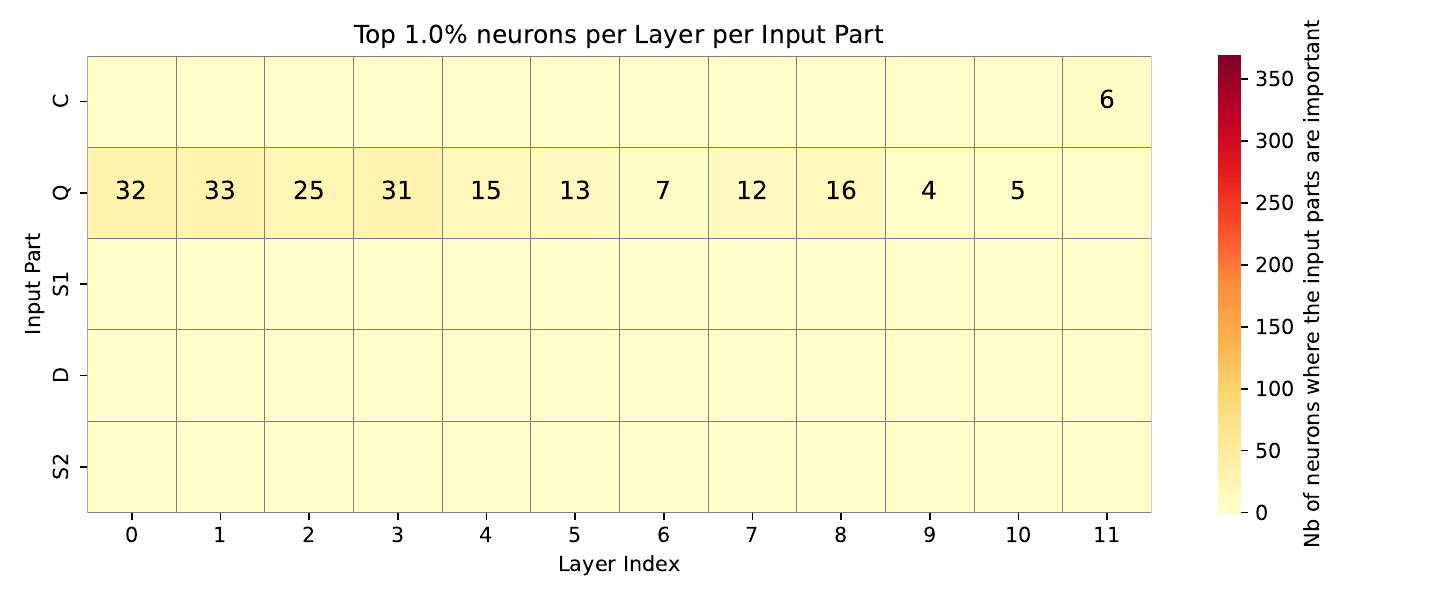}
        \caption{Ablation scheme $I_O$ for MonoBERT-base.}
    \end{subfigure}
    \begin{subfigure}[h]{\textwidth}
        \centering
        \includegraphics[width=\textwidth]{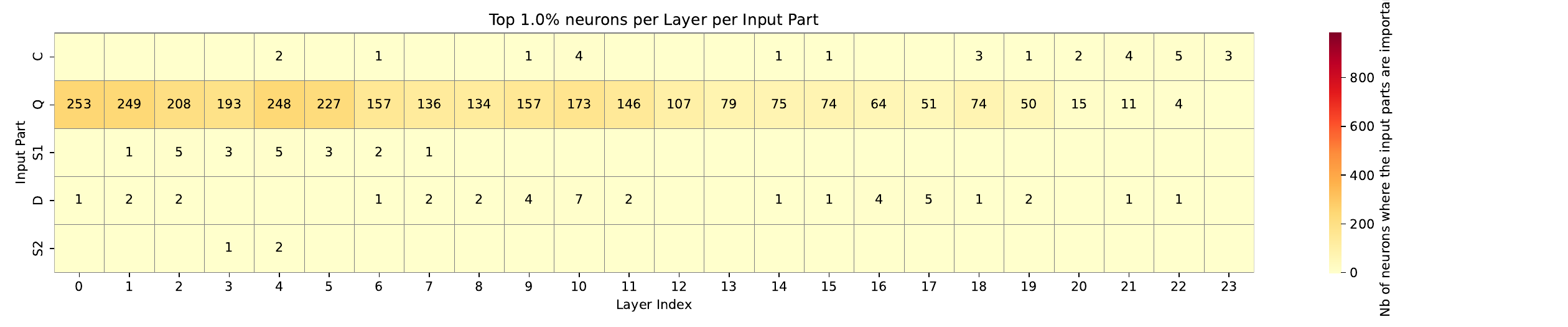}
        \caption{Ablation scheme $I_M$ for MonoBERT-large.}
    \end{subfigure}
    \begin{subfigure}[h]{\textwidth}
        \centering
        \includegraphics[width=\textwidth]{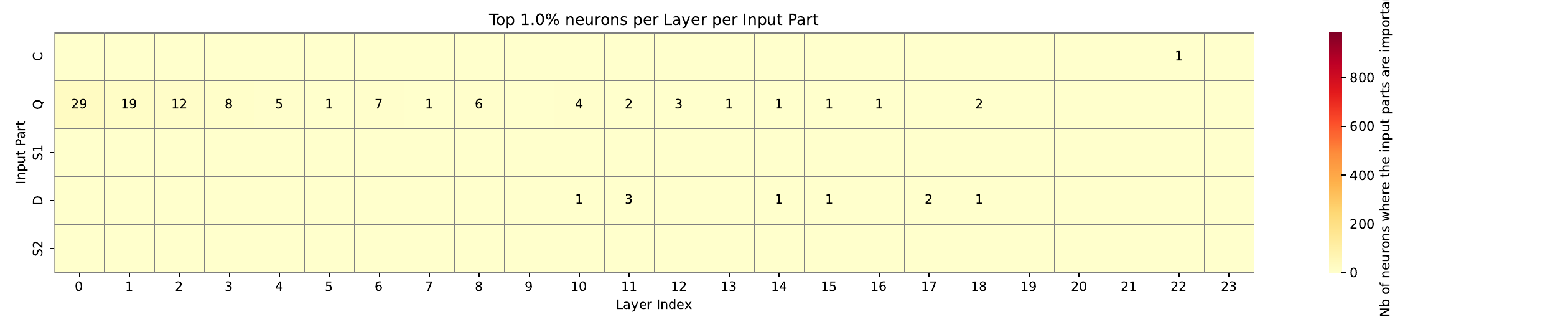}
        \caption{Ablation scheme $I_O$ for MonoBERT-large.}
    \end{subfigure}

    \caption[NIG results on the information flow (FFN)]{Distribution of the most important FFN neurons, relatively to each input part, per layer, in MonoBERT-base and MonoBERT-large following the ablation schemes obtained from the \textbf{Intersection} of the NIG scores over the TREC-DL and OOD datasets, at pruning percentage $t=1\%$.}
    \label{fig:nig_aggregation_ffn}
\end{figure}
\result{The \SEP{} tokens do not convey relevance information.}

For MiniLM, we can easily conclude on the limited importance of the \SEP{} tokens by observing Figures \ref{fig:nig_aggregation_attn} and \ref{fig:information_bottlenecks}. For both MonoBERT models, however, the \SEP{} tokens appear to be involved in a certain number of important transfers, with no obvious explanation as to why. The major difference between MiniLM and both MonoBERT models is their pre-training, but it seems unlikely that the MLM/NSP strategy used to pre-train BERT-base and BERT-large has led to a unique ability around these tokens in MonoBERT. Previous conclusions point to a higher consistency of the important parameters and of information transfers across domains in MiniLM compared to MonoBERT models. In addition, parameterizing the Information Bottleneck masks proved difficult: it was hard to maintain a comparable level of performance while maximizing sparsity, particularly for MonoBERT-large (see subfigure \ref{fig:monobert_mask_overview}), and to a lesser extent for MonoBERT-base. Together, these observations suggest that MonoBERT may be over-parameterized. With more parameters than necessary, the model may learn several ways of doing the same thing, splitting an otherwise clear signal across concurrent branches. To clarify this point, we further study NIG scores per input part of MonoBERT's FFN neurons in Figure \ref{fig:nig_aggregation_ffn}. If important information were indeed conveyed through the \SEP{} tokens, we would also expect important transformations to happen at their position in subsequent FFN layers. In contrast, we can see that the transformations applied at the \SEP{} tokens' position are not considered important compared to the transformations at the levels of the \CLS{}, query or document tokens. This suggests that the transfers towards the \SEP{} are not important for what they convey. This rather points towards their role as \emph{attention sinks}, i.e., to lower attention everywhere else and mitigate other transfers. 

This also suggests that NIG does not always identify the neurons that are truly meaningful for modeling relevance. We discuss more in-depth this implication in the Subsection \ref{subsection:design-implications}.

\result{The relevance prediction is decomposed into three consecutive stages, no matter the cross-encoders.}
%
%
In general, both approaches suggest a similar trend -- see Figures \ref{fig:nig_aggregation_attn} for NIG and 
\ref{fig:information_bottlenecks} for IB. Despite the mixed results of using IB on MonoBERT-large, perhaps because of the learning process, we note that the pattern resembles that obtained with NIG for $F_M$ and $F_O$. By analyzing the most important directions from NIG and the unmasked entries from IB, we can characterize the following steps of the relevance prediction process:
\begin{enumerate}[label=\textbf{\Roman*.}]
    \item In early to middle layers, the information flow is dominated by transfers between the queries and the documents. In particular, we identify two different types of transfers: self-contextualization, where the query (resp. document) sends information to itself, i.e., $Q \leftarrow Q$ (resp. $D \leftarrow D$), and cross-transfers, where the query (resp. document) sends information to the document (resp. query), i.e., $D \leftarrow Q$ (resp. $Q \leftarrow D$). We associate \textit{self-transfers} with the contextualization mechanism, typical of transformer-based models, and \textit{cross-transfers} with the detection of matching signals. Interestingly, we note that cross-transfers and self-transfers occur at the same layers. This suggests that cross-encoders can detect matching signals \textit{before} the contextualization of the query and document tokens, and can leverage both \emph{lexical} (almost at token-level) and \emph{semantic} matches.
    \item In later layers, important information transfers go mostly towards the query (clearly visible for MiniLM, at layers 8 and 9 in Figure \ref{fig:minilm_mask_overview}, layers 9 and 10 for MonoBERT-base in Figure \ref{fig:monobert_base_mask_overview}, and, for all models, in Figure \ref{fig:appendix_nig_attn_all}).
    \item Finally, in the last few layers, the information flows towards the \CLS{} token, mostly from the query in MiniLM (see Figure \ref{fig:minilm_mask_overview}), and from both the query and the document in both MonoBERT-base and MonoBERT-large (see Figures \ref{fig:monobert_base_mask_overview} and \ref{fig:monobert_mask_overview}). This difference further suggests that MiniLM integrates into the \CLS{} representation a signal that has already been refined in the query tokens, while the MonoBERT models further combine signals from the query and the document.
\end{enumerate}

This multi-step process echoes the process hypothesized by \citeauthor{zhan2020bertanalysis}~\cite{zhan2020bertanalysis}, but contradicts it on some key points. Their hypothesis originally describes the following steps:
\begin{enumerate}
    \item \textit{At first, query and document contextualize themselves without paying attention to each other.} Our results contradict this as we observe that cross-transfers and self-transfers occur jointly through early to middle layers and link this behavior to the two types of matching signals that cross-encoders can detect (lexical and semantic).
    \item \textit{Then interactions across query and document are important, and the model captures interaction signals to contextualize the tokens.} We confirm that behavior, but our experiments further suggest that the interactions in early-middle layers are not only used for contextualizing the tokens, but also to detect relevance signals (on this, see also the ablation results for $D+Q \nleftarrow D+Q$ in Figure \ref{fig:ablations}, Subsection \ref{subsection:ablations}).
    \item \textit{Finally, BERT relies heavily on the high-level representation to predict the relevance of the document to the query.} We confirm that behavior, while the importance of the direction $Q \leftarrow D$ in early layers (as well as the work of \citeauthor{lu-etal-2025-pathway}~\cite{lu-etal-2025-pathway} on the importance of lexical signals for cross-encoders) suggests that interactions across uncontextualized tokens also matter. We further underline that cross-encoders use query tokens to accumulate relevance signal(s) in later layers before transmitting them to the \CLS{}.
\end{enumerate}


\subsection{Analysis of the results from the ablation study}
\label{subsection:ablations}

A more direct -- but less granular -- way to study the information flow inside transformer-based models is to apply a perturbation-based method (see Section \ref{section:Related_works}). \citeauthor{zhan2020bertanalysis}~\cite{zhan2020bertanalysis} study the \emph{directions} followed by the information between input parts by conducting an ablation study. On MonoBERT-large, they show, for example, the role of the document in the contextualization of the query, or that \SEP{} tokens play a limited role in the prediction of relevance.

We extend their work by including two additional models and OOD datasets in our analysis. We follow their \textit{Attention Mask} approach, as we perturb specific parts of the attention matrices by \emph{setting the attention probability} to zero (which implies the attention might not sum to 1 anymore). In particular, we focus our study on the areas that are highlighted in Figure \ref{fig:attention_weight_segmented}. The perturbation is applied on all attention heads, from the first transformer layer to the $\ell$-th layer. We report the nDCG@10 result, on our set of datasets (both ID and OOD) and using the pool of queries and passages obtained after the \textbf{BM25 sampling} (Figure \ref{fig:dataset_construction}), of the masked models. For clarity, we report the percentage of the original performance that is obtained after ablation. For reference, Table \ref{tab:reranking_comparison} reports the performance of each model and dataset. We also report the expected performance of a random classifier as well as the worst possible one. As the pool of passages for each query in this experiment has been sampled by BM25, we note that this setup closely resembles the re-ranking task for which cross-encoders are used.

\begin{figure}[!htbp]
    \centering
    \begin{subfigure}[t]{0.48\textwidth}
        \centering
        \includegraphics[width=\textwidth]{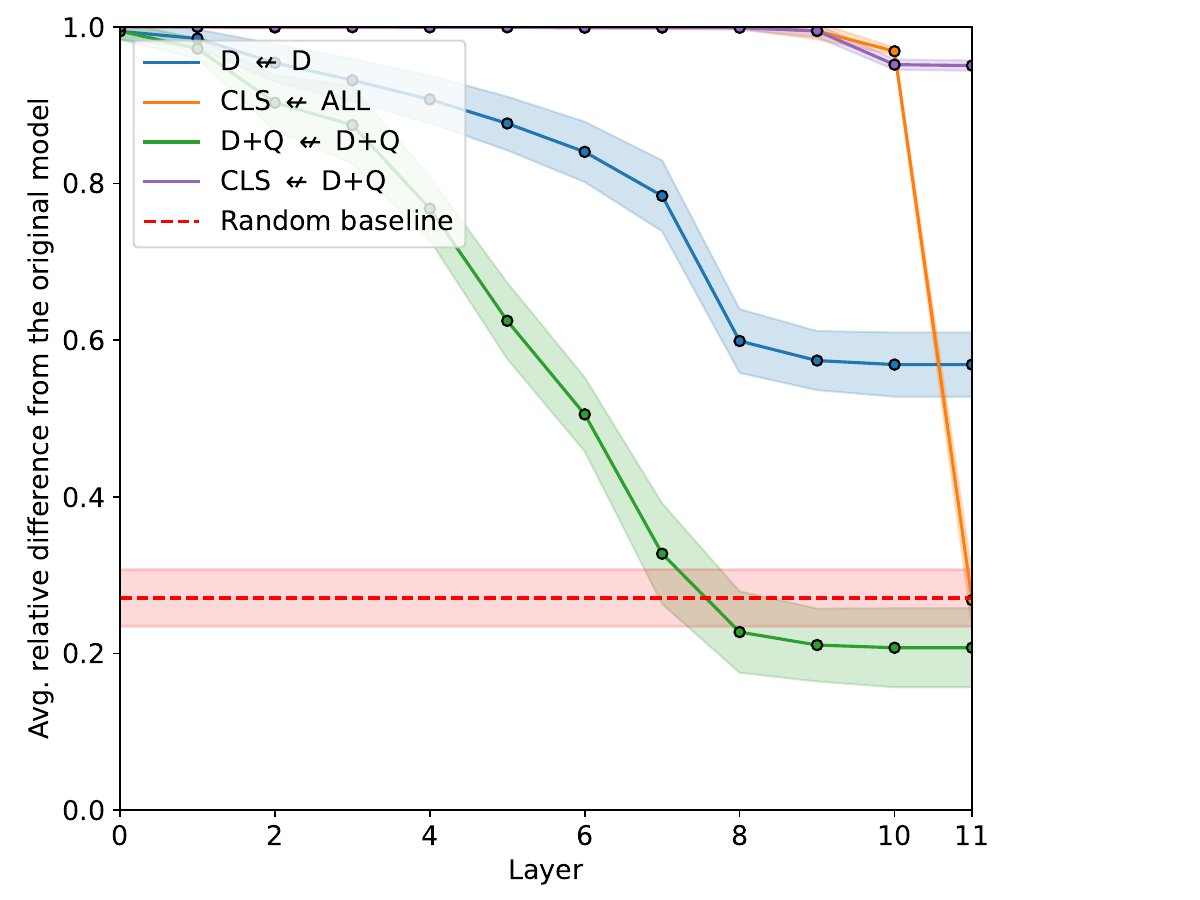}
        \caption{Ablations on the TREC-DL datasets for MiniLM.}
        \label{fig:minilm_msmarco_ablations}
    \end{subfigure}%
    \hfill%
    \begin{subfigure}[t]{0.48\textwidth}
        \centering
        \includegraphics[width=\textwidth]{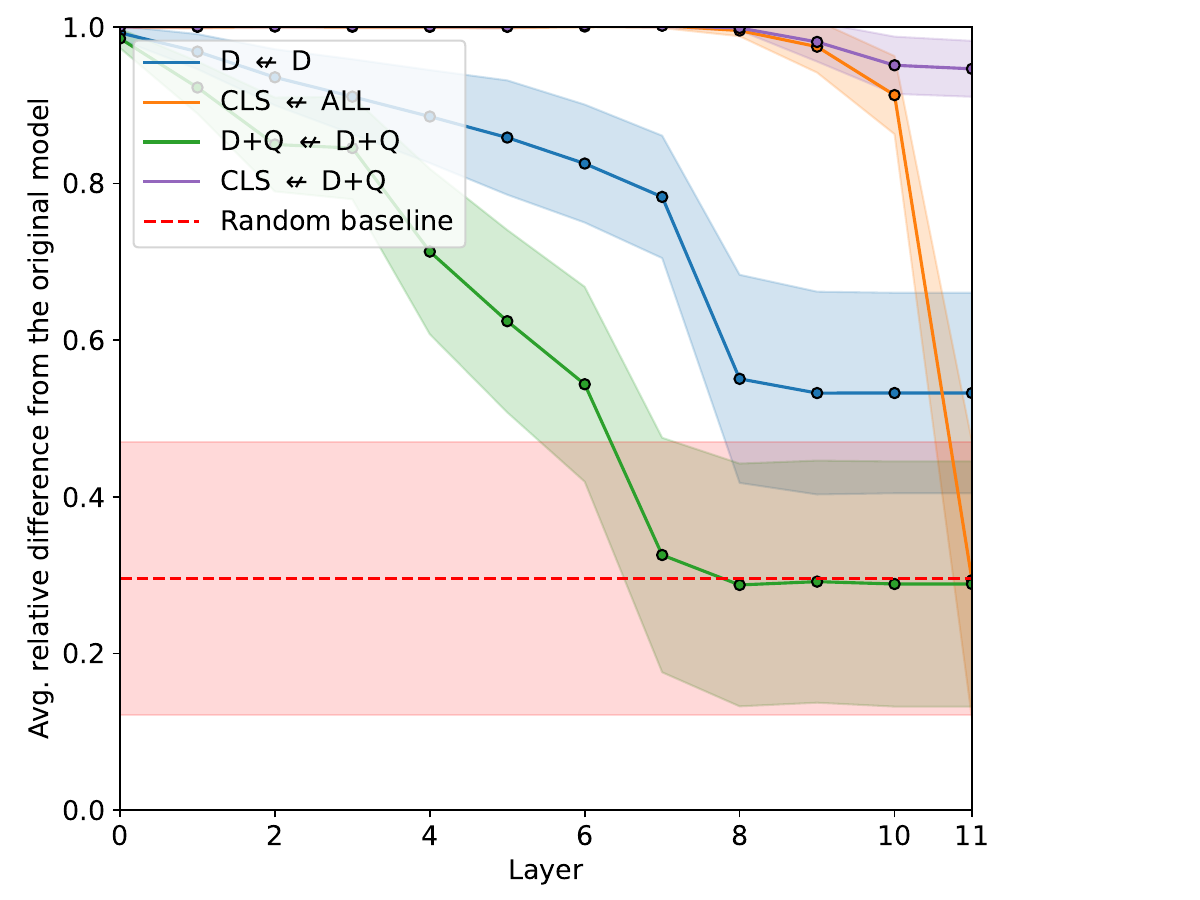}
        \caption{Ablations on the OOD datasets for MiniLM.}
        \label{fig:minilm_ood_ablations}
    \end{subfigure}
    \\ 
    \begin{subfigure}[t]{0.48\textwidth}
        \centering
        \includegraphics[width=\textwidth]{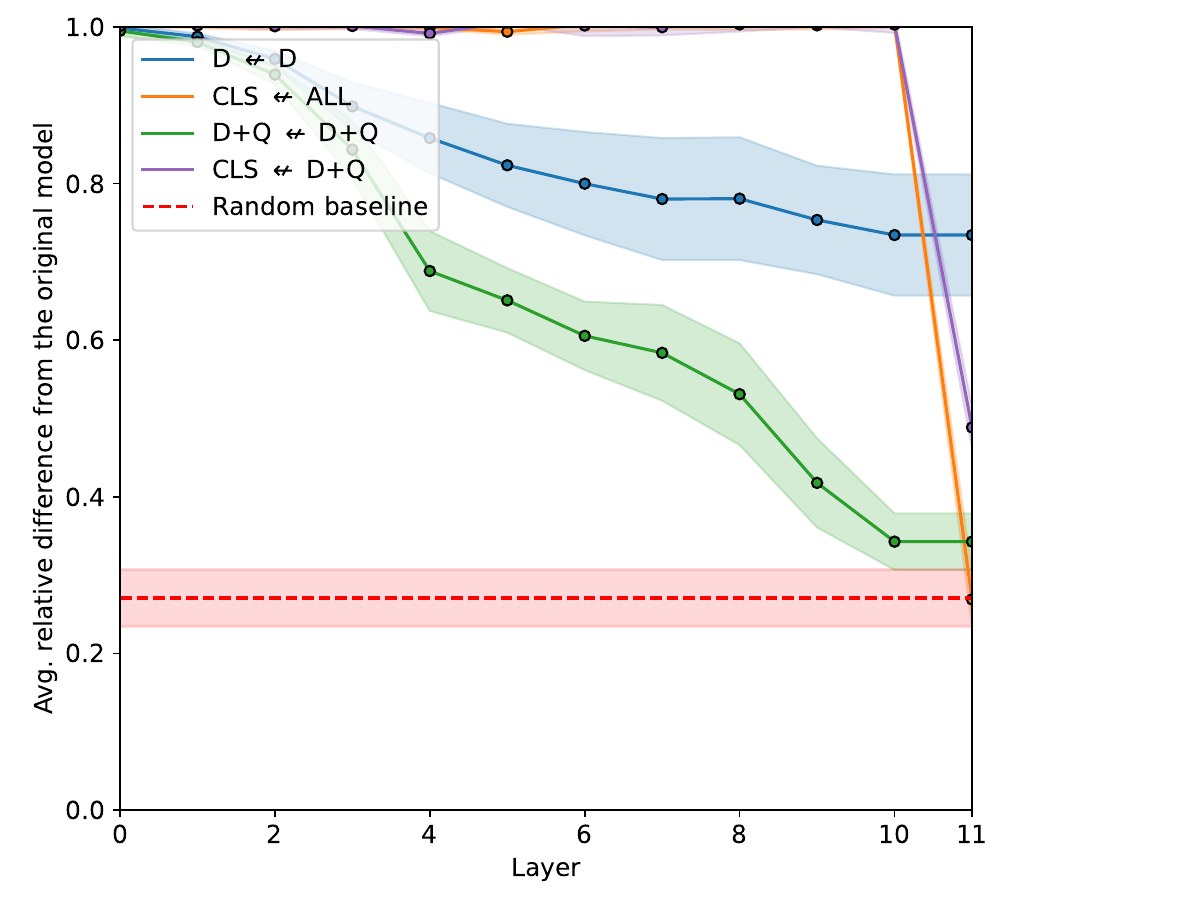}
        \caption{Ablations on the TREC-DL datasets for MonoBERT-base.}
        \label{fig:monobert_base_msmarco_ablations}
    \end{subfigure}%
    \hfill%
    \begin{subfigure}[t]{0.48\textwidth}
        \centering
        \includegraphics[width=\textwidth]{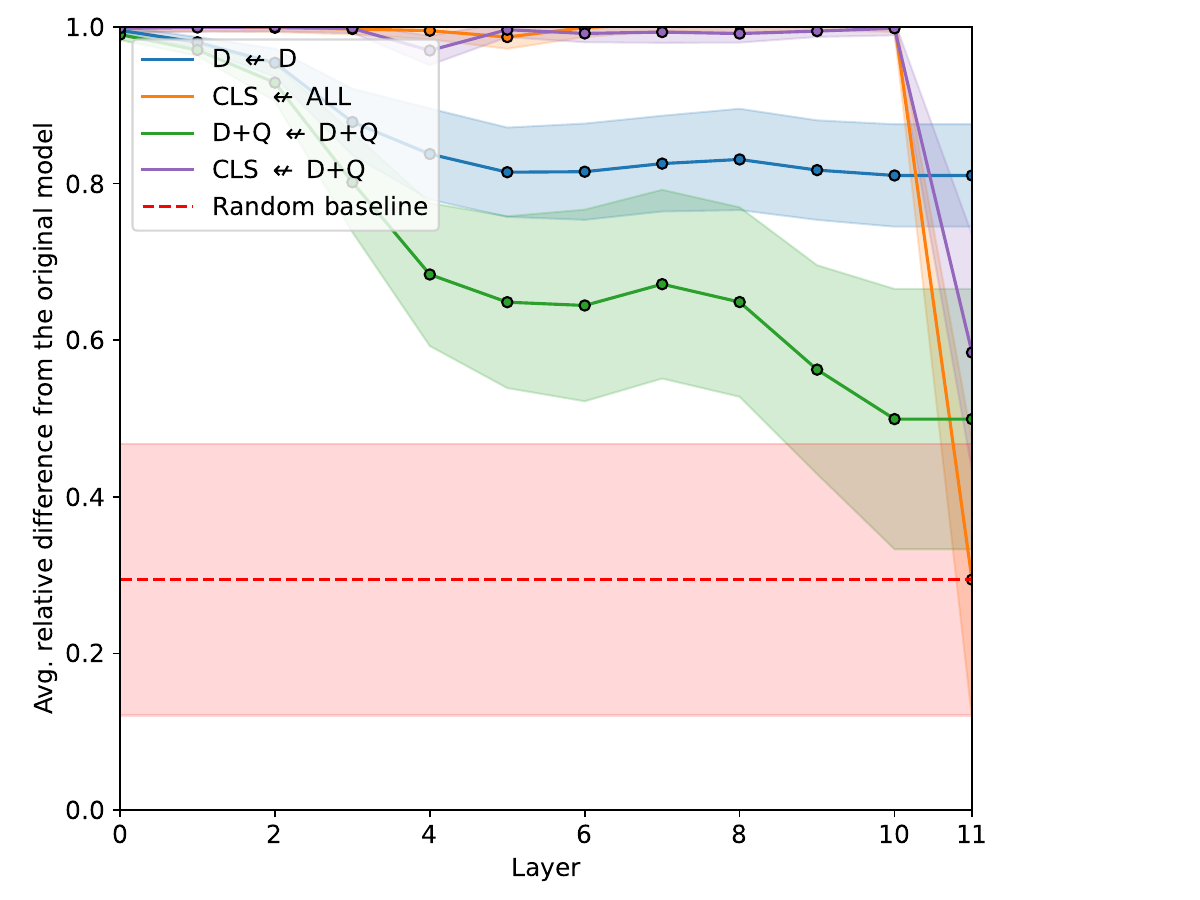}
        \caption{Ablations on the OOD datasets for MonoBERT-base.}
        \label{fig:monobert_base_ood_ablations}
    \end{subfigure}
    \\ 
    \begin{subfigure}[t]{0.48\textwidth}
        \centering
        \includegraphics[width=\textwidth]{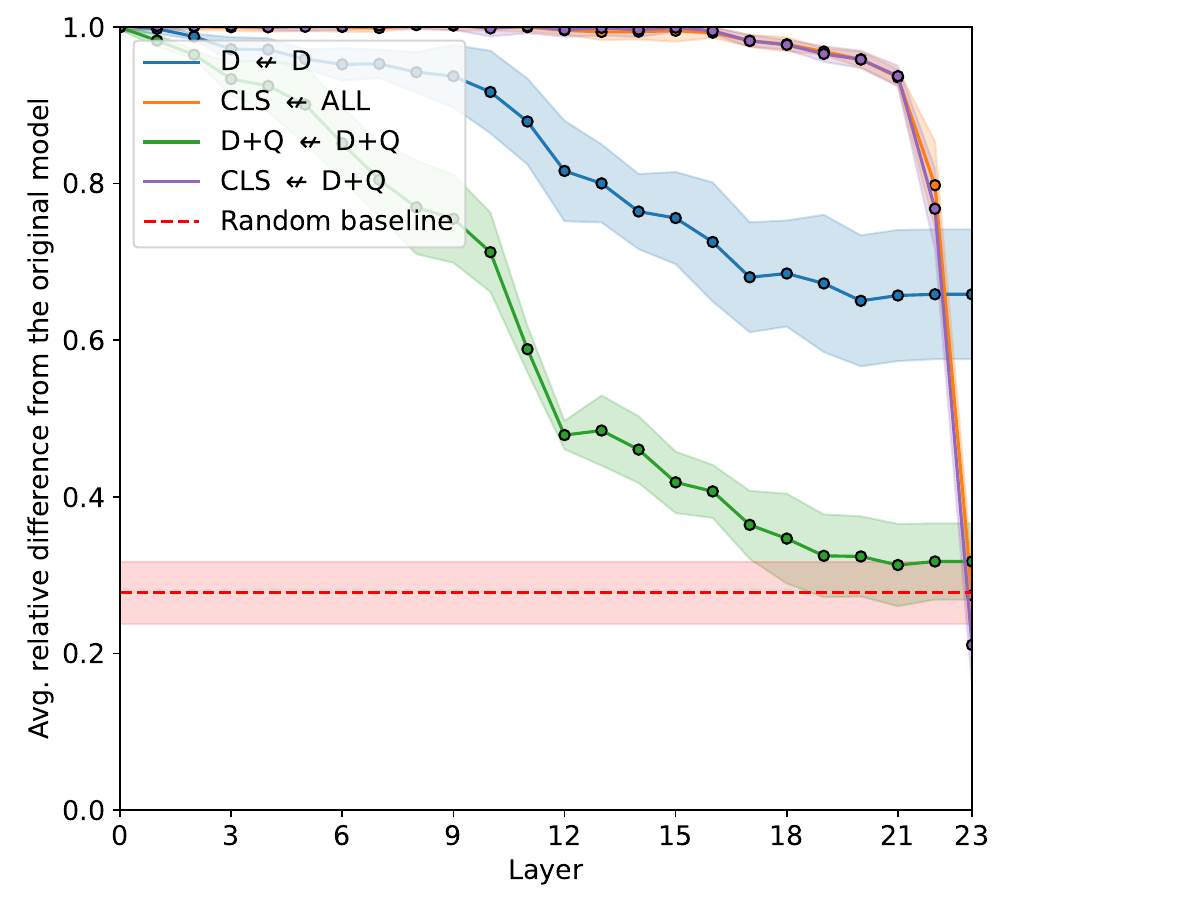}
        \caption{Ablations on the TREC-DL datasets for MonoBERT-large.}
        \label{fig:monobert_large_msmarco_ablations}
    \end{subfigure}%
    \hfill%
    \begin{subfigure}[t]{0.48\textwidth}
        \centering
        \includegraphics[width=\textwidth]{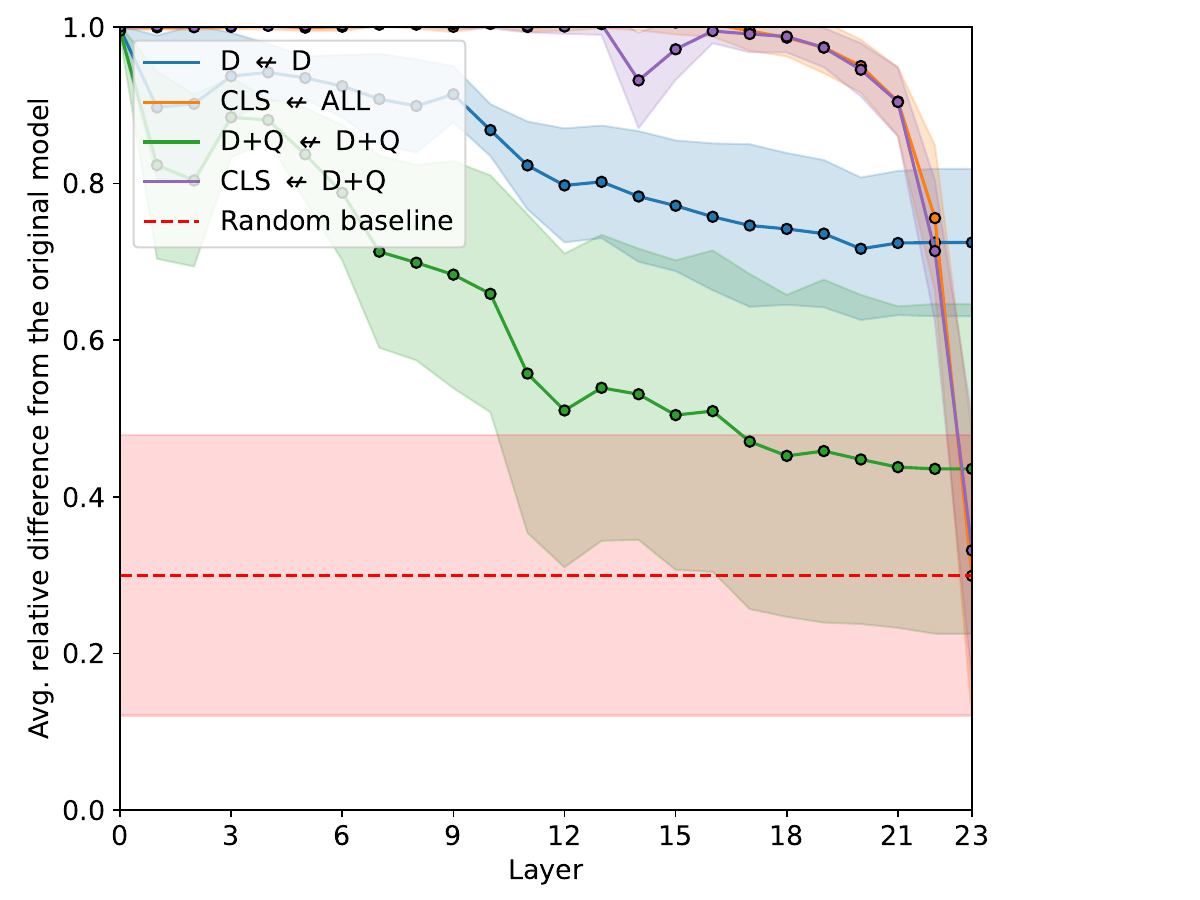}
        \caption{Ablations on the OOD datasets for MonoBERT-large.}
        \label{fig:monobert_large_ood_ablations}
    \end{subfigure}

    \caption[Ablation results on the information flow]{Average percentage of the original performance preserved after the ablations of directions inside the attention heads of the models up to the layer $\ell$. The areas around the dotted lines correspond to the variance. Averages and variances are computed over the TREC-DL datasets or the OOD datasets. The red dashed line and red area around it for the ``Random baseline'' come from Table~\ref{tab:reranking_comparison}.}
    \label{fig:ablations}

\end{figure}

\result{Cross-encoders do not rely on semantic matches only. MonoBERT-base and MonoBERT-large are able to capture matching signals even without explicit query-document interactions.}
The results of the ablations of the directions $D \nleftarrow D$, in blue, and $D+Q \nleftarrow D+Q$, in green, are reported in Figure \ref{fig:ablations} for both MonoBERT versions and MiniLM, on ID as well as OOD. Firstly, we remark that the impact of these two ablations, in both ID and OOD, is similar for the three models. Secondly, when preventing the document contextualization ($D \nleftarrow D$), we can observe a drop in performance around middle layers (7 or 9 for MiniLM and MonoBERT-large respectively, earlier for MonoBERT-base). This suggests that contextualization does not play an important role in the first layers, but gains importance as we progress through the layers. Implicitly, this means that the semantic matches should be captured at this level or after by cross-encoders. Also, we note that while masking $D \nleftarrow D$ throughout the layers of the models negatively impacts their performance, the final result is still far above the level of random rankings. Either cross-encoders do not rely only on semantic matches, or document information still reaches the model through the query tokens (recall the importance of $Q \leftarrow D$ in Subsection \ref{subsection:nig_information_flow}).

If we further prevent query contextualization and query and document from interacting with each other ($D+Q \nleftarrow D+Q$), we note that it always causes more damage to the performance than only blocking document contextualization. When this direction is masked across all the layers, the drop in performance reaches the level of random rankings in ID for all models. In OOD, both MonoBERT models partially maintain their performance, so they capture some matching signal even without explicit query-document interactions, presumably inside the \CLS{} (the previous section showed that the \SEP{} tokens are unlikely to convey it). Overall, this confirms the importance of query contextualization and query-document transfers, but also shows that MonoBERT models may have learned to use more intricate mechanisms than MiniLM to capture query-document matches.

\begin{figure}[t]
    \begin{subfigure}[t]{0.4\textwidth}
        \centering
        \includegraphics[width=\textwidth]{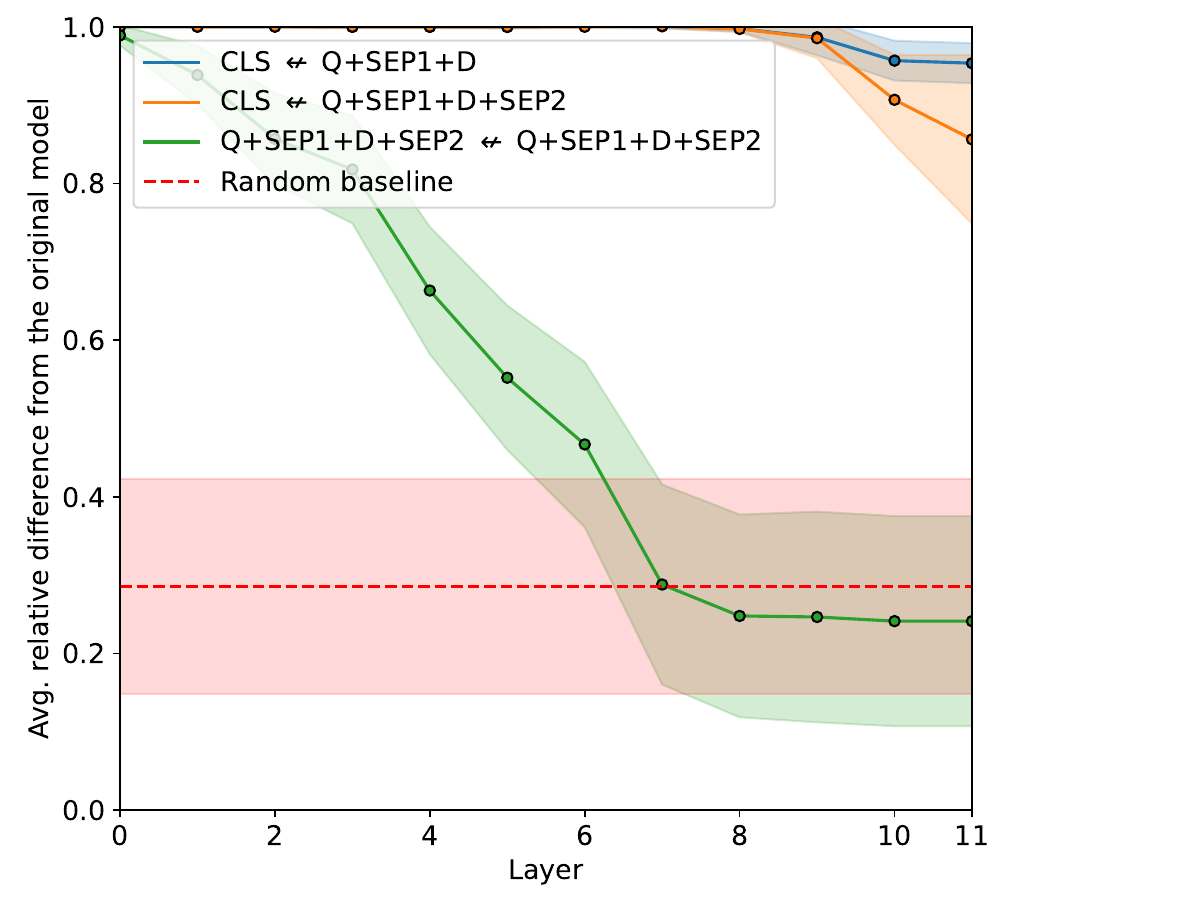}
        \caption{Ablations on all datasets for MiniLM.}
        \label{fig:complementary_ablations}
    \end{subfigure}%
    \begin{subfigure}[t]{0.6\textwidth}
        \centering
        \includegraphics[width=\textwidth]{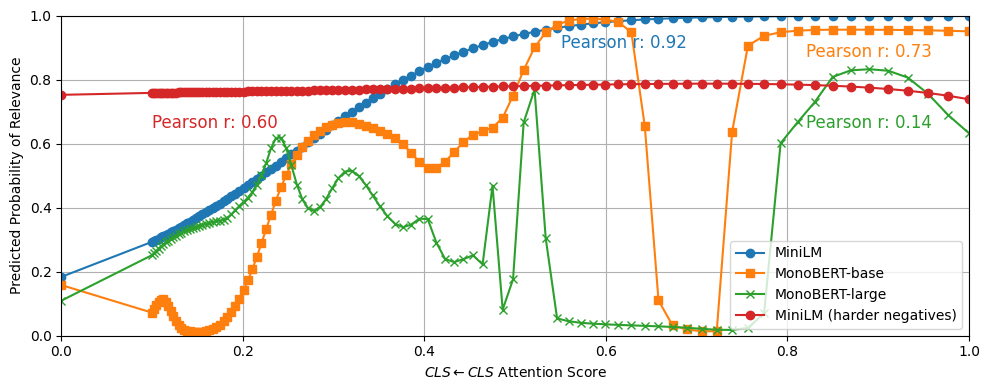}
        \caption{Evolution of the probability of relevance when masking all of the attention score matrices except the probability $\CLS \leftarrow \CLS$, and controlling its value.}
        \label{fig:controling_relevance}
    \end{subfigure}%
    \caption[Ablations to isolate the role of $\CLS \leftarrow \CLS$ transfers on the prediction of relevance]{Additional experiments to isolate the role of $\CLS \leftarrow \CLS$ transfers on the prediction of relevance, on MiniLM in particular.}
\end{figure}

\result{\SEP{} tokens carry little (MiniLM and MonoBERT-base) to no relevance (MonoBERT-large) information during the relevance prediction process inside cross-encoders.}
Looking back at the ablation curve of $D+Q \nleftarrow D+Q$, for ID, in Figure \ref{fig:ablations} for both MonoBERT cross-encoders, the drop of performance to the random classification level further reinforces the hypothesis that the relevance information does not transit through the tokens of \SEP{}. However, for MiniLM, we observe that the average percentage of preserved performance in ID and OOD drops even below the level of a random model, suggesting that the special tokens, contrary to MonoBERT, contain a signal that penalizes documents in the absence of relevance signals. To better isolate and understand the effect of the \SEP{} and \CLS{} tokens, we prevent the query and document from \emph{directly} sending information to \CLS{} ($\CLS \nleftarrow D+Q$, in purple). We observe that the results in that case differ significantly between the models. Indeed, we observe that while blocking these transfers of information across the whole model eventually causes the performance of MonoBERT-large to drop even below the level of random rankings (so \SEP{} and \CLS{} carry negative signals), this is not the case for MiniLM and MonoBERT-base.

For MiniLM, this ablation highlights that the model might have learned to use \SEP{} and/or \CLS{} to store relevance signals extracted from the query and document interactions. Although it echoes the presence of the first token \SEP{} as an important source of information in the ablation schemes $I_M$ and $I_O$ for MiniLM in Figure \ref{fig:nig_aggregation_attn}, this also calls for more research to understand their roles precisely.

Meanwhile, for MonoBERT-base, the analysis is more complex, as the performance of the model still drops significantly when the signal is pruned across all the layers. Yet, unlike its larger counterpart, the remaining transfers from the \SEP{} tokens to the \CLS{} are able to maintain the performance above the level of a random predictor. In the end, this suggests that MonoBERT-base may have learned to use the \SEP{} tokens to store some relevance information, but $\CLS \leftarrow \SEPone+\SEPtwo$ transfers are usually overridden by what is flowing from the query and the documents.

\result{MiniLM may base its prediction on signals of non-relevance, instead of relevance. When fine-tuned with harder negatives, this behavior may even be exacerbated.}
Following up on the role of the \SEP{} tokens inside MiniLM, we further prevent the \SEP{} tokens from sending information to the \CLS{} token in Figure \ref{fig:complementary_ablations} ($\CLS \nleftarrow Q+\SEPone+D$, in blue, and $\CLS \nleftarrow Q+\SEPone+D+\SEPtwo$, in orange). Surprisingly, even if the \CLS{} can only receive information from itself, across the entire model's layers, the drop in performance remains limited for MiniLM (around 80\% of the original performance is retained across all the datasets). This points to an information leak in our ablation setup (discussed in Section \ref{subsection:design-implications}). It also shows that MiniLM's predictions can be influenced simply by how much attention the \CLS{} pays to itself. This observation leads to a series of follow-up questions: is it possible to control the predicted probability of relevance of MiniLM by controlling the probability of attention from the \CLS{} to itself? To what extent? What about MonoBERT?

To answer these, we set up a small experiment in which we mask all the attention probability matrices of the cross-encoders, except the probability corresponding to the direction $\CLS \leftarrow \CLS$. To understand its impact, we observe the evolution of the predicted probability of relevance when progressively increasing this score, from 0.0 to 1.0, on a query and an unrelated document from MS MARCO. Figure \ref{fig:controling_relevance} shows that even though all the models originally predict a score close to 0, masking the attention score matrices can heavily impact the prediction. While there does not seem to be a correlation between the attention score of $\CLS \leftarrow \CLS$ and the probability of relevance predicted by both MonoBERT models (Pearson R correlation of $0.14$ for the large version and $0.73$ for the base, but non-monotonic curve), for MiniLM the correlation is much stronger ($0.92$). Note that in a real-world setup, an attention score of 0 for the direction $\CLS \leftarrow \CLS$ means the model is capturing information from the rest of the input inside \CLS{}, while a score of 1.0 means the model is focusing on what is already contained inside the token's representation. Here, we observe that the stronger MiniLM focuses the attention from the \CLS{} onto itself (and ignores the rest of the tokens), the more the probability of relevance increases. This experiment suggests that MiniLM does not scan the query and document tokens for ``relevance signals'' \emph{but rather for ``non-relevance signals''}. In their absence, its confidence in the relevance of the document to the query increases.

In Figure \ref{fig:controling_relevance}, we also report the same behavior but on a MiniLM trained using harder negatives, i.e., negatives sampled by ColBERT-v2 \cite{santhanam2022colbertv2}. Interestingly, this model exhibits a third type of behavior: when only $\CLS \leftarrow \CLS$ is allowed, no matter the weight of the attention, the prediction by default seems to be ``relevant''. This may indicate that when the model is fine-tuned with harder negatives, it relies even more on the detection of non-relevance signals for its prediction.

\result{\CLS{} only absorbs relevance signals in the very last layers of a cross-encoder.}
Returning to Figure \ref{fig:ablations}, the last ablation we tested is to prevent each token from sending information to \CLS{} ($\CLS \nleftarrow ALL$, in orange). We can see that when applied from the first layer up to the last layer, we logically end up with a random ranking performance. If we observe where the performance starts to drop, we note that the fall does not occur before the last few layers, especially in the case of MiniLM and MonoBERT-base. This suggests that the \CLS{} token only begins to integrate relevance signals from the query and the document tokens at the very end of the process. What is transmitted to the \CLS{} and the transformations they undergo through the last FFN layers are unclear at this stage and require more investigation.

With this, we now have a more complete understanding of the information flow across input parts inside different cross-encoders. Furthermore, the multi-stage process of relevance prediction in cross-encoders that we described suggests that they may be able to leverage both \emph{lexical} and \emph{semantic} interaction signals. Indeed, we observed that important interactions between a query and a document are captured by the models both before and after the contextualization of query and document tokens. After answering \textbf{RQ1} and \textbf{RQ2}, we now aim at characterizing the nature and origin of the signals that contribute to the prediction of relevance inside cross-encoders.

\section{RQ3: Characterizing the nature of the signals contributing to the prediction of relevance}
\label{section:understand_attention}

Among the attention heads we have found to play a crucial role in the process of modeling and predicting relevance in cross-encoder models, we can \textit{a priori} guess what the role of most of them is. Indeed, as highlighted by prior studies \cite{lu-etal-2025-pathway,zhan2020bertanalysis} as well as our findings in the previous Section \ref{section:information_flow}, attention heads in cross-encoder models play a key role in 1) contextualizing the query and/or document tokens (in particular for early to middle layers), 2) transferring information between the query tokens and document tokens (in particular for attention heads in early to middle/late layers) and 3) transferring information from the entire input to the \CLS{} (late layers). Roles 1) and 3) are as important as 2), but they are better understood. Role 2), usually seen as the \emph{detection} of query-document matches, remains opaque, both in the information it surfaces and in its mechanism.

In their work, \citeauthor{lu-etal-2025-pathway}~\cite{lu-etal-2025-pathway} use causal tracing \cite{causal_mediation_analysis} to investigate key components inside MiniLM and, in particular, the role of some attention heads in detecting query-document matches. Their approach reveals the causal chain leading to the relevance prediction, but it is hard to scale to multiple datasets, which limits conclusions about generalization (see \textbf{RQ5} and \textbf{RQ6}). We argue that more direct approaches can lead to comparable results while scaling more easily to multiple datasets. In particular, we believe that studying the attention scores between query and document tokens (red areas in Figure \ref{fig:attention_weight_segmented}), while considering the relevance of the document, can help us identify heads responsible for detecting matches (i.e., when most of the attention weight is focused on a single token) and to characterize their nature (lexical or semantic). To ensure each level of relevance is represented throughout our datasets, we employ the \emph{Diverse sampling} strategy (see Figure \ref{fig:dataset_construction}).

\subsection{Context and notations}

In this section, we aim to improve the understanding of the attention heads focusing on transferring information from the document (resp. query) to the query (resp. document), i.e., focusing on cross-attention (as opposed to the self-attention of an input part to itself, like in query/document contextualization).

To do so, we analyze the attention scores from query tokens to document tokens, named ``Q/D-attention'' (and vice-versa, i.e., ``D/Q-attention'') in the attention score matrices (see the two red areas on Figure \ref{fig:attention_weight_segmented}) and contrast the amount of cross-attention when a document is relevant versus when it is not relevant for the same query. We intend to differentiate the attention heads that have learned to detect \emph{relevant/important} matching signals from the attention heads that may be attending heavily to unimportant tokens, such as stop words in the context of the no-op operation \cite{clark-etal-2019-bert}.

Formally, for each attention head $h$ at layer $\ell$, relevance level $r$ and a query token $Q_i$, we compute the maximum attention probability over document tokens $D_j$ as $\max_j \patt(Q_i \leftarrow D_j)$.
We then compute the mean for a given query and denote this value $\maxatt{Q \leftarrow D/r}$. We define $\maxatt{D \leftarrow Q/r}$ for the attention of document tokens on query ones similarly.
To get as much contrast as possible, we consider the 3 following levels of relevance: ``relevant'' ($r=1$), ``hard negative'' ($r=0$), and ``easy negative'' ($r=-1$), 
and compare the differences, for each attention head, between $r=1$ and $r=-1$ as well as between $r=1$ and $r=0$, namely: 
\begin{formulas}
$\maxatt{Q\leftarrow D/1} - \maxatt{Q\leftarrow D/-1}$ 
and
$\maxatt{Q\leftarrow D/1} - \maxatt{Q\leftarrow D/0}$.
\end{formulas}

Other statistics than the maximum of attention can be computed from the same attention distributions. In Section~\ref{subsection:matching_signals}, we also use the Shannon entropy \cite{shannon_entropy} of the attention of a query token $Q_i$ over the document tokens,
$$E^{h,\ell}(Q_i \leftarrow D) = -\sum_{j} \patt(Q_i \leftarrow D_j) \log \patt(Q_i \leftarrow D_j),$$
summed over the query tokens and averaged over the queries of a dataset for a given relevance level $r$, which we denote $E^{h,\ell}_{Q \leftarrow D/r}$ ($E^{h,\ell}_{D \leftarrow Q/r}$ is defined similarly). Note that we thus average entropy values, and not attention distributions. A low entropy indicates that the head concentrates its attention on a few document tokens, whereas a high entropy indicates that its attention is spread over the document. As for the maximum of attention, we compare relevance levels through the differences $E^{h,\ell}_{Q\leftarrow D/1} - E^{h,\ell}_{Q\leftarrow D/-1}$ and $E^{h,\ell}_{Q\leftarrow D/1} - E^{h,\ell}_{Q\leftarrow D/0}$.

\subsection{Analyzing the attention process to find the heads responsible for detecting matches}
\label{subsection:matching_signals}

\begin{figure}[!htbp]
    \centering
    \begin{subfigure}[b]{0.47\textwidth}
        \centering
        \includegraphics[width=1\textwidth]{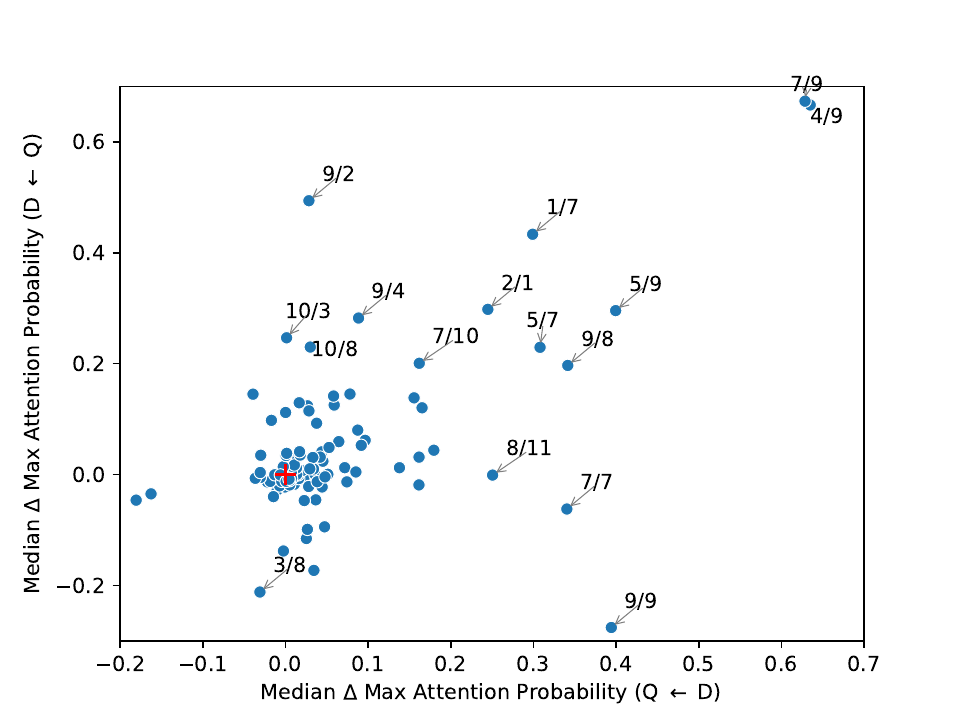}
        \caption{Results over the 4 TREC-DL datasets for MiniLM.}
        \label{fig:minilm_msm_attention}
    \end{subfigure}
    \hfill
    \begin{subfigure}[b]{0.47\textwidth}
        \centering
        \includegraphics[width=1\textwidth]{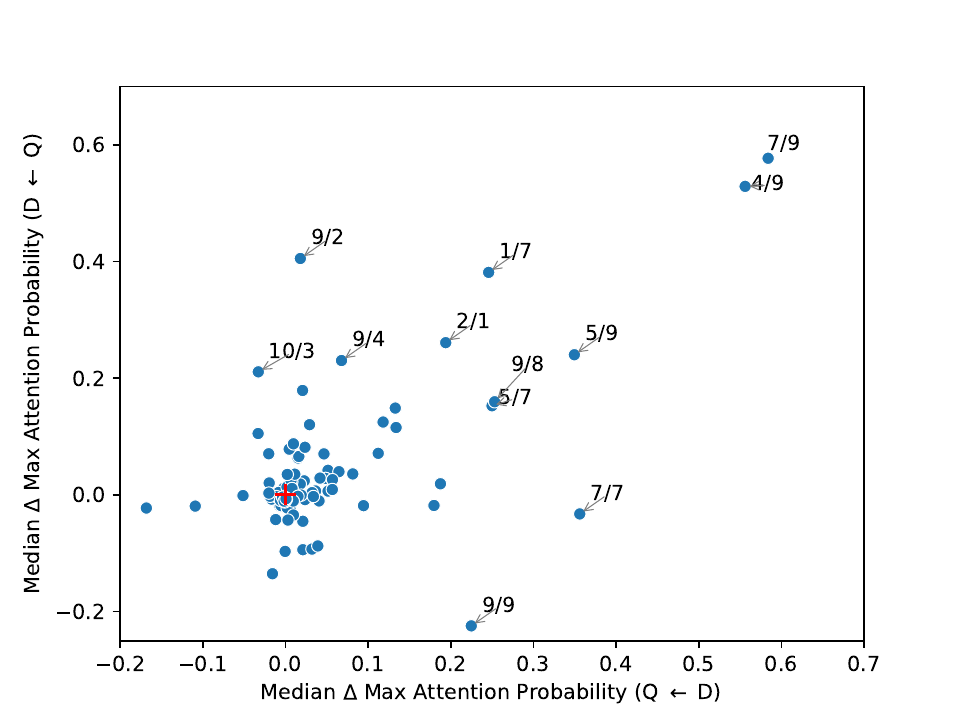}
        \caption{Results over the OOD datasets for MiniLM.}
        \label{fig:minilm_ood_attention}
    \end{subfigure}
    \centering
    \begin{subfigure}[b]{0.47\textwidth}
        \centering
        \includegraphics[width=1\textwidth]{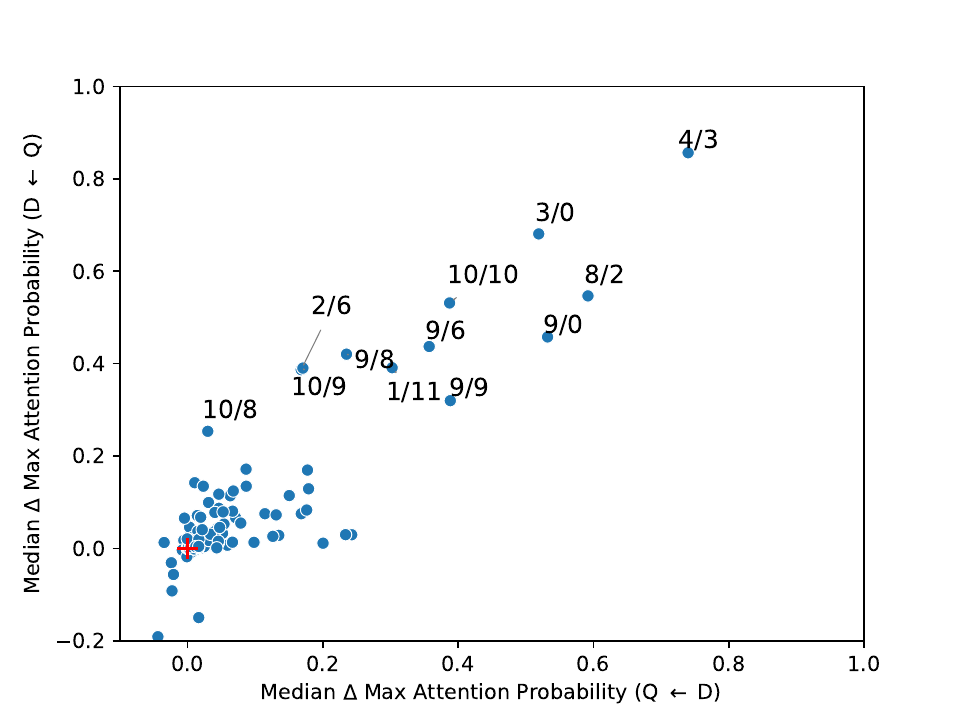}
        \caption{Results over the 4 TREC-DL datasets for MonoBERT-base.}
        \label{fig:monobert_base_msm_attention}
    \end{subfigure}
    \hfill
    \begin{subfigure}[b]{0.47\textwidth}
        \centering
        \includegraphics[width=1\textwidth]{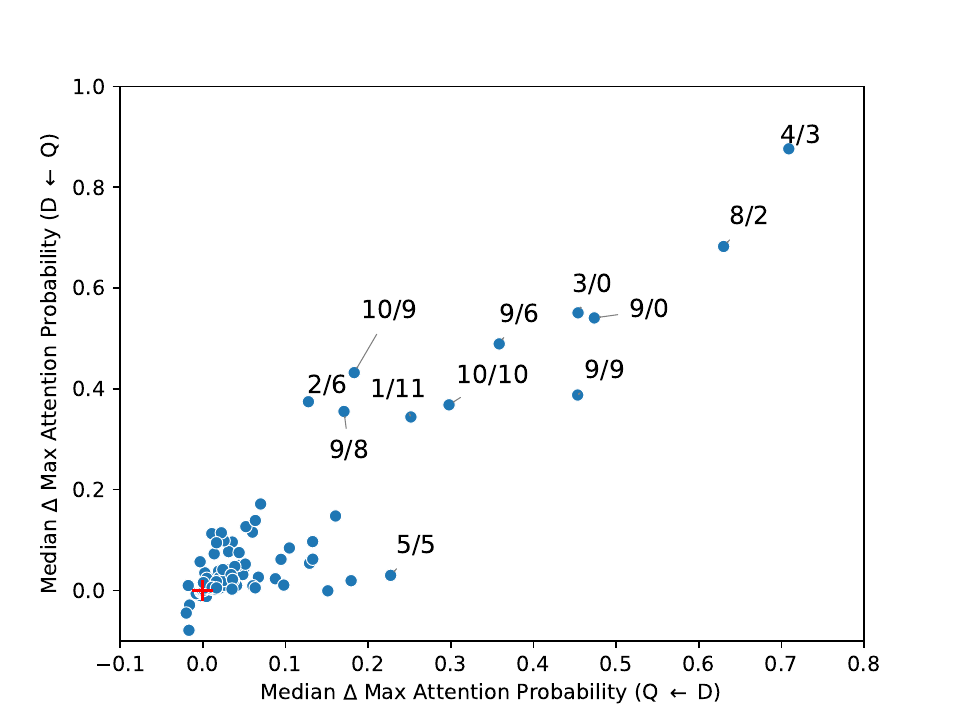}
        \caption{Results over the OOD datasets for MonoBERT-base.}
        \label{fig:monobert_base_ood_attention}
    \end{subfigure}
    
    \centering
    \begin{subfigure}[b]{0.47\textwidth}
        \centering
        \includegraphics[width=1\textwidth]{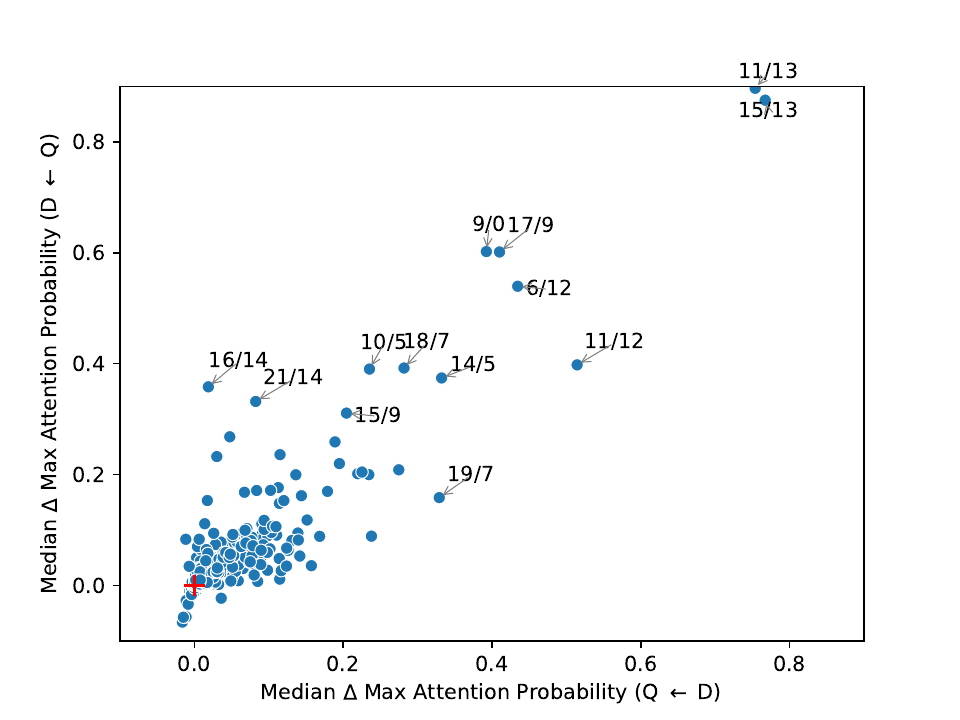}
        \caption{Results over the 4 TREC-DL datasets for MonoBERT-large.}
        \label{fig:monobert_msm_attention}
    \end{subfigure}
    \hfill
    \begin{subfigure}[b]{0.47\textwidth}
        \centering
        \includegraphics[width=1\textwidth]{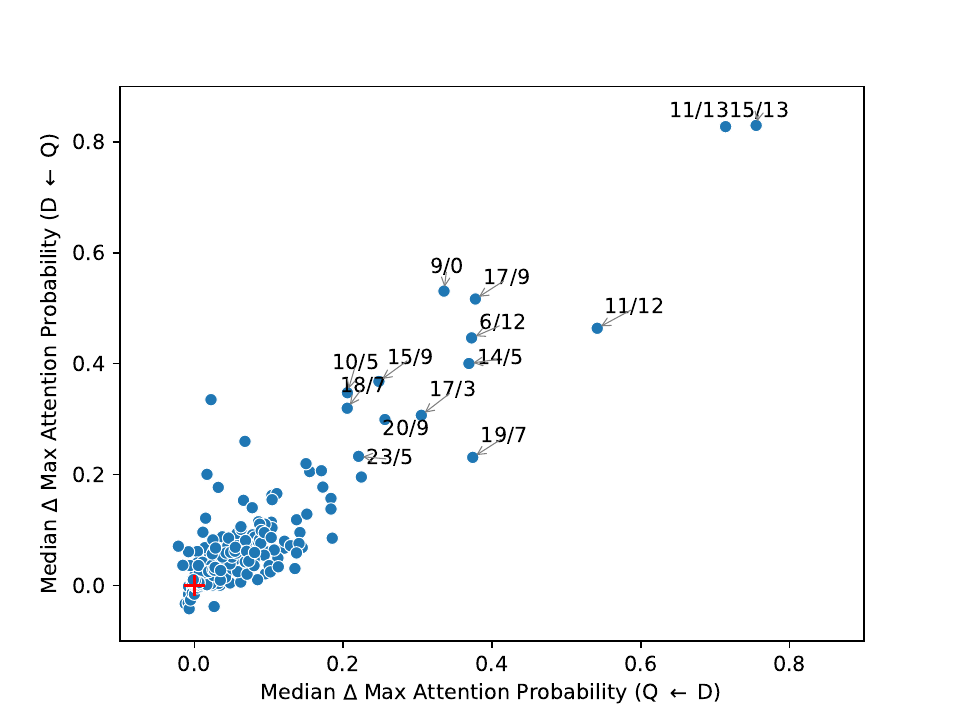}
        \caption{Results over the OOD datasets for MonoBERT-large.}
        \label{fig:monobert_ood_attention}
    \end{subfigure}
    \caption[Contrast of the attention patterns between relevant and easy negatives]{Relevant vs.\ easy negatives -- Median of the differences in the attention weights $\maxatt{Q \leftarrow D/1} - \maxatt{Q\leftarrow D/-1}$ and $\maxatt{D\leftarrow Q/1} - \maxatt{D\leftarrow Q/-1}$ for each attention head of all models. An attention head $h$ at a layer $\ell$ is denoted as $\ell/h$.}
    \label{fig:comparison_attention_weak}
\end{figure}

\begin{figure}[!htbp]
    \centering
    \begin{subfigure}[b]{0.47\textwidth}
        \centering
        \includegraphics[width=1\textwidth]{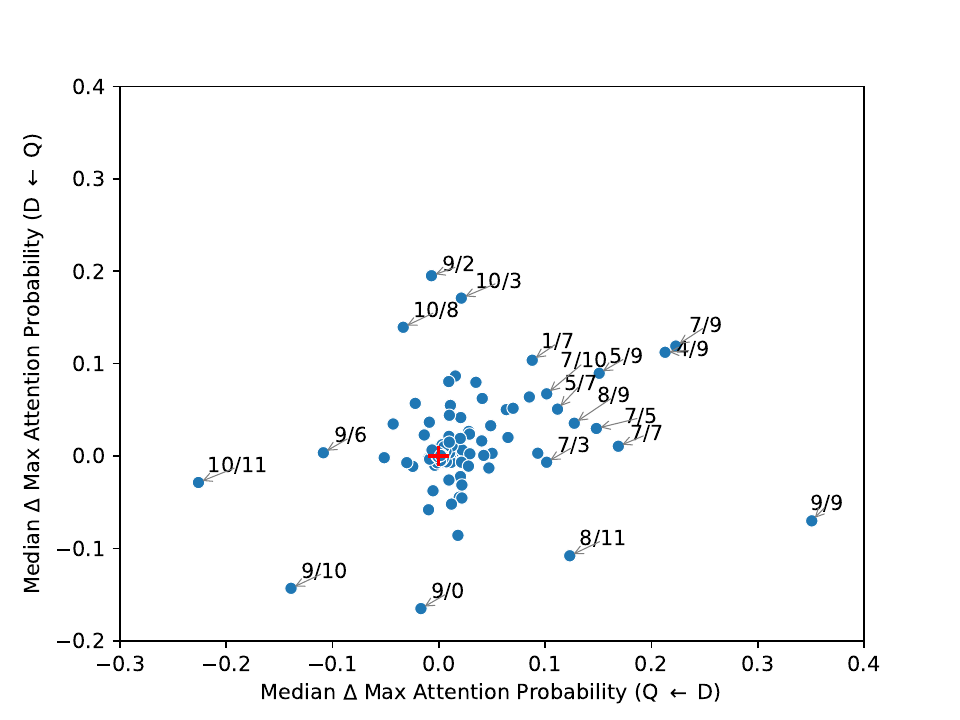}
        \caption{Results over the 4 TREC-DL datasets for MiniLM.}
        \label{fig:minilm_msm_attention_hard}
    \end{subfigure}
    \hfill
    \begin{subfigure}[b]{0.47\textwidth}
        \centering
        \includegraphics[width=1\textwidth]{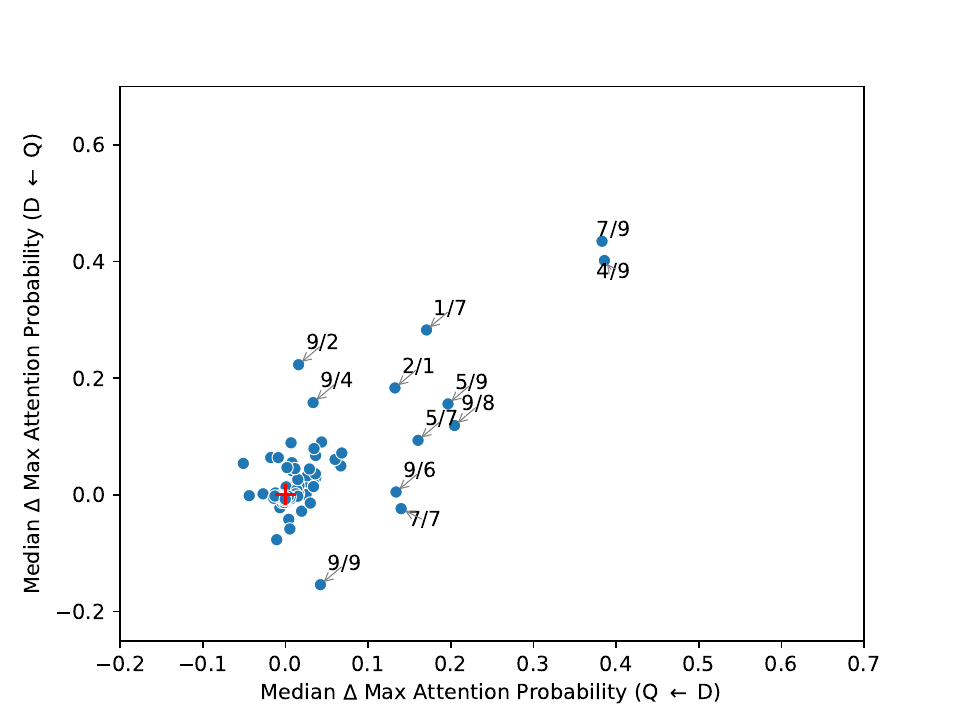}
        \caption{Results over the OOD datasets for MiniLM.}
        \label{fig:minilm_ood_attention_hard}
    \end{subfigure}
    \centering
    \begin{subfigure}[b]{0.47\textwidth}
        \centering
        \includegraphics[width=1\textwidth]{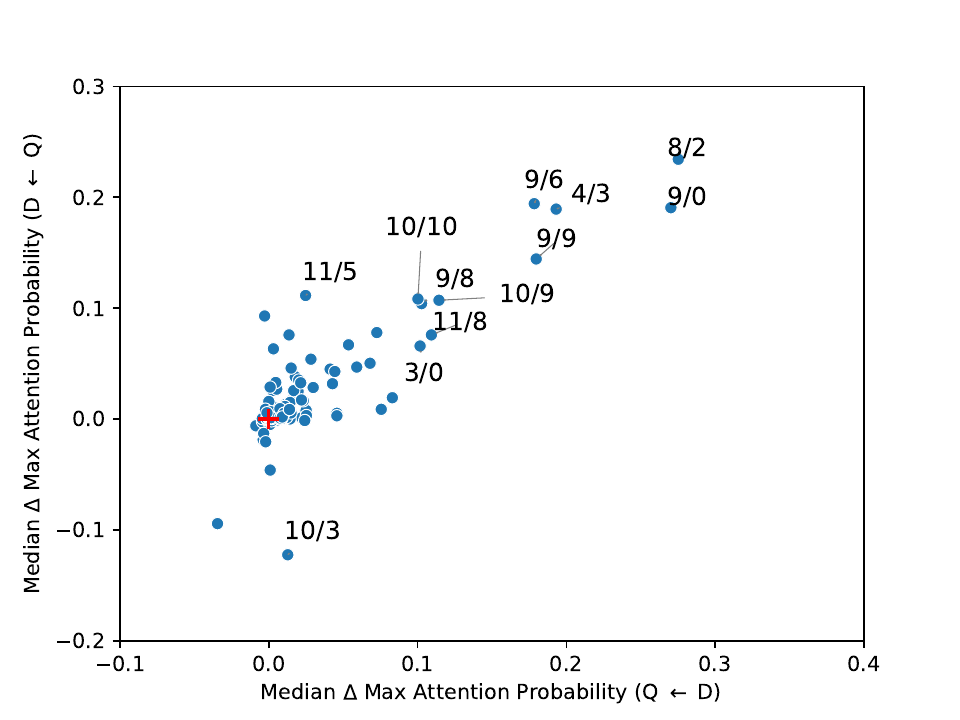}
        \caption{Results over the 4 TREC-DL datasets for MonoBERT-base.}
        \label{fig:monobert_base_msm_attention_hard}
    \end{subfigure}
    \hfill
    \begin{subfigure}[b]{0.47\textwidth}
        \centering
        \includegraphics[width=1\textwidth]{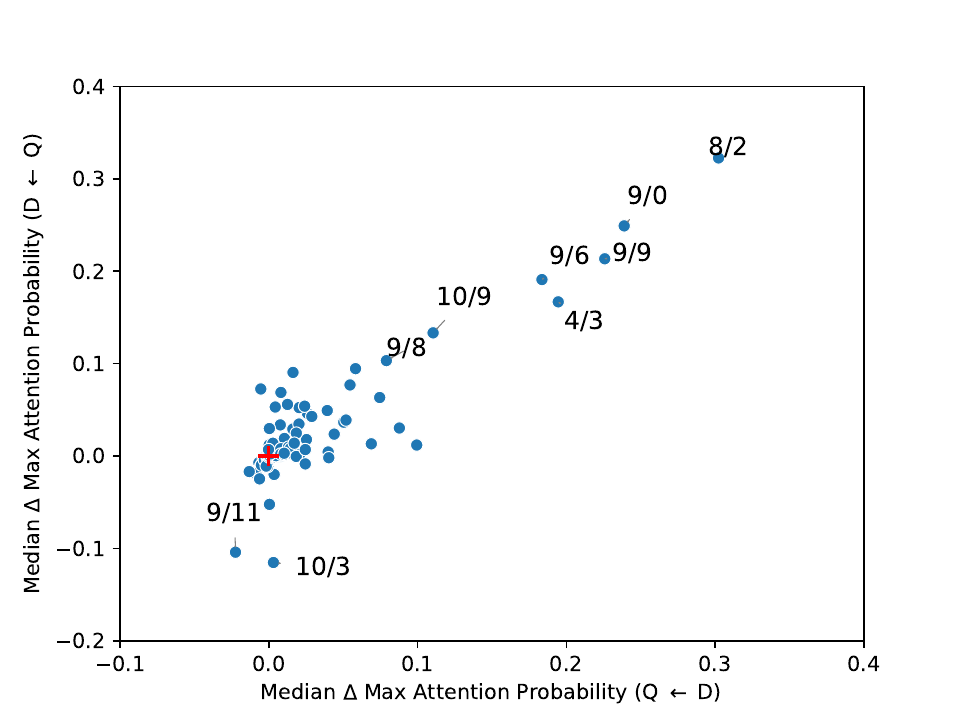}
        \caption{Results over the OOD datasets for MonoBERT-base.}
        \label{fig:monobert_base_ood_attention_hard}
    \end{subfigure}
    \centering
    \begin{subfigure}[b]{0.47\textwidth}
        \centering
        \includegraphics[width=1\textwidth]{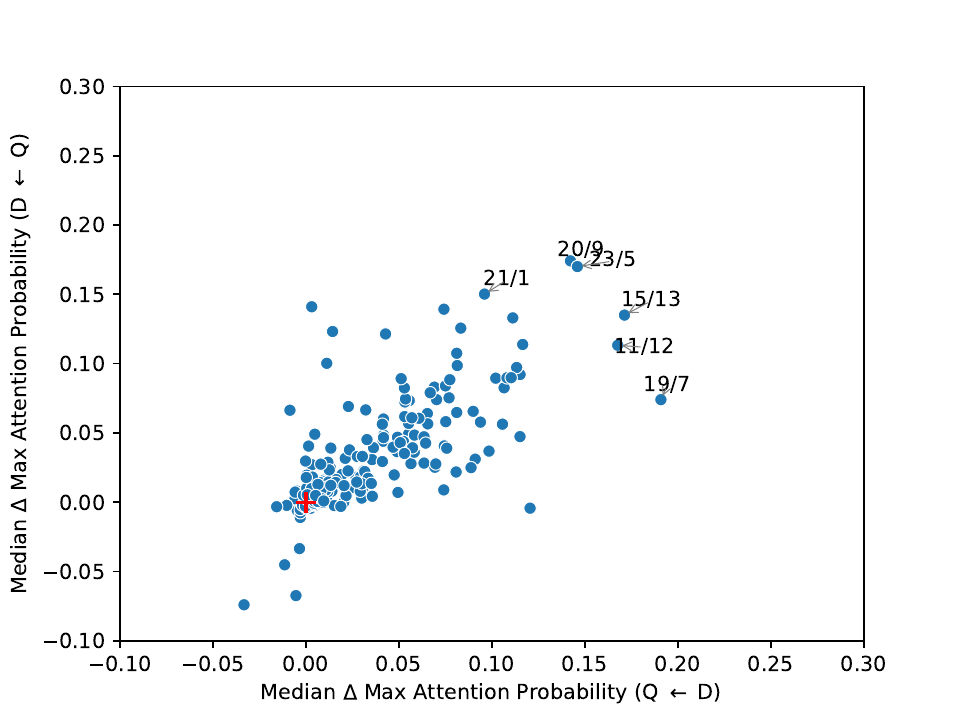}
        \caption{Results over the 4 TREC-DL datasets for MonoBERT-large.}
        \label{fig:monobert_msm_attention_hard}
    \end{subfigure}
    \hfill
    \begin{subfigure}[b]{0.47\textwidth}
        \centering
        \includegraphics[width=1\textwidth]{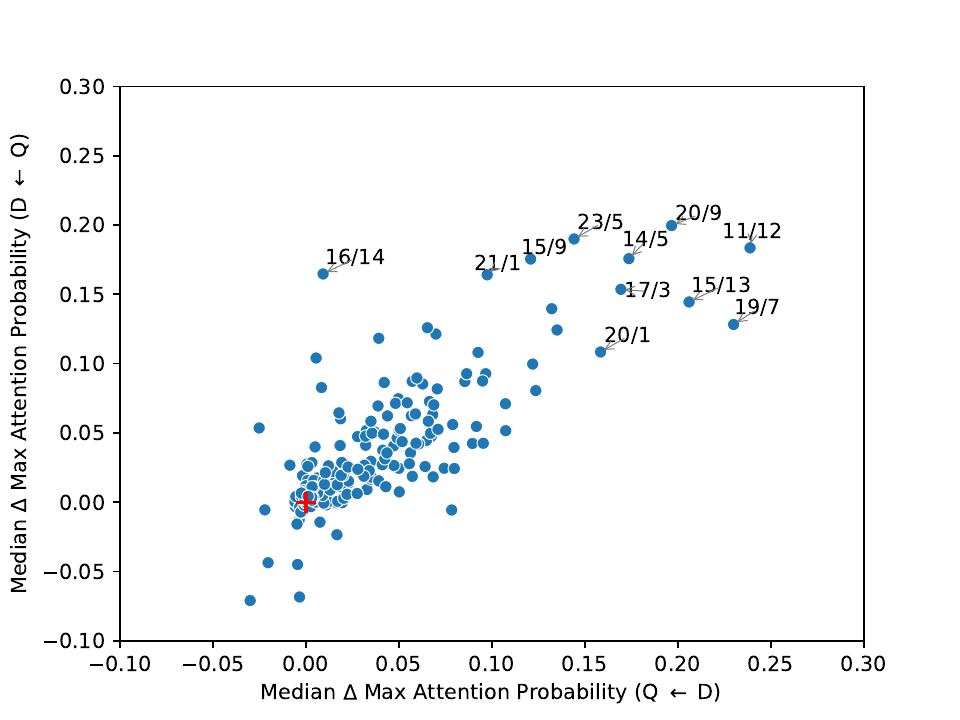}
        \caption{Results over the OOD datasets for MonoBERT-large.}
        \label{fig:monobert_ood_attention_hard}
    \end{subfigure}
    \caption[Contrast of the attention patterns between relevant and hard negatives]{Relevant vs.\ hard negatives -- Median of the differences in the attention weights $\maxatt{Q\leftarrow D/1}-\maxatt{Q\leftarrow D/0}$ and $\maxatt{D\leftarrow Q/1} - \maxatt{D\leftarrow Q/0}$ for each attention head of all models. An attention head $h$ at a layer $\ell$ is denoted as $\ell/h$.}
    \label{fig:comparison_attention_hard}
\end{figure}

Using the approach described above and by contrasting the maximum of Q/D-attention and D/Q-attention between two relevance levels across our different pools of datasets for our cross-encoder models, we can see if some attention heads are specialized in detecting \emph{relevant} matching signals.  Figure \ref{fig:comparison_attention_weak} describes the results for the contrast between $r=1$ and $r=-1$, and Figure \ref{fig:comparison_attention_hard} for $r=1$ and $r=0$.

\result{Cross-encoders have attention heads dedicated to detecting matching signals between a query and a relevant passage. For MiniLM and MonoBERT-base, these heads are relatively robust to high lexical proximity but for MonoBERT-large, it can lead to major changes in their matching patterns.}
Figure \ref{fig:comparison_attention_weak} shows that some heads have much higher cross-attention probability levels when the document is relevant compared to when it is not. Interestingly, we note that this is robust to domain shift for every model (\textbf{RQ6}, Figures \ref{fig:minilm_ood_attention}, \ref{fig:monobert_base_ood_attention} and \ref{fig:monobert_ood_attention}). The conclusion we can draw is that some attention heads in cross-encoders are indeed specialized in detecting \emph{relevant} matching signals between the query and the document. We refer to such attention heads as ``matching detection'' heads. Interestingly, we note that cross-attention is bi-directional, i.e., there is almost a linear relationship between the differences $\maxatt{Q \leftarrow D/1} - \maxatt{Q\leftarrow D/-1}$ and $\maxatt{D\leftarrow Q/1} - \maxatt{D\leftarrow Q/-1}$, meaning those heads detect matching tokens in the query (for the document) and in the document (for the query) at the same time. If each model has such heads, we note that they are located in almost all of the MiniLM and MonoBERT-base layers (from layer 1 to 10), compared to mostly middle to later layers (from layer 9 to 23) for MonoBERT-large.

Observing Figure \ref{fig:comparison_attention_hard}, contrasting relevant passages with hard negatives, we can observe similar patterns with lower differences compared to easy negatives (Figure \ref{fig:comparison_attention_weak}). For MiniLM and MonoBERT-base, it seems that the attention heads ($7/9$ and $4/9$ for MiniLM, and $4/3$ and $8/2$ for MonoBERT-base) that exhibit the most distinctive patterns in Figure \ref{fig:comparison_attention_weak} are relatively robust to this change, especially in OOD. Meanwhile, for MonoBERT-large, the equivalent two attention heads ($11/13$ and $15/13$) no longer stand out from the others. This comparison between the behaviors of the matching detection heads with different types of negative passages suggests that these heads are tricked by the lexical proximity between a query and a hard negative passage (sampled using BM25). We note that smaller BERT-based models, i.e., MiniLM and MonoBERT-base, seem more resilient to this issue than MonoBERT-large. 

\begin{figure}[t]
    \centering
        \centering
        \begin{subfigure}[b]{0.49\textwidth}
            \centering
            \includegraphics[width=\textwidth]{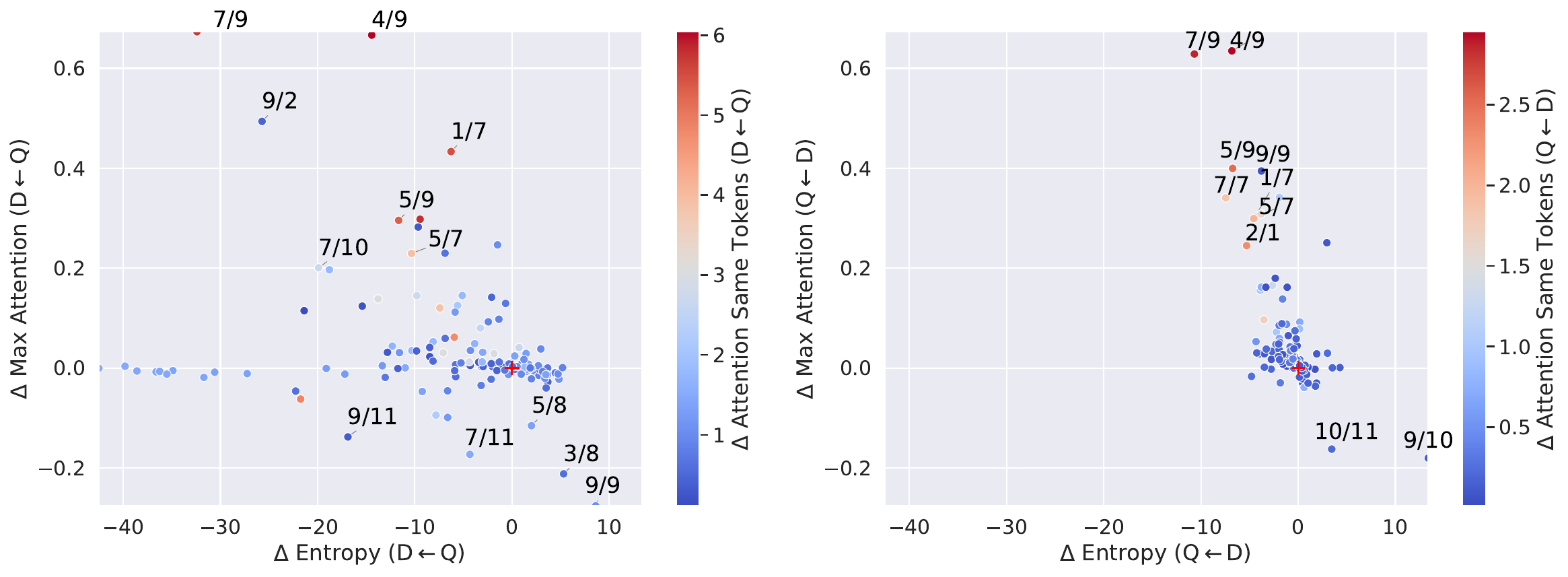}
            \caption{MiniLM on TREC-DL (easy).}
        \end{subfigure}
        \hfill
        \begin{subfigure}[b]{0.49\textwidth}
            \centering
            \includegraphics[width=\textwidth]{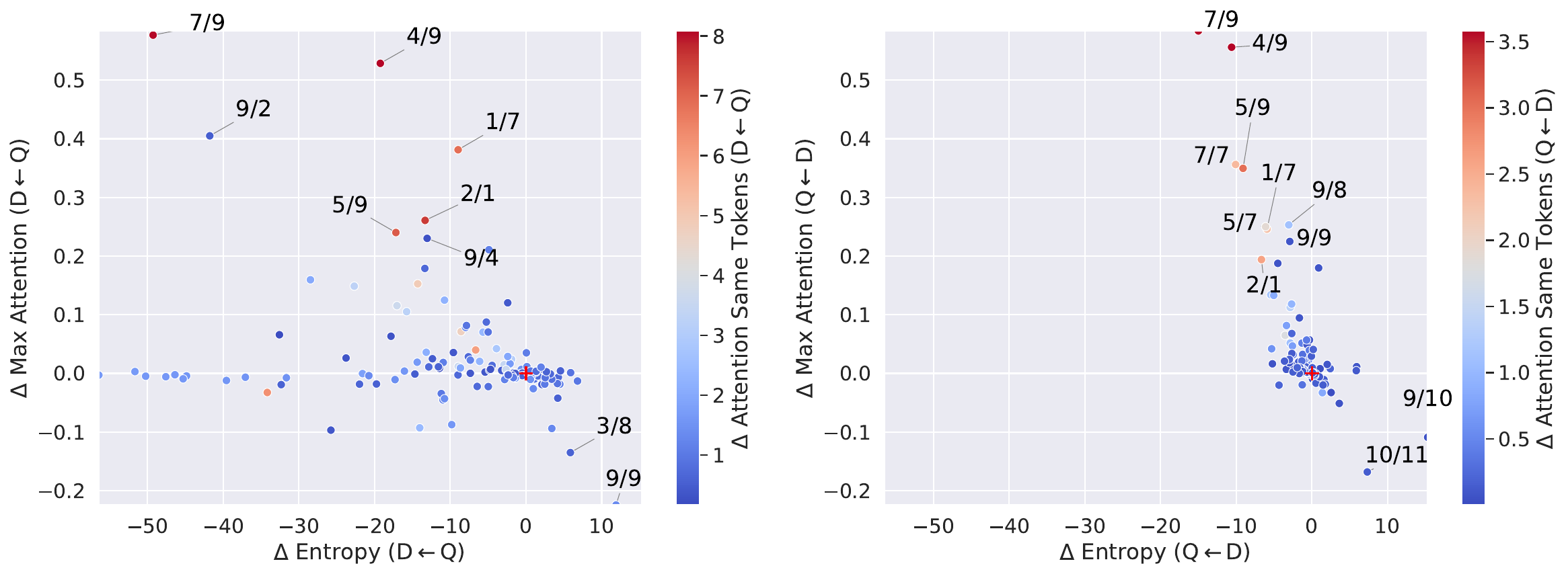}
            \caption{MiniLM on OOD (easy).}
        \end{subfigure}

        \begin{subfigure}[b]{0.49\textwidth}
            \centering
            \includegraphics[width=\textwidth]{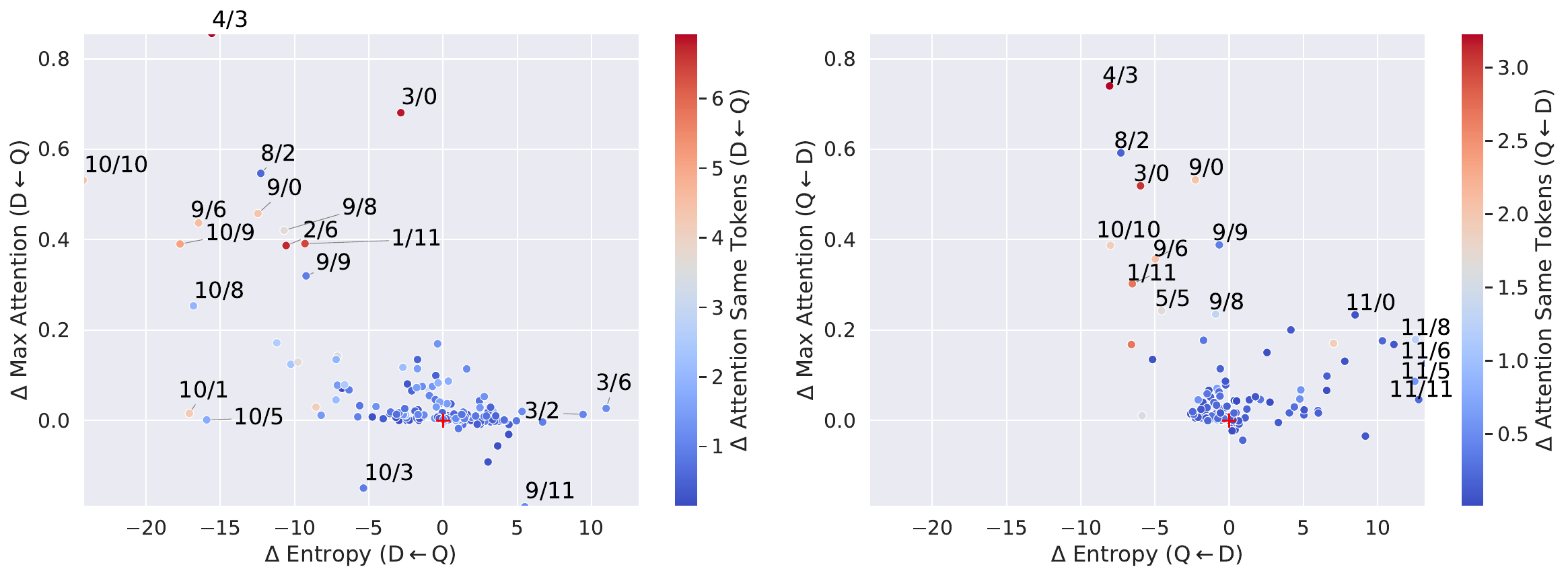}
            \caption{MonoBERT-base on TREC-DL (easy).}
        \end{subfigure}
        \hfill
        \begin{subfigure}[b]{0.49\textwidth}
            \centering
            \includegraphics[width=\textwidth]{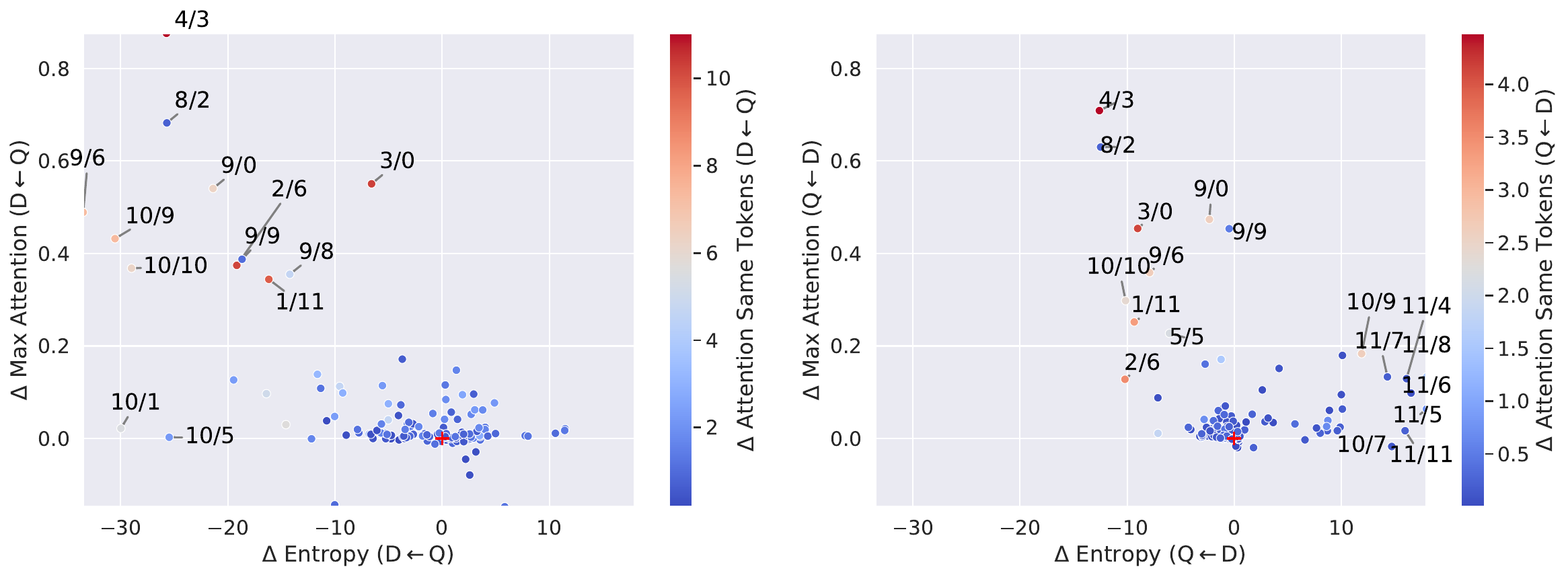}
            \caption{MonoBERT-base on OOD (easy).}
        \end{subfigure}
        
        \begin{subfigure}[b]{0.49\textwidth}
            \centering
            \includegraphics[width=\textwidth]{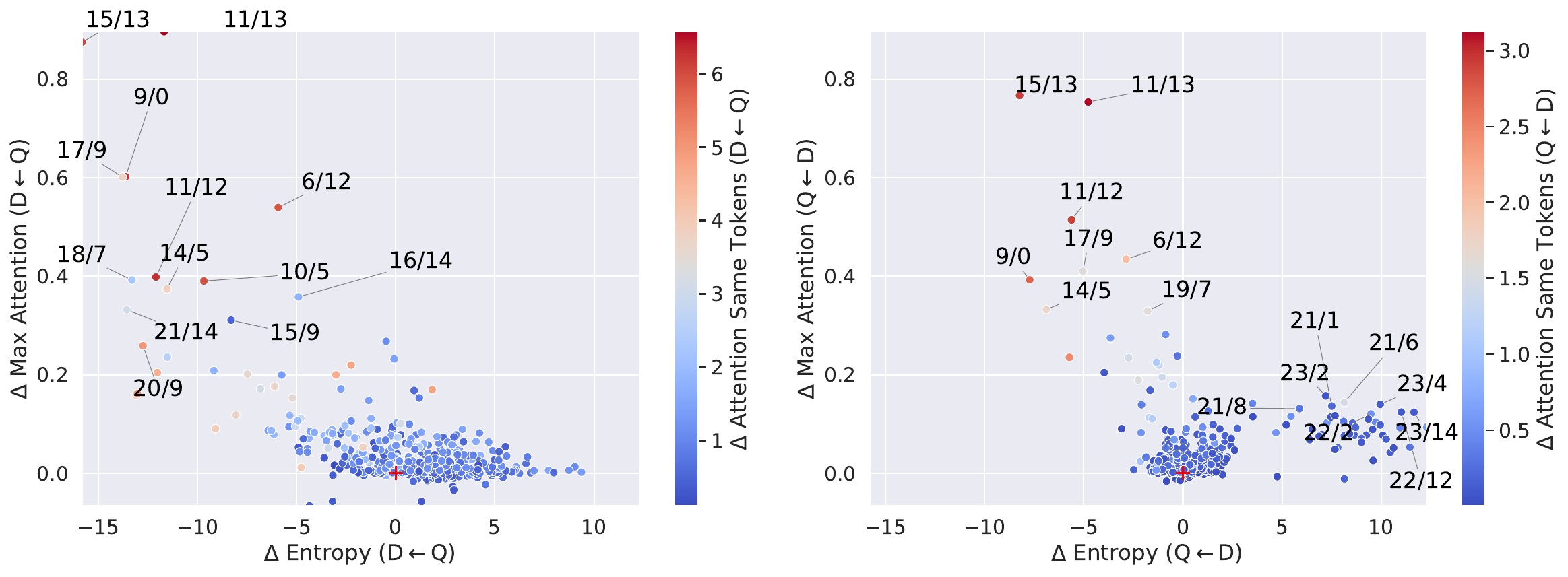}
            \caption{MonoBERT-large on TREC-DL (easy).}
        \end{subfigure}
        \hfill
        \begin{subfigure}[b]{0.49\textwidth}
            \centering
            \includegraphics[width=\textwidth]{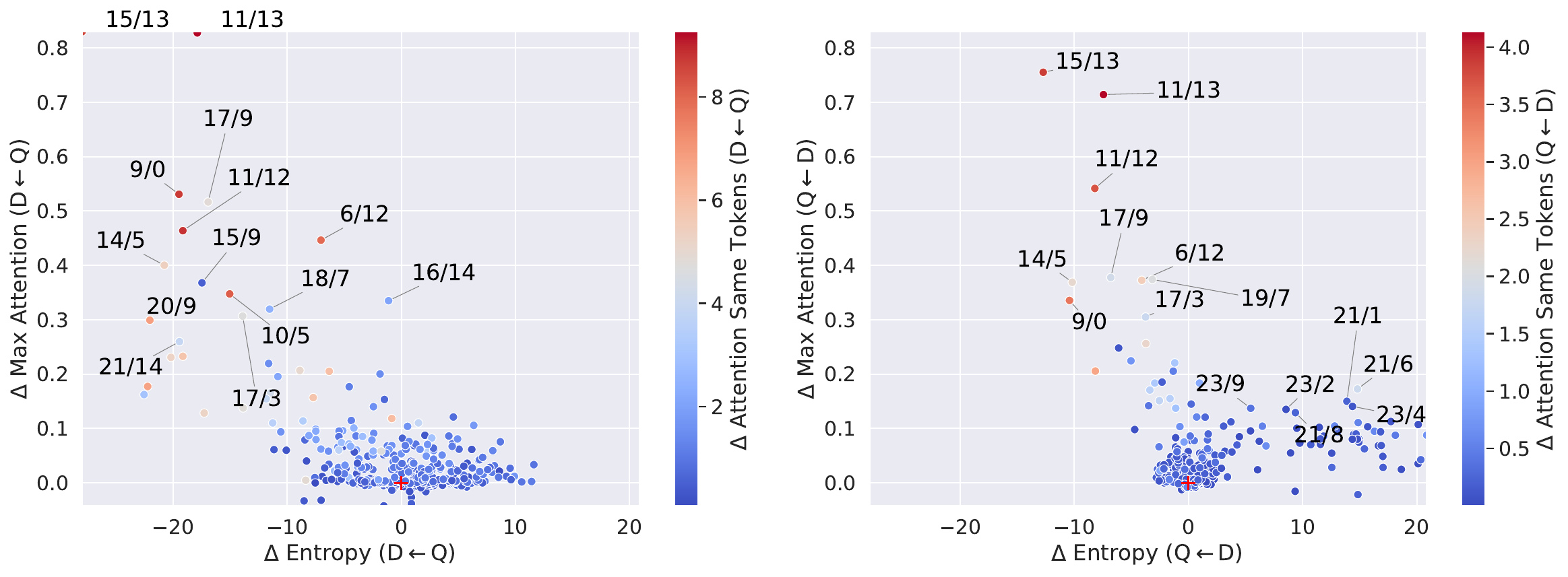}
            \caption{MonoBERT-large on OOD (easy).}
        \end{subfigure}

        \captionof{figure}[Comparison of the matching patterns between relevant and easy negatives]{Comparison of the matching patterns of the attention heads over the TREC-DL and OOD datasets, contrasting $r=1$ with $r=-1$ (easy negatives). Each point is an attention head $h$ at a layer $\ell$, denoted as $\ell/h$; the left panel corresponds to the $D \leftarrow Q$ direction and the right panel to $Q \leftarrow D$. x-axis: median of the differences in entropy ($E^{h,\ell}_{Q\leftarrow D/1} - E^{h,\ell}_{Q\leftarrow D/-1}$); y-axis: median of the differences in maximum attention ($\maxatt{Q\leftarrow D/1} - \maxatt{Q\leftarrow D/-1}$); color: median of the differences in the probability of attending to a token identical to the source token. The red cross marks the origin.}
        \label{fig:comparing_match_type_weak}
\end{figure}

\result{Matching detection heads can capture both exact matches and semantic matches.}
Beyond these high-level observations, it is difficult to determine the nature of the detected matches. To overcome this limitation, we take into account the source and target tokens of each match to differentiate exact matches (source and target tokens are identical) from semantic matches (source and target tokens are different). As we consider raw tokens and do not apply a pre-processing before studying them, we refer to a match between an identical source and target token as an \emph{exact}, and not a \emph{lexical} match, i.e., a match that would be detected by lexical approaches such as BM25 \cite{bm25}. Note, however, that we consider exact matches as a proxy for studying lexical matches, and that our approach tends to underestimate these so-called \emph{lexical} matches.
Figure \ref{fig:comparing_match_type_weak} provides details on the nature of query-document matches when comparing relevant passages ($r=1$) to easy negatives\footnote{Figure \ref{fig:comparing_match_type_hard} in the Appendix provides the same comparison between relevant passages and hard negatives ($r=0$).} ($r=-1$). This figure combines three differences per attention head: in entropy (x-axis), in maximum attention (y-axis, as in Figure~\ref{fig:comparison_attention_weak}), and in the probability of attending to a token identical to the source token (color). This last probability is conditioned on the event that a query token attends to the document (and symmetrically for $D \leftarrow Q$), so that it measures the fraction of the \emph{cross-attention} going to identical tokens. A head with a negative entropy difference and a positive maximum-attention difference focuses on a few specific document tokens when the passage is relevant, i.e., it detects matches; the color then tells whether these matches are exact (identical tokens) or semantic (different tokens).

If we first look at the results produced by contrasting easy negatives ($r=-1$) with relevant passages ($r=1$) in Figure \ref{fig:comparing_match_type_weak}, we see relatively similar patterns between ID and OOD situations (\textbf{RQ6}). For every model, the attention heads that are standing out in Figure \ref{fig:comparison_attention_weak}, i.e., $4/9$ and $7/9$ for MiniLM, $4/3$ and $3/0$ for MonoBERT-base and $15/13$ and $11/13$ for MonoBERT-large, are all focusing more on single tokens containing relevance signals (lower entropy when the passage is relevant than when it is non-relevant) and with the same identity (higher attention to identical tokens when the passage is relevant than when it is non-relevant). This observation suggests that these heads may be involved, more precisely, in the detection of \emph{exact matches}. At the same time, we also observe other attention heads, with high differences in the attention weights and low entropy, that do not focus on identical tokens like the heads $9/2$ for MiniLM, $8/2$ or $9/9$ for MonoBERT-base and $15/9$ or $17/9$ for MonoBERT-large. These heads may be involved in the detection of \emph{semantic matches}. In general, we observe a common trend across models and domains, where \emph{semantic} matching heads are located in later layers than \emph{exact} matching heads -- which corresponds to what is known about transformer-based models. We observe the same patterns when contrasting hard negatives ($r=0$) to relevant passages ($r=1$), though less pronounced (see Figure \ref{fig:comparing_match_type_hard} in Appendix).

\result{The aggregation of relevance signals is conveyed from the document tokens through the query tokens before eventually reaching the \CLS{} in MonoBERT-base and MonoBERT-large.}
In Figures \ref{fig:comparing_match_type_weak} and \ref{fig:comparing_match_type_hard}, we observe a very distinct pattern for the direction $Q\leftarrow D$ in both MonoBERT models, where, when the passage is relevant, the attention heads in the very last layers have higher entropy levels and a positive difference in the attention weights. This pattern echoes our description of the last step of the relevance prediction process in the previous Section \ref{section:information_flow}, with query tokens playing a pivotal role in the accumulation of relevance signal(s) from the entire input before transmitting them to the \CLS{}. We therefore interpret this difference with MiniLM as follows: both MonoBERT models aggregate \emph{raw relevance signals} from many document tokens into the query tokens, which explains both the higher attention and the higher entropy when the document is relevant.

\result{MiniLM has Q/D-attention heads that are more active when the passage is not relevant to the query. These attention heads are not detecting matching signals but non-relevance signals.}
Another difference, visible in Figures \ref{fig:comparison_attention_weak}, \ref{fig:comparison_attention_hard}, \ref{fig:comparing_match_type_weak} and \ref{fig:comparing_match_type_hard} (Appendix), is that some MiniLM heads cross-attend more when the document is \emph{not} relevant: matching detection heads ($3/8$ and $9/9$ in the first two figures) as well as other heads ($9/10$ and $10/11$ for the direction $Q \leftarrow D$ in the last two). Note that, while less pronounced, MonoBERT-base also partially exhibits this behavior only for the direction $D \leftarrow Q$ (see the heads $10/3$ and $9/11$), contrary to MonoBERT-large, which does not seem to possess the same mechanism.

From Figures \ref{fig:comparing_match_type_weak} and \ref{fig:comparing_match_type_hard} in the Appendix, focusing on the direction $D\leftarrow Q$, we remark that the attention heads $3/8$ and $9/9$ with higher cross-attention weights when the passage is non-relevant (in Figures \ref{fig:comparison_attention_weak} and \ref{fig:comparison_attention_hard}) also have a lower entropy and do not attend to identical tokens. Taken together, and echoing the experiment described in Figure \ref{fig:controling_relevance} in the previous section, the evidence suggests that these attention heads detect signals of non-relevance between the query and the document. Although we lack a clear understanding of this behavior, it appears both in-domain and out-of-domain, with the same heads consistently involved each time for both $D\leftarrow Q$ and $Q\leftarrow D$ for MiniLM, and solely for $D\leftarrow Q$ with MonoBERT-base.

At the same time, from the Figures \ref{fig:comparing_match_type_weak} and \ref{fig:comparing_match_type_hard}, but focusing on the direction $Q\leftarrow D$, we observe two attention heads in MiniLM, $9/10$ and $10/11$, located in the final layers of the model, exhibiting a comparable behavior to the attention heads in the final layers of MonoBERT for the same direction (see the previous point). However, instead of having higher attention scores for relevant passages, they have higher attention scores when passages are not relevant, but lower entropies. While we lack a clearer understanding of their roles, we consider this observation further evidence that MiniLM uses signals of non-relevance, and not relevance, to make its predictions (see the discussion in the subsection \ref{subsection:ablations}).

Looking back at Section \ref{section:NIG} where we measured the importance of different components inside the cross-encoders using NIG, the question we can ask ourselves is: are the most important attention heads, according to NIG, involved in query-document matches detection? In this section, we identified a small set of attention heads, scattered across the layers of every cross-encoder, whose behavior seems related to the detection of matches. Theoretically, these attention heads should be considered highly important for the IR task, even though their relative importance compared to the attention heads involved in the contextualization of the query and document tokens, or in the transfer of the relevant signal to the \CLS{} token, is unclear.
In Table \ref{tab:ranking_attention_heads}, we observe that most attention heads that exhibit an attention pattern representative of the detection of matching signals in Figures \ref{fig:minilm_msm_attention} and \ref{fig:monobert_msm_attention}, are also considered to be important in terms of their NIG score. We further remark the presence of additional attention heads, for which we currently lack a deeper understanding. This underlines the importance of mechanistic interpretability studies or similar works to unveil the role of key components. On this matter, Table \ref{tab:ranking_attention_heads} further reports the attention heads in MiniLM classified as ``Matching heads'' in \citeauthor{lu-etal-2025-pathway}~\cite{lu-etal-2025-pathway}. We observe a high overlap between the most important attention heads we identified with NIG, characterized as detecting query-document matching, and these ``Matching heads''. This provides further support for our findings and substantiates our claim that lightweight alternatives to causal patching exist.

\begin{table}[t]
\caption{Table detailing if the top 10 attention heads (layer $\ell$ / head $h$) in terms of NIG score are involved in detecting matching signals, i.e., if they are highlighted in one of the figures of this section, and if the Information Bottleneck (IB) agrees, i.e., if either $Q\leftarrow D$ or $D\leftarrow Q$ is preserved by the mask for the attention head. The ordering of the attention heads is from the highest NIG score to the lowest and follows the Fusion operation of aggregation.}
\centering
\begin{tabular}{cccc|ccc|ccc}
\multicolumn{4}{c|}{\textbf{MiniLM}} & \multicolumn{3}{c|}{\textbf{MonoBERT-base}} & \multicolumn{3}{c}{\textbf{MonoBERT-large}} \\
\toprule
\makecell{Attention\\heads} & In plots? & In IB? & \makecell{``Matching heads''\\in \cite{lu-etal-2025-pathway}?} & \makecell{Attention\\heads} & In plots? & In IB? & \makecell{Attention\\heads} & In plots? & In IB? \\
\midrule
4/9 & \checkmark & \checkmark & \checkmark & 3/0 & \checkmark & \checkmark & 1/9 & $\times$ & \checkmark\\
1/7 & \checkmark & \checkmark & \checkmark & 1/7 & $\times$ & \checkmark & 15/13 & \checkmark & \checkmark\\
2/1 & \checkmark & \checkmark & \checkmark & 2/6 & \checkmark & \checkmark & 9/0 & \checkmark & \checkmark\\
5/7 & \checkmark & \checkmark & \checkmark & 0/9 & $\times$ & \checkmark & 11/13 & \checkmark & \checkmark\\
7/9 & \checkmark & \checkmark & \checkmark & 0/10 & $\times$ & \checkmark & 11/12 & \checkmark & \checkmark\\
6/5 & $\times$ & \checkmark & \checkmark & 1/6 & $\times$ & $\times$ & 16/1 & $\times$ & \checkmark\\
5/9 & \checkmark & \checkmark & \checkmark & 9/5 & $\times$ & $\times$ & 14/5 & \checkmark & \checkmark\\
3/1 & $\times$ & \checkmark & \checkmark & 10/3 & \checkmark & \checkmark & 22/4 & $\times$ & \checkmark\\
2/10 & $\times$ & \checkmark & $\times$ & 9/11 & \checkmark & \checkmark & 10/10 & $\times$ & \checkmark\\
7/1 & $\times$ & $\times$ & $\times$ & 0/4 & $\times$ & \checkmark & 2/0 & $\times$ & $\times$\\
\bottomrule
\end{tabular}
\label{tab:ranking_attention_heads}
\end{table}

\section{RQ4: Deep dive into the subspace(s) of relevance and the construction of the relevance signal}
\label{section:subspace_analysis}

So far, we have seen that both attention heads and FFN layers play important roles inside cross-encoders (see Section~\ref{section:NIG}). These roles revolve mostly around the query and document tokens, their interactions, and ultimately the \CLS{} token (see Section~\ref{section:information_flow}). In particular, some attention heads contribute to detecting matching signals between query and document tokens while others contextualize tokens or transfer information to \CLS{} (see Section~\ref{section:understand_attention}). FFN layers and their ``important'' neurons process and refine information moved by the attention heads, integrating it into tokens' representation~\cite{geva_transformer_2021,geva2022transformer}. In cross-encoders, some of these neurons likely operate in a \emph{relevance subspace} where relevance signals, such as matching signals (Subsection~\ref{subsection:matching_signals}), are stored to support relevance prediction. This \emph{relevance subspace} is the focus of this section, and as in the previous section, we ensure that each level of relevance is represented by using the \emph{Diverse sampling} strategy (see Figure \ref{fig:dataset_construction}). 

\subsection{Context and Notations}

To study the \emph{relevance subspace}, we apply a Linear Discriminant Analysis (LDA) to three activations per layer (see Section \ref{subsection:architecture} for the notations): $\repA$, the attention output after LayerNorm, i.e., the residual stream\footnote{We refer here to the mechanistic interpretability paradigm, where the transformer layers are seen as modules that read and write from a \emph{residual stream} that goes all the way from the embedding layer to the final layer of the model \cite{elhage2021mathematical}.} after attention; $\repFFNr$, the raw FFN output; and $\repFFN$, the residual stream after the FFN and its LayerNorm. LDA is a classical method used to identify discriminative subspaces within high-dimensional representation spaces, in our case, the ``relevance subspace(s)'' of the hidden states representational space. We further compute activations with respect to each input part and the relevance label associated with the sample.\footnote{We provide more details on the different steps we follow to compute the LDA in Appendix \ref{appendix:lda}.} 

\subsection{Results analysis}

\begin{figure}[t]
    \centering
    \begin{subfigure}[t]{0.3\textwidth}
        \centering
        \includegraphics[width=1\textwidth]{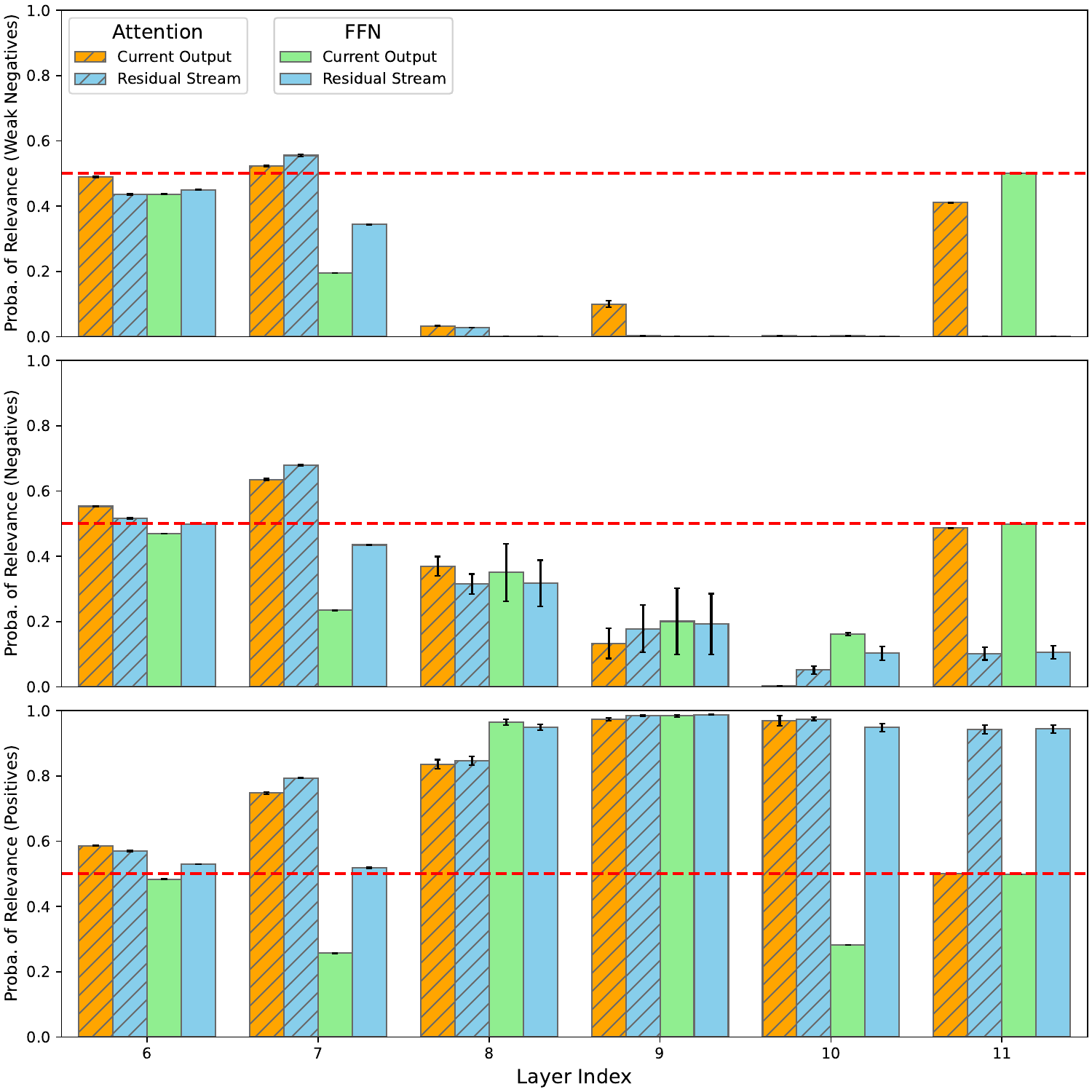}
        \caption{Details for MiniLM.}
        \label{fig:minilm_integration_relevance}
    \end{subfigure}
    \hfill
    \begin{subfigure}[t]{0.3\textwidth}
        \centering
        \includegraphics[width=1\textwidth]{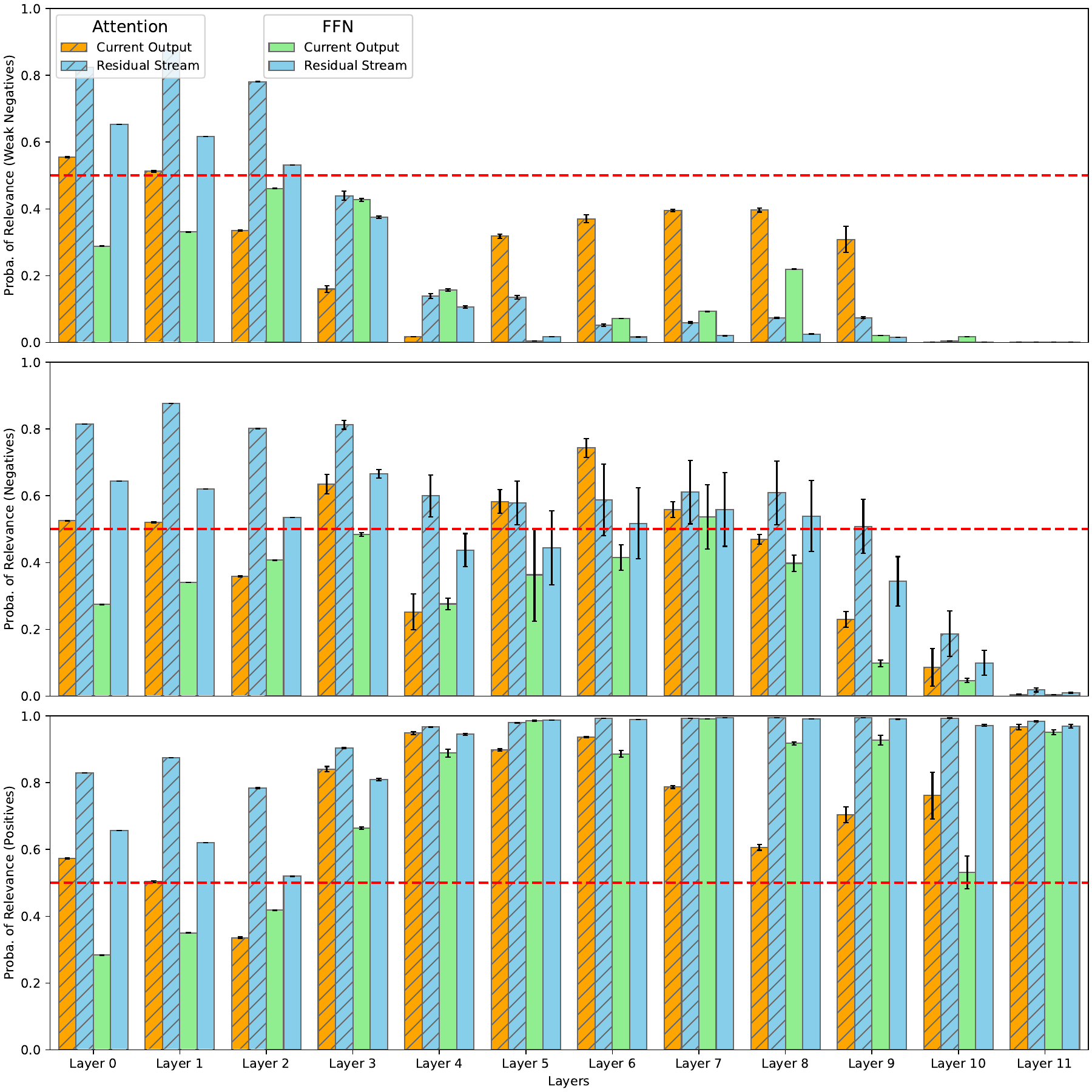}
        \caption{Details for MonoBERT-base.}
        \label{fig:monobert_base_integration_relevance}
    \end{subfigure}
    \hfill
    \begin{subfigure}[t]{0.3\textwidth}
        \centering
        \includegraphics[width=1\textwidth]{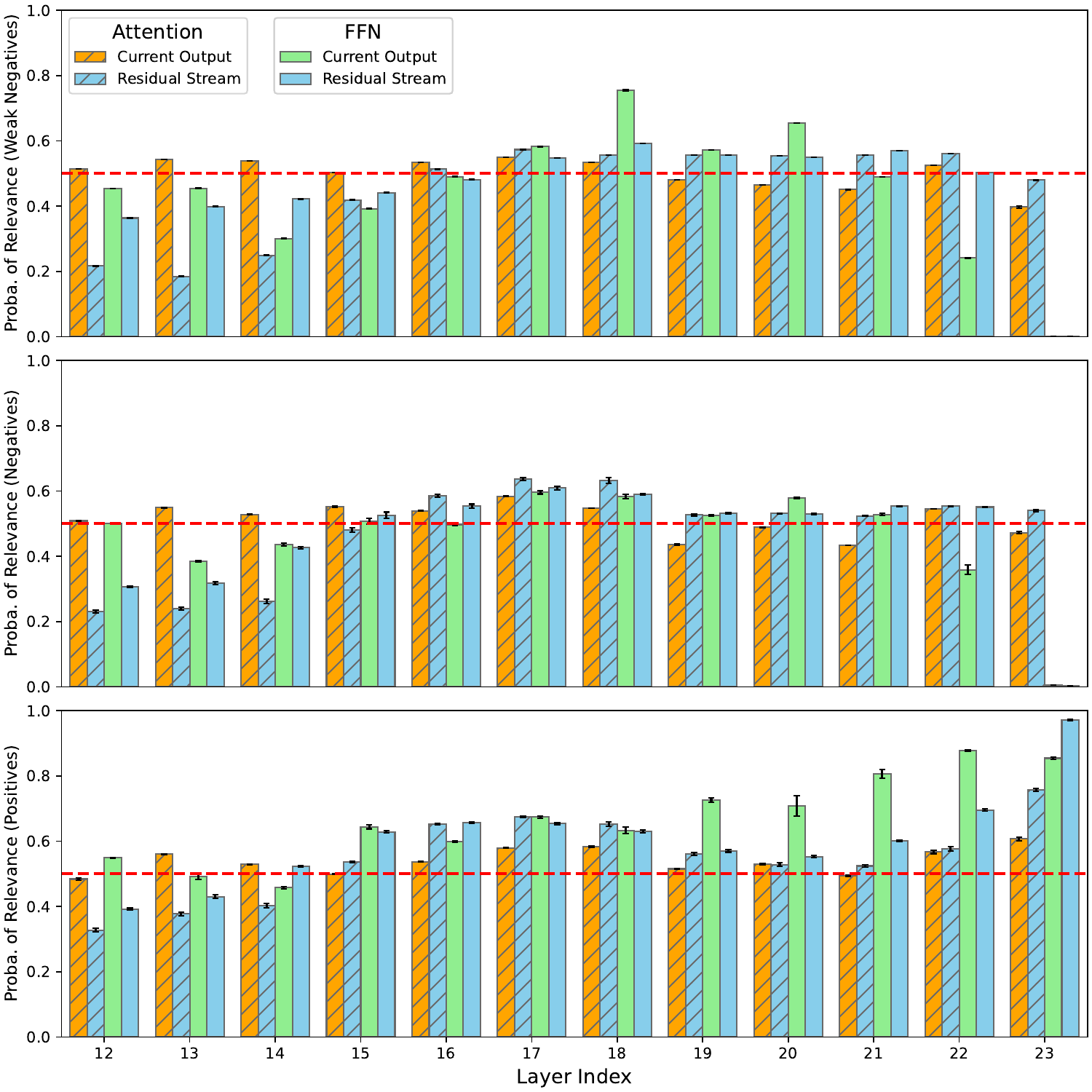}
        \caption{Details for MonoBERT-large.}
        \label{fig:monobert_integration_relevance}
    \end{subfigure}
    \caption[Evolution of the relevance prediction across the models' layers]{Evolution of the probability of relevance based on the \CLS{} internal representations, per layers, for $r \in \{-1, 0, 1\}$.}
    \label{fig:integration_relevance}
\end{figure}

\begin{figure}[t]
    \centering
    \begin{subfigure}[b]{0.9\textwidth}
        \centering
        \includegraphics[width=1\textwidth]{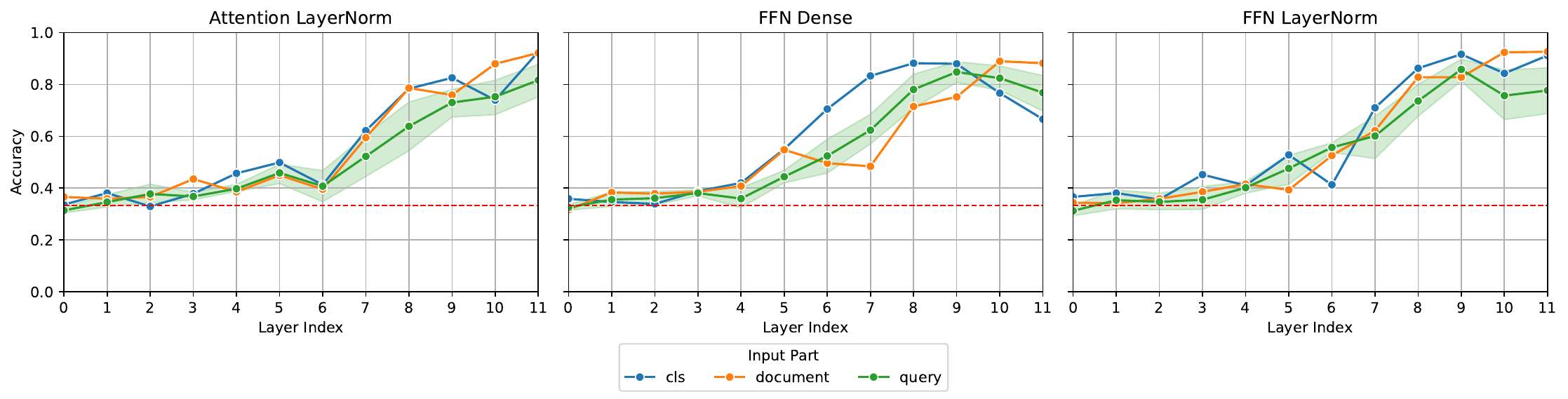}
        \caption{Details for MiniLM.}
\label{subfig:minilm_all_accuracy_msmarco}
    \end{subfigure}

    \centering
    \begin{subfigure}[b]{0.9\textwidth}
        \centering
        \includegraphics[width=1\textwidth]{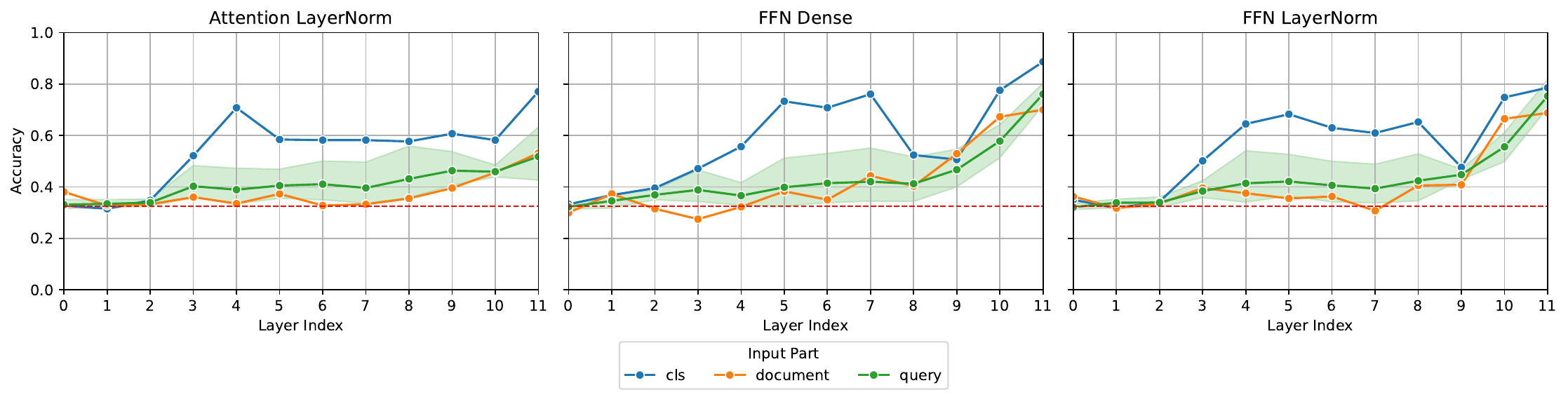}
        \caption{Details for MonoBERT-base.}
        \label{subfig:monobert_base_all_accuracy_msmarco}
    \end{subfigure}

    \centering
    \begin{subfigure}[b]{0.9\textwidth}
        \centering
        \includegraphics[width=1\textwidth]{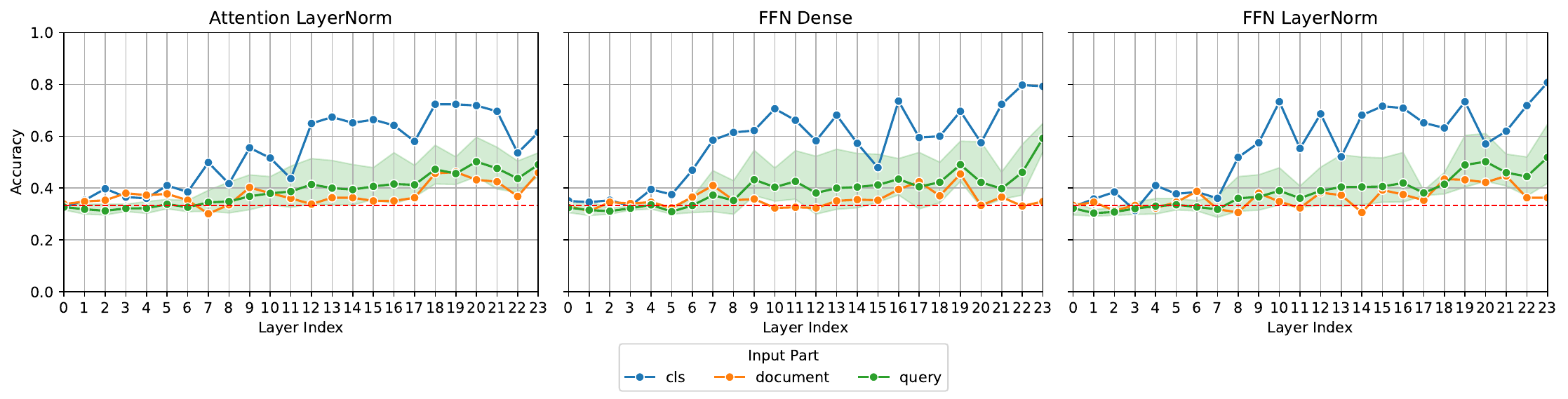}
        \caption{Details for MonoBERT-large.}
        \label{subfig:monobert_all_accuracy_msmarco}
    \end{subfigure}
    \caption[Evolution of the classification accuracy of the relevance subspace contained in the \CLS{}, query tokens and document tokens for TREC-DL datasets]{Evolution of the classification accuracy of the activations $\repA[p]$, $\repFFNr[p]$ and $\repFFN[p]$ projected along the LDA axis, by their original relevance label ($r \in \{-1,0,1\}$), per input part $p \in \{\CLS, Q_1,..., Q_k, \bar{D}\}, k=6$. LDA and classification are computed over samples from the TREC-DL datasets.}
    \label{fig:lda_accuracy}
\end{figure}

\result{A relevance subspace does not exist from the start but is constructed by cross-encoders to store and refine relevance signals.}
From the previous section, we know that matching signals, a necessary piece in the relevance prediction process \cite{lu-etal-2025-pathway}, are detected at different levels in cross-encoder models, from early layers until middle-later layers. From the conclusions we draw in Sections \ref{section:information_flow} and \ref{section:understand_attention} on the importance of ``matching detection'' heads, we hypothesize that if a ``relevance subspace'' exists, the attention mechanism must be gradually directing these matching signals inside it. Our study of the information flow, and specifically its late focus on the \CLS{} token representation, suggests that this subspace is centered around, or embedded within, the \CLS{} token's representation space. This, in turn, coupled with our remarks on the probable role in the no-op operation \cite{clark-etal-2019-bert} of the \CLS{} in early layers, implies that the relevance subspace only appears at the end of the processing.

To test this hypothesis, we plot in Figure \ref{fig:integration_relevance} the evolution of the confidence in the prediction ``Relevant'' of the three cross-encoders, per layers, based on the \CLS{} representation at each step of the processing. We apply a procedure similar to approaches using the unembedding matrix of LLMs on top of the residual stream, where they project an internal representation onto the vocabulary \cite{geva-etal-2023-dissecting}. In our work, we use the classification head of each cross-encoder on the internal representations of the \CLS{}. In our case, this projection takes as input the hidden state of the \CLS{} token at each layer, a vector of $\mathbb{R}^{\llmdim}$, before/after it is written to the residual stream (i.e. $\repAr[CLS]$, $\repA[CLS]$, $\repFFNr[CLS]$ and $\repFFN[CLS]$), and returns as an output a probability of relevance in $[0,1]$, based on the information currently stored inside it. We argue that, despite its obvious limitations (the classifier is not trained to predict relevance from these different representations), this approximation is reasonable and the trends it exhibits can still be interpreted as a proxy to monitor how much the \CLS{} representation points towards relevance vs non relevance.

From these Figures, we first see that, for MiniLM and MonoBERT-large, the predicted probabilities are highly similar across relevance levels for most of the models' layers (up to layer 7 for MiniLM and 18 for MonoBERT-large). This confirms that, before a critical point, the representation of the \CLS{} does not contain anything close to what the classification head of each model is expecting to predict the relevance. Meanwhile, we observe a dissimilar trend for MonoBERT-base, as this critical point occurs much earlier, around layer 3. After this layer, we observe that the model is already able to come close to its last layer's prediction, meaning the \CLS{} token starts relatively early to aggregate relevance signals.

\result{MiniLM and MonoBERT-base refine relevance signals directly inside query and document tokens before passing them to the \CLS{}, which ``composes'' them for the final prediction. In contrast, MonoBERT-large stores ``raw'' signals inside the \CLS{} token representation and, after that, refines them in place.}
This result is suggested first by observing the patterns in Figure \ref{fig:integration_relevance}, past the decision point where each model's behavior starts to change depending on the relevance levels. While the attention process in MiniLM and MonoBERT-base clearly integrates query/document relevance information into the \CLS{}, MonoBERT-large instead gradually builds this information, firstly inside the residual of the FFN, before transmitting it to the residual stream ($\repFFNr[CLS]$ in green at layers 20-21 first passes below or above 0.5, before $\repFFN[CLS]$, in blue and unhatched, in the next layers). For the smaller cross-encoders, this suggests that the ``raw'' information transferred to the \CLS{} token contains already, at least, a portion of the relevance signal. This could mean that either they can leverage ``raw'' matching signals, and perhaps other signals such as IDF information (already existing in the embedding matrix \cite{choiFindingInverseDocument2022}), as suggested in \citeauthor{lu-etal-2025-pathway}~\cite{lu-etal-2025-pathway}, or that the matching signals (and possibly others) are already processed when they are transferred to the \CLS{}. In contrast, MonoBERT-large seems to follow a much more incremental process, where the information that has been transferred to the \CLS{} token needs to be refined by the next few FFN layers. 

To investigate this observation further and lift the limitations of relying on the classification heads, we study the same phenomenon with an LDA, that further includes the query and document token representations. In Figures \ref{fig:lda_accuracy} and \ref{fig:lda_ood_accuracy} (in Appendix, as both figures depict similar patterns), we compare the evolution of the classification accuracy between the different relevance levels (easy and/or hard negatives and positives) of the internal representations from the \CLS{}, query and document tokens\footnote{For the queries, we consider independently the first $k=6$ tokens of each query. To get a single vector per document, we average its tokens' representations, per sample.}, after they are projected along the first (or first two if the three relevance levels are considered at the same time) axis of the LDA. 
Firstly, we note that, no matter the domain, the classification accuracy, from any input part, improves as the layer index increases for MiniLM, while the same pattern only happens for the \CLS{} token within MonoBERT-base and large. Secondly, if we now focus on the individual patterns of each module for each model, we can see that for MiniLM, the accuracy increases similarly at every level, while we observe differences between the attention and the FFN for MonoBERT-base (accuracy from query and document representations in the FFN increases towards the final layer, while they stay low in the attention). MonoBERT-large exhibits a somewhat similar pattern, but only for the query representations, not the document.

These trends can be seen more clearly if we only use the activations over query-passage pairs with relevance levels $r=-1$ (easy negatives) and $r=1$ (relevant passage) (see Figures \ref{fig:comparison_accuracy_minilm_strong_weak}, \ref{fig:comparison_accuracy_monobert_base_strong_weak} and \ref{fig:comparison_accuracy_monobert_strong_weak} in the Appendix).

Finally, this observation echoes the patterns observed in the Sections \ref{section:NIG} and \ref{section:information_flow} for the \textbf{RQ1} and \textbf{RQ2}, and reinforces the hypothesis that the relevance prediction processes of MiniLM, MonoBERT-base and MonoBERT-large are different. 

\result{Cross-encoders tend to rely more on exact matches when they encounter OOD samples. This behavior can lead them to be fooled more easily by hard negatives in this context.}
In the Appendix \ref{appendix:lda}, we provide additional figures detailing the evolution of the classification accuracy with only two relevance levels and over MS MARCO datasets or OOD datasets. As expected, we can clearly see that the classification accuracy is lower for relevance levels $r=1$ vs. $r=0$ than for $r=1$ vs. $r=-1$. Interestingly, we also note that the overall classification accuracy further diminishes if we switch from ID (Figures \ref{subfig:minilm_weak_negatives_accuracy_msmarco} and \ref{subfig:minilm_hard_negatives_accuracy_msmarco}, \ref{subfig:monobert_base_weak_negatives_accuracy_msmarco} and \ref{subfig:monobert_base_hard_negatives_accuracy_msmarco}, and \ref{subfig:monobert_weak_negatives_accuracy_msmarco} and \ref{subfig:monobert_hard_negatives_accuracy_msmarco}, all in Appendix) to OOD (Figures \ref{subfig:minilm_weak_negatives_accuracy_ood} and \ref{subfig:minilm_hard_negatives_accuracy_ood}, \ref{subfig:monobert_base_weak_negatives_accuracy_ood} and \ref{subfig:monobert_base_hard_negatives_accuracy_ood}, and \ref{subfig:monobert_weak_negatives_accuracy_ood} and \ref{subfig:monobert_hard_negatives_accuracy_ood}, all in Appendix). Furthermore, the characteristic increase in accuracy, observed in every setup, occurs later, in terms of layer index, with hard negatives and in OOD. It seems that intermediate representations in middle layers are more similar between hard negatives, sampled with BM25, and positives (making them harder to differentiate) than between easy negatives, sampled randomly, and positives. Given their respective nature, this confirms that cross-encoders rely heavily on both \emph{lexical} and \emph{semantic} matching, and also confirms that each type of matching is processed at a different stage: early to middle layers for lexical matches versus middle to later layers for semantic ones.

To conclude this Section and answer \textbf{RQ4}, it seems that MonoBERT-large progressively builds a single relevance signal within the representational space of the \CLS{} token by gathering information from the other tokens and then refining it. Meanwhile, MiniLM appears to be refining relevance information in place, across its query and document tokens, before transferring it to the \CLS{} token. Finally, MonoBERT-base exhibits a \textit{hybrid} behavior, where the \CLS{} starts to accumulate information on relevance relatively early (similarly to MonoBERT-large), but at the same time, the query and document tokens also seem to be refining additional signals in place (similarly to MiniLM). For MiniLM and MonoBERT-base, our results alone cannot tell whether the information sent to the \CLS{} in the last layers only needs to be filtered or weighted, or whether it is further refined inside the \CLS{}. Yet, from \citeauthor{lu-etal-2025-pathway}~\cite{lu-etal-2025-pathway} we know that it is possible to approximate, with a high correlation (80\%), the predictions of MiniLM by linearly combining signals that have been extracted and stored inside query tokens by so-called ``Relevance Scoring heads''. This insight corroborates our hypothesis about MiniLM and confirms that, contrary to MonoBERT-large, MiniLM, at least, has learned to integrate within its \CLS{} representation already refined relevance signals. For MonoBERT-base, we can suppose the model has already refined some signals inside the \CLS{} (see previous points), prior to gathering additional signals from the query and/or document tokens. 

\section{Discussion}
\label{section:discussions}

Before concluding, in this section we further discuss \textbf{RQ5} and \textbf{RQ6}, reflecting on the issues raised so far as well as discussing the limitations we identified in our study.

\subsection{RQ5: On the consistency of our findings across cross-encoders}\label{subsection:rq5}

In our work, we focused on three cross-encoders, MiniLM \cite{minilmv2} and MonoBERT \cite{monobert}, in both its base and large versions, that differ mostly by their pre-training regime (deep self-attention distillation \cite{minilmv2} for MiniLM versus Masked Language Modeling and Next Sentence Prediction \cite{bert} for MonoBERT) and their size (see Table \ref{tab:models-overview}). Their comparable performance on the IR task (see Table \ref{tab:reranking_comparison}) makes them interesting candidates to verify whether there is a single relevance prediction process towards which all cross-encoders have converged or not. 
Along with our first four Research Questions, we unveiled different insights suggesting that these cross-encoders possess their singularities. To answer \textbf{RQ5}: the stages of the relevance prediction process (contextualization and lexical matching, then semantic matching, then gathering of the relevance signal and construction of the relevance subspace) are similar across the studied models. This has been confirmed by our work around  \textbf{RQ2} and \textbf{RQ3}, in Sections \ref{section:information_flow} and \ref{section:understand_attention}, where we observed very similar flows of information across the input parts as well as attention heads specialized in detecting matching signals. Where the models differ, is principally in the details of these different steps.

For example, we identified one major difference in the stage of gathering relevance signals and constructing a relevance subspace in Section \ref{section:subspace_analysis}. MonoBERT-large gathers raw relevance signals from the document into the query, then into the \CLS{}, where it refines them and builds a relevance subspace. MiniLM and MonoBERT-base follow a different strategy. For MiniLM, the relevance signals gathered from the document to the query, are refined at the query-level
before being transferred to the \CLS{} token. Meanwhile, MonoBERT-base exhibits a ``hybrid'' pattern, as we observe, from a relatively early layer, the \CLS{} beginning to build a relevance subspace, and in later layers, relevance information also starts to appear in the query and document tokens, suggesting that some level of refinement also happens there.


Finally, and perhaps most importantly, we unveiled evidence in Subsection \ref{subsection:ablations} that MiniLM, contrary to MonoBERT, may not base its predictions on relevance signals, but on non-relevance signals. 
As described in the subsection \ref{subsection:ablations}, by training another MiniLM cross-encoder using harder negatives sampled from ColBERT-v2 \cite{santhanam2022colbertv2}, we further observed that this behavior may not be an artifact from the fine-tuning process or from the usage of random or ``easy'' negatives. Instead, it is perhaps tied to the pre-training regime of MiniLM (as we do not report the same trend for any of the two MonoBERT models), though the nature of the negatives seen during the fine-tuning also has an influence on its form. While early, this result strengthens the fact that not all cross-encoders share a unique relevance prediction process.

\subsection{RQ6: On the consistency of our findings across domains}

Overall, our findings indicate limited discrepancies in the models' behaviors in OOD compared to ID. Most of our observations on ID also verified themselves in OOD. This makes sense, as cross-encoders are considered to be relatively robust to domain shifts compared to other neural IR models like bi-encoders. However, previous works have suggested that there could exist some differences in MonoBERT's behavior in OOD \cite{Vast_2024}. In Section \ref{section:NIG}, we reproduced this finding but found it to be true mostly for MonoBERT (base and large), as MiniLM's most important neurons in both scenarios largely overlap. For MonoBERT, the role of these specific ``OOD'' neurons and attention heads remains opaque at this point. 
Through Sections \ref{section:information_flow} and \ref{section:understand_attention}, our observations and conclusions were consistent in both ID and OOD.
Finally, in Section \ref{section:subspace_analysis}, we uncovered that the construction of the relevance subspace(s) is hindered by domain shifts. We do not consider this to be a sign of a different mechanism but rather further evidence that transformer-based models, despite their reliance on lexical signals in information retrieval (IR), remain dependent on their domain knowledge and their semantic ability.

\subsection{Design implications}
\label{subsection:design-implications}

\paragraph{Impact of the query sampling}

Throughout this work, we relied on diverse sampling strategies to ensure a controlled setup but also to keep the computational cost of our study reasonable given its scale. We argue that as each sampling strategy is designed to fit the requirements of each experimental setup (reproducing an ``IR setup'', limiting the query-passage pairs or ensuring some levels of diversity in the qrels present in the final pools), and as we run our experiments on multiple datasets, their impact on our observations remains low, even for individual datasets such as NFCorpus or Natural Questions, which are arguably the most impacted datasets of our pool (see Table \ref{tab:sampling_results}).

We did not try to use a single sampling strategy across all our experiments, as none of them has the characteristics to fit all their requirements at once. For instance, while the BM25 sampling strategy is the closest to the IR task, and thus, perhaps the most relevant in the context of this study, we would not have been able to control the number of samples per relevance level for the Sections \ref{section:information_flow}, \ref{section:understand_attention} or \ref{section:subspace_analysis}. 

\paragraph{On the implications of using NIG to identify important neurons.}

While NIG attributes importance scores to every neuron in the model \emph{with regard to} the IR task, the roles of neurons with a score close to 0 or even largely negative are unclear at this point. By using triplets and back-propagating gradients from their contrast, we try to better target the core concept of ``relevance'' inside the model and attribute higher NIG scores to components closely related to it. However, this does not necessarily mean that the rest of the networks could be pruned without any impact on the model's performance. 
We have not explored the roles of neurons with negative NIG scores, and we do not have an explanation for the meaning of such scores. In our work, we restricted ourselves to exploring the roles of the most important neurons and attention heads, i.e., the ones with the highest NIG scores, and showed that our method allows us to identify key components in the relevance prediction process of cross-encoders.     

\paragraph{Information leakage in ``Attention Mask'' ablations.}

 Indeed, while it is more limited in the context of the other ablations, as we mask \textit{after} the softmax operation, following the Attention Mask procedure \cite{zhan2020bertanalysis}, the distribution of probabilities of each target token in each attention head is still updated based on the information available at this stage. In the case where we prevent all but one source token, as in the experiment where we blocked every transfer except the $\CLS \leftarrow \CLS$ in Section \ref{subsection:ablations} from sending information to the target token, its conditional probability of attention significantly impacts the target token. In that sense, although other tokens are hidden, their information can still influence the final results by modifying the conditional probability of attention of the last remaining source token. 

\paragraph{Is there a best tool to study the information flow?}

In Section \ref{section:information_flow}, we use 3 different approaches to study the information flow: the Neuron Integrated Gradients (NIG) \cite{sundararajan2017axiomatic,shrikumar2018computationally}, the Information Bottleneck (IB) \cite{Schulz2020Restricting} and an ablation study. Beyond the strengths and weaknesses listed in Table \ref{tab:ir-sota-overview}, we discuss here the insights and the granularity each of them brings.
From NIG, we obtain a relatively high-level picture of the information flow throughout the model. In addition, NIG results alone are hard to interpret in the context of the attention (see the analysis on the role of the \SEP{} in Section \ref{section:information_flow}), given the importance of the softmax operator, which greatly influences the roles of the tokens (attention sinks, no-op, etc.). While IB alleviates this limitation and can account for the interplay between the different components of the model, the learning procedure makes it harder to study multiple datasets. However, its success depends on finding the proper set of hyper-parameters (we present less accurate results for MonoBERT-large for example, for the same set of hyper-parameters). Finally, ablations can provide more causal results, informing us about the importance of a direction by measuring the impact of its absence on the final predictions. Ablations need no hyper-parameters, but a complete picture of every direction at every stage and across domains requires many targeted ablations, which is computationally intensive, and one must also make sure that no information leaks (see the previous point). 

It is hard to state whether one method is fundamentally better. From our study, we find it more interesting to combine their insights to provide a more complete picture and to make our conclusions more robust. 

\paragraph{Attention probability matrix and transformer-based model behavior.}

This question fits into the broader debate of using attention as an explanatory tool \cite{jain-wallace-2019-attention,wiegreffe-pinter-2019-attention}. In our case, we use the attention mechanism mostly in Sections \ref{section:information_flow} and \ref{section:understand_attention}. For Section \ref{section:information_flow}, we do not directly analyze the attention scores but instead analyze the flow of information with introspection techniques such as Neuron Integrated Gradients \cite{sundararajan2017axiomatic,shrikumar2018computationally}, ablations, or the Information Bottleneck approach \cite{Schulz2020Restricting}, which provide more meaningful results. 

In Section \ref{section:understand_attention}, we directly work on attention patterns and found that the results correlate with other results from the rest of our work, including methods that take into account the entire computational graph. For that reason, we believe that our method falls into the scope of the answer to the ``Attention is not Explanation'' paper \cite{jain-wallace-2019-attention} from \citeauthor{wiegreffe-pinter-2019-attention}~\cite{wiegreffe-pinter-2019-attention} so our results can be regarded as faithful and valuable because they are supported by other approaches.

Furthermore, we attempted to explain the matching behavior of the attention heads with the most distinctive patterns by analyzing the bilinear transformation defined by the Query-Key matrix $\left(W_Q^{h}\right)^{\top} W_K^{h}$ in Section \ref{subsection:matching_mechanism} in a mechanistic interpretability manner. As it captures most of the behaviors of the attention heads we identified, we also consider this experiment further validation of our analysis of the attention patterns.  

\subsection{On the impact of training strategies and models' depth}\label{discussion:design_choice_impact}

In addition to the discussions in Section \ref{subsection:rq5} about the consistencies of our findings across the models, we want to discuss more specifically the relationship between these differences and the pre-training strategies, as well as the size of the models. In our work, MonoBERT-base acts as ``pivot'' model, sharing both the number of layers of MiniLM and the pre-training strategy of MonoBERT-large. By comparing its behavior to those of MiniLM and MonoBERT-large, we can draw hypothesis on the source cause behind some of their traits.

In particular, from our analysis based on the Neuron Integrated Gradients in Sections \ref{section:NIG} and \ref{section:information_flow}, we noticed that MiniLM appears to be more robust to domain shift than both MonoBERT models, both in terms of what neurons are involved in the prediction process but also in terms of information flows. For the latter, this difference may be explained by the fact that the model has less ``space'' in its representations compared to MonoBERT-base or MonoBERT-large, but also by the specific pre-training of MiniLM \cite{minilmv2}. As the model has been pre-trained by distilling attention patterns from a teacher model, it is likely that this emphasis has led it to learn more consistent patterns. In contrast, models pre-trained on self-supervised tasks, such as Masked Language Modeling or Next Sentence Prediction, due to mechanisms such as the dropout which tends to add redundancy, may have learned more fuzzy attention patterns, which could have ultimately impacted their robustness. 
This hypothesis is further reinforced by the observation that MonoBERT-large is seemingly integrating relevance information directly inside the \CLS{} token, much more at least than MiniLM. This could also be a direct cause of the NSP pre-training, where the \CLS{} is used to predict if two sentences are consecutive or not. Interestingly, we have seen that MonoBERT-base shares a similar habit of integrating information of relevance inside the \CLS{}, but Section \ref{section:subspace_analysis} also revealed that it also uses the document and query tokens, just as MiniLM. Since depth is the only characteristic these two models share, our hypothesis is that building the relevance signal inside a single token, the \CLS{}, requires many layers. Consequently, smaller models may have learned to share this task across other tokens, in this case query and/or document tokens, and to only integrate ``refined'' signals inside the \CLS{}. 

From Section \ref{section:understand_attention}, we saw that the behaviors of the attention heads in MiniLM and MonoBERT-base are more robust to harder negatives and domain shifts than the ones of MonoBERT-large. We attribute this to the fact that MonoBERT-large has been fine-tuned with a simpler Binary Cross Entropy loss \cite{monobert}, while the cross-encoder based on MiniLM has been fine-tuned using a listwise loss\footnote{The exact loss used to fine-tune the MiniLM cross-encoder is hard to trace back from the documentation (available at \url{https://github.com/huggingface/sentence-transformers/tree/main/examples/cross_encoder/training/ms_marco}) but the authors indicate it could have been a LambdaLoss \cite{lambda_rank}.} and our MonoBERT-base with InfoNCE~\cite{infonce}. Figure \ref{fig:controling_relevance} underlines the impact of the negative sampling strategy on a model's behavior.

At the same time, and while all three models accumulate relevance inside the query tokens in the last few layers, the analysis of the matching patterns revealed distinct patterns between MiniLM and the two MonoBERT models. Indeed, while the difference in the attention weights is more pronounced when the document is relevant for both MonoBERT models (for the direction $Q \leftarrow D$), some attention heads in the last layers of MiniLM instead exhibit higher attention weights when the passage is not relevant. While there is no obvious link between this ability and the pre-training strategy of MiniLM (its most prominent unique trait), we do not have as of today, better explanation.

Some questions remain open: why MonoBERT-base relies less on its \CLS{} than MonoBERT-large despite the same pre-training recipe, and whether one strategy is more effective than the other.

\subsection{Implications of our findings for future work}

In general, interpretability studies constitute solid foundations to improve neural models. By revealing key mechanisms or how specific computations are implemented, it becomes possible to leverage architectural improvements and deduce the importance of design choices.
For instance in this work, we confirmed the prevalence of the following three steps in the three cross-encoders we studied, to predict the relevance of a document to a query:
\begin{enumerate*}
    \item Query and document independent contextualization across the first layers ($Q \leftarrow Q$ and $D \leftarrow D$);
    \item Query and document interactions, especially through the attention ($Q \leftarrow D$);
    \item Transmission of the relevance signals (``raw'' or ``refined'') to the \CLS{} token for the final prediction, from the query ($\CLS \leftarrow Q$).
\end{enumerate*}

In a recent work \cite{vast2026miceminimalinteractioncrossencoders}, we directly leverage this insight to design an optimized cross-encoder where only the most important directions, i.e., that preserve the previous steps, are kept. Thanks to the present interpretability study, we have been able to correctly identify and remove superfluous interactions from cross-encoders. This led to the improvement of the efficiency/effectiveness trade-off, with much lower latencies and also better OOD performance.

In addition, our work also raised some interesting questions, such as why MiniLM seems to have learned to search for ``non-relevance'' signals, instead of relevance ones, or how exactly attention heads learn to detect query-document matches. While we leave them unanswered at this stage, we believe they are interesting directions for future work.

\section{Limitations}

In this study, cross-encoders based on encoder-decoder \cite{monot5,rankt5} or decoder-only \cite{rankllama} architectures are not considered, which restricts the scope of our findings to encoder-only, and more specifically BERT-based, cross-encoders. Regarding the latter, and although comparisons would have been of interest, decoder-only models are not primarily designed for the IR task, as suggested by a recent study comparing encoder-only and decoder-only models of equal size across various NLP tasks \cite{ettin}. Regarding the former, while extending our experiments to encoder-decoder models such as RankT5 \cite{rankt5} would have yielded valuable insights, it would have simultaneously introduced additional layers of complexity into our analysis. First, prompt design is non-trivial and has been shown to heavily influence cross-encoder predictions \cite{marco_t5_interp}. Second, the cross-attention heads and the decoder side introduce an additional layer that requires explanation and remains largely under-explored. For these reasons, our efforts were focused on encoder-only cross-encoders.

The extension of this work to more recent encoders, such as DeBERTa-v3 \cite{he2021debertav3}, ELECTRA \cite{clark2020electra}, or ModernBERT \cite{modernbert}, is left for future work. Given the scale of our experimental setup, the decision was made to restrict our scope to BERT-based cross-encoders. Building on our discussion of the impact of pre-training and layer depth (see section \ref{discussion:design_choice_impact}), we believe the series of steps we identified is shared across encoders beyond BERT. Nevertheless, each model may implement a given step differently, and it therefore cannot be asserted that all encoder-only cross-encoders share a single, uniform behavior.

A further limitation, already noted in the Design Implications section \ref{subsection:design-implications}, concerns the accuracy of the Information Bottleneck results, particularly for the MonoBERT models. Because the method learns masks that depends on a carefully selected threshold to preserve performance while removing as many directions as possible, its results are hard to interpret in isolation. By combining IB with experiments based on NIG and ablations, and given that their results largely overlap, this limitation can be mitigated; however, further work is needed to design a more robust experimental plan specific to the Information Bottleneck.

\section{Conclusion}

In this work, we focused on better explaining the most effective IR models, namely the cross-encoders.
This lack of knowledge is a long-standing issue in the IR community that previous works have already attempted to tackle. After reviewing the current state of knowledge on interpreting cross-encoders, we identified that the main limitations are a restricted scope of study (one mechanism) and of applications (single model and single domain, often MS MARCO). In contrast, we comprehensively study the relevance prediction process inside three cross-encoders over multiple datasets, both in-domain and out-of-domain.

The different steps of this process have been roughly hypothesized in the past \cite{zhan2020bertanalysis}. Thanks to multiple experiments and the correlation of their results, we have been able to identify their nature and offer insights into their functioning:
\begin{enumerate}
    \item Initially, tokens lack contextualization, so the first mission of cross-encoders, similarly to transformer-based models, is to contextualize them to gain a semantic understanding of the query-passage pair. At this stage, self-contextualization (attention from the query or the document to itself) plays a major role, but we further showed that cross-contextualization also occurs at this stage (see Section \ref{section:information_flow}). By studying the attention process more in-depth (see Section \ref{section:understand_attention}), we showed that through early to middle layers, some attention heads are specialized in detecting \emph{lexical} matches between the query and the document.
    \item Moving forward in the relevance prediction process and the model's layers, we discovered other attention heads specialized in detecting query-document matches, but this time \emph{semantic}. This happens later in cross-encoders, as the input tokens need to be contextualized first.

    For both types of matching, we proposed an explanation for their detection (see subsection \ref{subsection:matching_mechanism}).
    \item Once these matching signals (and perhaps other signals) have been detected, cross-encoders refine and gather them inside a \emph{relevance subspace}. This step is decomposed into two different sub-steps, as cross-encoders first gather relevance signals in the query tokens, before transferring them to the \CLS{}. Importantly, we noticed that while MonoBERT models base their predictions on evidence that the passage is ``relevant'', MiniLM may instead be looking for the opposite, i.e., evidence that the passage is ``non-relevant''.

    At this stage, we noticed distinct strategies across the models, probably due to the different pre-training strategies. While MonoBERT models construct a relevance subspace inside the \CLS{} token and need multiple layers to accumulate the ``relevance'' signals and refine them, MiniLM constructs multiple relevance subspaces at the level of the query and document tokens (see Section \ref{section:subspace_analysis}). It suggests that MiniLM refines the ``relevance'' signals it has detected before sending them to the \CLS{}, while MonoBERT gathers inside it ``raw'' signals. Note that unlike its larger counterpart, MonoBERT-base also tends to construct relevance subspaces at the level of the query and document tokens, in a similar fashion to MiniLM. Since both models have the same depth, we hypothesize that the number of layers is a bottleneck for building a relevance subspace inside the \CLS{} alone, which smaller models avoid by spreading the task across tokens.  
    \item Finally, the ``relevance'' signals are refined/aggregated inside the \CLS{} before the final relevance prediction. 
\end{enumerate}

By comparing three BERT-based cross-encoders, of different sizes (from 33M to 300M) and trained following different strategies, we reinforced the hypothesis that this multi-stage process is shared by most BERT-based cross-encoders (even if they differ in their final processing). In addition, we also showed that this process is consistent across domains, which might explain their better generalization ability compared to other neural IR models. Our work, however, underlined how difficult it is for these models to deal with hard negatives, especially in OOD. It suggests that cross-encoders may over-rely on \emph{lexical} matches when they fail to capture the semantics of a new domain. 

Following this in-depth comparison of MiniLM and MonoBERT, we can assert that they have comparable performance, despite major differences in their behaviors. Some of our findings further indicate that MonoBERT (large and base) might be over-parameterized (see Section \ref{section:NIG}). Given its slightly better performance (see Table \ref{tab:reranking_comparison}) and the greater robustness of its mechanism across domains, MiniLM-L12-v2 is a more reasonable choice than MonoBERT-base or large, for both the efficiency-effectiveness trade-off and generalization.

Overall, our work provides a robust foundation for future interpretability studies in IR. We demonstrated the importance of including various datasets and considering more than a single model to put results into perspective and strengthen conclusions. Some of the insights we unveiled, such as the fact that MiniLM may rely on ``non-relevance'' signals and how some attention heads can detect query-document matches, also constitute interesting directions for future work.


\begin{acks}
The authors acknowledge the ANR – FRANCE (French National Research Agency) for its financial support of the GUIDANCE project n°ANR-23-IAS1-0003 as well as the CLUSTER IA project n°ANR-23-IACL-0007.

Furthermore, this work was granted access to the HPC resources of IDRIS under the allocation 2024-AD011014444R1 made by GENCI.
\end{acks}

\bibliographystyle{ACM-Reference-Format}
\bibliography{biblio}

\newpage 
\appendix



\listoffigures

\section{Models}

\subsection{MonoBERT-base details}\label{appendix:monobert_base_details}

Given that MonoBERT-large and the cross-encoder based on MiniLM do not share the same training loss (binary cross-entropy for MonoBERT-large, and an unknown loss for MiniLM\footnote{Sentence-transformers, who trained the model, does not detail the loss that was used.}), but also to keep the number of experiments reasonable, we decided to follow recent studies training cross-encoders \cite{squeezing_water,schlatt:2025} and to use an InfoNCE loss, with 7 negatives and 1 positive per query. The model's re-ranking performance can be found in \autoref{tab:reranking_comparison}.

\subsection{Overview of the models}

\begin{table}[h]
\caption{Details of the models in terms of layer numbers and structures, as well as overall parameter counts.}
\begin{tabular}{|c|c|c|c|c|}
\hline
\textbf{Model} & \textbf{\# layers} & \textbf{\# attention heads per layer} & \textbf{Hidden size} & \textbf{\# params} \\ \hline \hline
MiniLM-L12-v2 \cite{minilmv2} & 12 & 12 & 384 & 33M \\ 
\hline 
MonoBERT-base \cite{monobert} & 12 & 12 & 768 & 110M \\\hline 
MonoBERT-large \cite{monobert} & 24 & 16 & 1024 & 340M \\ \hline 
\end{tabular}
\label{tab:models-overview}
\end{table}

\subsection{Re-ranking performance of the models}
\label{annex:baselines}

Table \ref{tab:reranking_comparison} details the performance on the re-ranking task of the cross-encoders with our pool of datasets. We report their performance when re-ranking the top 100 passages retrieved by BM25 (using default values for the hyper-parameters $k_1=0.9$ and $b=0.4$) in \textit{Anserini} \cite{anserini}, along with the performance of BM25.

Furthermore, Table \ref{tab:reranking_comparison} also provides references for the ranking performance of a \textbf{Random} classifier and for the \textbf{Worst} (possible) classifier, on each dataset. One can observe that in the case of the \textbf{Worst} classifier, some datasets have an above 0 nDCG@10 value, in particular TREC-DL19 \cite{craswell2019overview} (0.02), NFCorpus \cite{nfcorpus} (0.12) and TREC-Covid \cite{trec-covid} (0.01). This is mostly due to the composition of the datasets themselves: 1) The unexpectedly high ``worst'' nDCG@10 value for NFCorpus comes from the distributions of the annotations in the dataset. Table \ref{tab:datasets-passages} highlights that there are only around 10 passages per query with a relevance level of 0. By construction of the ``worst'' possible ranking, this leads to a higher proportion of relevant passages in the top-100 passages and potential leaks in the top-10, which then explains the high nDCG@10 value. Similarly, TREC-Covid contains both a very high number of relevant passages per query and very few queries according to Table \ref{tab:datasets-passages}. This could explain that for a few of them, one relevant passage managed to make its way to the top-10 and then explains the above 0 nDCG@10. The same explanation may also apply to TREC-DL19: dividing the number of passages annotated with a relevance level of 1, 2 or 3 by the 43 queries of this track yields around 150 relevant passages per query. Note also that the annotation process of TREC-DL19 was supported by BM25, which further increases the probability that marginally relevant passages could leak into the top-10. 

\begin{table}[ht]
\caption{Performance across datasets in nDCG@10. \textbf{Random} corresponds to the nDCG@10 value obtained after averaging the score over 1000 random rankings. \textbf{Worst} corresponds to the nDCG@10 value obtained after re-ordering the top passages retrieved by BM25 in the increasing order of their relevance level (least relevant first).}
\begin{tabular}{l|cccc|cccccc|c}
\toprule
\textbf{Model} & \textbf{DL 19} & \textbf{20} & \textbf{21} & \textbf{22} & \textbf{NFCorp.} & \textbf{FiQA} & \textbf{T.-Covid} & \textbf{Rob.-04} & \textbf{NQ} & \textbf{BioASQ} & \textbf{Avg.} \\
\midrule
BM25 & 0.51 & 0.48 & 0.45 & 0.27 & 0.32 & 0.24 & 0.59 & 0.45 & 0.23 & 0.52 & 0.41 \\
\texttt{MiniLM} & 0.74 & 0.68 & 0.65 & 0.49 & 0.35 & 0.31 & 0.74 & 0.46 & 0.36 & 0.50 & 0.53 \\
\texttt{MonoBERT-base} & 0.73 & 0.68 & 0.64 & 0.49 & 0.35 & 0.30 & 0.73 & 0.47 & 0.36 & 0.50 & 0.52 \\
\texttt{MonoBERT-large} & 0.71 & 0.68 & 0.63 & 0.49 & 0.35 & 0.32 & 0.70 & 0.45 & 0.35 & 0.47 & 0.52 \\
\midrule
\midrule
\texttt{Random} & 0.24 & 0.17 & 0.24 & 0.18 & 0.23 & 0.04 & 0.40 & 0.19 & 0.07 & 0.06 & 0.18 \\
\texttt{Worst} & 0.02 & 0.00 & 0.00 & 0.00 & 0.12 & 0.00 & 0.01 & 0.00 & 0.00 & 0.00 & 0.02 \\
\bottomrule
\end{tabular}
\label{tab:reranking_comparison}
\end{table}

\newpage

\section{Datasets}

\subsection{Overview}

\begin{table}[h]
\centering
\caption{Details of the datasets, including the task associated with it, their respective number of topics and the distribution of the annotations, per relevance-level. Relevance levels in bold correspond to the label ``relevant'' after binarization. ``MS MARCO'' corresponds to the aggregation of the 4 tracks of TREC-DL.}
\begin{tabular}{ccccc}
\toprule
\textbf{Datasets} & \textbf{\# Topics} & \textbf{Relevance Levels} & \textbf{\# Qrels} & \textbf{Total} \\
\midrule
MS MARCO \cite{MSMARCO,craswell2019overview,craswell2020overview,craswell2021overview,craswell2022overview} & 226 &  0 (not relevant) & 303K & - \\
(Web Retrieval) & & 1 (related) & 59K & - \\
         & & \textbf{2 (relevant)}& 51K & - \\
         & & \textbf{3 (perfectly relevant)} & 4K & 417K \\
\addlinespace \hline
\makecell{FiQA \cite{fiqa} \\(Question Answering)} & 648 & \textbf{1 (relevant)} & 57K & 57K \\
\addlinespace \hline
Natural Questions & 6.5K & -1 (easy negatives) & 326K & - \\
(Question Answering) & & 0 (hard negatives) & 603K & - \\
                  & & 1 (contains the answer, retrieved by BM25) & 45K & - \\
                  & & \textbf{2 (relevant)} & 6.5K & 980K \\
\addlinespace \hline
\makecell{BioASQ  \cite{bioasq}\\(BioMedical-IR)} & 500 & \textbf{1 (relevant)} & 2359 & 2359 \\
\addlinespace \hline
TREC-Covid \cite{trec-covid} & 50 & 0 (not relevant) & 43K & - \\
(BioMedical-IR) & & 1 (partially relevant) & 11K & - \\
            & & \textbf{2 (relevant)} & 16K & 69K \\
\addlinespace \hline
NFCorpus \cite{nfcorpus} & 323 & 0 (doc. is marginally relevant to a related query) & 3.5K & - \\
(BioMedical-IR) & & 1 (doc. is relevant to a related query) & 11.8K & - \\
        & & \textbf{2 (directly relevant)} & 576 & 15.8K \\
\addlinespace \hline
TREC-Robust04 \cite{robust04} & 249 & 0 (not relevant) & 294K & - \\
(News Retrieval) & & \textbf{1 (relevant)} & 16K & - \\
            & & \textbf{2 (highly relevant)} & 1K & 311K \\
\bottomrule
\end{tabular}
\label{tab:datasets-passages}
\end{table}

\subsection{Pre-processing} 
\label{appendix:sampling_strategies}

\begin{table}[htbp]
\centering
\caption{Proportion of the original set of queries remaining after the different sampling strategies per dataset. Note that for Natural Questions, the reported ratios are below one because of the upper bound on the number of queries per dataset, set to 1000 ($\frac{1}{3.4}\approx0.29$). For NFCorpus, however, the low ratio for Diverse Sampling is due to the filtering of the queries without ``relevant'' passages after the binarization (see Table \ref{tab:datasets-passages}) or because the model does not recognize them as ``relevant''.}
\begin{tabular}{l|ccc}
\toprule
\textbf{Datasets} & \textbf{Top Sampling} & \textbf{Diverse Sampling} & \textbf{BM25 Sampling} \\  
\midrule
\textbf{TREC-DL 19} & 1 & 1 & 1 \\
\textbf{TREC-DL 20} & 0.96 & 1 & 1 \\
\textbf{TREC-DL 21} & 0.96 & 1 & 1 \\
\textbf{TREC-DL 22} & 0.99 & 1 & 1 \\
\textbf{NFCorpus} & 0.13 & 0.37 & 0.95 \\
\textbf{FiQA} & 0.46 & 1 & 1 \\
\textbf{TREC-Covid} & 0.98 & 1 & 1 \\
\textbf{Robust-04} & 0.57 & 1 & 1 \\
\textbf{NQ} & 0.08 & 0.29 & 0.29 \\
\textbf{BioASQ} & 0.78 & 1 & 1 \\
\bottomrule
\end{tabular}
\label{tab:sampling_results}
\end{table}

\begin{figure}[H]
    \centering
    \includegraphics[width=0.8\linewidth]{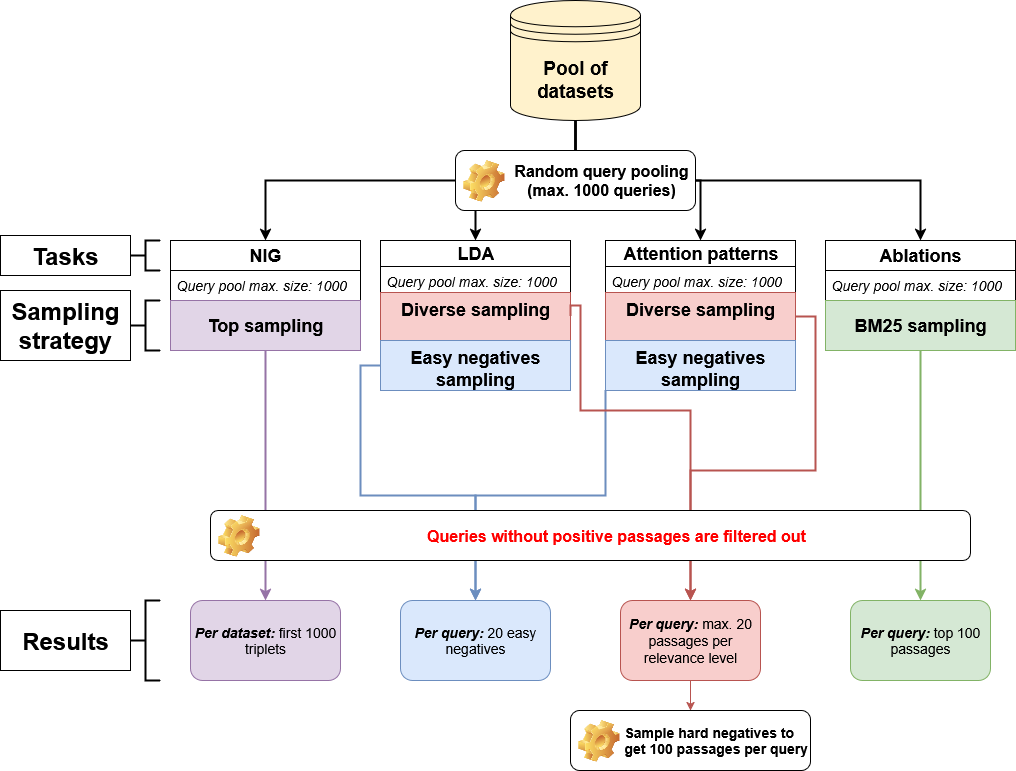}
    \caption[Description of the sampling strategies]{Description of the different pre-processing steps we follow for each of our experiments. \textbf{Easy negatives sampling} indicates that, in addition to the pool of 100 passages for each query, we further randomly sample 20 passages, not already in the pool of assessments for that query, in the dataset to serve as \emph{easy negatives}.}
    \label{fig:dataset_construction}
\end{figure}

To fit our needs in the different experiments that we run in this work, we applied different pre-processing strategies to our datasets. Figure \ref{fig:dataset_construction} illustrates the different steps of each strategy we performed to 1) limit the maximum number of queries considered and 2) limit the number of passages per query.
The first step, shared across all tasks, consists of randomly sampling a maximum of 1000 queries (\textbf{Random Query Pooling} in Figure \ref{fig:dataset_construction}) per dataset (only impacts Natural Questions as shown in Table \ref{tab:datasets-passages}). Then, for this subset of queries, we sample the passages depending on the analysis we wish to perform, giving rise to three different \textbf{sampling strategies}: 
\begin{itemize}
    \item[1-] \textbf{Top sampling} means we sample 1000 query-passage pairs of each dataset. We apply it only to the experiments based on NIG (see Sections \ref{section:NIG} and \ref{section:information_flow}) to follow the work done in \citeauthor{Vast_2024}~\cite{Vast_2024}.
    \item[2-] \textbf{Diverse sampling} means we ensure the different relevance levels are present (when available in the annotations) in the pool of passages of each query. For this strategy, we fix both a maximum number of passages per relevance level and a target number of passages per query. If the latter is not satisfied after sampling passages in the annotations, we use BM25\footnote{For pre-processing, BM25 is used with the same default values for each dataset, i.e., $k_1=0.9$ and $b=0.4$.} \cite{bm25} and sample hard negatives up to the target number of passages. Once we sampled the target number of passages per query, we sampled easy negatives (see \textbf{Easy negatives sampling} on Figure \ref{fig:dataset_construction}). This strategy is used for both LDA (see Section \ref{section:subspace_analysis}) and the analysis of the attention patterns (see Sections \ref{section:information_flow} and \ref{section:understand_attention}) to ensure a certain level of variety in the distribution of the data. Note that to simplify our analysis of the results based on this strategy, we considered a passage as ``relevant'' (resp. ``non-relevant'') if, and only if, it is annotated as such in the original datasets \emph{and} the model has classified it as such.
    \item[3-] \textbf{BM25 sampling} means the selection of k (usually 100) passages per query using BM25. We use this strategy for the ablation study (see Section \ref{section:information_flow}) and evaluations, as it mimics a typical IR setup and allows us to get meaningful results before and after the ablations that we can compare.
\end{itemize} 

As we pre-processed each dataset in different ways to fit our needs in this work, we detail in Table \ref{tab:sampling_results} the proportion of queries (from the original set) still present after each sampling and filtering strategy.  

\section{Ablation study}

\begin{figure}[H]
    \centering
    \begin{subfigure}[t]{0.5\linewidth}
        \centering
        \includegraphics[width=\textwidth]{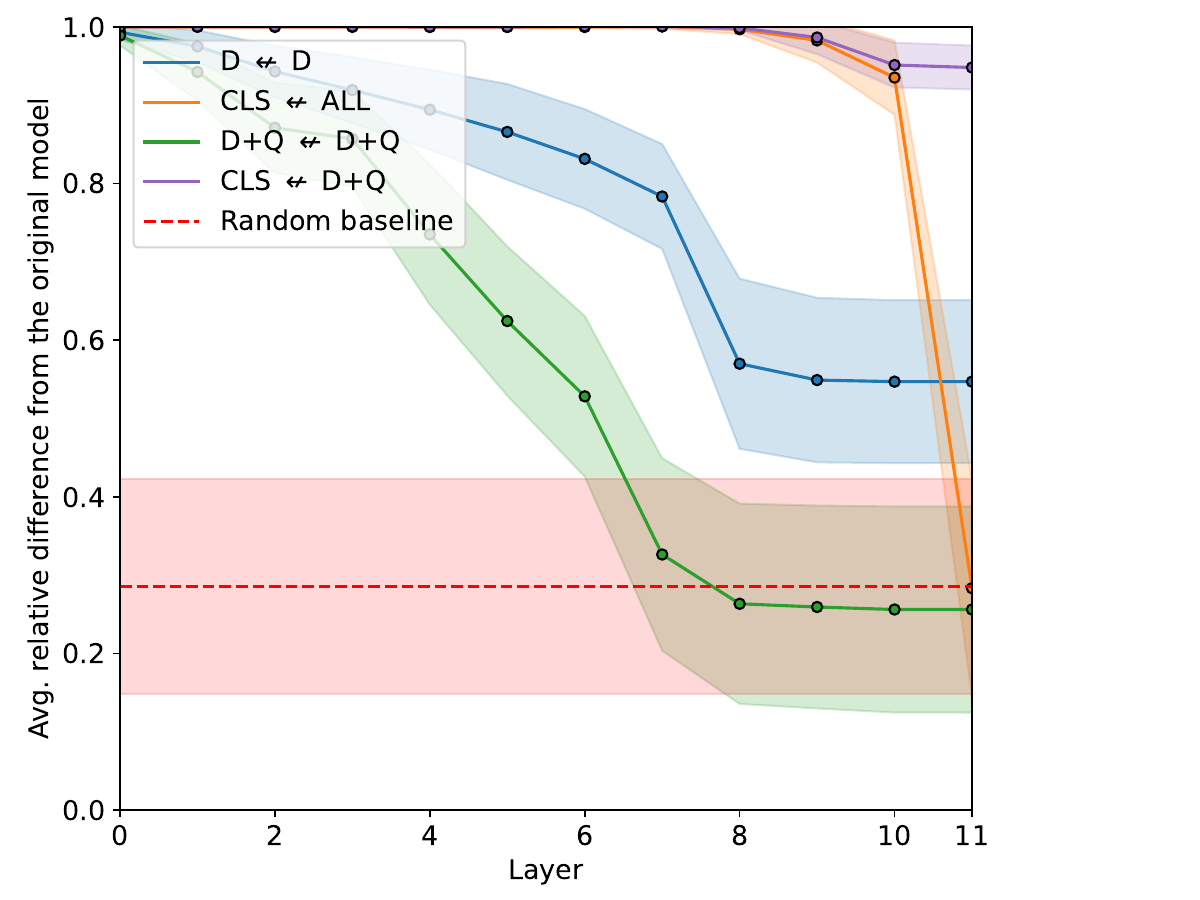}
        \caption{Average percentage of the original performance preserved on MiniLM.}
        \label{fig:minilm_all_ablations}
    \end{subfigure}%
    \begin{subfigure}[t]{0.5\linewidth}
        \centering
        \includegraphics[width=\textwidth]{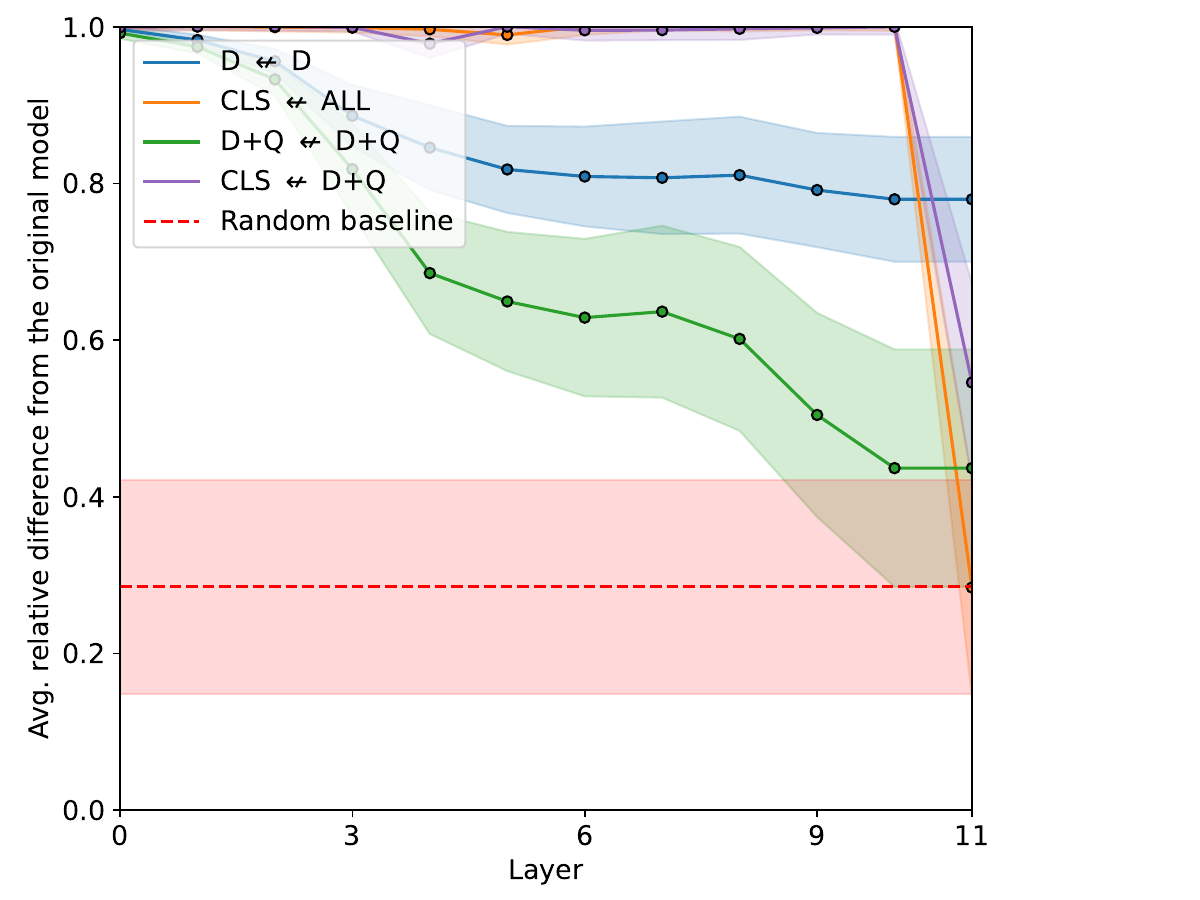}
        \caption{Average percentage of the original performance preserved on MonoBERT-base.}
        \label{fig:monobert_base_all_ablations}
    \end{subfigure}
    
    \begin{subfigure}[t]{0.5\linewidth}
        \centering
        \includegraphics[width=\textwidth]{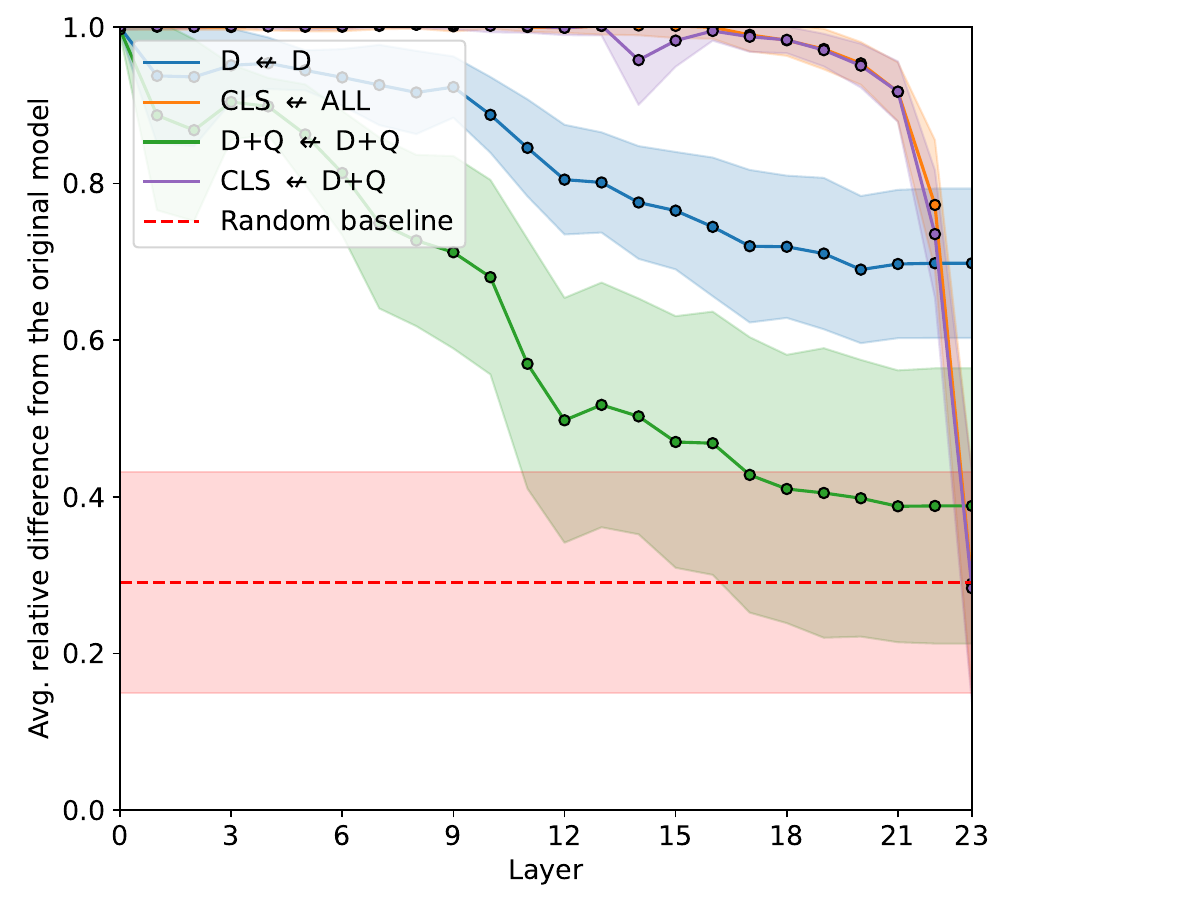}
        \caption{Average percentage of the original performance preserved on MonoBERT-large.}
        \label{fig:monobert_all_ablations}
    \end{subfigure}%
    \caption[Summary of the ablation results across all the datasets]{Average percentage of the original performance preserved after the ablations of directions inside the attention heads of the models up to the layer $\ell$. The average is computed over all the datasets.}
    \label{fig:all_ablations}
\end{figure}

\newpage 
\section{Neuron Integrated Gradients}
\label{appendix:nig}

\begin{figure}[h]
    \centering    
    \begin{subfigure}[h]{0.44\textwidth}
        \centering
        \includegraphics[width=\textwidth]{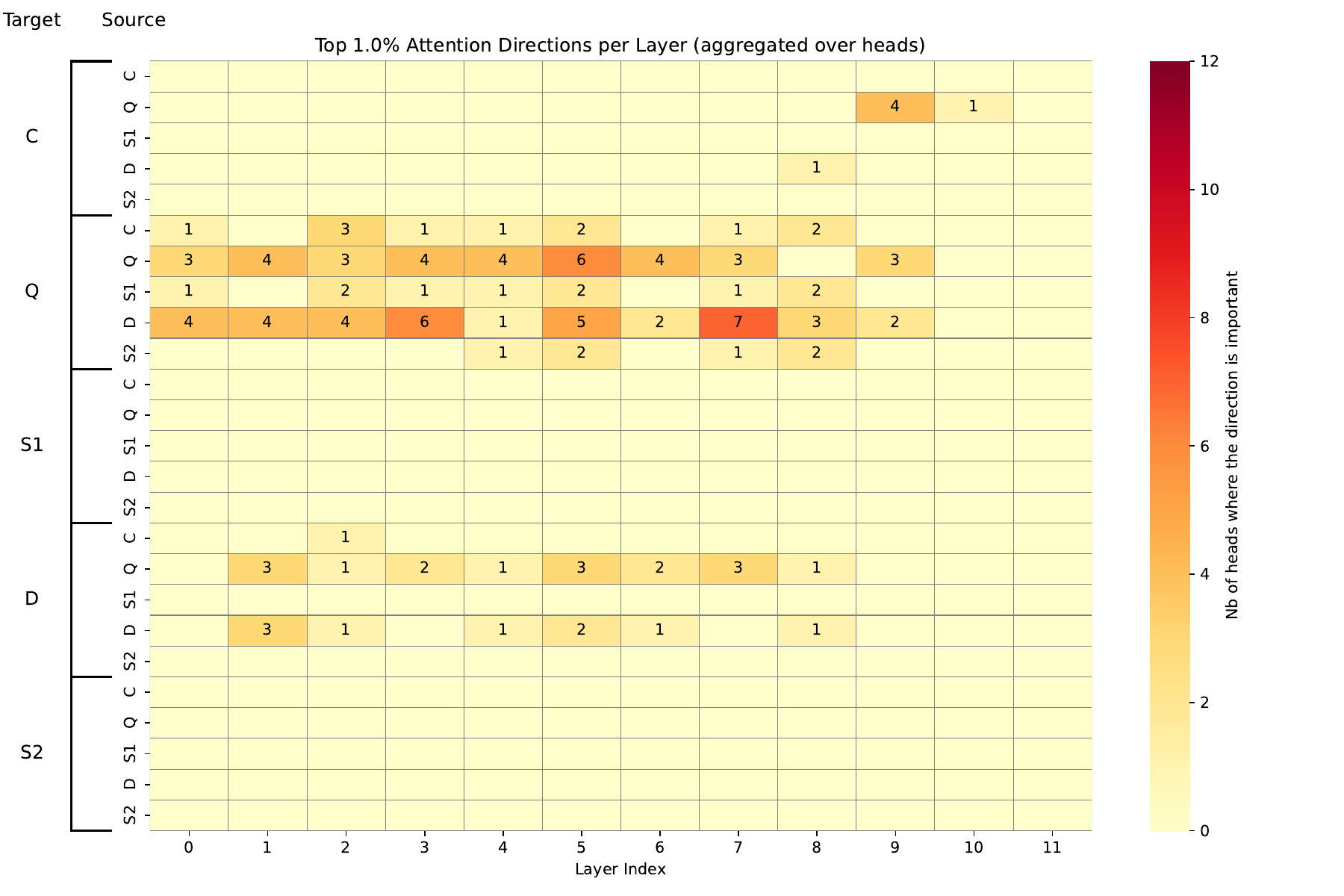}
        \caption{Ablation scheme $I_M$ for MiniLM.}
    \end{subfigure}
    \hspace{0.02\textwidth} 
    \begin{subfigure}[h]{0.47\textwidth}
        \centering
        \includegraphics[width=\textwidth]{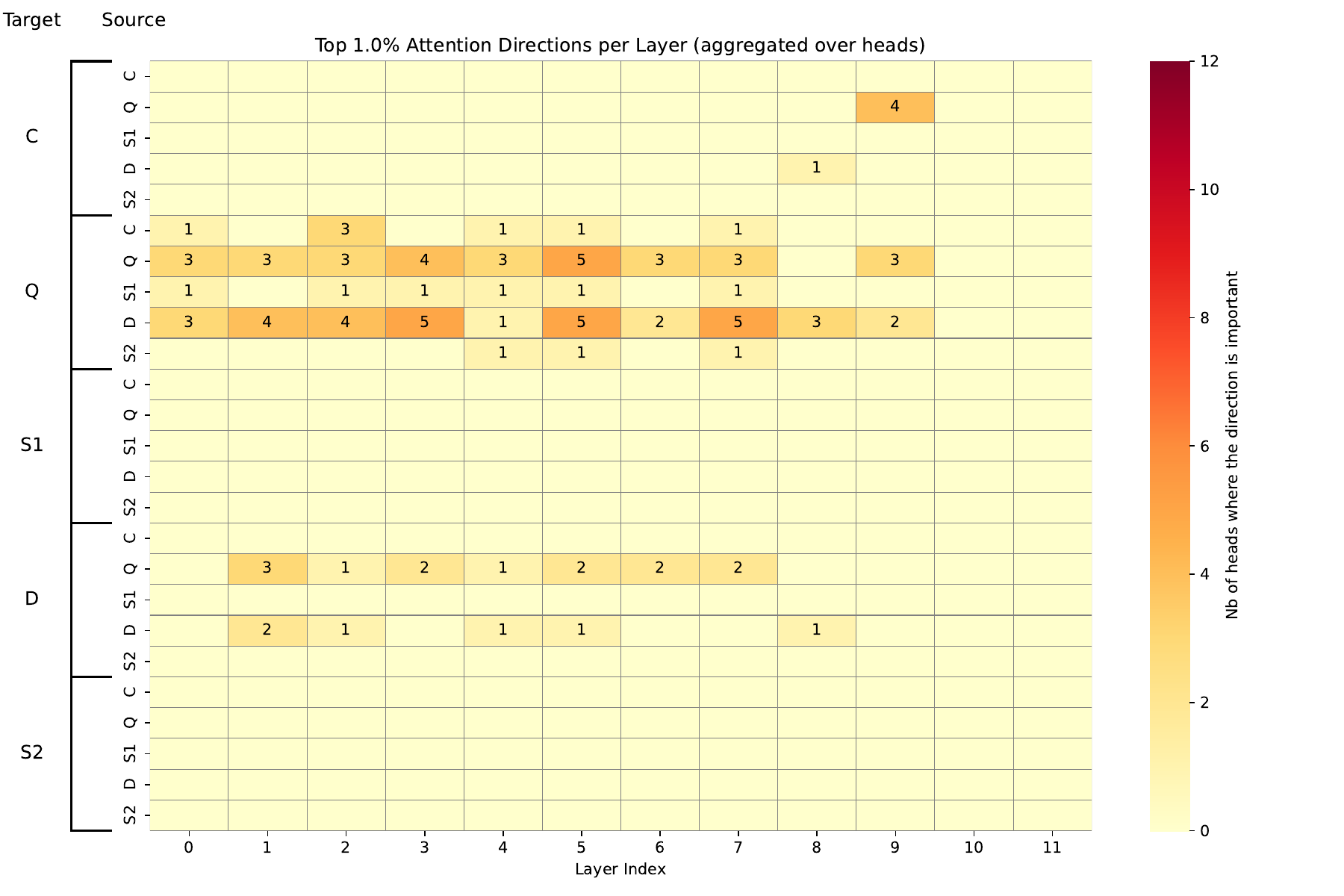}
        \caption{Ablation scheme $F_M$ for MiniLM.}
    \end{subfigure}

    \begin{subfigure}[h]{0.44\textwidth}
        \centering
        \includegraphics[width=\textwidth]{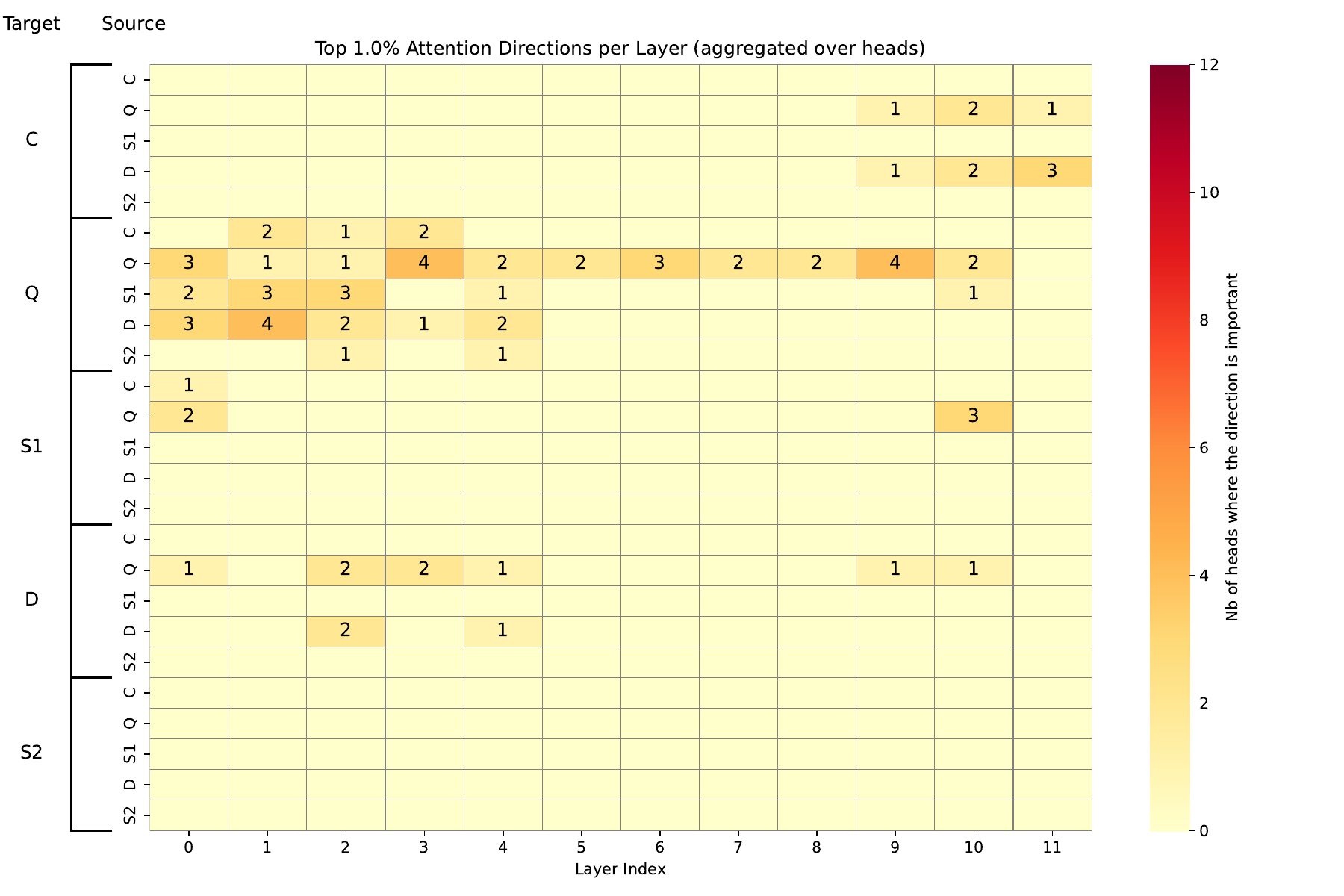}
        \caption{Ablation scheme $I_M$ for MonoBERT-base.}
    \end{subfigure}
    \centering
    \hspace{0.02\textwidth} 
    \begin{subfigure}[h]{0.47\textwidth}
        \centering
        \includegraphics[width=\textwidth]{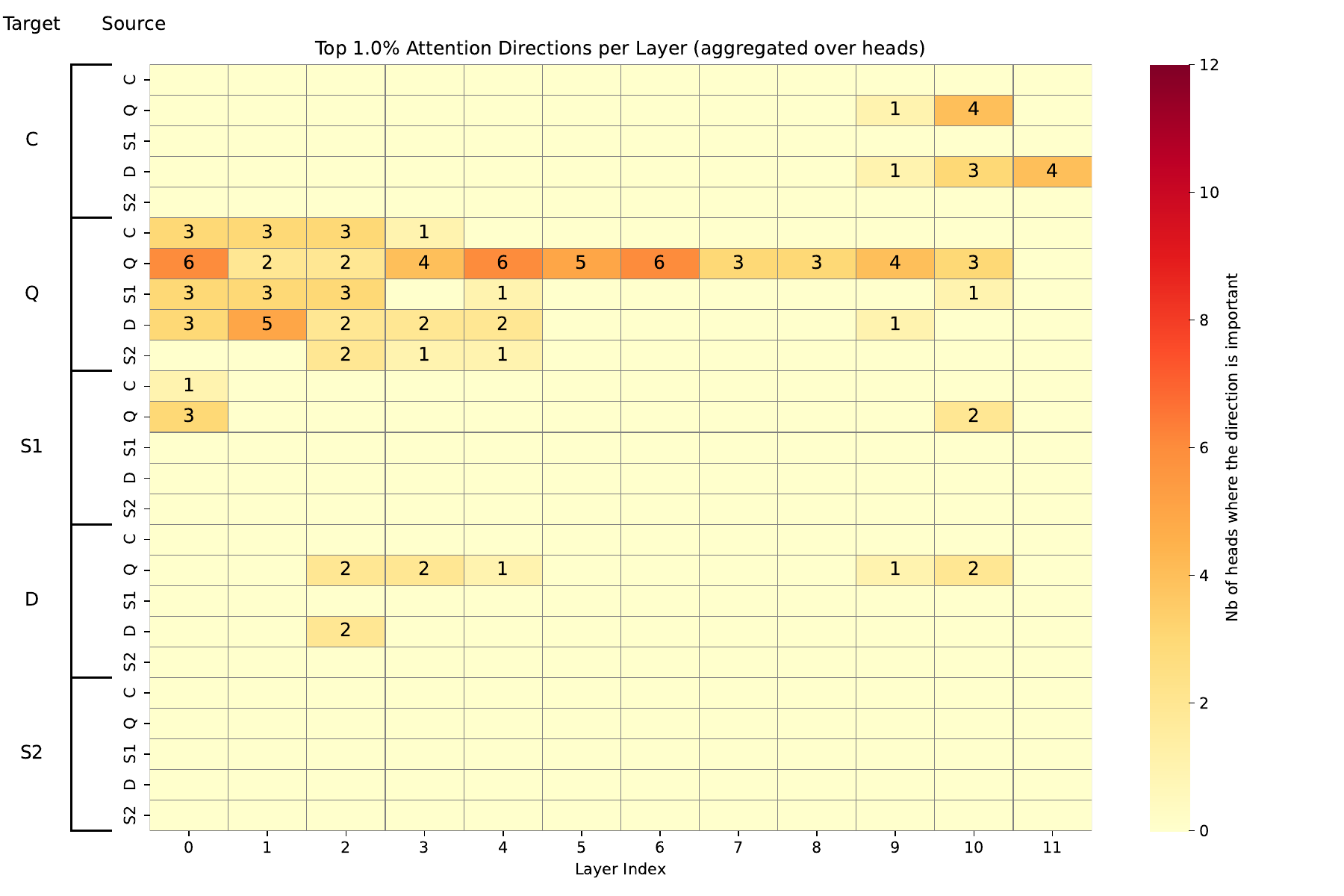}
        \caption{Ablation scheme $F_M$ for MonoBERT-base.}
    \end{subfigure}
    
    \begin{subfigure}[h]{0.44\textwidth}
        \centering
        \includegraphics[width=\textwidth]{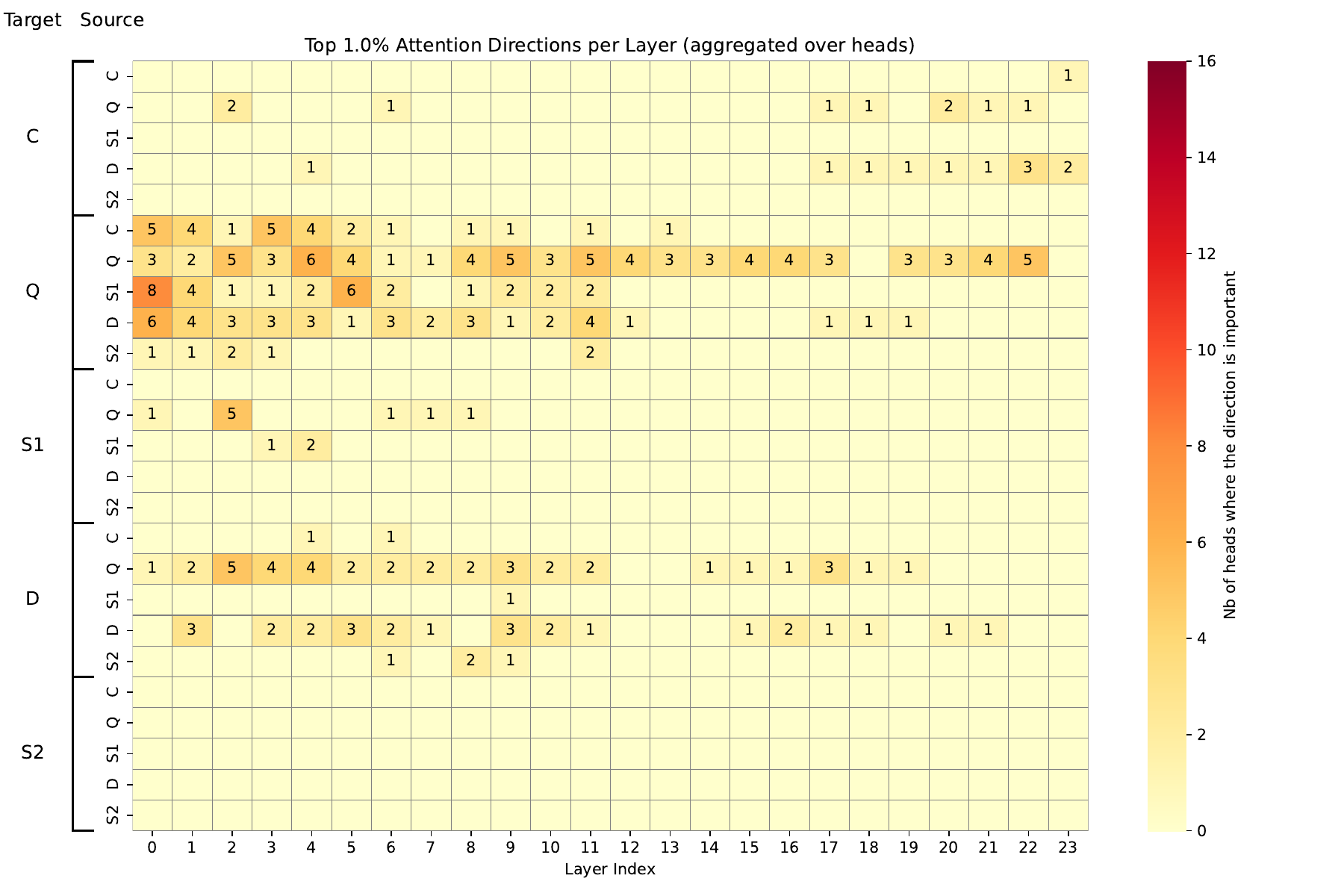}
        \caption{Ablation scheme $I_M$ for MonoBERT-large.}
    \end{subfigure}
    \centering
    \hspace{0.02\textwidth} 
    \begin{subfigure}[h]{0.47\textwidth}
        \centering
        \includegraphics[width=\textwidth]{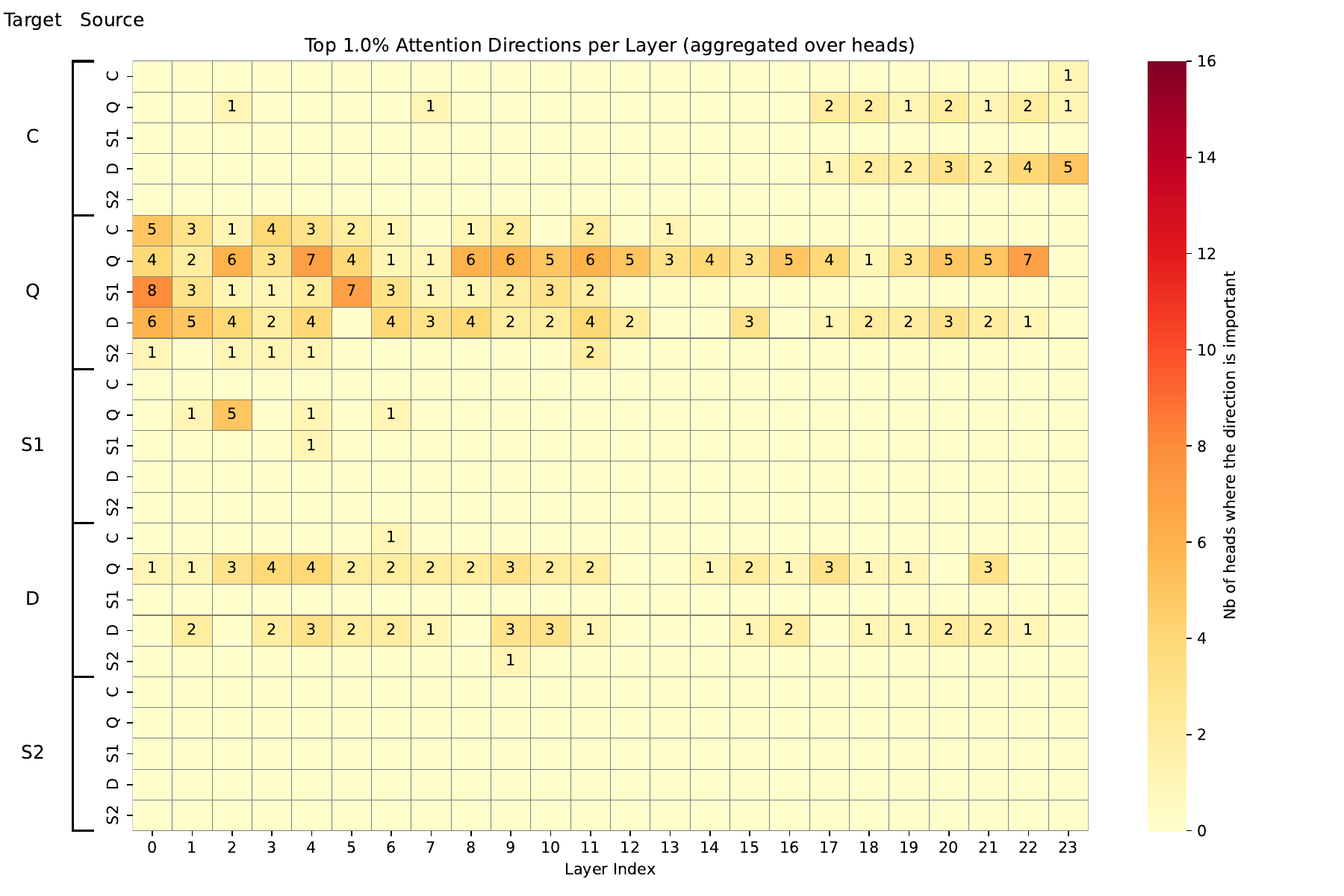}
        \caption{Ablation scheme $F_M$ for MonoBERT-large.}
    \end{subfigure}
    
    \caption[NIG results on the information flow for $I_M$ and $F_M$]{Distribution of the most important attention heads, relative to each input part, per layer, following the ablation schemes $I_M$ and $F_M$ at pruning percentage $t=1\%$.}
    \label{fig:appendix_nig_attn_msmarco}
\end{figure}

\begin{figure}[h]
    \centering
    \begin{subfigure}[h]{0.47\textwidth}
        \centering
        \includegraphics[width=\textwidth]{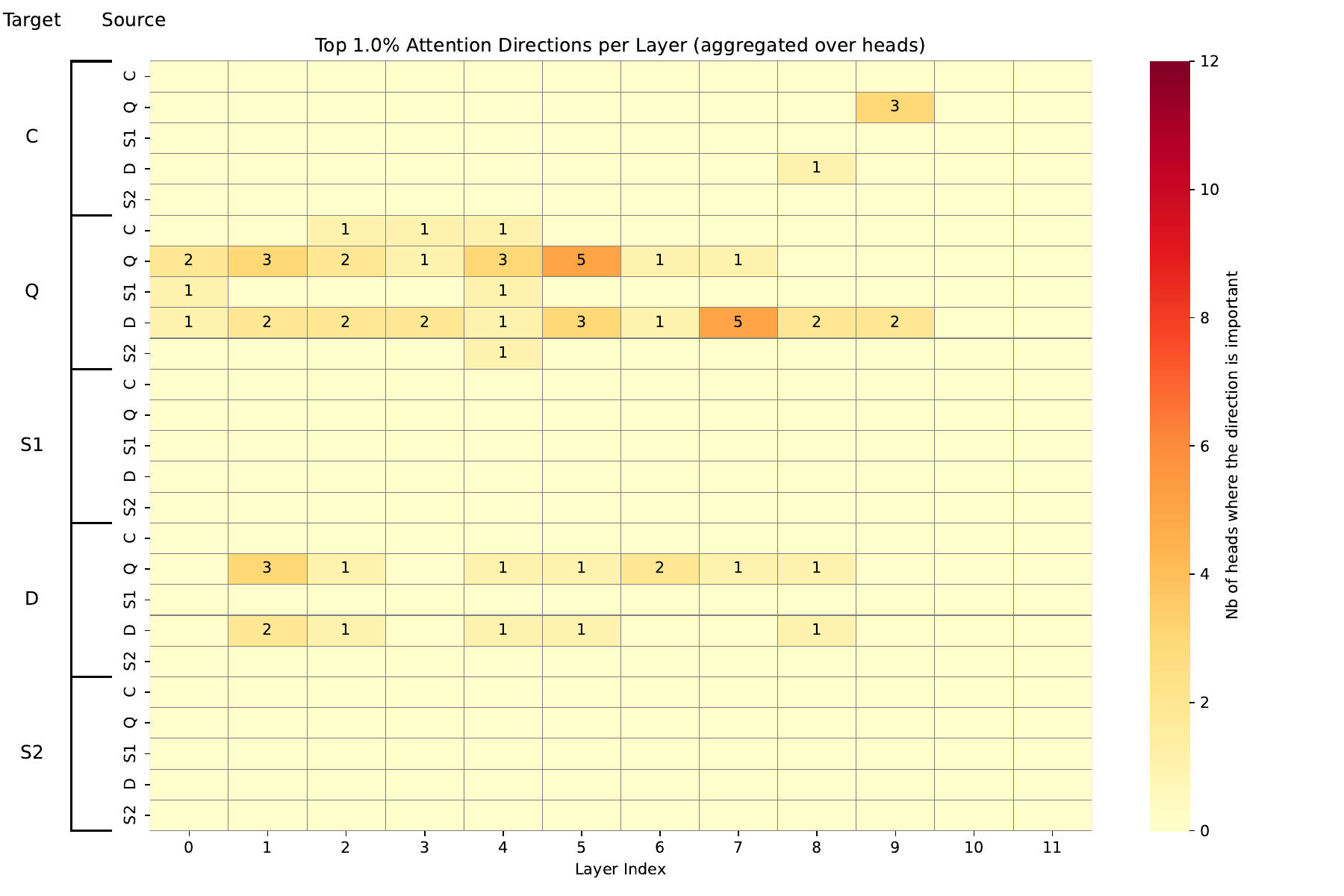}
        \caption{Ablation scheme $I_A$ for MiniLM.}
    \end{subfigure}
    \hspace{0.02\textwidth} 
    \begin{subfigure}[h]{0.47\textwidth}
        \centering
        \includegraphics[width=\textwidth]{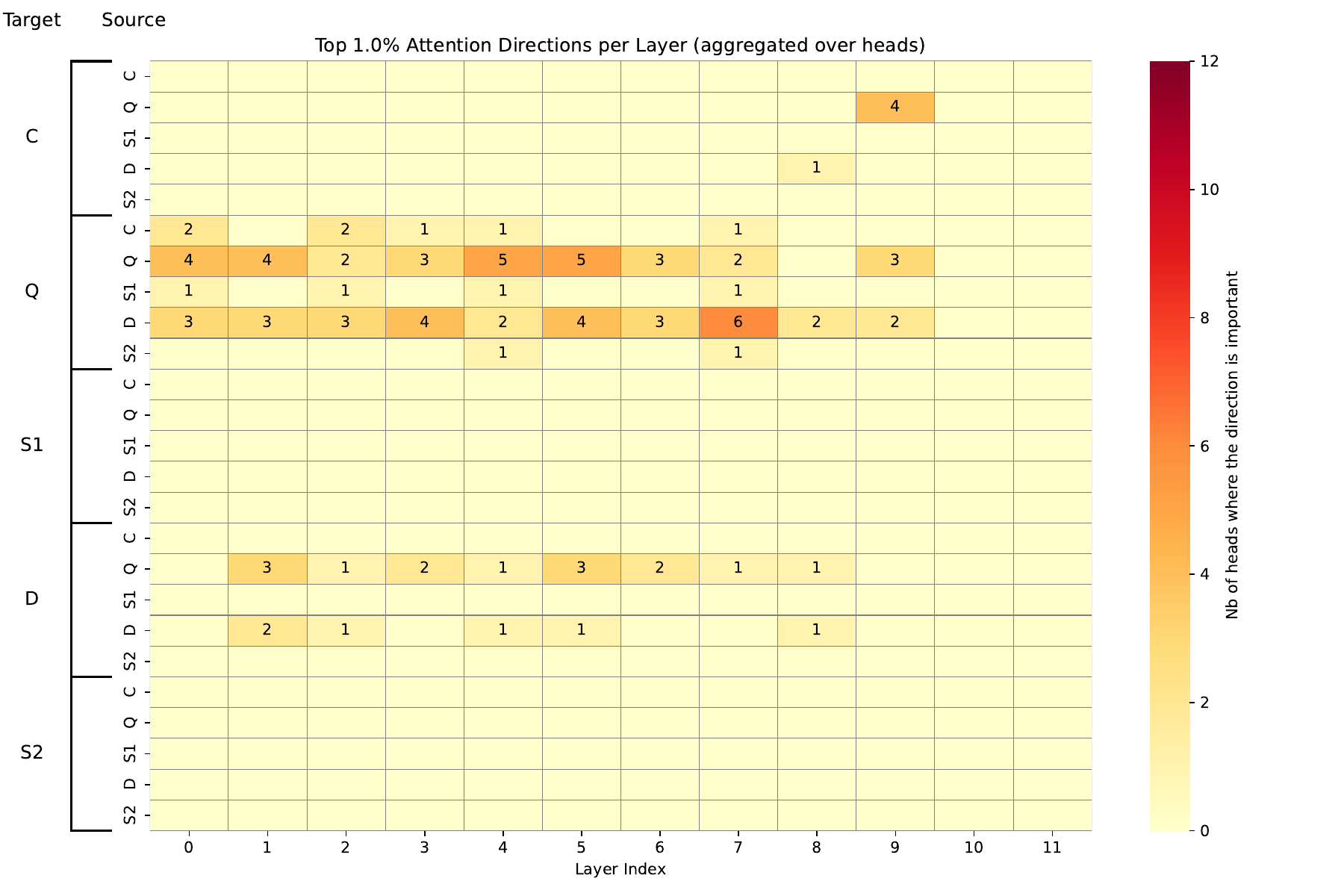}
        \caption{Ablation scheme $F_A$ for MiniLM.}
    \end{subfigure}

    \begin{subfigure}[h]{0.44\textwidth}
        \centering
        \includegraphics[width=\textwidth]{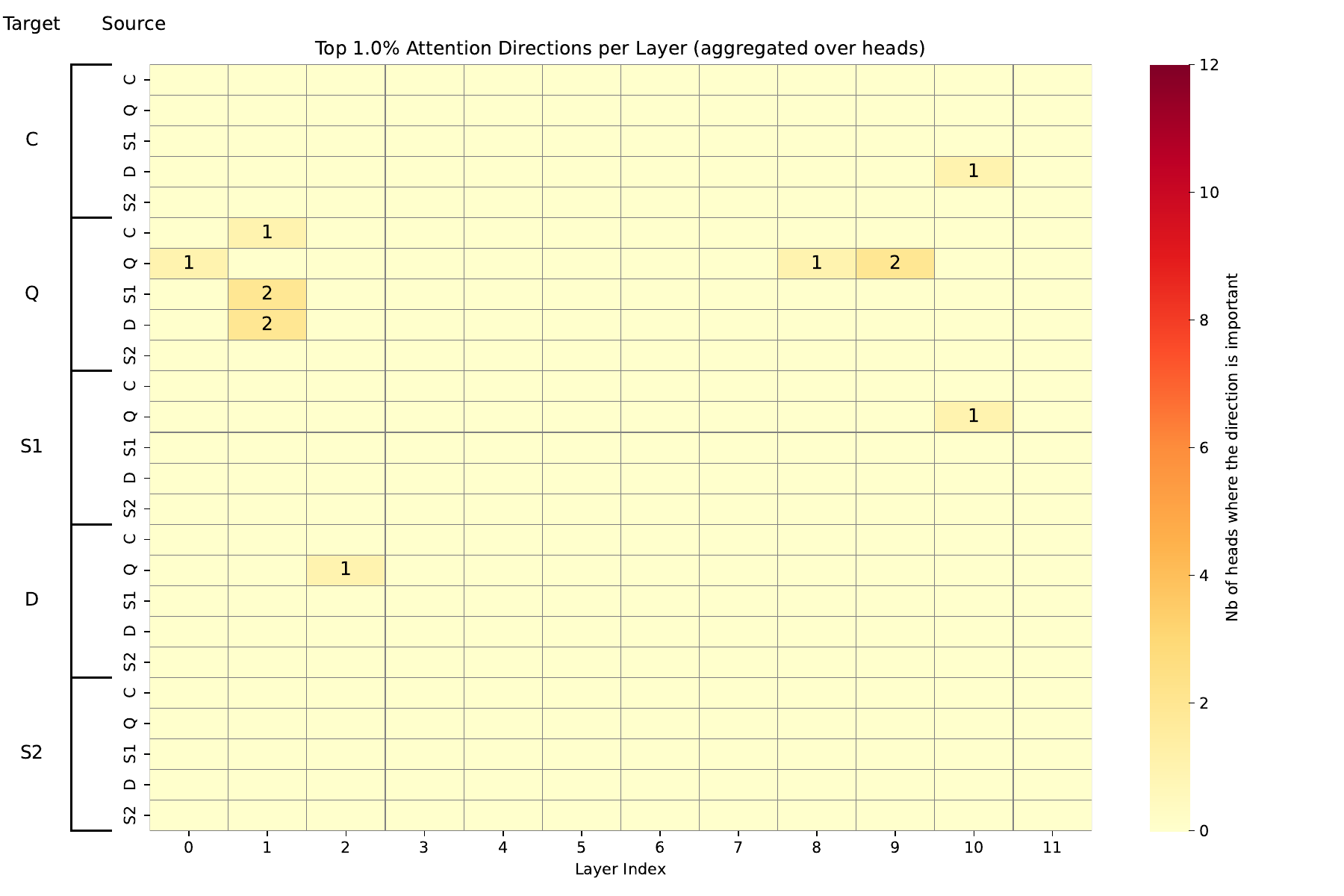}
        \caption{Ablation scheme $I_A$ for MonoBERT-base.}
    \end{subfigure}
    \centering
    \hspace{0.02\textwidth} 
    \begin{subfigure}[h]{0.47\textwidth}
        \centering
        \includegraphics[width=\textwidth]{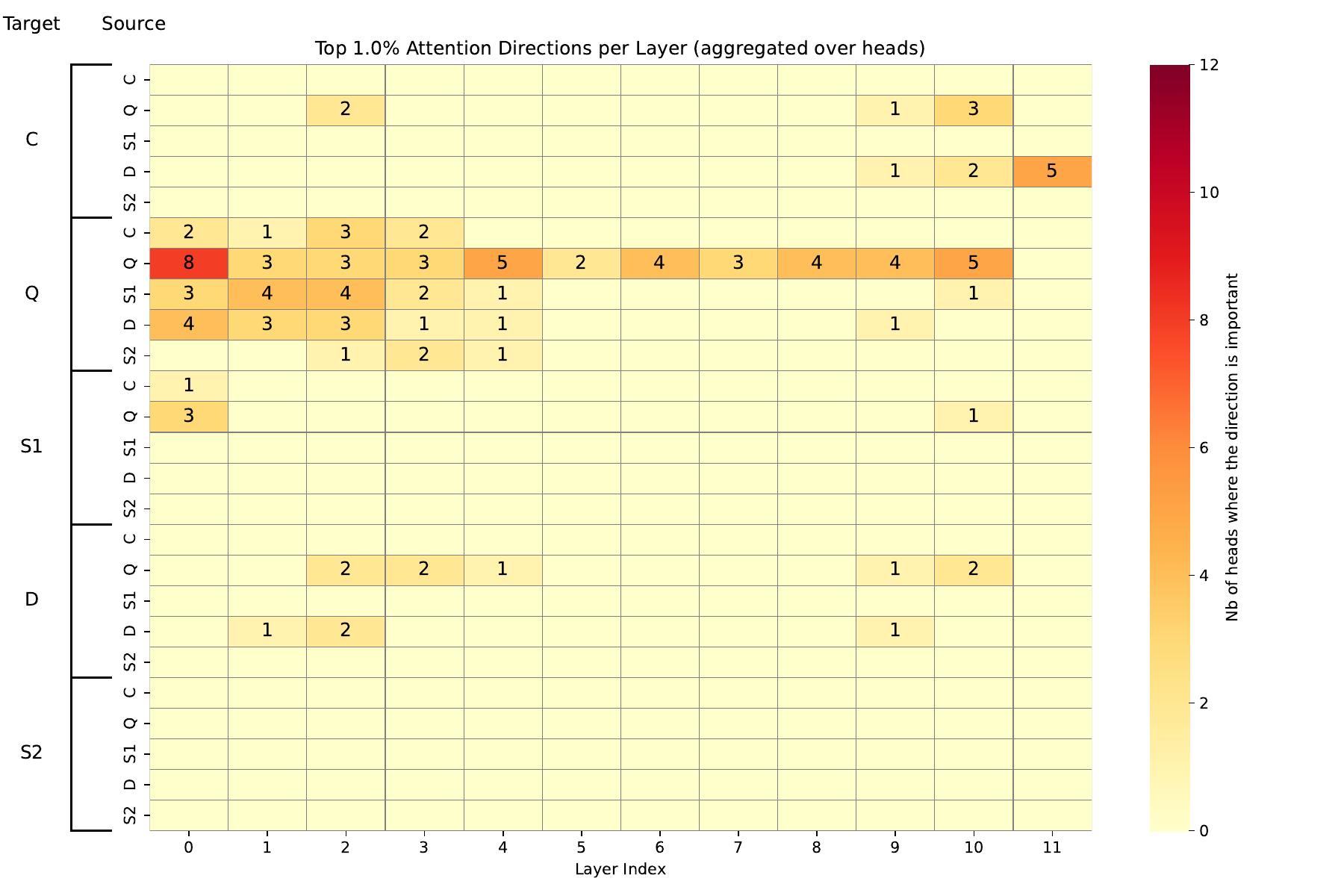}
        \caption{Ablation scheme $F_A$ for MonoBERT-base.}
    \end{subfigure}
    
    \begin{subfigure}[h]{0.47\textwidth}
        \centering
        \includegraphics[width=\textwidth]{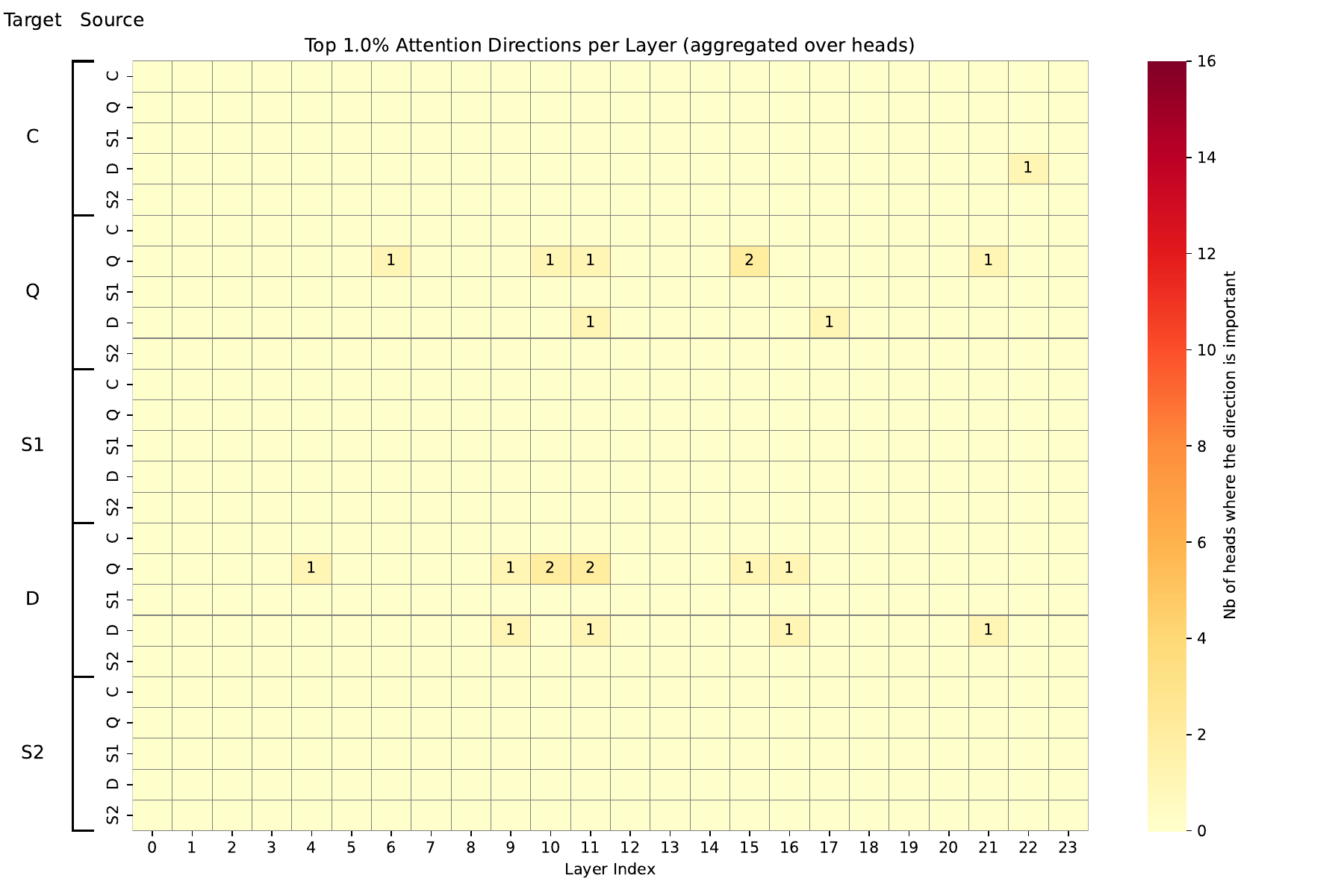}
        \caption{Ablation scheme $I_A$ for MonoBERT-large.}
    \end{subfigure}
    \hspace{0.02\textwidth} 
    \begin{subfigure}[h]{0.47\textwidth}
        \centering
        \includegraphics[width=\textwidth]{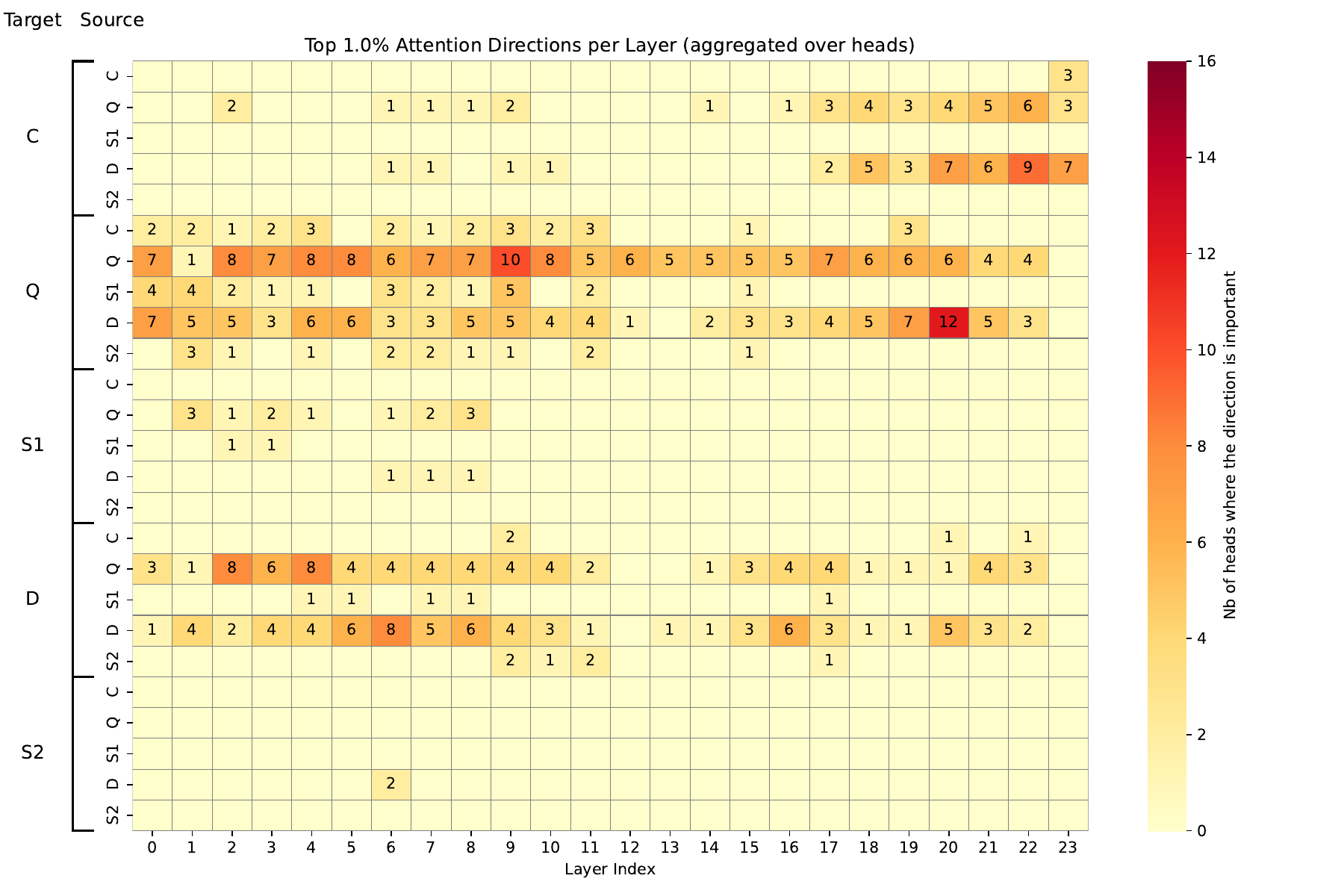}
        \caption{Ablation scheme $F_A$ for MonoBERT-large.}
    \end{subfigure}
    
    \caption[NIG results on the information flow for $I_A$ and $F_A$]{Distribution of the most important directions across attention heads per layer for ablation schemes $F_A$ and $I_A$ at pruning percentage $t=1\%$.}
    \label{fig:appendix_nig_attn_all}
\end{figure}

\section{Background on the Information Bottleneck approach}\label{appendix:information_bottleneck}

In the Information Bottleneck framework \cite{Schulz2020Restricting}, the objective is to restrict as much as possible the flow of information using so-called ``bottlenecks'' without hindering the performance of the model too much. By design, these bottlenecks are masks, used to prevent the model from using some dimensions from its hidden states. From \cite{jiang-etal-2021-bert}, the information bottleneck in IR can be summarized as follows:
\begin{equation}
    \label{equ:Information_Bottleneck}
    T = \mu \odot X+(\mathbf{1}-\mu)\odot\epsilon
\end{equation}
where $T$ is the output representation, $X \in \mathbb{R}^{o}$ represents an output of the model of dimension $o$, $\epsilon$ represents the noise, $\mathbf{1} \in \mathbb{R}^{o}$ is the all-one vector and $\mu \in \mathbb{R}^{o}$ is the weight vector controlling the balance between noise and the original signal. 
They constrain $\mu_i\in [0,1]$ by setting $\mu_i = \sigma(\alpha_i)$, where $\sigma$ is the sigmoid function and the $\alpha_i$ are the learned parameters. The training process starts from a frozen re-ranker such as MonoBERT, with $\alpha_i$ initialized to 5 for each dimension $i$, so that $\mu_i = 0.99$ and $T \approx X$ at the beginning of the training.

\subsection{Application to our study}
Unlike them, we are not interested in optimizing $\alpha$ for the hidden dimensions but instead for the attention score matrices. Following the setup introduced in the previous parts, we study the different directions between the 5 input parts and try to learn which directions, in every attention head, can be masked without hindering the performance of the models on the IR task. To do so, we reformulate the experimental framework of \citeauthor{jiang-etal-2021-bert}~\cite{jiang-etal-2021-bert} to have, for each attention head $h$ at the layer $\ell$ in the model, a vector $\mu^{h,\ell} \in \mathbb{R}^{25}$, where the 25 rows correspond to the 25 possible directions\footnote{In practice, there is 25 + 1 dimensions in total, to account for \PAD{} tokens, but we omit it in our explanations for the sake of simplicity.}, i.e., $\mu^{h,\ell}_{t\leftarrow s}\in \mathbb{R}$, with $s$ and $t$ respectively the source and target input parts, in an attention score matrix for the IR task (see Figure \ref{fig:attention_weight_segmented}). The equation \ref{eq:alpha_masks} below illustrates the relationship that exists between $\mu^{h,\ell}$ and the different sections of the corresponding attention scores matrix $A^{h,\ell}$.

\begin{equation*}
A^{h,\ell}_{final}= A^{h,\ell}_{original} \bigodot
\begingroup 
\setlength\arraycolsep{1.5pt}
\begin{pmatrix}
\textcolor{red}{\mu^{h,\ell}_{\CLS\leftarrow \CLS}} & \textcolor{blue}{\mu^{h,\ell}_{\CLS\leftarrow Q}} & \cdots & \textcolor{blue}{\mu^{h,\ell}_{\CLS\leftarrow Q}} & \textcolor{green}{\mu^{h,\ell}_{\CLS\leftarrow \SEPone}} & \textcolor{orange}{\mu^{h,\ell}_{\CLS\leftarrow D}} & \cdots & \textcolor{orange}{\mu^{h,\ell}_{\CLS\leftarrow D}} & \textcolor{purple}{\mu^{h,\ell}_{\CLS\leftarrow \SEPtwo}}\\
\textcolor{cyan}{\mu^{h,\ell}_{Q\leftarrow \CLS}} & \cdots & & \ddots & & & & & & \\
\vdots & & & & & &  & & \vdots \\ 
\textcolor{cyan}{\mu^{h,\ell}_{Q\leftarrow \CLS}} & \cdots & & & & \ddots & & &  \\
\vdots & & & & & &  & & \vdots  \\ 
 & & & \cdots & & & & & \textcolor{brown}{\mu^{h,\ell}_{\SEPtwo\leftarrow \SEPtwo}} 
\end{pmatrix}
\endgroup
\label{eq:alpha_masks}
\end{equation*}

\noindent We fix the maximum input length to 128 tokens in the experiments. Importantly, we do obtain the vector $\mu^{h,\ell}$ by using the Gumbel-Softmax trick on $\alpha^{h,\ell}$. Finally, our loss is composed of a distillation loss term (in our case based on MSE), where the teacher and student are the frozen re-ranker itself, with or without the masks $\mu$, and of a regularization term over the values of $\alpha$ to enforce sparsity:

\begin{equation}
\mathcal{L}
      \;=\;
      \underbrace{
        \frac{1}{N}\,
        \left\lVert
          \mathbf{z}
          \;-\;
          \mathbf{z_\mu}
        \right\rVert_{2}^{2}
      }_{\displaystyle \mathcal{L}_{\mathrm{distill}}}
      \;+\lambda\;
      \underbrace{
       \sum_{\ell \in L} \sum_{h \in H}\, 
        \left\lVert \boldsymbol{\sigma}(\alpha^{h,\ell}) \right\rVert_{1}
      }_{\displaystyle \mathcal{L}_{\mathrm{reg}}}
\label{eq:distill_l1_loss}
\end{equation}
with $z$ the original score, $z_\mu$ the score when the masks are applied, $\lambda$ the coefficient of the regularization term, and $L$ and $H$ respectively the maximum number of layers in the model and the maximum number of attention heads per layer.

\subsection{Experimental details} 

We apply the same set of hyper-parameters to every model. This includes the weight of the regularization term set to $\lambda=10$ for every model. In particular, for every dimension $i,j$, $\alpha^{h,\ell}_{i,j}$ is initialized to 3 ($\sigma(3) \approx0.95$) and the learning rate is 1. The mask $\mu^{h,\ell}$ is computed as $\mu^{h,\ell} = \operatorname{GS}(\alpha^{h,\ell}, \tau)$, with $\tau=1.0$ \footnote{When testing the models, we remove the Gumbel noise and apply a softmax function on the weights $\alpha^{h,\ell}_{i,j}$. To binarize the masks, we manually determined a threshold $t$ above which the direction is kept to maximize the sparsity while conserving most of the original model performance. We set $t=0.95$, $t=0.97$ and $t=0.99$ respectively for MiniLM, MonoBERT-base and MonoBERT-large.}. 
To stay consistent with the dataset used during their fine-tuning, we rely on the train set of MS MARCO \cite{MSMARCO}. 

In practice, the learned masks preserve only $\approx9\%$ of the directions for MiniLM, $\approx15\%$ for MonoBERT-base, and $\approx21\%$ for MonoBERT-large.

\newpage
\section{Attention patterns}

\subsection{Attention patterns analysis over all the datasets}

\begin{figure}[H]
    \centering
    \begin{subfigure}[t]{0.32\textwidth}
        \centering
        \includegraphics[width=1\textwidth]{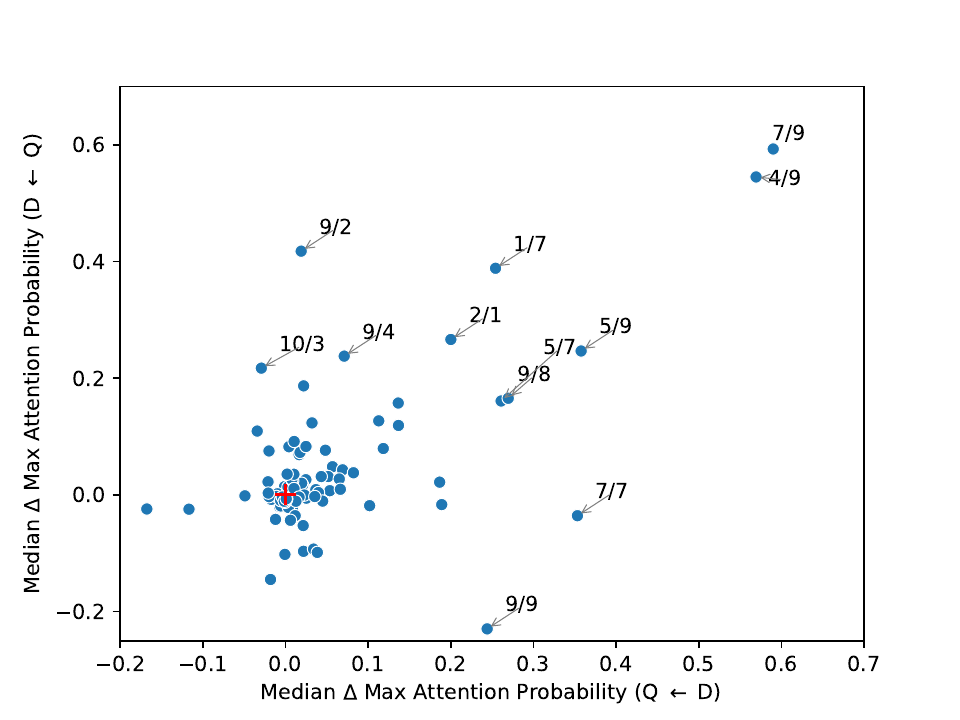}
        \caption{Results over all the datasets for MiniLM.}
        \label{fig:minilm_all_attention}
    \end{subfigure}
    \hfill
    \begin{subfigure}[t]{0.32\textwidth}
        \centering
        \includegraphics[width=1\textwidth]{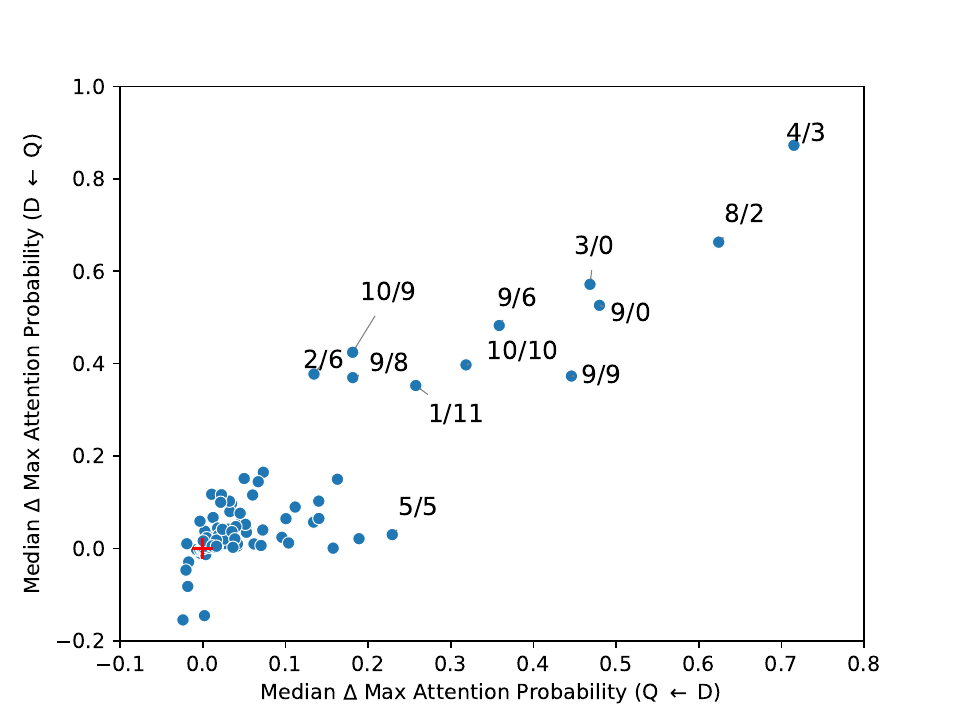}
        \caption{Results over all the datasets for MonoBERT-base.}
        \label{fig:monobert_base_all_attention}
    \end{subfigure}
    \hfill
    \begin{subfigure}[t]{0.32\textwidth}
        \centering
        \includegraphics[width=1\textwidth]{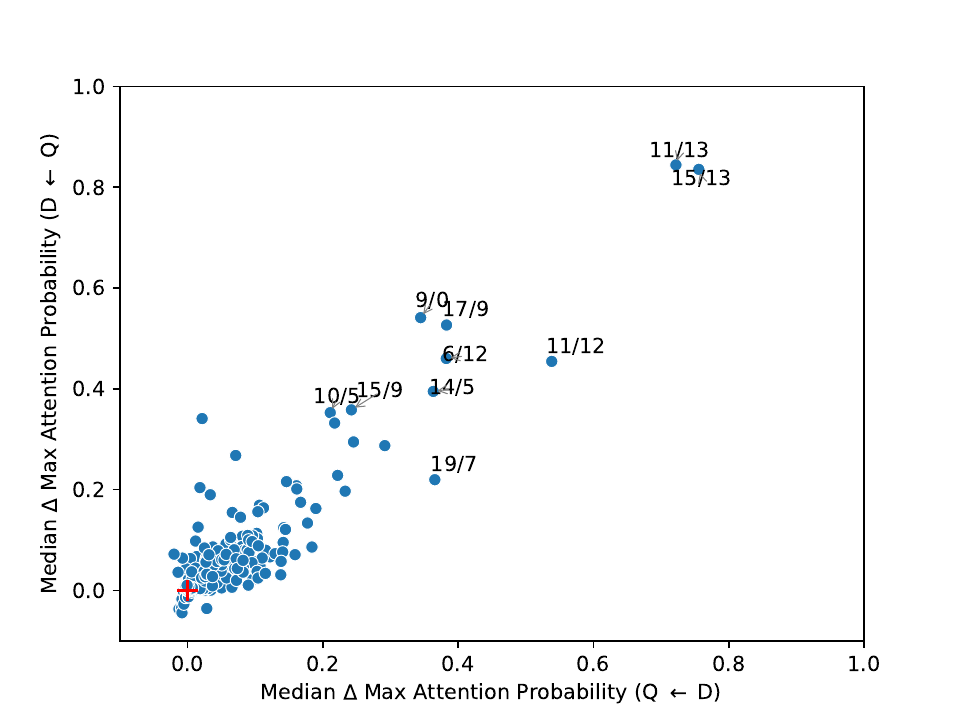}
        \caption{Results over all the datasets for MonoBERT-large.}
        \label{fig:monobert_all_attention}
    \end{subfigure}
    \caption[Summary of the matching patterns across all the datasets for relevant and easy negatives]{Median of the differences in the attention weights $\maxatt{Q,D/1} - \maxatt{Q,D/-1}$ and  $\maxatt{D,Q/1} - \maxatt{D,Q/-1}$ for each attention head.}
    \label{fig:attention_all_datasets}
\end{figure}

\begin{figure}[H]
    \centering
    \begin{subfigure}[t]{0.32\textwidth}
        \centering
        \includegraphics[width=1\textwidth]{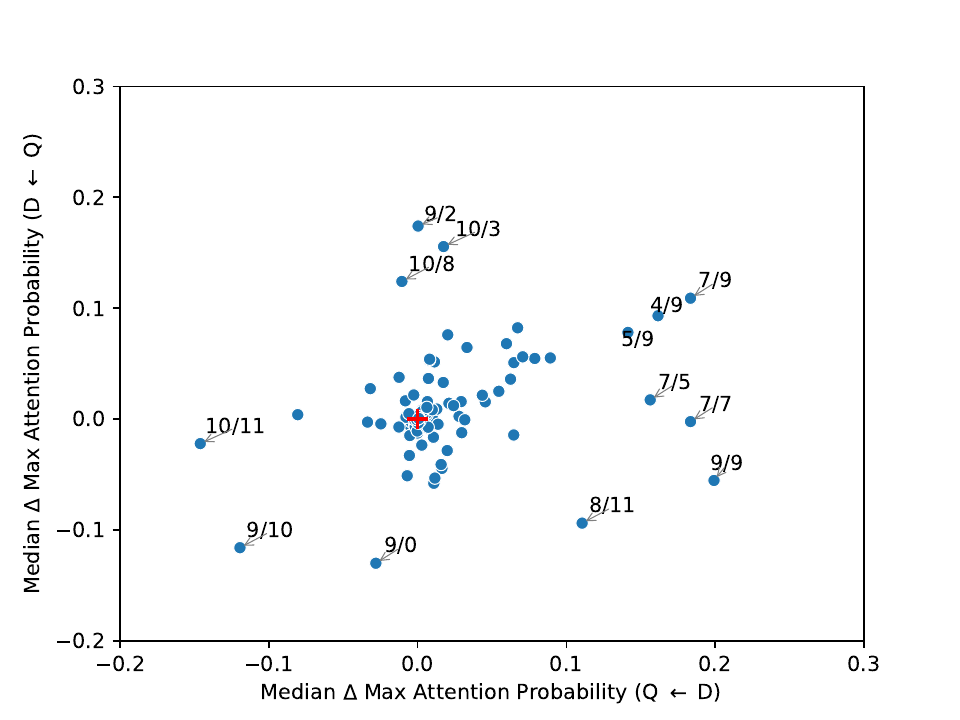}
        \caption{Results over all the datasets for MiniLM.}
        \label{fig:minilm_all_attention_hard}
    \end{subfigure}
        \hfill
    \begin{subfigure}[t]{0.32\textwidth}
        \centering
        \includegraphics[width=1\textwidth]{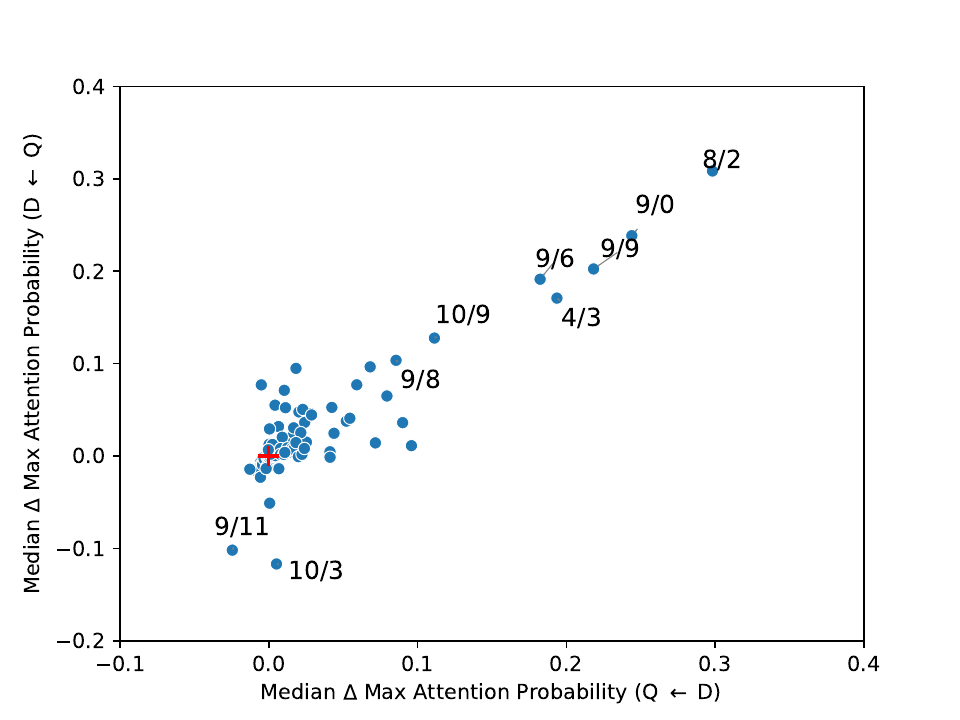}
        \caption{Results over all the datasets for MonoBERT-base.}
        \label{fig:monobert_base_all_attention_hard}
    \end{subfigure}
    \hfill
    \begin{subfigure}[t]{0.32\textwidth}
        \centering
        \includegraphics[width=1\textwidth]{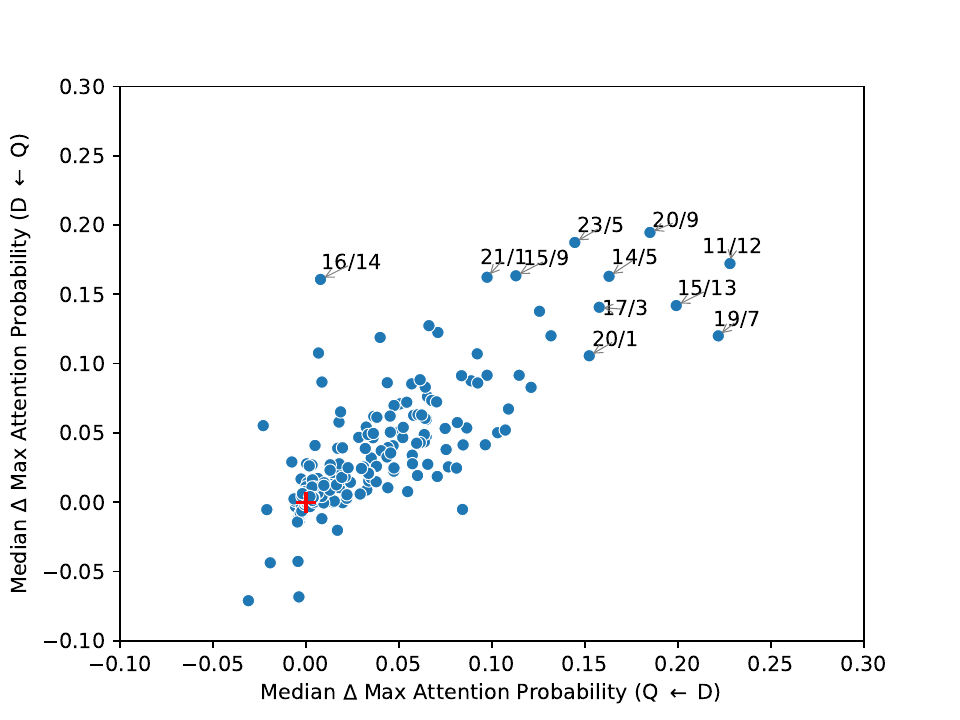}
        \caption{Results over all the datasets for MonoBERT-large.}
        \label{fig:monobert_all_attention_hard}
    \end{subfigure}
    \caption[Summary of the matching patterns across all the datasets for relevant and hard negatives]{Median of the differences in the attention weights $\maxatt{Q,D/1} - \maxatt{Q,D/0}$ and  $\maxatt{D,Q/1} - \maxatt{D,Q/0}$ for each attention head.}
    \label{fig:attention_all_datasets_hard}
\end{figure}

\newpage

\subsection{Analysis of the matching signals over hard negatives}

\begin{figure}[htbp]
        \centering
        \begin{subfigure}[b]{0.49\textwidth}
            \centering
            \includegraphics[width=\textwidth]{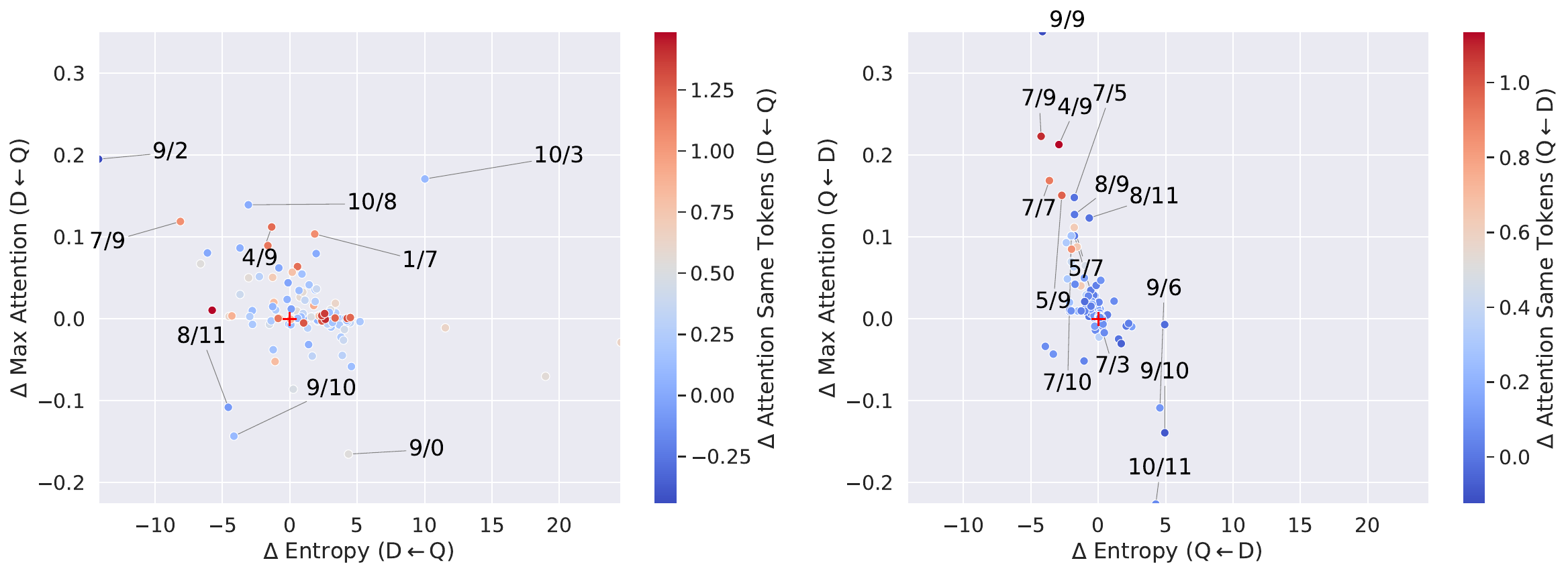}
            \caption{MiniLM on TREC-DL (hard).}
        \end{subfigure}
        \hfill
        \begin{subfigure}[b]{0.49\textwidth}
            \centering
            \includegraphics[width=\textwidth]{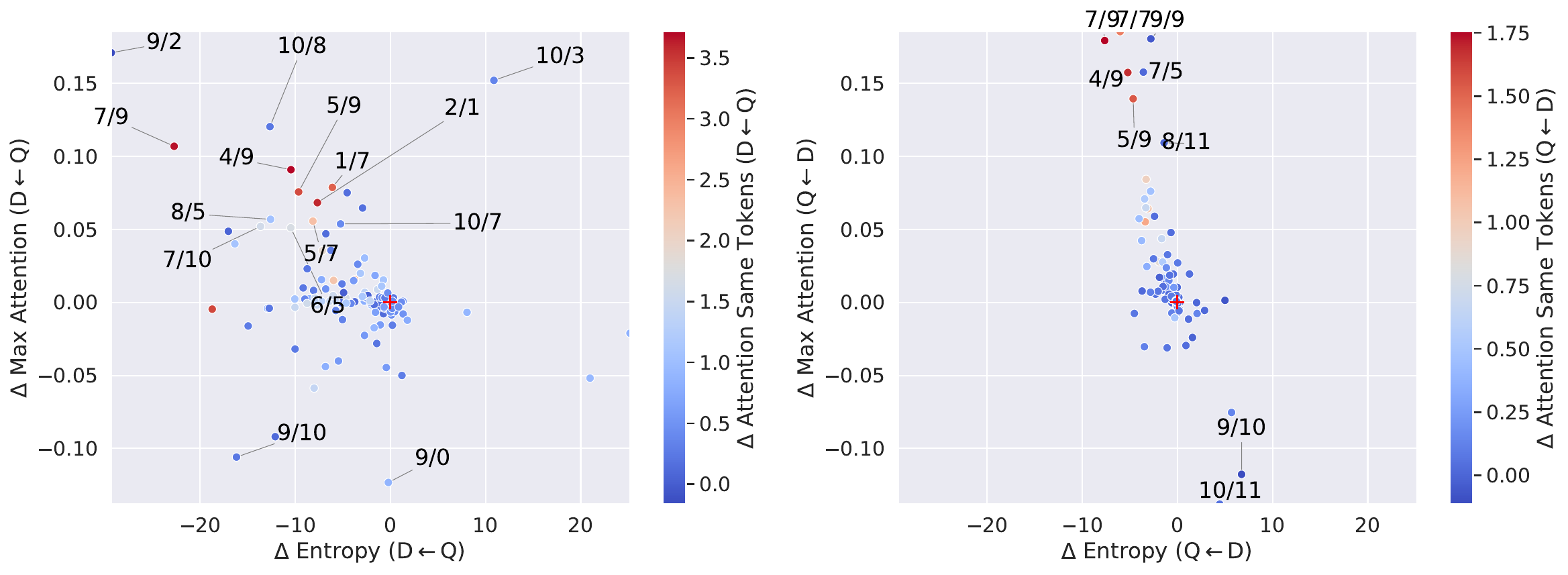}
            \caption{MiniLM on OOD (hard).}
        \end{subfigure}

        \begin{subfigure}[b]{0.49\textwidth}
            \centering
            \includegraphics[width=\textwidth]{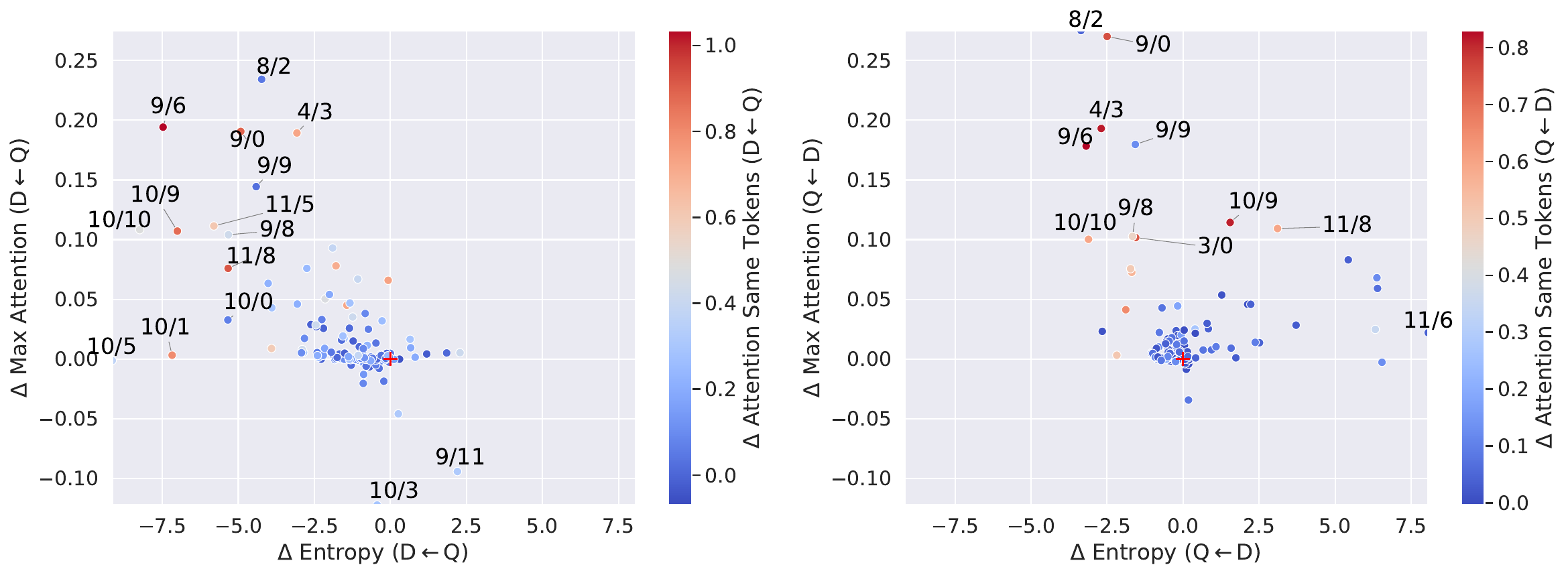}
            \caption{MonoBERT-base on TREC-DL (hard).}
        \end{subfigure}
        \hfill
        \begin{subfigure}[b]{0.49\textwidth}
            \centering
            \includegraphics[width=\textwidth]{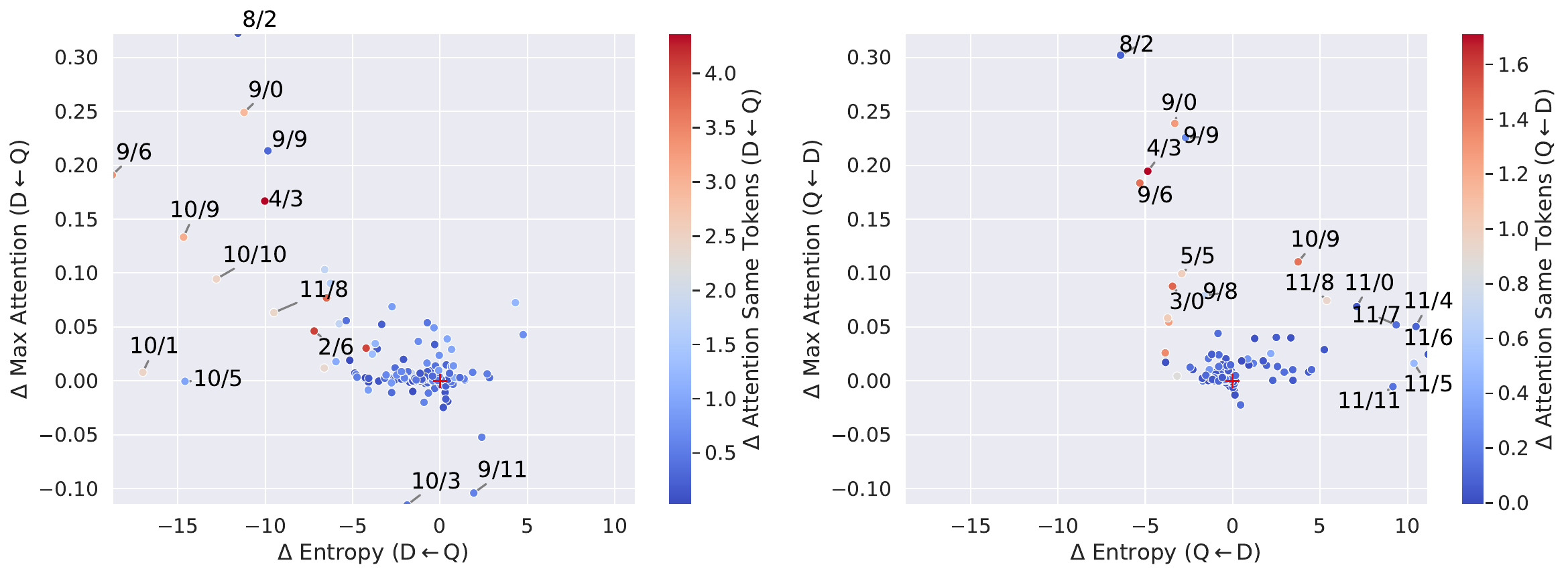}
            \caption{MonoBERT-base on OOD (hard).}
        \end{subfigure}
        
        \begin{subfigure}[b]{0.49\textwidth}
            \centering
            \includegraphics[width=\textwidth]{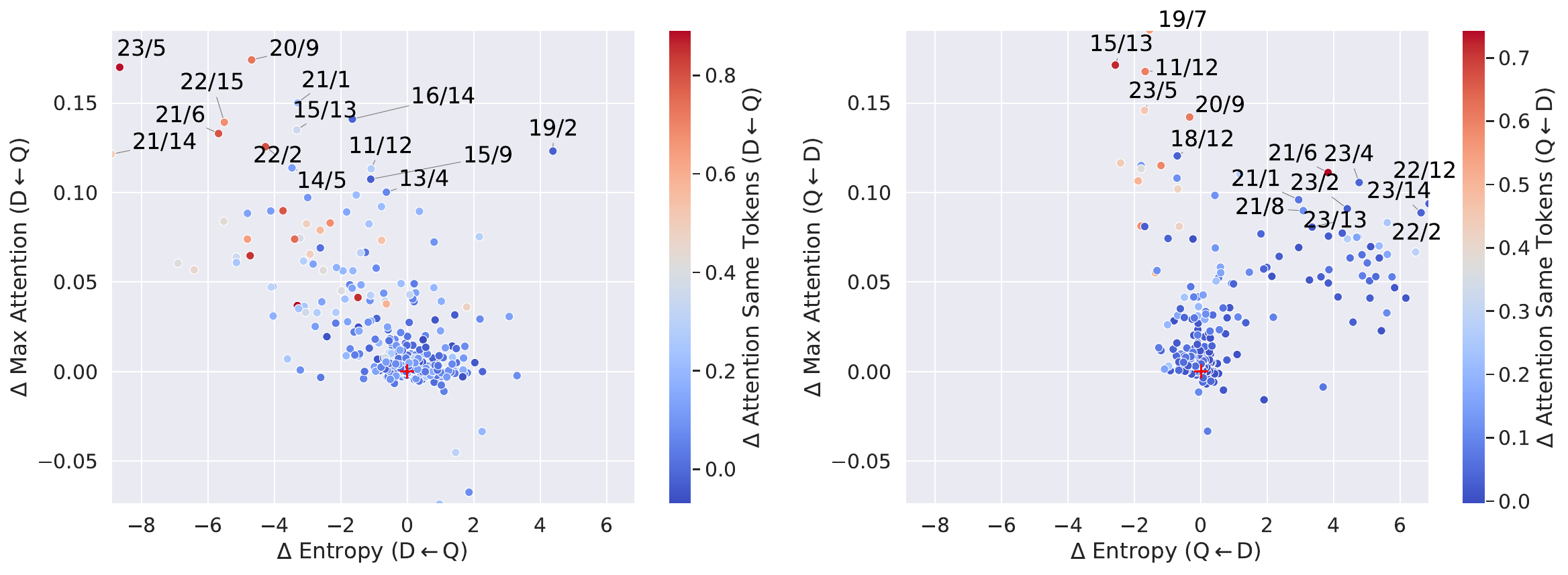}
            \caption{MonoBERT-large on TREC-DL (hard).}
        \end{subfigure}
        \hfill
        \begin{subfigure}[b]{0.49\textwidth}
            \centering
            \includegraphics[width=\textwidth]{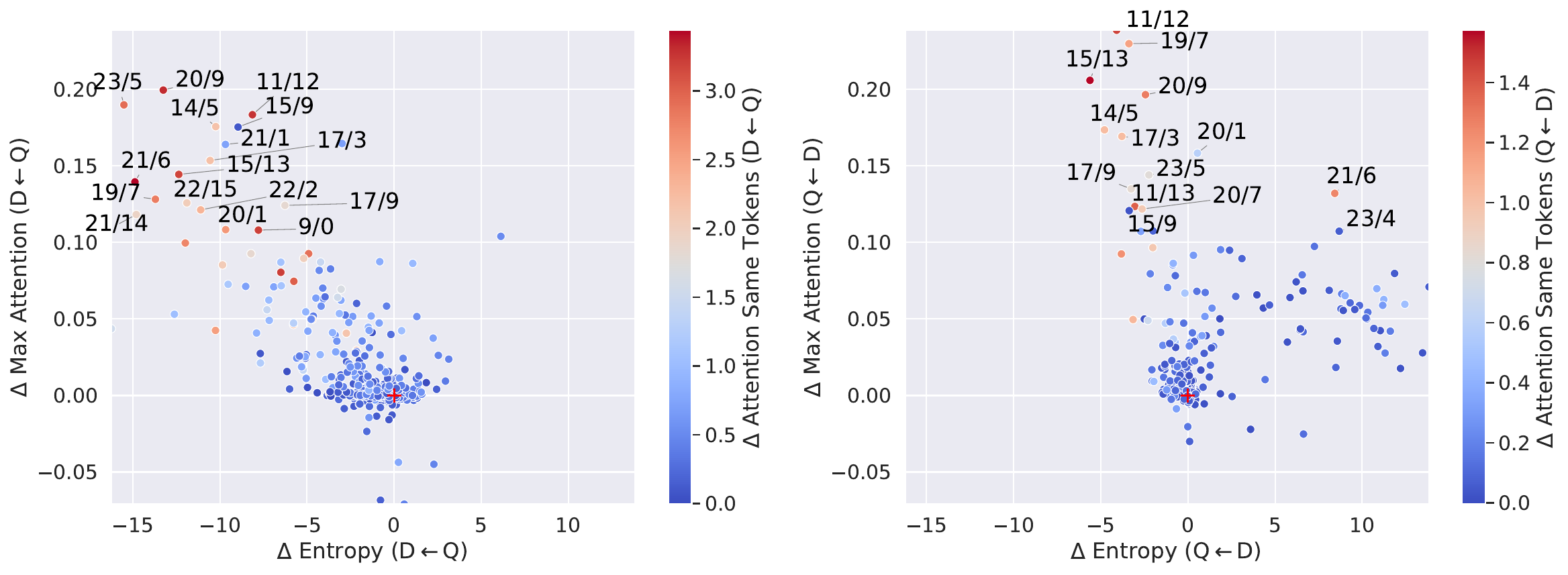}
            \caption{MonoBERT-large on OOD (hard).}
        \end{subfigure}

        \caption[Comparison of the matching patterns between relevant and hard negatives]{Comparison of the matching patterns of the attention heads over the TREC-DL and OOD datasets, contrasting $r=1$ with $r=0$ (hard negatives). Same axes as Figure \ref{fig:comparing_match_type_weak}: each point is an attention head $\ell/h$; left panel $D \leftarrow Q$, right panel $Q \leftarrow D$; x-axis: median of the differences in entropy; y-axis: median of the difference in maximum attention; color: median of the difference in the probability of attending to a token identical to the source token.}
        \label{fig:comparing_match_type_hard}
\end{figure}

\subsection{Unveiling the mechanism behind matching attention heads}
\label{subsection:matching_mechanism}

\begin{figure}[!htbp]
    \centering
    \begin{subfigure}[b]{0.49\textwidth}
        \centering
        \includegraphics[width=1\textwidth]{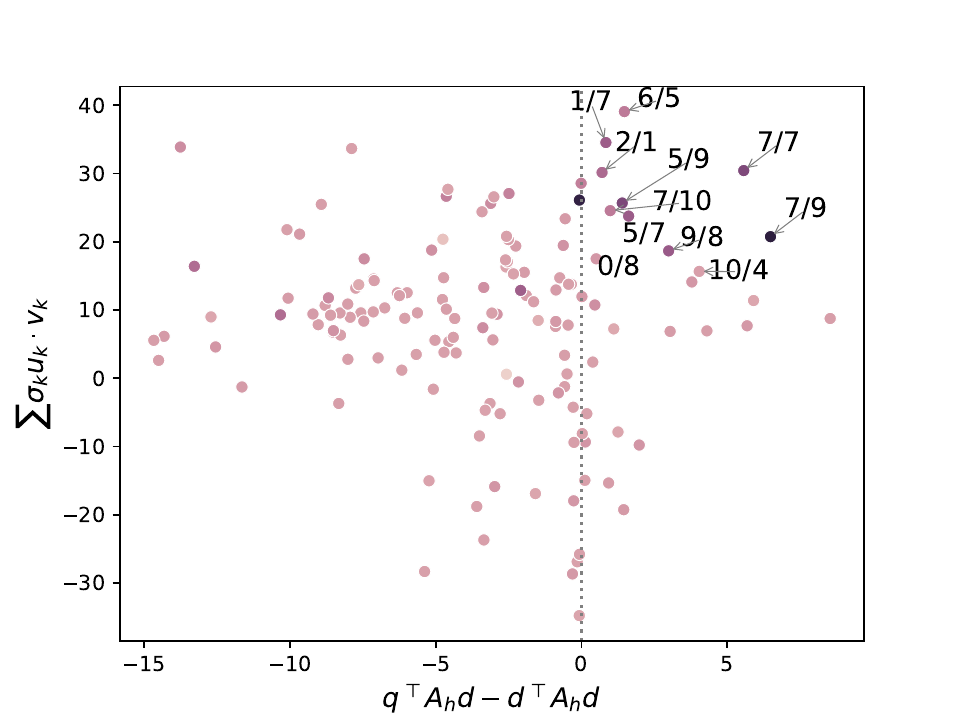}
        \caption{Details for MiniLM from $Q \leftarrow D$.}
    \end{subfigure}
    \begin{subfigure}[b]{0.49\textwidth}
        \centering
        \includegraphics[width=1\textwidth]{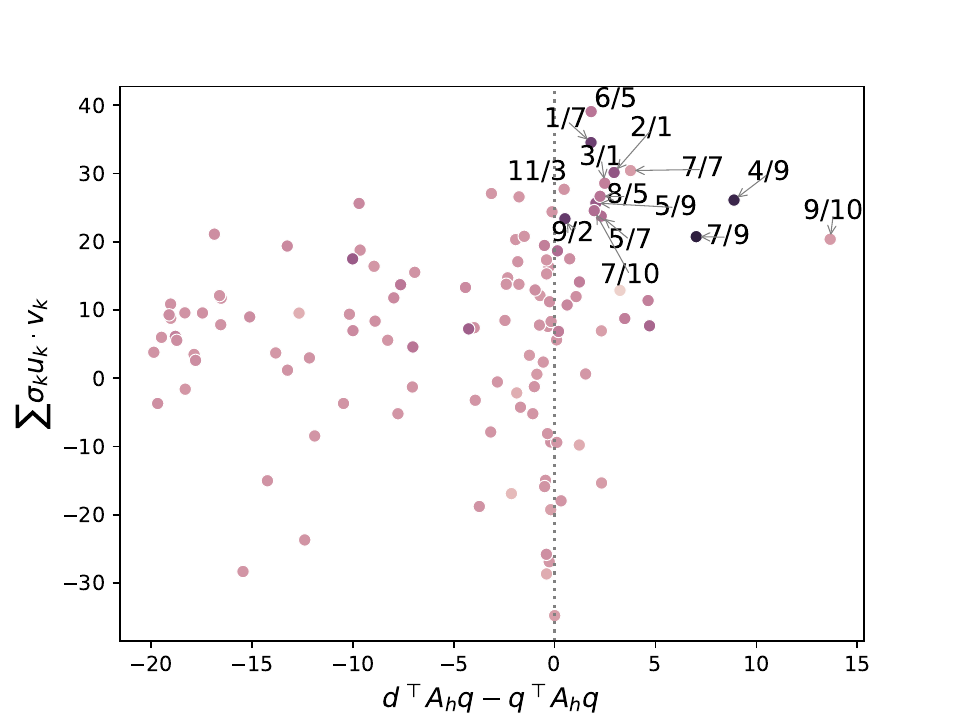}
        \caption{Details for MiniLM from $D \leftarrow Q$.}
    \end{subfigure}
    \centering
    \begin{subfigure}[b]{0.49\textwidth}
        \centering
        \includegraphics[width=1\textwidth]{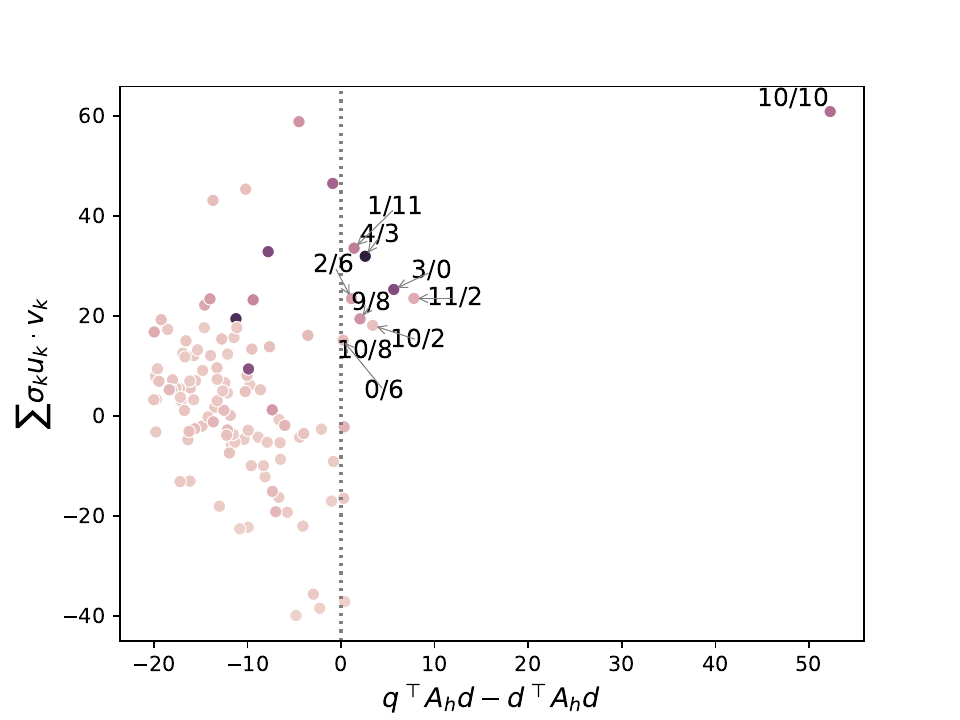}
        \caption{Details for MonoBERT-base from $Q \leftarrow D$.}
    \end{subfigure}
    \begin{subfigure}[b]{0.49\textwidth}
        \centering
        \includegraphics[width=1\textwidth]{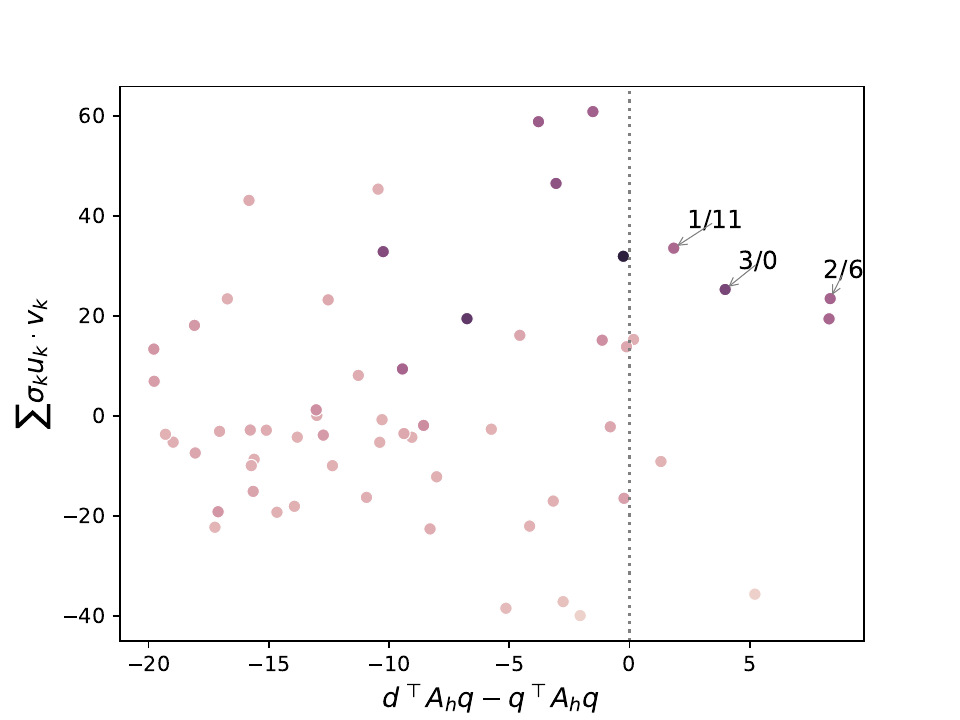}
        \caption{Details for MonoBERT-base from $D \leftarrow Q$.}
    \end{subfigure}
    \vspace{-0.2cm}
    \centering
    \begin{subfigure}[b]{0.49\textwidth}
        \centering
        \includegraphics[width=1\textwidth]{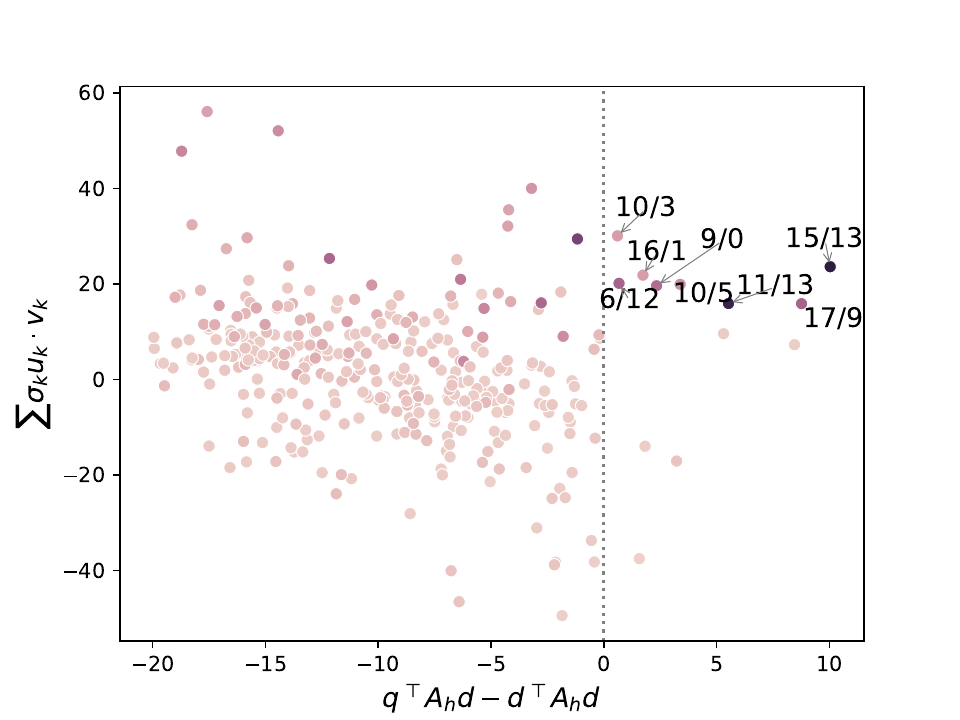}
        \caption{Details for MonoBERT-large from $Q \leftarrow D$.}
    \end{subfigure}
    \begin{subfigure}[b]{0.49\textwidth}
        \centering
        \includegraphics[width=1\textwidth]{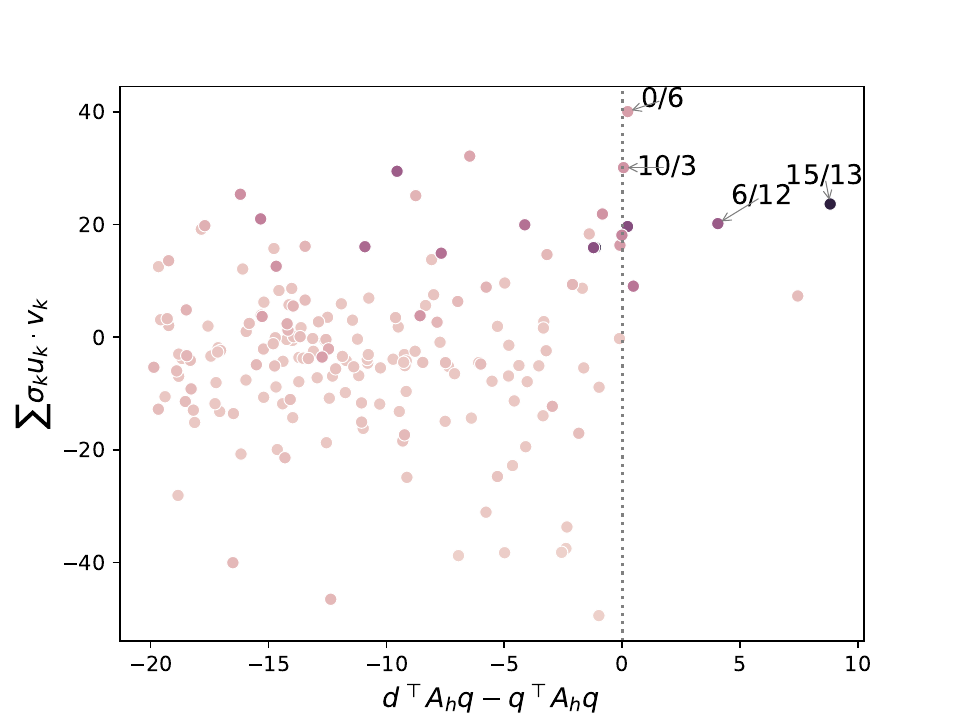}
        \caption{Details for MonoBERT-large from $D \leftarrow Q$.}
    \end{subfigure}
    \caption[Mechanistic interpretations of the detection of the matching signals in the attention of BERT-based cross-encoders]{Distribution of the attention heads (layer $\ell$ / head $h$) based on their Singular Value Decomposition and their ability to do cross-contextualization. Color levels relate to the contrasts between $r=1$ and $r=-1$ on all the datasets (see Figures \ref{fig:minilm_all_attention} and \ref{fig:monobert_all_attention}). In the axis labels, $A_h$ stands for the Query-Key matrix $W_{QK}^{h}$, and $q$ and $d$ for the mean query and document representations $\bar{q}$ and $\bar{d}$.}
    \label{fig:matching_heads}
    \vspace{-0.5cm}
\end{figure}

From Section \ref{section:understand_attention}, we now understand the role of the most important attention heads better, yet the mechanism(s) underlying the detection of matching signals remain elusive. From our analysis on the attention weight matrices, we can hypothesize that for a query token (resp. document token) to be matched with a document token (resp. query token), attention heads' sub-operations need to meet certain criteria. Following the mechanistic interpretability target to break down and explain, in a human-understandable manner, the behavior of each component inside neural networks, we study the mechanisms hidden behind the detection of query-document matching by attention heads. To do so, we follow the work of \citeauthor{elhage2021mathematical}~\cite{elhage2021mathematical} and decompose the operations inside each attention head into the Query-Key matrix and the Output-Value matrix.

From the self-attention mechanism \cite{attention}, we get that for any attention head $h$ at a layer $\ell$, the attention logits are obtained from the bilinear form defined by the Query-Key matrix $W_{QK}^{h,\ell} = \left(W_Q^{h}\right)^{\top} W_K^{h}$ (see Section~\ref{subsection:architecture}). The logit of the probability that token $i$ in the input attends to token $j$ is $E_i^\top W_{QK}^{h,\ell} E_j$, where $E_i$ and $E_j$ denote the representations of these two tokens at the input of the layer.
 Suppose now that the representation of the $i^{th}$ token of the query and the $j^{th}$ token of the document can be expressed as $E_i = \bar{q} + x_i + a_i$ and $E_j = \bar{d} + y_j + b_j$, where $\bar{q}$ and $\bar{d}$ respectively represent the part of the embedding of queries and documents shared by every query and document\footnote{In practice, $\bar{q}$ and $\bar{d}$ are the average representations of a query and a document, computed over multiple queries and passages.}, $x_i$ and $y_j$ their semantic/lexical information, and $a_i$ and $b_j$ some residual information. Then, the attention logit for $E_i$ on $E_j$ ($Q \leftarrow D$) can be reformulated as:
\begin{align}
    (E_i)^\top W_{QK}^{h,\ell} E_j &= (\bar{q} + x_i + a_i)^\top W_{QK}^{h,\ell} (\bar{d} + y_j + b_j) \\
    &= \bar{q}^\top W_{QK}^{h,\ell} \bar{d} + x_i^\top W_{QK}^{h,\ell} y_j + \epsilon
\end{align}
\noindent with $\epsilon$, a residual that, hopefully, is close to 0 for ``matching detection'' heads. Note that we can obtain a similar reformulation for the attention logit $E_j$ on $E_i$ ($D \leftarrow Q$).
In the following, we develop our reasoning based on the direction $Q\leftarrow D$, but it also applies to $D \leftarrow Q$.

From that formulation, we hypothesize that for $Q \leftarrow D$ matches to happen, $\bar{q}^\top W_{QK}^{h} \bar{d} - \bar{d}^\top W_{QK}^{h} \bar{d} > 0$.
To validate this hypothesis, we plot the difference $\bar{q}^\top W_{QK}^{h} \bar{d} - \bar{d}^\top W_{QK}^{h} \bar{d}$ (which should be greater than 0 when query-document matching is sought) against the propensity of matching tokens with the same meaning, i.e., where $x_i$ and $y_j$ are close. We evaluate the second condition by measuring the similarity of the left and right singular vectors of the Singular Value Decomposition of $W_{QK}^{h}$. Specifically, with $W_{QK}^{h}=\sum\sigma_ku_kv_k^\top $, we compute $\sum_k \sigma_k u_k \cdot v_k$. From Figure \ref{fig:matching_heads}, we can see that the heads identified in Table~\ref{tab:ranking_attention_heads} tend to follow our hypothesis, i.e., they favor cross-attention rather than self-attention, and their left and right singular vectors are more aligned.

Yet, we can see that our hypothesis is not able to correctly predict the behavior of all the ``matching heads'' (not all of them appear in the upper right area of each plot), so more work is needed to refine it. 

\newpage
\section{Linear Discriminant Analysis}
\label{appendix:lda}

\subsection{Description of the computation process}

In practice, to compute the LDA, we start by computing the outputs of the attention and FFN modules across the layers of the cross-encoders while distinguishing the sets of query-passage pairs $\mathcal{S}_r$ with $r \in \{-1,0,1\}$:
\begin{equation}
    \Sigma_r^{\ell,m}(p) = \frac{1}{|\mathcal{S}_r|} \sum_{s \in \mathcal{S}_r} \left(P(s)-0.5\right) \, \mathbf{x}_p^{(\ell,m)}(s) \, \mathbf{x}_p^{(\ell,m)}(s)^\top
\end{equation}
with $P(s)$ the probability of relevance predicted for the input $s$, $p \in \{\CLS, Q, \SEPone, D, \SEPtwo\}$ the input part, $m \in \{a, \tilde f, \emptyset\}$ the activation (see Section~\ref{subsection:architecture}), $\ell$ the layer it belongs to, and $\mathbf{x}_p^{(\ell,m)}(s)$ the value of this activation for the sample $s$. In the formula, we also re-weight the product by the predicted probability of relevance of each query-passage pair, centered at 0.5. Note that for the query $Q$, we work at the level of its first $k=6$ tokens, separately, $Q_1, ..., Q_k$, while for the document $D$, we work with an averaged representation, per sample, $\bar{D}$, and not directly with their individual token representations.

We then compute the between-class and within-class covariance matrices over the relevance levels and use them to solve a generalized version of the eigenvalue problem\footnote{Given that we are trying to solve a generalized version of the eigenvalue decomposition with two matrices A and B, B different from the identity matrix, we use the Scipy implementation which supports this version of the eigenvalue problem: \url{https://docs.scipy.org/doc/scipy/reference/generated/scipy.linalg.eigh.html}.}. As a result, we order the eigenvectors by their corresponding eigenvalues to maximize the class separability. In our case, and for the sake of visualization, we proceed to retain only the first two axes and project onto them the activations previously generated to visualize a potential structure underlying the cross-encoder models' internal representations. 
\newpage

\subsection{Evolution of the classification accuracy for OOD datasets}

\begin{figure}[!htbp]
    \centering
    \begin{subfigure}[b]{0.9\textwidth}
        \centering
        \includegraphics[width=1\textwidth]{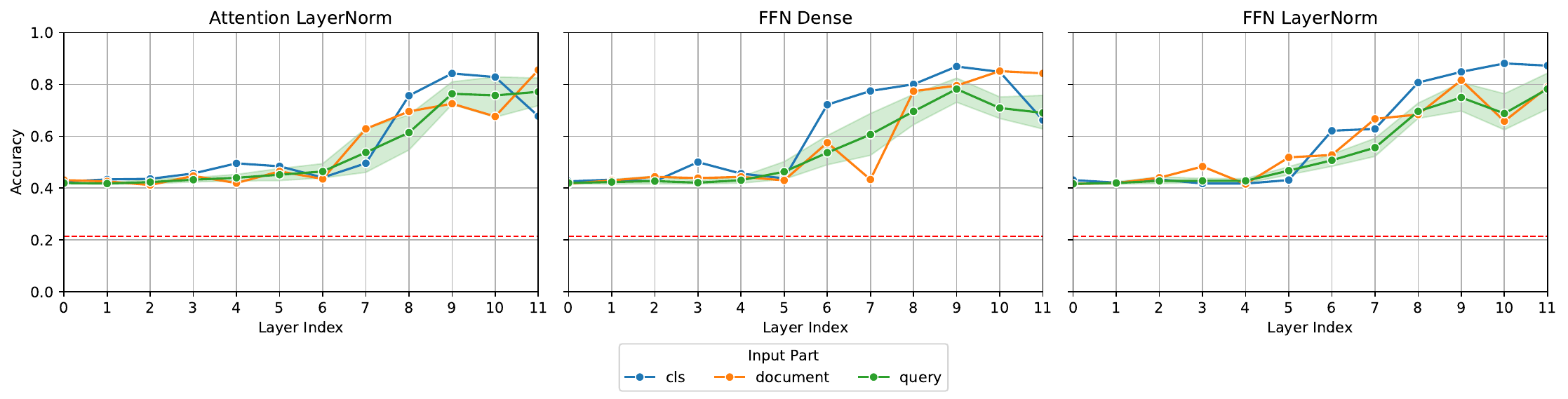}
        \caption{Details for MiniLM.}
        \label{subfig:minilm_all_accuracy_ood}
    \end{subfigure}

    \centering
    \begin{subfigure}[b]{0.9\textwidth}
        \centering
        \includegraphics[width=1\textwidth]{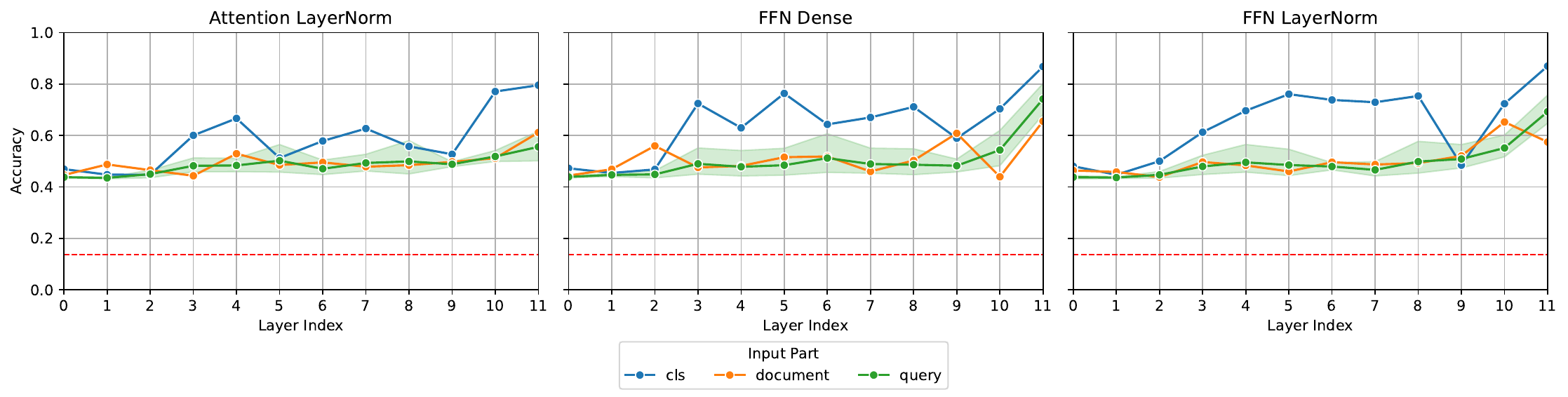}
        \caption{Details for MonoBERT-base.}
        \label{subfig:monobert_base_all_accuracy_ood}
    \end{subfigure}

    \centering
    \begin{subfigure}[b]{0.9\textwidth}
        \centering
        \includegraphics[width=1\textwidth]{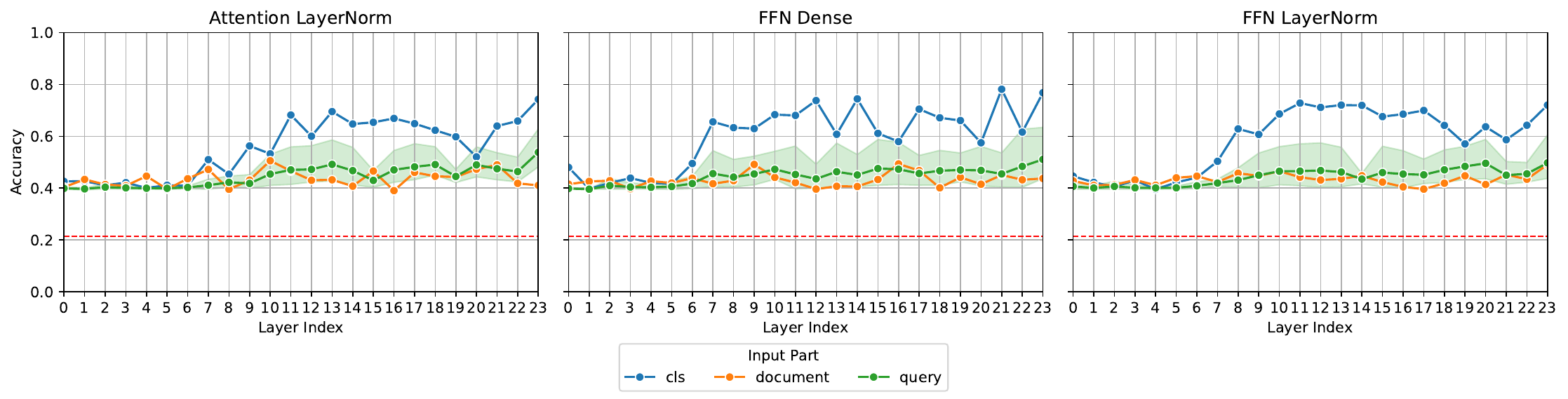}
        \caption{Details for MonoBERT-large.}
        \label{subfig:monobert_all_accuracy_ood}
    \end{subfigure}
    \caption[Evolution of the classification accuracy of the relevance subspace contained in the \CLS{}, query tokens and document tokens for OOD datasets]{Evolution of the classification accuracy of the activations $\repA[p]$, $\repFFNr[p]$ and $\repFFN[p]$ projected along the LDA axis, by their original relevance label ($r \in \{-1,0,1\}$), per input part $p \in \{\CLS, Q_1,..., Q_k, \bar{D}\}, k=6$. LDA and classification are computed over samples from the OOD datasets.}
    \label{fig:lda_ood_accuracy}
\end{figure}

\newpage
\subsection{Additional figures}

\begin{figure}[t]
    \centering
    \begin{subfigure}[b]{1\textwidth}
        \centering
        \includegraphics[width=1\textwidth]{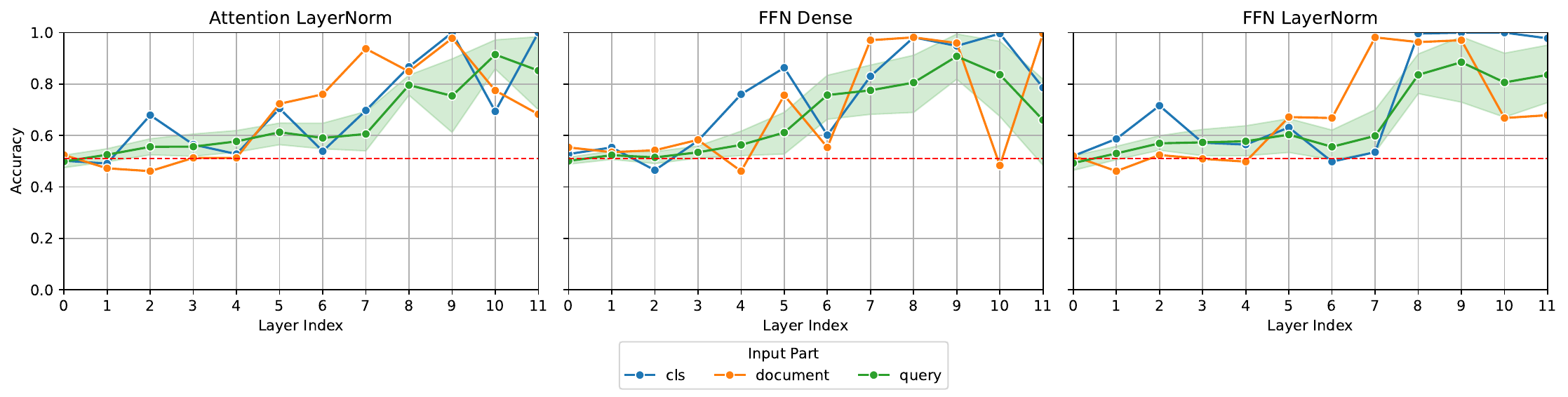}
        \caption{Details on MS MARCO datasets, relevance levels $r=-1$ and $r=1$.}
        \label{subfig:minilm_weak_negatives_accuracy_msmarco}
    \end{subfigure}
    \begin{subfigure}[b]{1.0\textwidth}
        \centering
        \includegraphics[width=1\textwidth]{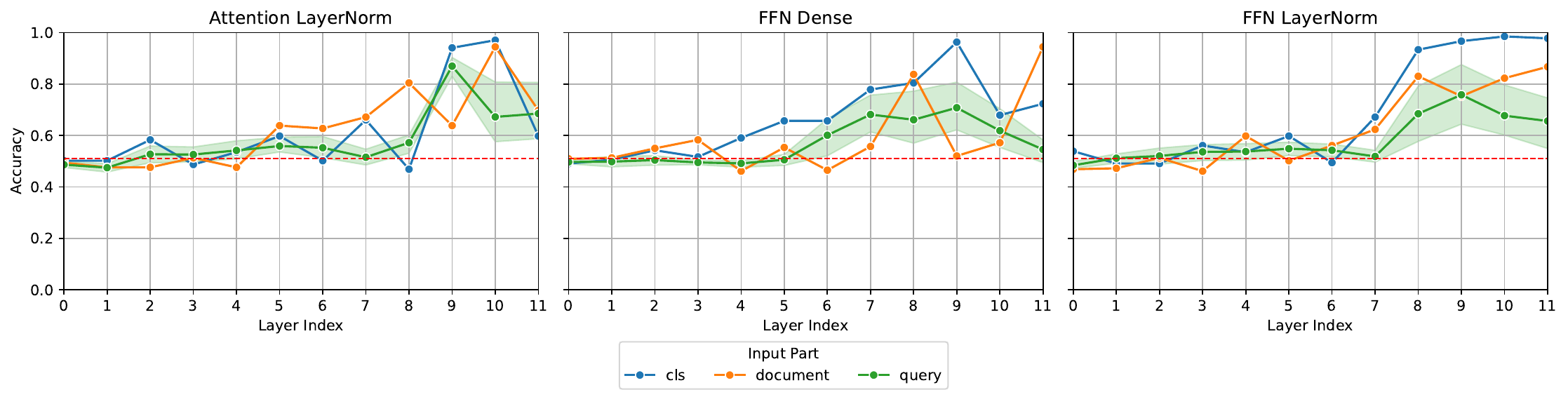}
        \caption{Details on MS MARCO datasets, relevance levels $r=0$ and $r=1$.}
        \label{subfig:minilm_hard_negatives_accuracy_msmarco}
    \end{subfigure}
    
    \centering
    \begin{subfigure}[b]{1\textwidth}
        \centering
        \includegraphics[width=1\textwidth]{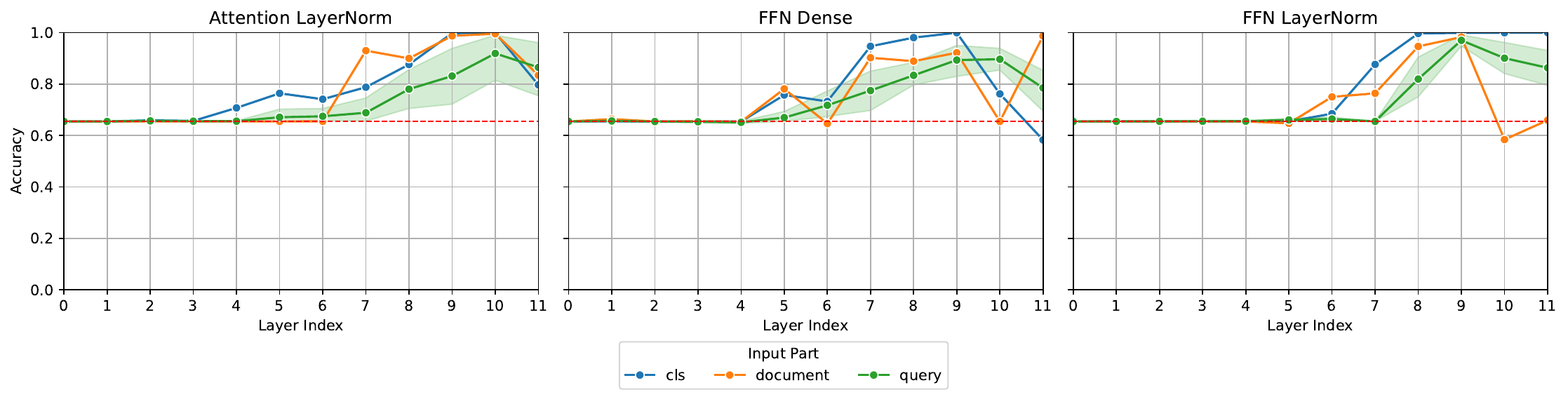}
        \caption{Details on OOD datasets, relevance levels $r=-1$ and $r=1$.}
        \label{subfig:minilm_weak_negatives_accuracy_ood}
    \end{subfigure}
    \begin{subfigure}[b]{1.0\textwidth}
        \centering
        \includegraphics[width=1\textwidth]{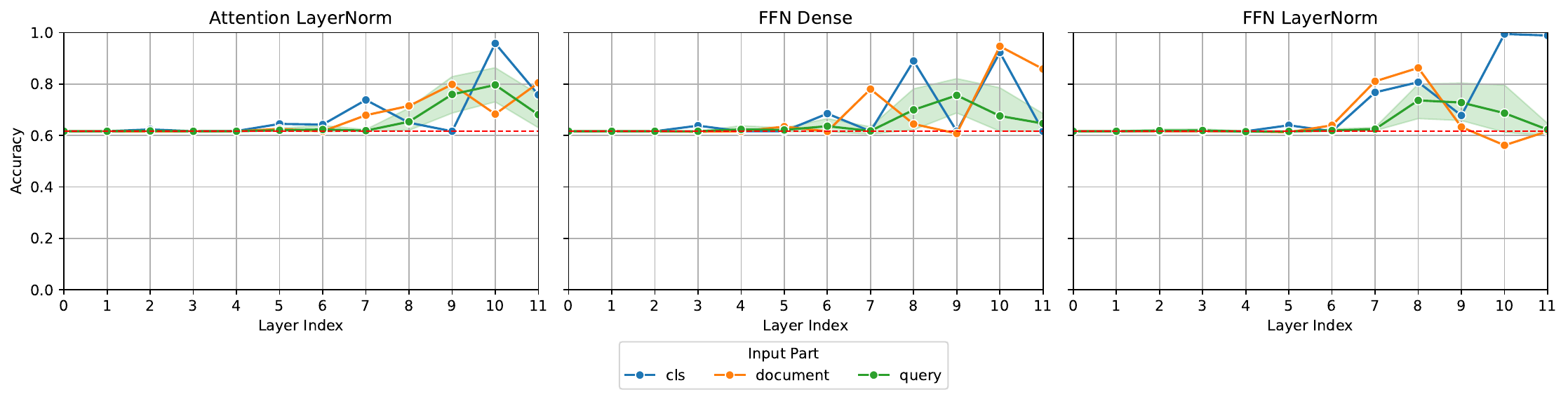}
        \caption{Details on OOD datasets, relevance levels $r=0$ and $r=1$.}
        \label{subfig:minilm_hard_negatives_accuracy_ood}
    \end{subfigure}
    \caption[Evolution of the binary classification accuracies for MiniLM]{Evolution of the classification accuracy, per input part and modules of MiniLM, over either the MS MARCO or the OOD datasets, for relevance levels $r=1$ and $r\in\{-1,0\}$. On each plot, the red line corresponds to the proportion of the largest class.}
    \label{fig:comparison_accuracy_minilm_strong_weak}
\end{figure}

\begin{figure}[t]
    \centering
    \begin{subfigure}[b]{1\textwidth}
        \centering
        \includegraphics[width=1\textwidth]{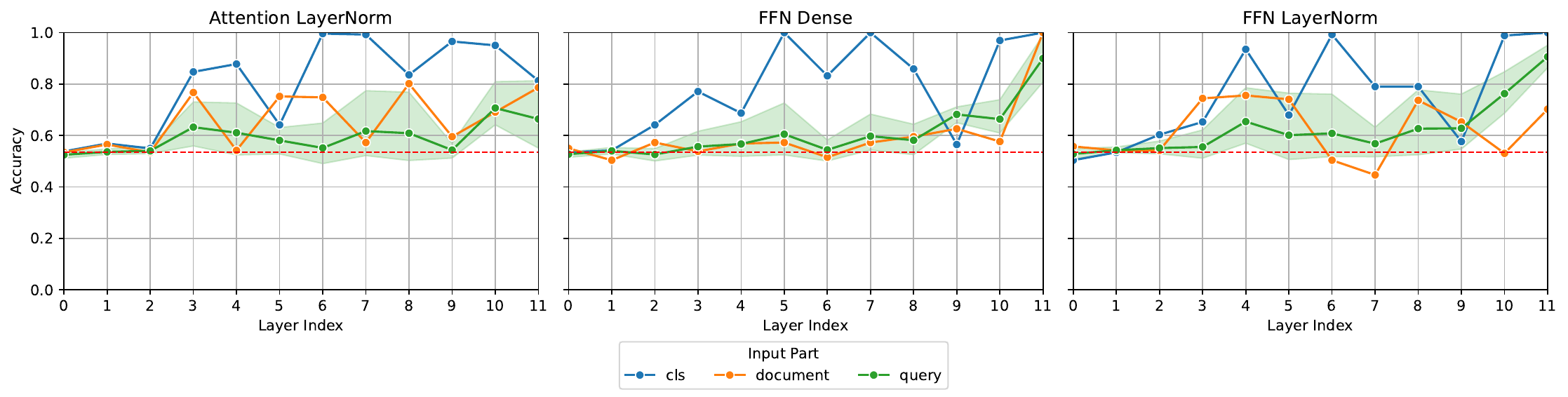}
        \caption{Details on MS MARCO datasets, relevance levels $r=-1$ and $r=1$.}
        \label{subfig:monobert_base_weak_negatives_accuracy_msmarco}
    \end{subfigure}
    \begin{subfigure}[b]{1.0\textwidth}
        \centering
        \includegraphics[width=1\textwidth]{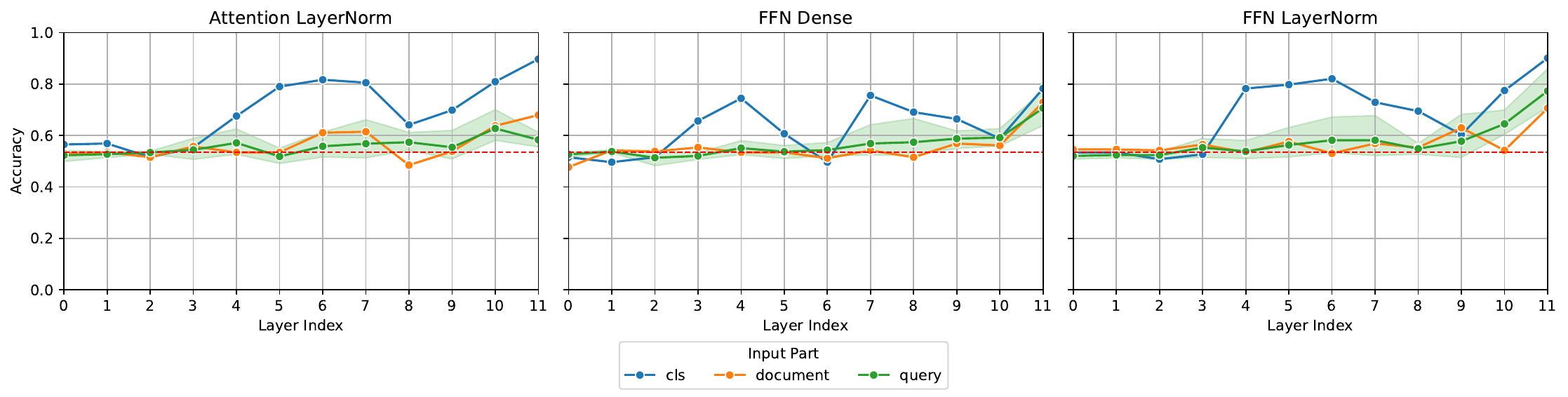}
        \caption{Details on MS MARCO datasets, relevance levels $r=0$ and $r=1$.}
        \label{subfig:monobert_base_hard_negatives_accuracy_msmarco}
    \end{subfigure}
    
    \centering
    \begin{subfigure}[b]{1\textwidth}
        \centering
        \includegraphics[width=1\textwidth]{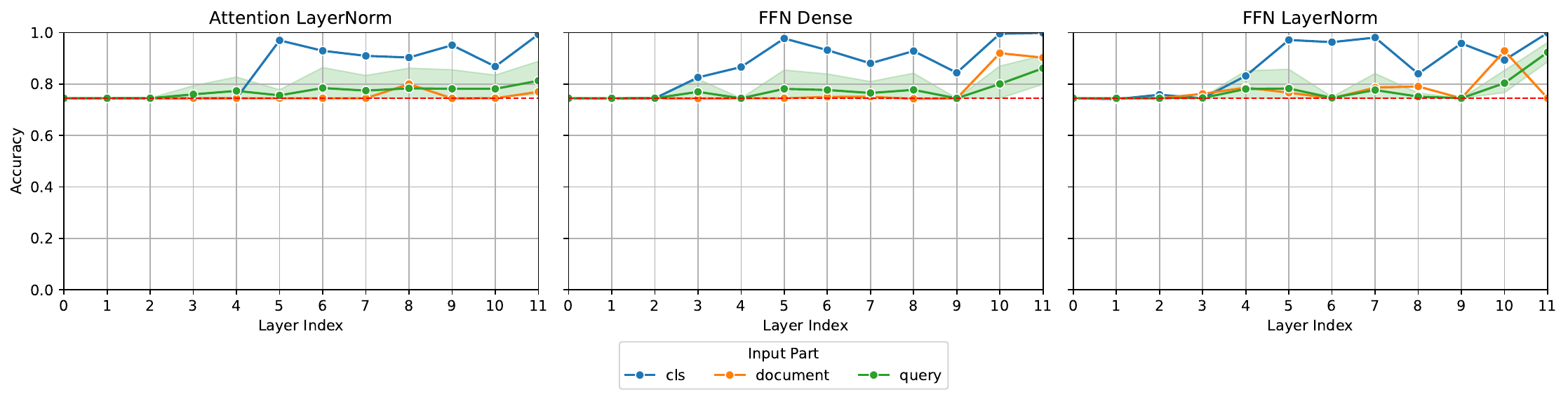}
        \caption{Details on OOD datasets, relevance levels $r=-1$ and $r=1$.}
        \label{subfig:monobert_base_weak_negatives_accuracy_ood}
    \end{subfigure}
    \begin{subfigure}[b]{1.0\textwidth}
        \centering
        \includegraphics[width=1\textwidth]{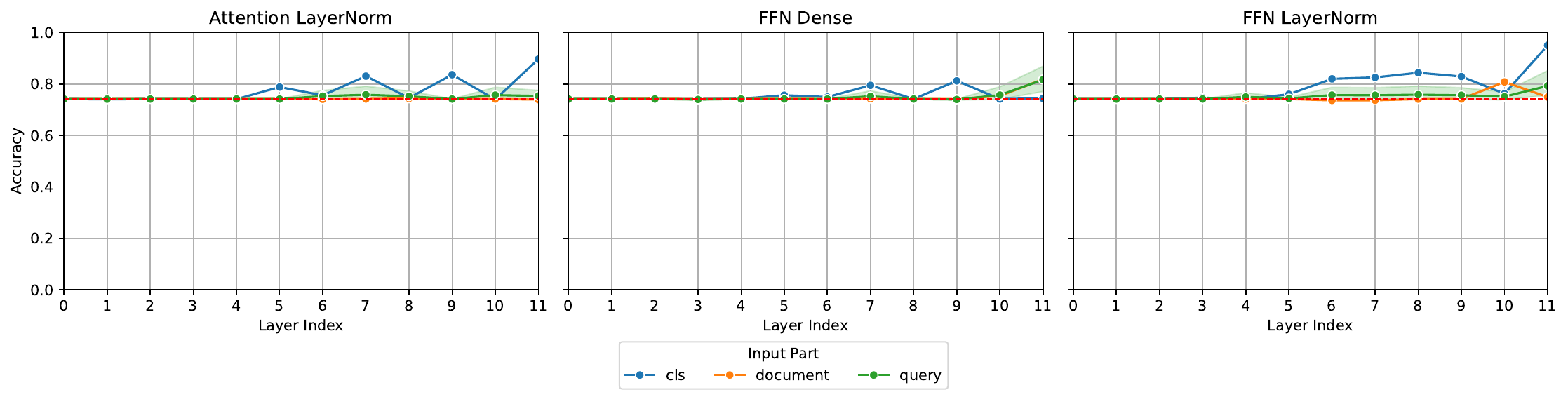}
        \caption{Details on OOD datasets, relevance levels $r=0$ and $r=1$.}
        \label{subfig:monobert_base_hard_negatives_accuracy_ood}
    \end{subfigure}
    \caption[Evolution of the binary classification accuracies for MonoBERT-base]{Evolution of the classification accuracy, per input part and modules of MonoBERT-base, over either the MS MARCO or the OOD datasets, for relevance levels $r=1$ and $r\in\{-1,0\}$. On each plot, the red line corresponds to the proportion of the largest class.}
    \label{fig:comparison_accuracy_monobert_base_strong_weak}
\end{figure}

\begin{figure}[t]
    \centering
    \begin{subfigure}[b]{1\textwidth}
        \centering
        \includegraphics[width=1\textwidth]{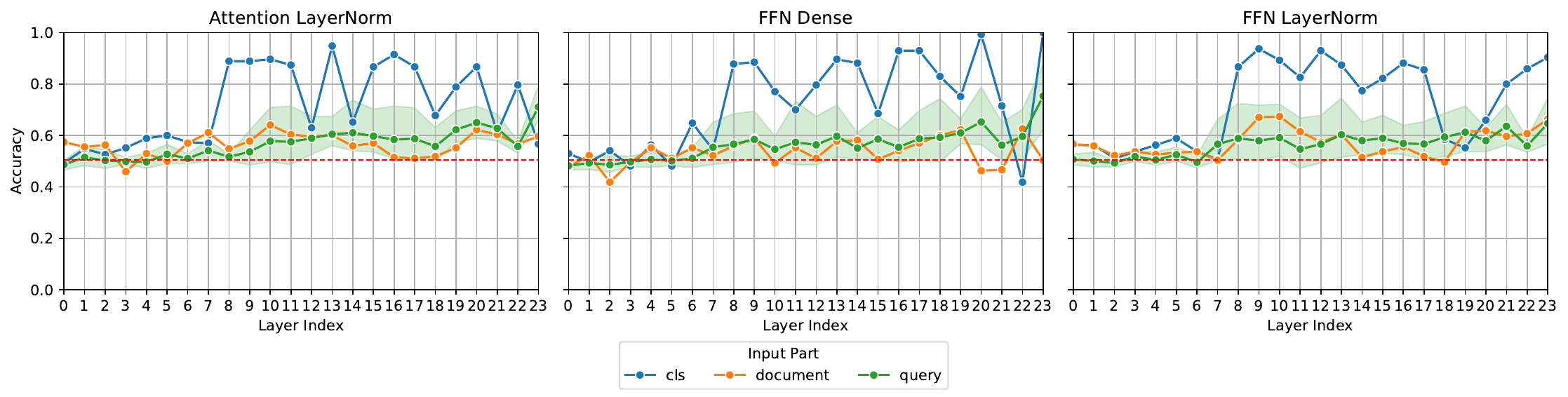}
        \caption{Details on MS MARCO datasets, relevance levels $r=-1$ and $r=1$.}
        \label{subfig:monobert_weak_negatives_accuracy_msmarco}
    \end{subfigure}
    \begin{subfigure}[b]{1.0\textwidth}
        \centering
        \includegraphics[width=1\textwidth]{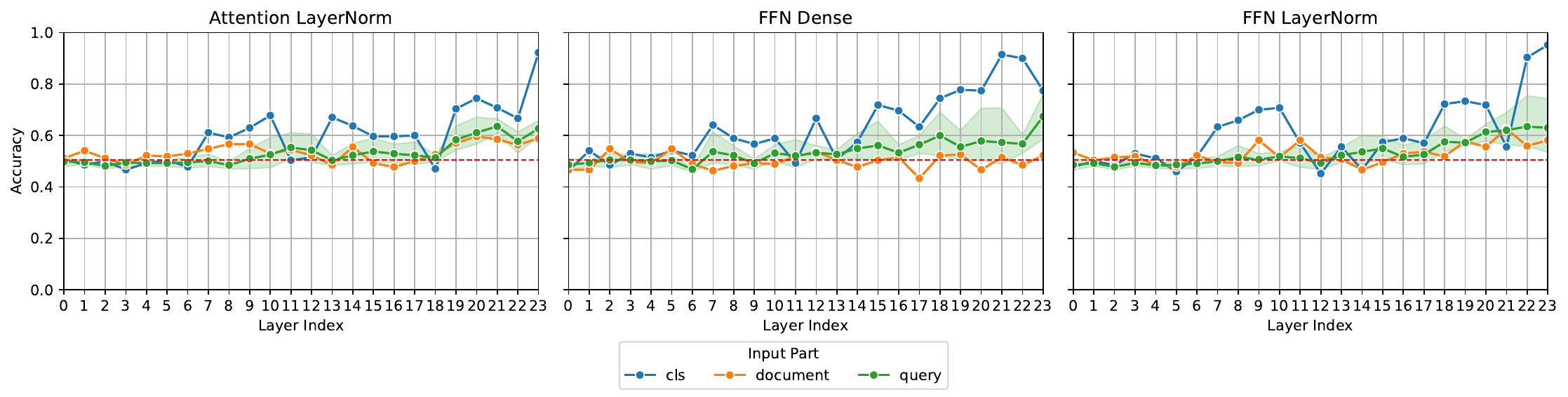}
        \caption{Details on MS MARCO datasets, relevance levels $r=0$ and $r=1$.}
        \label{subfig:monobert_hard_negatives_accuracy_msmarco}
    \end{subfigure}
    
    \centering
    \begin{subfigure}[b]{1\textwidth}
        \centering
        \includegraphics[width=1\textwidth]{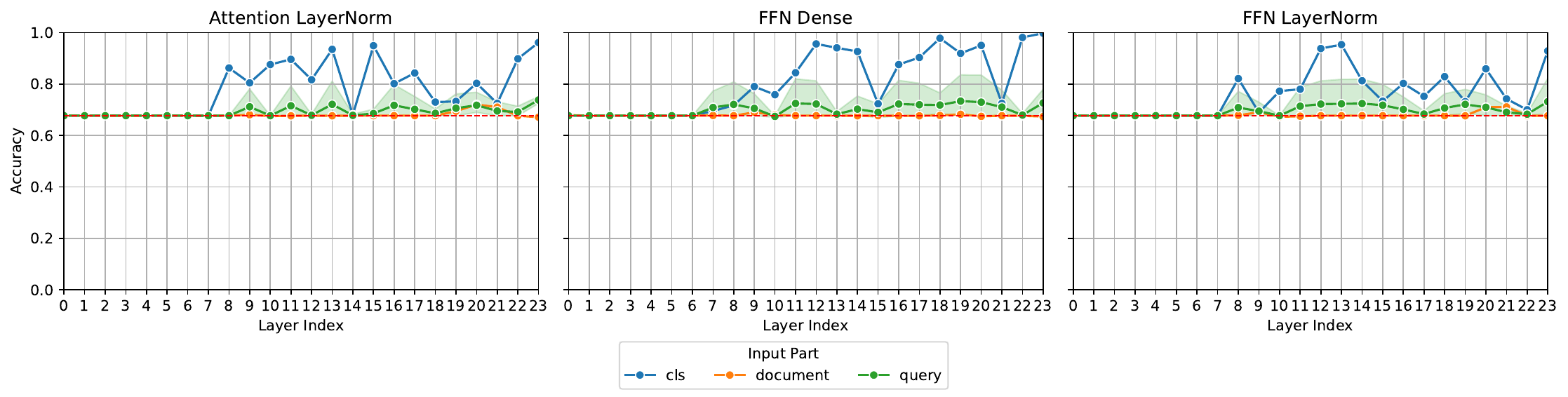}
        \caption{Details on OOD datasets, relevance levels $r=-1$ and $r=1$.}
        \label{subfig:monobert_weak_negatives_accuracy_ood}
    \end{subfigure}
    \begin{subfigure}[b]{1.0\textwidth}
        \centering
        \includegraphics[width=1\textwidth]{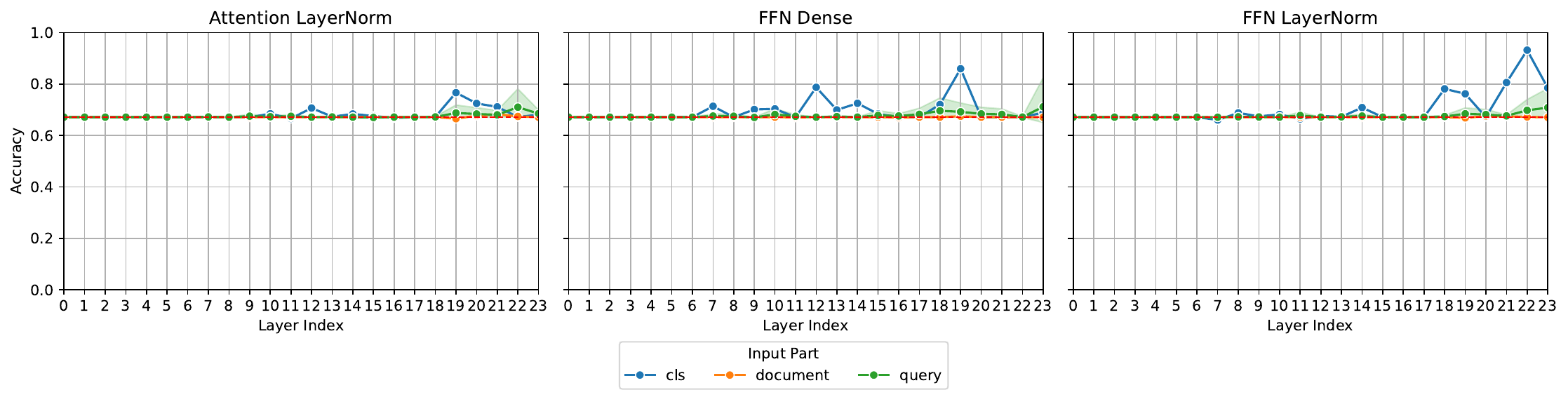}
        \caption{Details on OOD datasets, relevance levels $r=0$ and $r=1$.}
        \label{subfig:monobert_hard_negatives_accuracy_ood}
    \end{subfigure}
    \caption[Evolution of the binary classification accuracies for MonoBERT-large]{Evolution of the classification accuracy, per input part and modules of MonoBERT-large, over either the MS MARCO or the OOD datasets, for relevance levels $r=1$ and $r\in\{-1,0\}$. On each plot, the red line corresponds to the proportion of the largest class.}
    \label{fig:comparison_accuracy_monobert_strong_weak}
\end{figure}

We provide here, for reference, the evolution of the classification accuracy when we limit the LDA (see Section \ref{section:subspace_analysis}) to two classes: $r=1$ and either $r=-1$ or $r=0$ for both MiniLM (Figure \ref{fig:comparison_accuracy_minilm_strong_weak}) and MonoBERT (Figures \ref{fig:comparison_accuracy_monobert_base_strong_weak} and \ref{fig:comparison_accuracy_monobert_strong_weak}).

We first observe that, for all three models, distinguishing the activations of query-relevant passage pairs from pairs with easy negatives seems easier than from pairs with hard negatives. In addition, we observe that when the domain is different from the training domain, this difference is exacerbated (Figures \ref{subfig:minilm_hard_negatives_accuracy_ood} and \ref{subfig:monobert_hard_negatives_accuracy_ood}).

\end{document}